\documentclass[a4paper,12pt,oneside]{smartthesis}
\pdfoutput=1
\usepackage{times}
\usepackage{here}
\usepackage{epsfig}
\usepackage{rotating}
\usepackage{enumerate}
\usepackage{natbib}
\usepackage{setspace}
\usepackage{amssymb,amsmath,euscript}
\usepackage{floatflt}
\usepackage{mathrsfs}

\usepackage{epsf,latexsym}
\usepackage{color,graphicx,graphics,subfigure}
\usepackage[sf,small,hang]{caption}
\usepackage{pstcol}
\bibpunct{[}{]}{~,}{~a}{~,}{~,}
\newcommand{\gam} {{$\gamma$-ray}~}
\newcommand{\gams} {{$\gamma$-rays}~}

\begin{document}
\frontmatter
\thispagestyle{empty}
\Huge
\begin{center}
{TeV Gamma-Ray Observations of Mrk421 and 
H1426+428 with TACTIC Imaging Telescope}\\

\vspace{1.5 cm}

\Large

{A Thesis submitted\\
to the}\\
University of Mumbai \\
for the \\
{\bf Ph. D. (SCIENCE) Degree\\
in PHYSICS} \\

\vspace{0.8 cm}
\LARGE
{Submitted By} \\

\vspace{0.1 cm}
\Huge
{\bf  Kuldeep Kumar Yadav} \\
\Large
\vspace{1.0 cm}
{Under the Guidance of} \\
\LARGE
{Dr. R. C. Rannot} \\
\Large		
\vspace{1.2 cm}

{Astrophysical Sciences Division} \\
{Bhabha Atomic Research Centre} \\
{Trombay, Mumbai 400 085} \\ 
\LARGE
{\bf October 2010}
\end{center}
\normalsize

\newpage
\chapter{}

\Huge
\begin{center}
\vspace{5 cm}
{\bf \textit{Dedicated}} \\
{\bf \textit{to}} \\
{\bf \textit{My Parents}}\\
\end{center}
\newpage
\Large
\begin{center}

{\bf \underline{STATEMENT BY THE CANDIDATE}}

\vspace{2 cm}

\end{center}

\large

\noindent As required by the University Ordinance 770, I wish to state that the work embodied in this thesis titled `` \textbf{TeV Gamma-Ray Observations of Mrk421 and 
H1426+428 with TACTIC Imaging Telescope}'' forms my own contribution to the research work carried out under the guidance of \textbf{Dr. R.C. Rannot} at \textbf{Astrophysical Sciences Division, Bhabha Atomic Research Centre, Trombay, Mumbai}. This work has not been submitted for any other degree of this or any other University. Whenever reference have been made to previous works of others, it has been clearly indicated as such and included in the Bibliography. 
\vspace{1.5 cm}
\begin{flushright}
----------------------------------- \\
Signature of candidate \hspace{0.5cm}{  }   \\
    {Name : \bf Kuldeep Kumar Yadav} \hspace{1cm}{} \\
\end{flushright}

\vspace{1.0 cm}
\begin{flushleft}
Certified by \\
\vspace{1.0cm}
-----------------------------------\\
\hspace{.5cm} Signature of Guide\\
{Name : \bf Dr. R. C. Rannot} \\
\end{flushleft}

\normalsize

\newpage
\newpage
\chapter{Abstract}
We have observed extragalactic sources Mrk421 ($z$=$0.30$) and H1426+428 ($z$=$0.129$) in very high energy \gam region using TACTIC telescope at Mount Abu ($24.6^\circ $N, $72.7^\circ$E, $1300\,m$ asl), Rajasthan, India. Both of these sources belong to the blazar sub-class of active galactic nuclei whose jets are closely aligned to our line of sight. Blazars are most extreme and powerful variable sources of radiation of energies ranging from the radio to the \gam regimes. Mrk421 was observed during 07 December 2005 to 18 April 2007 for a total of $\sim$$331\,hours$ while H1426+428 was observed for $\sim$$244\,hours$ between 22 March 2004 and 11 June 2007. Detailed analysis of Mrk421 data revealed the presence of a \gam signal with the excess of $\sim$$(951\pm 82)$ \gam like events corresponding to a statistical significance of $\sim$$12.0\sigma$ in $\sim$$97\,hours$ during 27 December 2005 to 07 February 2006. The combined average flux was measured to be $\sim$$(1.04\pm0.14)$ Crab units. Obseved time averaged differential energy spectrum was found to be consistent with both a pure power law $d\Phi/dE=(4.66\pm0.46)\times 10^{-11}\,E^{(-3.11\pm0.11)}\,\,cm^{-2}\,s^{-1}\,TeV^{-1}$ and power law with an exponential cutoff functions $d\Phi/dE=f_0E^{-\Gamma}exp(-E/E_0)$ with $f_0=(4.88\pm0.38)\times10^{-11}\,cm^{-2}\,s^{-1}\,TeV^{-1}$, $\Gamma=2.51\pm0.26$ and  $E_0=(4.7\pm2.1)TeV$. The later function fits the observed spectrum well with lower value of reduced $\chi^2$. However, during the observation period from 18 December 2006 to 18 April 2007, the source was found to be in a low emission state and we have placed an upper limit of $1.44\times\,10^{-12}\,photons\,cm^{-2}\,s^{-1}$ at  $3\sigma$ level on the integrated \gam flux above $1\,TeV$. In addition, the TACTIC light curves have also been compared with those obtained with the RXTE/ASM and $Swift$/BAT satellite-based experiments. Further, we have obtained the SED of Mrk421 in the VHE range using TACTIC telescope and compared it with those obtained by the HESS, MAGIC, VERITAS, Whipple, HEGRA and CAT telescopes. The detection of $TeV$ photons from the distant object H1426+428 have important implications for estimating the EBL, in addition to the understanding of the particle acceleration and $\gamma$-ray production mechanisms in the AGN jets. Detailed analysis of H1426+428 data does not indicate the presence of any statistically significant $TeV$ \gam signal. Accordingly we have placed an upper limit of $\leq1.18\times10^{-12}\,photons\,cm^{-2}\,s^{-1}$ on the integrated  $\gamma$-ray flux above $1\,TeV$ at 3$\sigma$ significance level.




\chapter{List of Publication}

{\bf {\Large{Publications in refereed journals}}}
\begin{enumerate}

\item Search for TeV $\gamma$-rays from H1426+428 during 2004-07 with the TACTIC telescope;\\
\textbf{ K.K. Yadav}, R.C. Rannot, P. Chandra, A. K. Tickoo, K. Venugopal, N. Bhatt, S. Bhattacharyya, K. Chanchalani, V.K. Dhar,  H.C. Goyal, R.K. Kaul, M. Kothari, S. Kotwal, M.K. Koul, R. Koul, B.S. Sahaynathan, M. Sharma, S. Thoudam;\\
\textbf{\textit{J. Phys. G: Nucl. Part. Phys., 36 (2009) 085201}}.

\item ANN-based energy reconstruction procedure for TACTIC gamma-ray telescope and its comparison with other conventional methods;\\
V. K. Dhar, A.K. Tickoo, M.K. Koul, R.C. Rannot, \textbf{K.K. Yadav}, P. Chandra, B. P. Dubey, R. Koul;\\
\textbf{\textit{Nucl. Instrum. and meth. A., 606 (2009) 795}}.

\item Very High Energy $\gamma$-ray observations of Mrk 501 using TACTIC imaging $\gamma$-ray telescope during 2005-06;\\
S. V. Godambe, R.C. Rannot, P. Chandra, \textbf{K. K. Yadav}, A. K. Tickoo, K. Venugopal, N. Bhatt, S. Bhattacharyya, K. Chanchalani, V. K. Dhar,  H. C. Goyal, R. K. Kaul, M. Kothari, S. Kotwal, M.K. Koul, R. Koul, B. S. Sahaynathan, M. Sharma, S. Thoudam;\\
\textbf{\textit{J. Phys. G: Nucl. Part. Phys., 35 (2008) 065202}}.

\item Observations of TeV $\gamma$-rays from Mrk421 during Dec.2005 to May2006 with the TACTIC telescope;\\
\textbf{K.K. Yadav}, P. Chandra, A.K. Tickoo, R.C. Rannot, S. Godambe, M.K. Koul, V.K. Dhar, S. Thoudam, N. Bhatt, S. Bhattacharyya, K. Chanchalani, H.C. Goyal, R.K. Kaul, M. Kothari, S. Kotwal, R. Koul, S. Sahayanathan, M. Sharma, K. Venugopal;\\
\textbf{\textit{Astroparticle Physics, 27 (2007) 447}}.

\item Very High Energy $\gamma$ -ray and Near Infrared observations of 1ES2344+514 during 2004-05;\\
S.V. Godambe, R.C. Rannot,  K.S. Baliyan, A.K. Tickoo, S. Thoudam, V.K. Dhar, P. Chandra, \textbf{K.K. Yadav}, K. Venugopal, N. Bhatt, S. Bhattacharyya, K. Chanchalani, S. Ganesh, H.C. Goyal, U.C. Joshi, R.K. Kaul, M. Kothari, S. Kotwal, M.K. Koul, R. Koul, S. Sahaynathan, C. Shah, M. Sharma;\\
\textbf{\textit{J. Phys. G: Nucl. Part. Phys., 34 (2007) 1683}}.

\item The TACTIC atmospheric Cherenkov Imaging telescope;\\
R. Koul, A.K. Tickoo, S.K. Kaul, S.R. Kaul, N. Kumar, \textbf{K.K. Yadav}, N. Bhatt, K. Venugopal, H.C. Goyal, M. Kothari, P. Chandra, R.C. Rannot, V.K. Dhar, M.K. Koul, R.K. Kaul, S. Kotwal, K. Chanchalani, S. Thoudam, N. Chouhan, M. Sharma, S. Bhattacharyya, S. Sahayanathan;\\
\textbf{\textit{Nucl. Instrum. and meth. A., 578 (2007) 548}}.

\item A generalized ray-tracing procedure for an atmospheric Cherenkov imaging telescope and optical characteristics of the TACTIC light collector;\\
A.K. Tickoo, R.L. Suthar, R. Koul, M.L. Sapru, N. Kumar, C.L. Kaul, \textbf{K.K. Yadav}, S. Thoudam, S.K. Kaul, K. Venugopal, M. Kothari, H.C. Goyal, P. Chandra, V.K. Dhar, R.C. Rannot, M.K. Koul, S.R. Kaul;\\
\textbf{\textit{Nucl. Instrum. and meth. A., 539 (2005) 177}}.

\item Real time data acquisition and control system for the 349-pixel TACTIC atmospheric Cerenkov imaging telescope;\\
\textbf{K.K. Yadav}, R. Koul, A. Kanda, S.R. Kaul, A.K. Tickoo, R.C. Rannot, P.Chandra, N. Bhatt, N. Chouhan, K. Venugopal, M. Kothari, H.C. Goyal, V.K. Dhar and S.K. Kaul;\\
\textbf{\textit{Nucl. Instrum. and Meth.A 527 (2004) 411}}.

\item Programmable Topological Trigger Generator for the 349-pixel Imaging Camera of the TACTIC telescope;\\
S.R. Kaul, R. Koul, I.K. Kaul, A.K. Tickoo, \textbf{K.K. Yadav}, M.L. Sapru and C.L. Bhat;\\
\textbf{\textit{Nucl. Instrrum. and  Meth.A, 496 (2003) 400.}}

\item On a single-count rate stabilization scheme employed in the imaging camera of  the TACTIC gamma-ray telescope;\\
N. Bhatt, \textbf{K.K. Yadav}, A. Kanda, R. Koul, S.R. Kaul, H.C. Goyal, V.K. Dhar, S.K. Kaul, K. Venugopal, A.K. Tickoo, I.K. Kaul, M. Kothari and C.L. Bhat;\\
\textbf{\textit{Meas.Sci \& Tech, 12(2001) 167-171.}}

\end{enumerate}

\newpage
{\bf {\Large{Publications in conferences}}}
\begin{enumerate}
\item Feasibility of operating TACTIC telescope during partial moonlit conditions;\\
K.K. Singh, A.K. Tickoo, \textbf{K.K. Yadav}, R.C. Rannot, V.K. Dhar, R. Koul;\\
\textit{Presented at “ 16th National Space Science Symposium ( NSSS-2010)", Rajkot, India, (2010).}

\item Recent TeV observations of Mrk 421 with the TACTIC gamma-ray telescope;\\
R.C. Rannot, P. Chandra, \textbf{K.K. Yadav}, A.K. Tickoo, K. Chanchalani, A. Goyal, H.C. Goyal, M. Kothari, S. Kotwal, 
N. Kumar, P. Marandi, K. Venugopal, C.K. Bhat, N. Bhatt, S. Bhattacharyya, V.K. Dhar, M.K. Koul, R. Koul, 
S.Sahayanathan, M.Sharma and K.K.Singh;\\
\textit{Presented at “16th National Space Science Symposium ( NSSS-2010)", Rajkot, India, (2010).}
 
\item VHE gamma-ray observations of 3C279 during 2008-09 using TACTIC telescope;\\
P. Chandra, R.C. Rannot, \textbf{K.K. Yadav}, A.K. Tickoo, K. Venugopal, K. Chanchalani, A. Goyal, H.C. Goyal, M. Kothari, 
S. Kotwal, N. Kumar, P. Marandi, K.K. Singh, C.K. Bhat, N. Bhatt, S. Bhattacharyya, V.K. Dhar, M.K. Koul, R. Koul, 
S. Sahayanathan and M. Sharma;\\
\textit{Presented at “16th National Space Science Symposium ( NSSS-2010)", Rajkot, India, (2010).}

\item Fractal and Wavelet analysis approach for detectiong gamma-rays from Cerenkov images using Imaging Cerenkov Telescope;\\
C.K. Bhat, \textbf{K.K. Yadav}, R. Koul;\\
\textit{Presented at “16th National Space Science Symposium ( NSSS-2010)", Rajkot, India, (2010).}
 
\item Pulse profile digitization system for TACTIC gamma-ray telescope;\\
N. Chouhan, R.C. Rannot, S.R. Kaul, A.K. Tickoo, \textbf{K.K. Yadav}, P. Chandra, S. Kotwal, K. Venugopal, K. Chanchalani, M. Kothari, H.C. Goyal, P. Marandi, N. Kumar, R. Koul;\\
\textit{Presented at "National Symposium on Nucl. Ins.(NSNI-2010)", Mumbai, India, (2010).}

\item TACTIC and MACE gamma-ray telescopes;\\
\textbf{K.K. Yadav} for the HIGRO collaboration		
\textit{proceedings to "44th Rencontres de Moriond", La Thuile (Val d'Aosta, Italy) on "Very High Energy Phenomena in the Universe", February 1-8, 2009}

\item TeV gamma-ray observations of Mrk421 during 2007-2008 with the TACTIC telescope;\\
\textbf{K.K. Yadav}, R.C. Rannot, P. Chandra, A.K. Tickoo, V.K. Dhar, M.K. Koul, S. Thoudam, K. Venugopal, S. Sahayanathan, M. Sharma, C.K. Bhat, N. Bhatt, S. Bhattacharyya, K. Chanchalani, S.V. Godambe, H.C. Goyal, M. Kothari, S. Kotwal, R. Koul;\\
\textit{XXVII meeting of the astronomical society of India (ASI-2009) February 18-20, 2009\\
Indian Institute of Astrophysics, Bangalore}

\item TeV gamma-ray observations of the Crab Nebula with the TACTIC telescope;\\
A.K. Tickoo, R.C. Rannot, P. Chandra, \textbf{K.K. Yadav}, V.K. Dhar, M.K. Koul, S. Thoudam, M. Sharma, C.K. Bhat, N. Bhatt,  S. Bhattacharyya, K. Chanchalani, S.V. Godambe, H.C. Goyal, M. Kothari, S. Kotwal, R. Koul, S. Sahayanathan;\\
\textit{XXVII meeting of the astronomical society of India (ASI-2009) February 18-20, 2009\\
Indian Institute of Astrophysics, Bangalore}

\item The TACTIC gamma-ray telescope;\\
R. Koul, A.K. Tickoo, S.K. Kaul, S.R. Kaul, N. Kumar, \textbf{K.K. Yadav}, N. Bhatt, K. Venugopal, H.C. Goyal, M. Kothari, P. Chandra, R.C. Rannot, V.K. Dhar, M.K. Koul, R.K. Kaul, S. Kotwal, K. Chanchalani, S. Thoudam, N. Chouhan, M. Sharma, S. Bhattacharyya, S. Sahayanathan;\\
\textit{National Symposium on Gamma-Ray Astronomy, IIA Bangalore. Nov 23-24, 2007}

\item Observations of Mrk421 and Mrk501 during 2005-2006 with the TACTIC telescope;\\
\textbf{K.K. Yadav}, P. Chandra, A.K. Tickoo, R.C. Rannot, S. Godambe, M.K. Koul, V.K. Dhar, S. Thoudam, N. Bhatt, S. Bhattacharyya, K. Chanchalani, H.C. Goyal, R.K. Kaul, M. Kothari, S. Kotwal, R. Koul, S. Sahayanathan, M. Sharma, K. Venugopal;\\
\textit{National Symposium on Gamma-Ray Astronomy, IIA Bangalore. Nov 23-24, 2007}

\item A search for TeV gamma-rays from 1ES2344+514 and H1426+428 using TACTIC telescope;\\
S.V. Godambe, R.C. Rannot, A.K. Tickoo, S. Thoudam, V.K. Dhar, P. Chandra, \textbf{K.K. Yadav}, K. Venugopal, N. Bhatt, S. Bhattacharyya, K. Chanchalani, H.C. Goyal, R.K. Kaul, M. Kothari, S. Kotwal, M.K. Koul, R. Koul, S. Sahayanathan, M. Sharma;\\
\textit{National Symposium on Gamma-Ray Astronomy, IIA Bangalore. Nov 23-24, 2007}

\item A CAMAC-based multi-channel digital delay generator for TACTIC $\gamma$-ray telescope;
S. R. Kaul, \textbf{K.K. Yadav}, A.K. Tickoo, R. Koul;\\
\textit{XXV meeting of the astronomical society of India (ASI-2007) February 7-9, 2007
Department of Astronomy, Osmania University, Hyderabad}

\item TeV observations of Mrk501 with the TACTIC gamma-ray Telescope during 2005-06; \\
S. V. Godambe , P. Chandra, R.C. Rannot, \textbf{K.K. Yadav}, A.K. Tickoo, K., Venugopal, N. Bhatt,S. Bhattacharyya, V. K. Dhar, H.C. Goyal, R. K. Kaul, M. Kothari,S. Kotwal, M.K. Koul, R.Koul, S. Sahaynathan, M. Sharma, S. Thoudam;\\
\textit{XXV meeting of the astronomical society of India (ASI-2007) February 7-9, 2007\\
Department of Astronomy, Osmania University, Hyderabad }

\item TACTIC observations of Mrk421 at TeV energies during Dec. 2005 to April 2006;\\
\textbf{K. K. Yadav}, P. Chandra, A. K. Tickoo, R.C. Rannot, S. V. Godambe, N. Bhatt, S. Bhattacharyya, M.K. Koul, V. K. Dhar, S. Thoudam,H.C. Goyal, R. K. Kaul, M. Kothari, S. Kotwal, R.Koul, S. Sahaynathan, M. Sharma, K. Venugopal; \\
\textit{XXV meeting of the astronomical society of India (ASI-2007) February 7-9, 2007 \\
Department of Astronomy, Osmania University, Hyderabad}

\item Recent TeV observations of Mrk 501 with the TACTIC gamma-ray telescope; \\
S. V. Godambe , R.C. Rannot, P. Chandra, \textbf{K. K. Yadav}, A. K. Tickoo, K., Venugopal, N. Bhatt, S. Bhattacharyya, K. Chanchalani,V. K. Dhar, H. C. Goyal, R. K. Kaul, M. Kothari, S. Kotwal, M.K. Koul, R. Koul, S. S.Sahaynathan, M. Sharma, S. Thoudam;\\
\textit{30$^{th}$ International Cosmic Ray Conference 2007 Merida Yucatan Mexico} 

\item Very High Energy observations of 1ES2344+514 and Mrk501;\\
S.V. Godambe, S. Thoudam, R.C. Rannot, P. Chandra, \textbf{K.K. Yadav}, A. K. Tickoo, S. Sahaynathan, M. Sharma, K. Venugopal, N. Bhatt, S. Bhattacharyya, V.K. Dhar, H.C. Goyal, R.K. Kaul, M. Kothari, S.Kotwal, R. Koul;\\
{\it XIV National Science Symposium, Andhra University, Visakhapatnam February. 9-12, 2006.}

\item TeV $\gamma$-ray observations of the Blazar Markarian 421 from January to April 2004 with TACTIC Imaging Element; \\             
R.C. Rannot, P. Chandra, S. Thoudam, \textbf{K.K. Yadav}, K. Venugopal, N. Bhatt, S. Bhattacharyya, V.K. Dhar, H.C. Goyal, S. Godambe, R.K. Kaul, M. Kothari,  S. Kotwal, R. Koul, A.K. Tickoo, S. Sahayanathan, M.L. Sapru, M.K. Koul, M.Sharma;\\
 {\it Proc.``29 $^{th}$ ICRC"; Pune, India, (2005), OG 2.3}

\item VHE observations of H1426+428 using TACTIC imaging telescope: 2004 observations; \\
S. Thoudam, \textbf{K.K. Yadav}, R.C. Rannot,  S. Sahyanathan, M. Sharma, K. Venugopal, N. Bhatt, S. Bhattacharyya, P. Chandra, V.K. Dhar, H.C. Goyal, S. Godambe, R.K. Kaul, M. Kothari, S. Kotwal, R. Koul, A.K.Tickoo;\\
 {\it Proc.``29$^{th}$ ICRC"; Pune, India, (2005), OG 2.3}

\item Very High $\gamma$-ray and Near Infrared observations of 1ES2344+514 with TACTIC and MIRO telescopes;\\ 
S.V. Godambe,  S. Thoudam, R.C. Rannot, P. Chandra, S. Sahyanathan, M. Sharma, K. Venugopal, N. Bhatt, S. Bhattacharyya, V.K. Dhar, H.C. Goyal, R.K. Kaul, M. Kothari, S. Kotwal, R. Koul, \textbf{K.K. Yadav}, A.K. Tickoo, K.S. Baliyan, U.C. Joshi, S. Ganesh, C. Shah, A. Ohlan;\\
 {\it Proc.``29$^{th}$ ICRC"; Pune, India, (2005), OG 2.3}

\item ANN  based energy  estimation procedure and energy spectrum of the Crab Nebula as measured by the TACTIC $\gamma$-ray telescope;\\             
V.K. Dhar, M.K. Koul,A.K. Tickoo, \textbf{K.K. Yadav}, S. Thoudam, B.P. Dubey, K. Venugopal, N. Bhatt, S. Bhattacharyya, P. Chandra, H.C. Goyal, R.K. Kaul, M. Kothari, S. Kotwal, R. Koul, R.C. Rannot, S. Sahyanathan, M.Sharma;\\
{\it Proc.``29$^{th}$ ICRC"; Pune, India, (2005), OG 2.2}

\item Design  and performance of the data acquisition and control system of the TACTIC $\gamma$-ray telescope;\\  
\textbf{K.K. Yadav}, N. Chouhan, R. Koul, S.R. Kaul,A.K. Tickoo, R.C. Rannot, P. Chandra, N. Bhatt, K. Venugopal, M. Kothari, H.C. Goyal, V.K. Dhar, S.K.Kaul;\\
{\it  Proc.``29$^{th}$ ICRC"; Pune, India, (2005), OG 2.7 }

\item Study of TeV photons from Mrk421 with the TACTIC $\gamma$-ray telescope - 2004 observations;\\            
R.C. Rannot, P. Chandra, S. Thoudam, \textbf{K.K. Yadav},  M. Sharma  K. Venugopal, N. Bhatt, S. Bhattacharyya,  V.K. Dhar, H.C. Goyal, S. Godambe, R.K. Kaul, M. Kothari, S. Kotwal, R. Koul, A.K.Tickoo, S.Sahyanathan;\\
{\it Proc.``23$^{rd}$ ASI meeting"; Nainital, India: Bull. Astron. Soc. India, 33 (2005) 403}

\item Recent TeV observations of  1ES2344+514 with the TACTIC telescope;\\ 
S. Godambe, S. Thoudam, R.C. Rannot, P. Chandra, A.K. Tickoo, S. Sahyanathan, 
M. Sharma, K. Venugopal, N. Bhatt, S. Bhattacharyya, V.K. Dhar, H.C. Goyal, R.K. Kaul, M. Kothari, S. Kotwal, R. Koul, \textbf{K.K.Yadav}; \\
{\it Proc.``23$^{rd}$ ASI meeting"; Nainital, India: Bull. Astron. Soc. India, 33 (2005) 404}

\item VHE observations of H1426+428 using TACTIC imaging telescope - 2004 observations; \\  
S. Thoudam, \textbf{K.K. Yadav}, R.C. Rannot,  S. Sahyanathan,  M. Sharma  K. Venugopal, N. Bhatt, S. Bhattacharyya, P. Chandra, V.K. Dhar, H.C. Goyal, S. Godambe, R.K. Kaul, M. Kothari, S. Kotwal, R. Koul, A.K. Tickoo;\\
{\it Proc.``23$^{rd}$ ASI meeting"; Nainital, India: Bull. Astron. Soc. India, 33 (2005) 403}

\item TeV energy spectrum of the Crab Nebula as measured by the TACTIC $\gamma$-ray telescope;\\
A.K. Tickoo, \textbf{K.K. Yadav}, M.K. Koul, S. Thoudam, V.K. Dhar, K. Venugopal, N. Bhatt, S. Bhattacharyya, P. Chandra, H.C. Goyal, R.K. Kaul, M. Kothari, S. Kotwal, R. Koul, R.C. Rannot,  S. Sahyanathan, M. Sharma;\\
{\it Proc.``23$^{rd}$ ASI meeting"; Nainital, India: Bull. Astron. Soc. India, 33 (2005) 389}

\item Remote Control and Monitoring System for the TACTIC Gamma - Ray Telescope;\\
J.J. Kulkarni, N. Chouhan, L.P. Babu,\textbf{K.K. Yadav}, U.C. Lad, P.S. Dhekne and R. Koul;\\
\textit{Proc. National Symposium on Nuclear Instrumentation, IGCAR, 2004,648.} 
      
\item Improved data analysis program for TACTIC Imaging element;\\
S. Thoudam, \textbf{K.K. Yadav}, P. Chandra and R.C. Rannot;\\
\textit{Presented at XIII NSSS-2004, Feb 17 - 20, 2004,Kottayam.}

\item Energy sensitive search for pulsed photons from Crab pulsar using the TACTIC Imaging element;\\
R.C. Rannot, \textbf{K.K. Yadav},V.K. Dhar, A.K. Tickoo, N. Bhatt, S. Bhattacharyya, P. Chandra, H.C. Goyal, C.L. Kaul, R.K. Kaul, M. Kothari, S. Kotwal,R. Koul, S. Sahayanathan, M.L. Sapru, M. Sharma, K. Venugopal and C.L. Bhat;\\
\textit{Proc.XXII ASI meeting , Bull. Astr. Soc. India, 31(2003) 357.} 

\item Artificial Neural Network-based image cleaning method for atmospheric Cerenkov imaging telescopes;\\
V.K. Dhar, A.K. Tickoo, M.K. Koul, C.L. Kaul, R.C. Rannot, \textbf{K.K. Yadav}, B.P. Dubey and R. Koul;\\
\textit{Proc.XXII ASI meeting , Bull. Astr. Soc. India, 31(2003) 495.}

\item Multinode data acquisition and control system for the 4-element  TACTIC telescope array;\\
\textbf{K.K. Yadav}, N. Chouhan, S.R. Kaul and R. Koul;\\
\textit{Bull. Astr. Soc. India , 30 (2002), 403.} 

\item Towards absolute gain calibration of the TACTIC imaging element;\\
A.K. Tickoo, V.K. Dhar, K. Venugopal,  \textbf{K.K. Yadav}, S.K. Kaul, R. Koul, M. Kothari, H.C. Goyal, P. Chandra, N. Bhatt and C.L. Bhat;\\
\textit{Bull. Astr. Soc. India, 30 (2002), 381.}

\item Active mirror alignment control system for the MACE telescope;\\
S.V. Kulgod, Lizy Pious, V.K. Chadda, S.R. Kaul, \textbf{K.K. Yadav} and R. Koul;\\
\textit{Bull. Astr. Soc. India, 30 (2002), 321.}

\item Recent results on Crab Nebula and Mkn-421 observations with TACTIC imaging element;\\
N. Bhatt, N.K. Agrawal, C.K. Bhat, S. Bhattacharyya, S.K. Charagi, P. Chandra, N. Chouhan, V.K. Dhar, A. Goyal, H.C. Goyal, K. Kamath, C.L. Kaul, I.K. Kaul, R.K. Kaul, S.K. Kaul, S.R. Kaul, M. Sharma, M. Kothari, S.V. Kotwal, D.K. Kaul, M.K. Koul, R.C. Rannot, A.K. Razdan, S. Sahayanathan, M.L. Sapru, N. Satyabhama, A.K. Tickoo, K. Venugopal, \textbf{K.K. Yadav} and C.L. Bhat;\\
\textit{Bull. Astr. Soc. India, 30 (2002), 385.}
      
\item Search for pulsed TeV gamma - rays from Crab pulsar with the TACTIC Imaging $gamma$-ray Telescope;\\
R.C. Rannot, \textbf{K.K. Yadav}, V.K. Dhar, A.K. Tickoo, N. Bhatt, S. Bhattacharyya, P. Chandra, H.C. Goyal, C.L. Kaul, R.K. Kaul, M. Kothari, S.V. Kotwal, R. Koul, S. Sahayanathan, M.L. Sapru, M. Sharma, K. Venugopal and C.L. Bhat;\\
\textit{Bull. Astr. Soc. India, 30 (2002) 669.}

\item Recent observations of 1ES2344+514 and PSR0355+54 with   TACTIC imaging element;\\
A.K. Tickoo, V.K. Dhar, \textbf{K.K. Yadav}, R.C. Rannot, S. Bhattacharyya,   P. Chandra, H.C. Goyal, C.L. Kaul, R.K. Kaul, M. Kothari, S.V. Kotwal, R. Koul, S. Sahayanathan, M.L. Sapru, M. Sharma, K. Venugopal, N. Bhatt and C.L. Bhat;\\
\textit{Bull. Astr. Soc. India, 30 (2002) 775.}

\item Computer Network based data acquisition and control system for the  4 - element TACTIC telescope array;\\
\textbf{K.K. Yadav}, A. Kanda, N. Bhatt, R. Koul and S.R. Kaul;\\
\textit{Proc. INIT-2001, 6-9 Feb,2001, Mumbai.}

\item A programmable trigger-generator for the 349 pixel imaging camera of the TACTIC telescope;\\
S.R. Kaul, R. Koul, I.K. Kaul, A. Kanda, A.K. Tickoo, \textbf{K.K. Yadav} and  C.L. Bhat;\\
\textit{Proc. INIT-2001, 6-9 Feb, 2001, Mumbai.}

\item Observations of the Crab Nebula, Mrk 501 and Mrk 421 using the TACTIC Imaging Element;\\
N. Bhatt, N.K. Agrawal, C.K. Bhat, S. Bhattacharyya, V.K. Dhar, A. Goyal, H.C. Goyal, C.L. Kaul, D.K. Koul, I.K. Kaul, R.K. Kaul, S.K. Kaul, S.R. Kaul, M.K. Koul, R. Koul, M. Kothari, R.C. Rannot, A.K. Razdan, S. Sahayanathan, M.L. Sapru, N. Sathyabhama, A.K. Tickoo, K. Venugopal, \textbf{K.K.Yadav}, C.L. Bhat;\\
\textit{Pro. 26th ICRC,Aug.1999,OG 2,4,20.}


\item Development of LAWA Cerenkov radiation detector for the MYSTIQUE gamma ray telescope;\\
V.K. Dhar, \textbf{K.K. Yadav}, S. Bhattacharyya, A.M. Wani, R.K. Kaul, C.L. Bhat,  R. Koul, S.R. Kaul, S.K. Kaul;\\
\textit{Bull. Astr. Soc. India (1999), 27, 305-308.}

\end{enumerate}

\newpage
\chapter{Acknowledgement}
\noindent I wish to express my sincere gratitude to my thesis guide Dr. R. C. Rannot for his advice, encouragement and deep involvement at all stages of this work. I am also grateful to him for critically going through the manuscript of this thesis.

\noindent I express my gratitude to Sh. R. Koul, Head, Astrophysical Sciences Division, BARC, who has always been a source of inspiration, encouragement and guidance to me.

\noindent I gratefully acknowledge Dr. A. K. Tickoo, who first introduced me to data analysis techniques and helped me in understanding the details of the TACTIC experiment.

\noindent I am thankful to Sh. M. K. Koul, Sh. V. K. Dhar, Sh. P. Chandra and Sh. K. K. Singh for their involvement in the simulation and data analysis aspects of the TACTIC telescope.

\noindent I am also thankful to Dr. S. Bhattacharyya and Sh. N. Bhatt for their help with the word processing aspects of the manuscript. 

\noindent The contribution of my colleagues Sh. H. C. Goyal, Sh. M. Kothari, Sh. K. Venugopal, Sh. K. Chanchalani, Sh. S. Kotwal, Sh. P. Marandi and Sh. N. Kumar towards the observation aspects of the TACTIC telescope is gratefully acknowledged.

\noindent I also acknowledge Sh. S. K. Koul, Sh. S.R. Kaul, Sh. N. Chouhan and Sh. N. Agrawal for their support in resolving the hardware related issues of the telescope.

\noindent I wish to thank Dr. A. K. Mitra, Dr. A. K. Razdan, Dr. D. K. Koul, Dr. C. K. Bhat, Sh. S. Sahayanathan and Sh. M. Sharma for their constant encouragement.

\noindent I also take this opportunity to express my gratitude to all my family members and relatives specially my father, my beloved mother, my eldest brother and my father-in-law for their unstinted support and encouragement.

\noindent Finally, I would like to express my appreciation towards my wife Chitra and our children Srishti (10 years) and Shaurya (3 years) for allowing me to concentrate on my thesis programme and willingly sacrifice the precious moments which were rightfully theirs.

\newpage
\tableofcontents
\listoffigures
\listoftables
\newpage
\chapter{List of Acronyms and Abbreviations}
\begin{table}[h]
\begin{tabular}{lll}
1ES && First Einstein Survey\\
ACE &&Atmospheric Cherenkov Event\\
ACR &&Atmospheric Cherenkov Radiation\\
ACT &&Atmospheric Cherenkov Technique\\
ADC &&Analog to Digital Convertor\\
AGC &&Absolute Gain Calibration\\
AGN &&Active Galactic Nuclei\\
ANN  &&Artificial Neural Network\\
asl && above sea level\\
ASM && All Sky Monitor\\
BAT && Burst Alert Telescope\\
BeppoSAX && Satellite per astronomia X\\
BL Lac &&BL Lacerate\\
BTV && Binary TeV\\
CAMAC&& Computer Automated Measurement and Control\\
CANGAROO && Collab. of Australia and Nippon for Gamma-Ray Observatory in the Outback\\
CAT && Cherenkov Array at Themis\\
CCR &&Chance Coincidence Rate\\
CDC &&Charge to Digital Converter\\
CELESTE && Cherenkov Low Energy Sampling and Timing Experiment\\
CGRO && Compton Gamma Ray Observatory\\
CMBR && Cosmic Microwave Background Radiation\\
COG && Center Of Gravity\\
CORSIKA && COsmic Ray SImulations for KASCADE\\     
CPC &&Compound Parabolic Concentrator\\
CR &&Cosmic Ray \\
CTA &&Cherenkov Telescope Array\\
CU  &&Crab Unit\\
DAC &&Data Acquisition and Control\\
DDG  &&Digital Delay Generator\\
EAS &&Extensive Air Shower\\
EBL &&Extragalactic Background Radiation\\
EGRET && Energetic Gamma-Ray Experiment Telescope\\
EH &&Event Handler\\
\end{tabular} 
\end{table}
\begin{table}[H]
\begin{tabular}{lll}
EHE && Extremely High Energy\\
ERC &&External Radiation Compton \\
eV && electron volt\\
FEM&& Finite Element Method\\
FR && Fanaroff-Riley\\
FSRQ &&Flat Spectrum Radio Quasars\\
FTP&& File Transfer Protocol\\
FWHM && Full Width at Half Maximum\\
GPS&& Global Positioning System\\ 
GRB &&Gamma Ray Bursts\\
GUI&& Graphics User Interface\\
GZK && Greizen-Zatsepin -Kuz'min\\
HBL &&High frequency peaked BL Lac objects   \\
HE &&High Energy\\
HEAO 1&& High Energy Astrophysics Observatory 1\\
HEGRA && High Energy Gamma Ray Astronomy\\
HESS && High Energy Stereoscopic System\\
HiRes && High Resolution Fly's Eye\\
HV &&High Voltage\\
IACT &&Imaging Atmospheric Cherenkov Telescope\\
IBL &&Intermediate frequency peaked BL Lac objects   \\
IC &&Inverse Compton\\
IPC  &&Inter Process Communication\\
IR && Infra Red\\
KASCADE && KArlsruhe Shower Core and Array DEtector\\
kpc && killo parsec\\
LAT &&Large Area Telescope\\
LBL &&Low frequency peaked BL Lac objects  \\
LED&& Light Emitting Diode\\ 
LONS &&Light Of Night Sky\\
LST && Local Sidereal Time\\ 
MACE && Major Atmospheric Cherenkov Experiment\\
MAGIC && Major Atmospheric Gamma-ray Imaging Cherenkov\\
MJD && Modified Julian Day\\
Mrk && Markarian\\
3NCT && Nearest Neighbour Non-collinear Triplets\\
3NCQ && Nearest Neighbour Non-Collinear  Quadruplets\\
\end{tabular} 
\end{table}
\begin{table}[H]
\begin{tabular}{lll}
NKG && Nishimura-Kamata-Greisen\\
NNP && Nearest Neighbour Pairs\\
NIM&& Nuclear Instrumentation Module\\
NSB &&Night Sky Background\\
OVV &&Optically Violently Variable\\
PACT && Pachmarhi Array of Cherenkov Telescopes\\
PC&& Personal Computer\\ 
PCA && Proportional Counter Array\\
PCR &&Prompt Coincidence Rate\\
pe  &&photo electron\\
PKS && Parkes Catalog of radio Sources\\
PMT &&PhotoMultiplier Tube\\
PWN && Pulsar Wind Nebula \\
RAM&& Random Access Memory\\
RAP && Radio Active Pulser\\
RGC && Relative Gain Calibration\\
RMS && Root Mean Square\\
RQQ &&Radio Quit Quasars\\
RTOS &&Real Time Operating System\\
RXTE && Rossi X-Ray Timing Explorer\\
SCR &&Single Channel Rate\\
SSC &&Synchrotron Self Compton\\
SED &&Spectral Energy Distribution\\
SMBH &&SuperMassive Black Hole\\
SNR &&SuperNova Remnants\\
SSRQ &&Steep Spectrum Radio Quasars\\
TACTIC && TeV Atmospheric Cherenkov Telescope with Imaging  Camera\\
TTG &&TACTIC Trigger Generator\\
TTL&& Transistor-Transistor Logic\\
UHE && Ultra High Energy\\
UV && Ultra Violet\\
VDN &&Voltage Divider Network\\
VERITAS && Very Energetic Radiation Imaging Telescope Array System\\
VHE &&Very High Energy \\
XRT && X-Ray Telescope\\
\end{tabular} 
\end{table}

\newpage

%
\mainmatter
\chapter{Very high energy gamma-ray astrophysics}
\section{Introduction}
Cosmic \gams represent the highest energy band of the electromagnetic spectrum and provide insight into some of the most dynamic processes in the Universe. The window to the high energy universe was opened a few decades ago with the advent of X-ray astronomy and recent advances have given us access to observe photons of still higher energy. Although the distinction between X-rays and \gams is not very sharp, often X-rays are considered to be photons produced by atomic, thermal or non-thermal processes, while \gams are produced in nuclear or non-thermal processes. Above $1\,MeV$ all photons are \gams but this very wide energy range needs some subdivision for clarity. Weekes~\cite{Weekes1988} defined ``High Energy'' (HE) gamma rays as those between $30\,MeV$ to $10\,GeV$, ``Very High Energy'' (VHE) as photon energies from $10\,GeV$ to $100\,TeV$ ($10^{10}$ to $10^{14}\,eV$), ``Ultra High Energy'' (UHE) from $100\,TeV$ to $100\, PeV$ ($10^{14}$ to $10^{17}\,eV$) and ``Extremely High Energy'' (EHE) from $100\,PeV$ to $100\,EeV$ ($10^{17}$ to $10^{20}\,eV$). These energy ranges overlap the regions of the spectrum historically covered by the different detection techniques: satellite based detectors (HE), atmospheric Cherenkov detectors (VHE), air shower arrays (UHE) and nitrogen fluorescence detectors or very large air shower arrays (EHE). In this thesis I shall follow the convention of Very High Energy (VHE) \gam astronomy as the study of astrophysical sources of photons $\geq10\,GeV$.
\par
In this chapter, we discuss the scientific motivation for pursuing \gam astronomy, production and absorption processes of VHE \gams along with their propagation in the interstellar medium. We also discuss various types of galactic and extragalactic sources of \gams.
\section{Brief history}
While \gam astronomy in the $100\,MeV$ energy region started with the seminal paper of Morrison~\cite{Morrison1958}, astronomy at $TeV$ energies seems to date from the prescient prediction by Cocconi~\cite{Cocconi1959} in 1959 of the detectability of $TeV$ \gams from the Crab Nebula. Although the model for \gam emission adopted by him was not correct (as it over estimated the detected flux by orders of magnitude) he sowed the seeds for the first serious atmospheric Cherenkov experiments to detect VHE \gams from cosmic sources. Cocconi's prediction was based on the detection of air showers from \gams by existing particle arrays. The detection of Atmospheric Cherenkov Radiation (ACR) from Cosmic Ray (CR) air showers~\cite{Galbraith-Jelley1953} opened the path for ground-based studies of \gams. The first systematic VHE \gam observations using the Atmospheric Cherenkov Technique (ACT), were made in Crimea by a group at the Lebedev Institute in 1960~\cite{Chudakov1965}. Operating at an energy threshold of $\sim4\,TeV$, no statistically significant fluxes were detected from any source directions by this telescope. However, they laid the ground work for the future development of the ACT. This was followed by the development of Whipple \gam telescope in 1968, with a large optical reflector ($10\,m$ diameter), at Mount Hopkins in Southern Arizona. While larger collection area of the Whipple telescope led to the reduction in energy threshold, it did not immediately result in a significant improvement in the flux sensitivity. However, observations with the Whipple telescope resulted in the first detection of \gams from the Crab Nebula~\cite{Fazio1972}. More details of the early history of \gam astronomy can be found in a review paper by Weekes~\cite{Weekes1988}.
\par
In the first couple of decades of observations with marginal detections, it was clear that the sensitivity of ACT would need to be substantially improved by an efficient means of rejecting the CR background. Now it is known that even for the strongest \gam sources, the CR background events outnumber the \gams by a factor of about 1000 or so. With the advances in computers, rigorous Monte Carlo simulations of shower development revealed some characteristic differences in the electromagnetic (initiated by \gams) and the hadronic showers. In order to get the best method for background rejection, a number of discriminants were attempted e.g. differences in the spectral contents exploited by the early Whipple and the Crimean groups, lateral distribution differences utilized by CELESTE~\cite{Naurois2002}, STACEE~\cite{Williams2004} and Pachmarhi group~\cite{Bhat1997}. It was the idea originally recognised by Jelley and Porter~\cite{Jelley-Porter1963} that dramatically enhanced the performance of atmospheric Cherenkov telescopes. The image of ACR incident on the telescope could be used to reject the CR background by significantly improving the angular resolution of the telescope and the shape of image could be used as an independent discriminant~\cite{Hillas1985}. Demonstrating the power of the imaging atmospheric Cherenkov technique, the Whipple group firmly established the detection of $TeV$ \gam emission from the Crab Nebula with a statistical significance of $9.0\sigma$ in 1989~\cite{Weekes1989}.
\par
The stereoscopic approach (detection of ACR from the shower in different projections) allows precise determination of the arrival direction of primary \gams on an event by event basis and effective suppression of the background light due to CR showers and from local muons, therefore provides further improvement in the performance of the imaging atmospheric Cherenkov technique. The observations of the Crab Nebula with HEGRA~\cite{Daum1997} confirmed the predictions about the stereoscopic Imaging Atmospheric Cherenkov Telescope (IACT) arrays~\cite{Aharonian1993}. The collective experience of all the historical experiments pointed towards a larger collection area of the IACT, small pixel size of the camera and the power of stereoscopic technique as the parameters with the potential for improving the sensitivity of the IACT.
\begin{figure}[t]
\centering
\includegraphics*[width=0.92\textwidth,height=0.61\textheight,bb= 0 0 567 709]{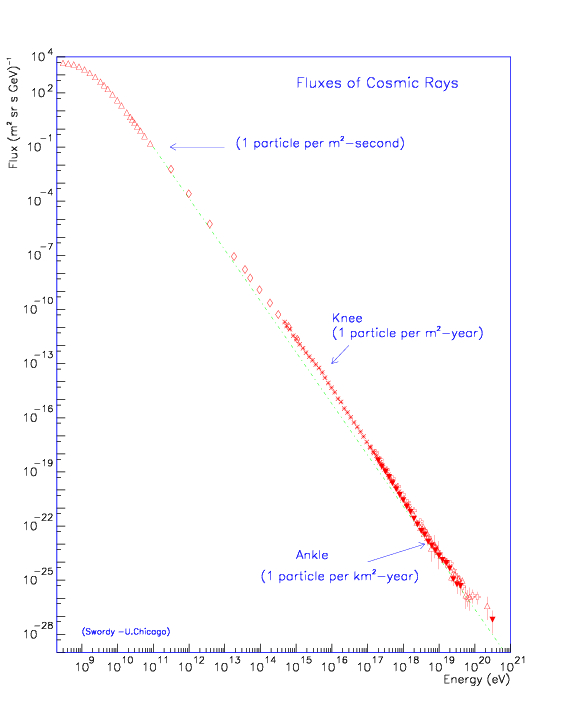}
\caption{\label{all-particle-spectrum}The all-particle spectrum of CRs measured by different experiments. This figure has been adapted from S. Swordy~\cite{S.Swordy}.}
\end{figure} 
\section{Cosmic rays}
Ever since their discovery by the Austrian physicist Victor Hess, nearly a century ago in 1912, CRs have been puzzling physicists and astrophysicists. He observed that an ionizing radiation impinges at the top of the Earth's atmosphere continuously. The origin of CRs (exact sites and processes of their production), their composition over the entire observed energy range and their propagation in space basically still remains unknown after so many years of research. The CR spectrum measured by different experiments from a few $GeV$ energy up to $10^{20}\,eV$ is shown in Fig.~\ref{all-particle-spectrum}. Up to energies $<10^{13}\,eV$ CRs can directly be measured by satellite based experiments, however at higher energies their small flux makes the direct detection by these experiments difficult. As shown in Fig.~\ref{all-particle-spectrum} the CR flux decreases from $\sim1\,particle\,\,m^{-2}\,s^{-1}$ at energies around $100\,GeV$ to below $1\,particle\,\,m^{-2}\,year^{-1}$ for the energy $10^{16}\,eV$. For energies above $10^{12}\,eV$ it is possible to detect the CR initiated atmospheric cascade of secondary particles called Extensive Air Shower (EAS), with ground based detectors at high altitudes. The detection of charged secondary particles and the measurement of ACR, produced by an air shower are the two established techniques to observe EAS. The secondary particles produced by lower energy primaries do not survive till the ground level while ACR produced by them can still be measured. The energy spectrum in the region beyond $1\,GeV$ is unaffected by the magnetic field of the Earth and the propagation of CR charged particles to the Earth through the solar wind. The differential flux $dN/dE$, in this energy region can be described by a power law, $dN/dE \propto E^{-\alpha}$. Upto the knee point located at $\simeq5\times10^{15}\,eV$ the spectral index has a value of $\alpha=2.7$ while the spectrum steepens from the knee and the index assumes a value $\alpha=3.3$ upto a point called the ankle (at $\simeq3\times10^{18}\,eV$). From the ankle onwards and upto $\simeq3\times10^{19}\,eV$ the differential spectral index has a value of $\alpha=2.6$. More recent experiments HiRes~\cite{Bergman2007} and Auger~\cite{Auger2010} have shown the suppression of CR spectrum by a factor of two in comparison to the power law extrapolation above $3\times10^{19}\,eV$.
\par
The suppression at the highest energy end in CR spectrum is presumably due to the Greizen-Zatsepin-Kuz'min (GZK) cutoff which is a theoretical upper limit on the energy of CRs from distant sources. Independently computed in 1966 by Greisen~\cite{Greisen1966}, and Kuz'min and Zatsepin~\cite{Zat-Kuz1966}, this limit is caused due to CR energy losses in their interaction with Cosmic Microwave Background Radiation(CMBR). They predicted that CRs with energies over the threshold energy of $5\times10^{19}\,eV$ would interact with CMBR photons to produce pions via the Δ resonance
\begin{equation*}
 \gamma_{CMBR} + p \rightarrow \bigtriangleup^{*} \rightarrow p + \pi^{0}
\end{equation*}
\begin{equation*}
 \gamma_{CMBR} + p \rightarrow \bigtriangleup^{*} \rightarrow n + \pi^{+}
\end{equation*}
This process continues until the CR energy falls below the pion production threshold. Due to the mean path associated with this interaction, extragalactic CRs traveling over distances larger than $50\,Mpc$ and with energies greater than this threshold should never be observed on Earth. This distance is also known as \textbf{GZK horizon}. The existence of GZK cutoff has been a much debated subject. A number of observations have been made by the AGASA experiment that appeared to show CRs from distant sources with energies above this limit (called ultra-high-energy CRs, or UHECRs). The observed existence of these particles was the so-called GZK paradox or cosmic ray paradox.
\par
The chemical composition of CR has been measured directly up to the knee point and found to comprise mainly charged particles, about $98\%$ of these particles are protons and nuclei, whilst $2\%$ are electrons. Of the protons and nuclei, about 87$\%$ are protons, 12$\%$ are helium nuclei and the remaining 1$\%$ are heavier nuclei~\cite{Longair2004}. The spectrum of electrons ( and positrons) is steeper than the one of protons and nuclei. The \gam component is just a tiny fraction $<10^{-4}$. It is generally believed that CRs with the energy below the knee value mostly originate from galactic accelerators and they have been confined inside our Galaxy by its magnetic field which is of the order of 1$\mu$G. The region between knee and ankle is still under debate, while at higher energies above $10^{17}\,eV$ particles are believed to be mostly of extragalactic origin, since the galactic magnetic field is unable to trap them within the galaxy\footnote{Rigidity  $R= B \times r_{g}$ where $B$ is Magnetic field and $r_{g}$ is the gyroradius\\ gyroradius $r_{g} = mv_{\perp}/qB$}. More recently the Auger collaboration has announced that CRs of energy above $10^{19}\,eV$ appear to come from Active Galactic Nuclei (AGN) located within $75\,Mpc$ of our Galaxy~\cite{Auger2007}. The value of extragalactic magnetic field is of the order of $5\times10^{-10}\,G$ within a radius of $75\,Mpc$ from the solar system~\cite{Angelis2008}.
\par
The energy spectrum of CR shows very few variations which can be interpreted as the evidence that the same mechanism and the same class of sources are responsible for most of their production. The best known possible sites of production of CR includes Supernova Remnants (SNRs of shell type), Plerions, Pulsars, Microquasars (X-ray binary systems) and Young stars ( and young open star clusters) within our galaxy while extragalactic ones are AGN, Gamma Ray Bursts (GRBs), Starburst galaxies and Clusters of galaxies.
\par
Despite being the most abundant component of CRs, protons do not retain the directional information about their origin as they get deflected by the magnetic fields. For a proton of energy $E$ crossing a magnetic field $B$, the deflection radius is approximately given by~\cite{Angelis2008ebl}:
\begin{equation*}
 R/1 pc \,\simeq \,(0.01)\, \frac{E/1 TeV} {B/ 1\mu G} 
\end{equation*}
This means that a proton of energy below $300\,PeV$\footnote{$PeV = 10^{15}\,eV$}, coming from the galactic center ($\sim8\,kpc$ from the Earth) will lose its directional information. \gams, neutrinos and neutrons are possible tracers for search and investigation of CR sources because they are not deflected by magnetic fields and thus point to the place of their creation. Mean decay time of a neutron being $886\,s$, implies that only the highest energy neutrons can be detected over large distances\footnote{Neutrons from the center of the Galaxy can only be detected if it has energy $\gtrsim 10^{18}\,eV$}. The detection of neutrinos from AGN allows to distinguish between the two major competing classes of models of \gam production mechanisms as they favor the hadronic one (discussed in next chapter) and therefore neutrino observations are complementary to \gam astronomy. However detection of neutrinos is very difficult as they have small interaction cross section, necessitating very large size detectors ($\gtrsim\,1\,km^{3}$). Currently \gam astronomy is the most suitable and efficient probe to study CR sources. 
\section{Scientific motivation}It is known that VHE astrophysical processes produce relativistic particles and the associated gamma radiation over a wide range of energy. \gams provide one of the best windows into the non-themal universe  and a means of probing fundamental physics beyond the reach of terrestrial accelerators. The energy region of \gams has been explored so far rather inhomogeneously. Neverthless, unprecedented progress has been made during the last decade in this field which has the following scientific motivations:
\subsection{Origin of cosmic rays}
As the cosmic \gams reach us without changing their original direction, they provide information about their sources. CRs are remarkable for a variety of reasons. Some of their important characteristics are:
\begin{itemize}
 \item The energy range of CRs is enormous i.e. $10^{6}\,eV$ to $10^{20}\,eV$. The presence of VHE CRs indicates the evidence of powerful astrophysical accelerators. 
\end{itemize}
 \begin{itemize}
  \item The sources which power CRs have a total luminosity of $\>10^{41}\,erg\,s^{-1}$. There is evidence that CRs have existed for as long as the galaxy itself ($\sim10^{10}\,yr$) but the estimated galactic containment time is only $\sim10^{7}\,yr$~\cite{Garcia-Munoz1977}. Hence CRs must continuously be replenished during the lifetime of the galaxy. 
 \end{itemize}
\begin{itemize}
 \item CRs are abundant and play an important role in the energy balance of the galaxy. They have an energy density of $\sim1 eV$ $cm^{-3}$ comparable to that contained in the galactic magnetic field or in the CMBR.
\end{itemize}
\begin{itemize}
 \item CRs are the largest source of material reaching the Earth from outside the Solar System. Their chemical composition (modulated by propagation effects) reflects the nucleosynthetic processes occurring at their origin.
\end{itemize}
\subsection{Astrophysical considerations} While the typical energy of a visible photon is $\sim1\,eV$, a VHE \gam has an energy more than 10 orders of magnitude greater than this. Based on this energy difference the thermal radiation as the underlying production mechanism for high energy photons is not applicable for $\gamma$-rays. It is expected that \gams are produced via particle interactions in very powerful accelerators. \gam astronomy therefore probes astrophysics not accessible by conventional astronomical techniques.
\par
As discussed later in this chapter, \gams can also be used as probes of Extragalactic Background Light (EBL), especially VHE photons which interact with optical to infrared background. The measured VHE spectra are used to set upper limits on EBL density. The $TeV$ catalog of extragalactic sources contains a large fraction of nearby sources, which is possibly due to the absorption of \gams by the EBL. It is expected that the number of extragalactic sources will increase significantly if the analysis threshold energy of a VHE \gam telescopes is reduced significantly below $100\,GeV$, as at these energies photons penetrate deeper into space.
\section{Gamma rays}
\subsection{Production mechanisms of \gams}
\begin{itemize}
\item 
\textbf{Bremsstrahlung}\\ When a charged particle is deflected in the electric field of a nucleus it emits electromagnetic radiation called Bremsstrahlung. Classically, the acceleration produced by a nucleus (Ze) on a particle of charge e and mass m is proportional to Ze$^{2}$/m. The complete quantum mechanical treatment of electron bremsstrahlung by an atom is complex because of the effects of screening by the atomic electrons and the finite nuclear radius~\cite{Evans1955}. This process is particularly important for relativistic electrons in the presence of atomic and molecular material wherein electrons will be deflected and emit \gams via the bremsstrahlung~\cite{Harwit1988} mechanism. In astrophysical situations, the production of \gams depends on the electron energy distribution and the gas density. The spectrum of emitted photons has a power law form with the same spectral index as that of the parent accelerated particles. To produce a \gam photon of a given energy, the electron should have more than twice that energy~\cite{Ramana-wolfendale1993}. Bremsstrahlung in the field of atomic electrons is also important when the numbers of e$^{\pm}$ are much higher than those of the nuclei as in regions where temperatures are more than $10^{10}\,K$.
\end{itemize}
\begin{itemize}
\item
\textbf{Synchrotron radiation}\\ Synchrotron radiation is the process in which radiation is emitted by relativistic charged particles in the presence of a magnetic field. In a uniform magnetic field a non-relativistic electron moves in a spiral path at a constant pitch angle $\theta$. Its velocity along the field lines is constant while it gyrates about the magnetic field direction at the angular frequency of Larmour precession
\begin{equation}
  \omega_{L}=\frac{e\,B}{m_e\,c} 
\end{equation} 
where $m_e$ is rest mass of electron and B is magnetic field intensity normal to the velocity vector of the electron. At relativistic energies, radiation is beamed into a cone of angle $\theta= m_{e}c^{2}/E$~\cite{Weekes2003} and occurs as a continuum spectrum. The critical frequency at which the maximum power is emitted is
\begin{equation}
 \omega_{c}=\frac{3}{2}\frac{e\,B}{m_e\,c}\,\gamma^2\,sin\theta 
\end{equation}
where $\gamma$ is Lorentz factor of electron. The energy loss rate is given by 
\begin{equation}
 -\frac{dE}{dx} = \frac{1}{c}\, \frac{dE}{dt} = (\frac{2\,e^{4}}{3\, m_e^{2}\,c^{4}})\,\gamma^{2}\,B^{2} \qquad ergs \,\, cm^{-1}
\end{equation}
where $E$ is in $ergs$ and $B$ in $Gauss$\cite{Lang1980}.\\
Because of higher mass of proton $(m_{p}/m_{e} \simeq 1836)$ the energy loss rate appears $(m_{p}/m_{e})^{4} \simeq 10^{13}$ times slower, characteristic frequency of synchrotron radiation is $(m_{p}/m_{e})^{3} \simeq 6\times10^{9}$ times smaller and time needed for radiation of synchrotron $\gamma$-ray photon is $(m_{p}/m_{e})^{5/2} \simeq 1.5\times10^{8}$ times longer as compared to an electron of the same energy and therefore proton synchrotron radiation is considered as an inefficient process~\cite{Aharonian-book2004}.\\
If the radiating electrons have a power law energy spectrum of the form
\begin{equation}
 N_{e}(E_{e}) = K\, E_{e}^{-\Gamma}
\end{equation}
the synchrotron photon number spectrum will be of the form
 \begin{equation}
 G_{\gamma}(E_{\gamma}) \propto E_{\gamma}^{-(\Gamma+1)/2}
\end{equation}
which is flatter than that of the parent electron spectrum for values of $\Gamma > 1$~\cite{Ramana-wolfendale1993}. It should be noted here that in other windows of electromagnetic radiation, spectra are generally expressed in terms of the power radiated. In order to find the radiated power spectrum$(W m^{2}Hz^{-1})$ one has to multiply the equation (1.5) by $E_{\gamma}$ leading to synchrotron power spectrum~\cite{Ramana-wolfendale1993}
 \begin{equation}
 J_{s}(E_{\gamma}) \propto  E_{\gamma}^{-(\Gamma-1)/2}.
\end{equation}
Synchrotron radiation of accelerated electrons is one of the most important processes in the non-thermal universe and in the context of VHE $\gamma$-rays, this is the usual process for generating the seed photons for Inverse Compton (IC) scattering (discussed below). However, UHE CRs (electrons and protons) can directly emit synchrotron radiation at VHE range. Quantitatively, to produce $1.0$ $ TeV $ $\gamma$-ray in the presence of a magnetic field of $1\,G$, the electron should have an energy of $\sim 7.2 \times 10^{15}\,eV$.
\end{itemize}
\begin{itemize}
\item 
\textbf{Inverse Compton (IC) scattering}\\ In the IC scattering a relativistic electron or positron of energy $E_{e}$ scatters off a low energy photon of energy $\epsilon$ and transfers part of its energy to the photon. The energy of Compton boosted photon is given by~\cite{Ramana-wolfendale1993}
\begin{equation}
 E_{\gamma} \simeq \epsilon\, \gamma^{2}\qquad  for \qquad \gamma\,\epsilon \ll m_{e}\,c^{2}    
\end{equation} 
and 
\begin{equation}
 E_{\gamma} \simeq E_{e}  \qquad for \qquad \gamma \,\epsilon \gg m_{e}\,c^{2}.    
\end{equation} 
The corresponding cross sections are given by~\cite{Ramana-wolfendale1993} 
 \begin{equation}
  \sigma_{c} = \sigma_{T}\left(1-\frac{2\,\gamma \,\epsilon}{m_{e}\,c^{2}}\right)
 \end{equation} 
and 
\begin{equation}
  \sigma_{c} = \left(\frac{3}{8}\right)\, \sigma_{T}\,\left(\frac{m_{e}\,c^{2}}{\gamma\,\epsilon}\right)\left[ln\left(\frac{2\,\gamma \,\epsilon}{m_{e}\,c^{2}}\right)+1/2\right]
 \end{equation} 
where $\sigma_{T} = 8\pi r_{e}^{2} / 3 $ is the Thomson cross section and has a value of $\sim6.65 \times 10^{-25}$ $cm^{2}$. In order to produce a $1.0$ $TeV$ photon against microwave background which has mean energy of $7\times10^{-4}$ $eV$ requires an electron of energy $\sim 1.7\times10^{13}$ $eV$, while against starlight ($\epsilon \sim 1.0$ $eV$) and X-rays ($\epsilon \sim 10$ $keV$) nearly $1.0$ $TeV$ electron can produce the photon of similar energy~\cite{Ramana-wolfendale1993}.
\par
Considering the differential energy spectrum of electrons given by equation (1.4) the exponent of the $\gamma$-ray energy spectrum will be given by $-(\Gamma+1)/2$.
\end{itemize}
\begin{itemize}
\item 
\textbf{Curvature radiation}\\ Very high energy particles ($e^{\pm}$) radiate curvature radiation in their direction of motion~\cite{Manchester&Taylor1977} when forced to move along the curved magnetic field lines in very intense magnetoshere of pulsars. The characteristic energy of the radiation is given by
\begin{equation}
 E_{c}\,(eV) \approx \frac {3\,h\,c\,\gamma^{3}}{4\,\pi \,\rho_{c}} = \frac{2.96 \times 10^{-5}\, \gamma^{3}}{\rho_{c}\,(cm)}
\end{equation}
where $\rho_{c}$ is the radius of curvature of the magnetic field line and $\gamma = E_{e}/m_{e} c^{2}$. A $10^{13}\,eV$ electron moving along a field line with a curvature of $10^{8}\,cm$, typical for a pulsar, emits photons of energy $\approx 2.5\,GeV$.
\end{itemize}

\begin{itemize}
\item 
\textbf{$\pi^{0}$ decay}\\Protons can produce $\pi$ mesons either in inelastic collisions with matter or interacting with ambient radiation, e.g. the microwave background. Both charged ($\pi^{+}$, $\pi^{-}$) and neutral pions ($\pi^{0}$) are produced with the same probability in the nucleonic cascades and are the lightest ($m_{\pi^{0}} = 135\, MeV$, $m_{\pi^{\pm}} = 140\,MeV$) mesons. Neutral pions with a mean life of $0.83 \times 10^{-16} \,s \times \gamma_{\pi}$ (where $\gamma_{\pi}$ is the Lorentz factor of the pion) decay into \gams.\\
While interacting with matter, the threshold kinetic energy, for producing a particle of mass m in a proton-proton collision is given by
\begin{equation}
 T_{th} = 2\,m\,c^{2}(1+\frac{m}{m_{p}})
\end{equation} 
where $m_{p}$ is the proton mass. To produce a $\pi^{0}$ meson, the threshold kinetic energy is $279.6\,MeV$. This process produces a very weak \gam flux due to the steep spectrum of the CRs ($\approx E^{2.7}$) and its cross section ($\sigma_{pp} \approx2.5\,m\,barns$).\\
The second process of VHE \gam generation is the interaction of proton with radiation which has relatively small cross section ($\approx 250\,\mu\,barns$) but is more promising. The threshold energy for the production of $\pi^{0}$ from photons of $2.7\,K$ microwave background is $\sim10^{20}\,eV$~\cite{Ramana-wolfendale1993} and is around $10^{14}\,eV$ for $1\,keV$ photons. In the vast intergalactic distances this process leads to sufficient energy degradation and hence to a steepening of the primary CR proton spectrum at energies $\ge 10^{20} \,eV$. The resulting \gams have energies in excess of $\approx 10^{19} \,eV$ and interact with the same microwave background photons to produce $e^\pm$ pairs. These particles further produce high energy \gams by IC scattering process. The resulting electromagnetic cascade gives \gams of lower energies\cite{wdowczyk1972,stecker1973}.
\end{itemize}
\subsection{Absorption of VHE \gams} The absorption mechanisms of $\gamma$-rays can be divided into two parts: $\gamma$-matter interaction and $\gamma$-$\gamma$ interaction.
\begin{itemize}
\item 
\textbf{Gamma-matter interaction}\\There are three processes by which \gams can interact with matter.
(\textit{i}) Photoelectric effect: this process has importance at low energy (hard X-ray region).
(\textit{ii}) Compton scattering: it is the dominant process in the \gam energy range from few hundred $keV$ to $10\,MeV$.
(\textit{iii}) Pair production: this is the most important energy loss mechanism for HE and VHE \gam with matter. The incident \gam is completely annihilated, i.e. $h\nu \rightarrow e^{+} + e ^{-}$. The interaction takes place in the electric field of a nucleus and threshold for the interaction is $>1.02\,MeV$ (i.e. $2\,m_{e}\,c^{2}$). It can also take place in the field of an electron (a rare phenomenon) but the cross section is much less and the threshold is higher ($4\,m_{e}\,c^{2} = 2.044\,MeV$). Note that the kinetic energies are not equally shared between electron and positron and the initial trajectory is not necessarily the mean of their emission angles. The pair production becomes a dominant process above energy of about $30\,MeV$. It is the key process in HE space telescopes and also plays a vital role in the development of the atmospheric electromagnetic cascades that make VHE \gam astronomy possible. The mean distance that a \gam photon travels before undergoing pair production is 
\begin{equation}
\lambda_{pp} = \frac{1}{N\sigma_{pp}}
\end{equation} 
where N is the number of target nuclei per unit volume. $\sigma_{pp}$ is the cross section for pair production whose value is given by
\begin{equation}
 \sigma_{pp} = \sigma_{0}\,Z^{2}\,\left[\frac{28}{9}\,\,ln\,\left(\frac{183}{Z^{1/3}}\right) - \frac{2}{27}\right] \qquad  cm^{2} \,\,atom^{-1}
\end{equation}
where $\sigma_{0}\,=(1/137)\,(e^{4}/(m^{2} _{e}\, c^{4})) = 5.8 \times 10^{-28}\, cm^{2}/nucleus = 0.58 \,m\,barn$~\cite{Weekes2003}. At very high energies the cross section is independent of energy. The mean free path for pair production is related to the radiation length, $X_{0}$[discussed later] by
\begin{equation}
\lambda_{pp} = \frac{9}{7 X_{0}}
\end{equation} 
\end{itemize}
\begin{itemize}
\item 
\textbf{Gamma-gamma interaction}\\This process requires very unusual combination of high energy photons and a high density of low energy photons. \gams are absorbed by photon-photon pair production ($\gamma + \gamma \to e^{+} + e^{-}$) on background photon fields provided the center of mass energy of the two photon system is more than the square of rest mass energy of electron. A \gam photon of energy $\epsilon_{1}$ collides with another photon of lower energy $\epsilon_{2}$ and gives rise to a pair of particles, each of mass m, if $\epsilon_{1}$ is greater than a threshold value $\epsilon_{t}$, given by~\cite{Ramana-wolfendale1993}
\begin{equation}
 \epsilon_{t} = \frac{2\, m^{2}\,c^{4}}{\epsilon_{2}\,(1-cos\theta)}
\end{equation} 
where $\theta$ is the angle between the photon trajectories. Taking a case of head on collisions, $\epsilon_{2}\epsilon_{t} = 0.26 \times 10^{12}$ (all energies are in $eV$). Following are the cases with astronomically important radiation fields, with their mean energy and required threshold energy of incident photon~\cite{Ramana-wolfendale1993}\\\\\\
----------------------------------------------------------------------------------------\\
2.7 K CMBR \,$(6\times 10^{-4}\,eV) \qquad \epsilon_{t} \approx 4 \times 10^{14}\,eV$ \\
starlight \qquad \, $(2\,eV) \qquad \qquad \, \, \, \, \, \, \epsilon_{t} \approx 10^{11}\, eV$ \\
X-ray \qquad \, \, \, $(1\,keV) \qquad \qquad \, \, \, \, \, \, \epsilon_{t} \approx 3 \times 10^{11}\, eV$. \\
------------------------------------------------------------------------------------------\\
The cross section for this process is not very small but such collisions are not very frequent because of the low densities of target photons. Nevertheless, even with low photon densities these collisions lead to significant attenuation of photons traveling over typical extragalactic distances. The effect is particularly important where the presence of EBL, limits the distance to which VHE \gam telescopes can detect sources.
\end{itemize}
\begin{figure}[t]
\centering
\includegraphics*[width=0.65\textwidth,height=0.3\textheight,angle=0,bb=26 41 657 514]{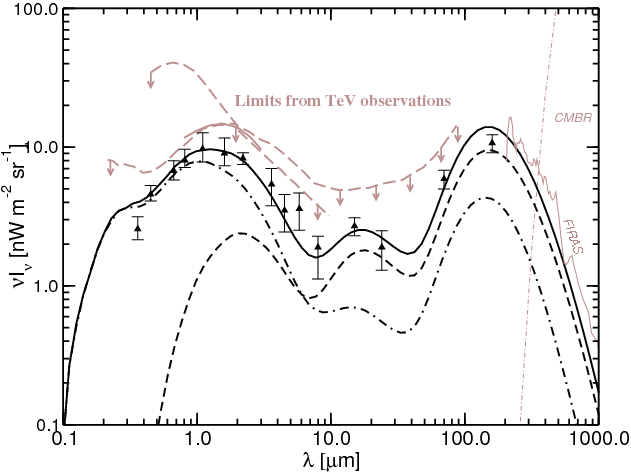}
\caption{\label{ebl-sed}Spectral Energy Distribution of EBL. This figure has been adapted from~\cite{Kneiske2010}.}
\end{figure} 
\begin{figure}[t]
\centering
\includegraphics*[width=0.70\textwidth,height=0.28\textheight, angle=0,bb= 0 0 841 595]{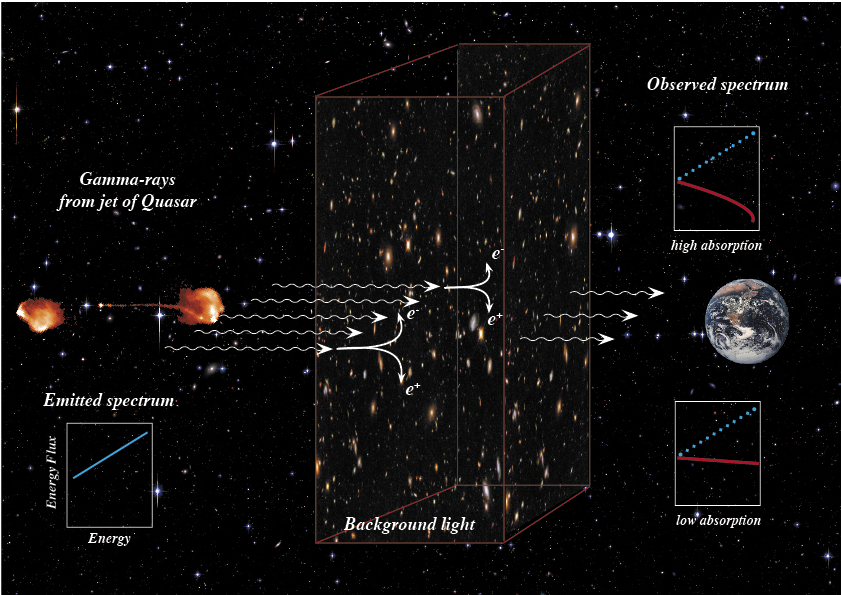}
\caption{\label{abs-illus} Illustration of absorption of extragalactic VHE photons by EBL traveling over large distances. This picture has been taken from~\cite{ebl}.}
\end{figure}
\begin{figure}[t]
\centering
\includegraphics*[width=0.60\textwidth,height=0.3\textheight, angle=0,bb=0 0 440 349]{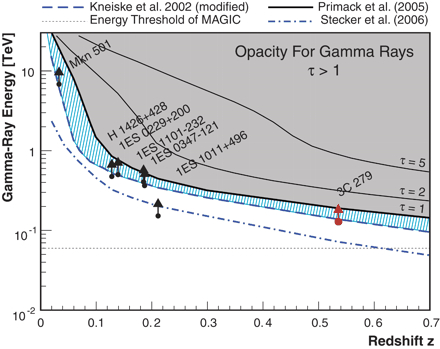}
\caption{\label{hor}Illustration of \gam horizon which can be constrained by observations upto z=0.536. The prediction range of EBL models is illustrated by thick solid black line and dashed-dotted blue line. Dashed blue line represents an upper limit of EBL based on 3C279 data taken with MAGIC telescope. Grey area indicates $\tau>1$ i.e. flux of \gams is strongly suppressed. The image has been adapted from ~\cite{Albert2008-ebl}.}
\end{figure}
\begin{figure}[t]
\centering
\includegraphics*[width=0.9\textwidth,height=0.5\textheight, angle=0,bb= 0 0 973 717]{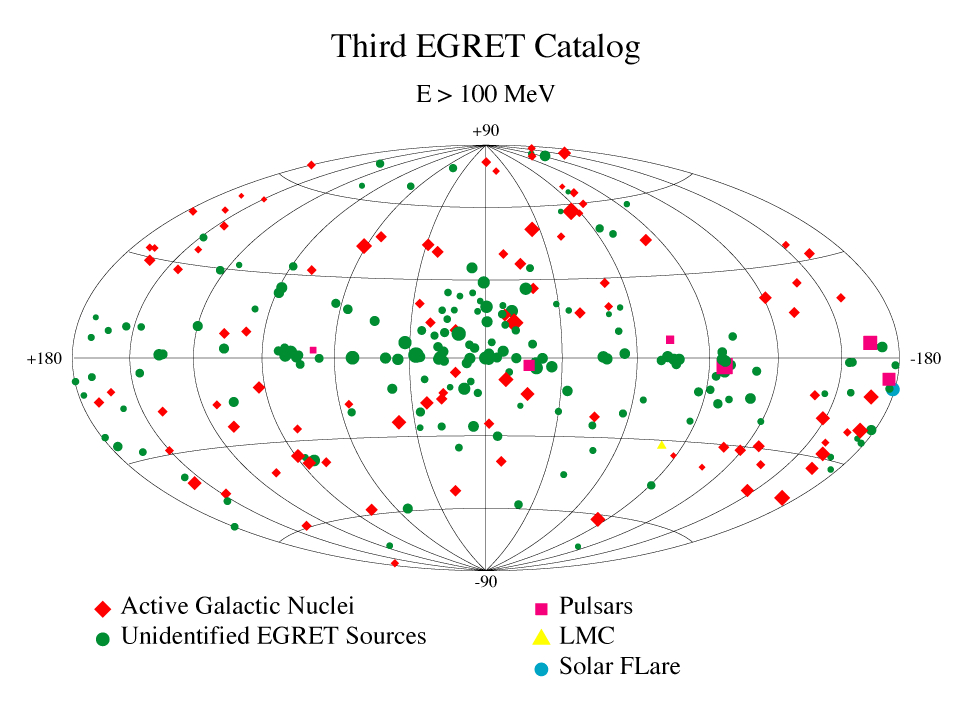}
\caption{\label{EGRET}Third EGRET catalog (1999) at energy more than 100 MeV. This figure has been adapted from~\cite{nasa}.}
\end{figure} 
\begin{figure}[t]
\centering
\includegraphics*[width=0.95\textwidth,height=0.38\textheight, angle=0,bb= 0 0 567 267]{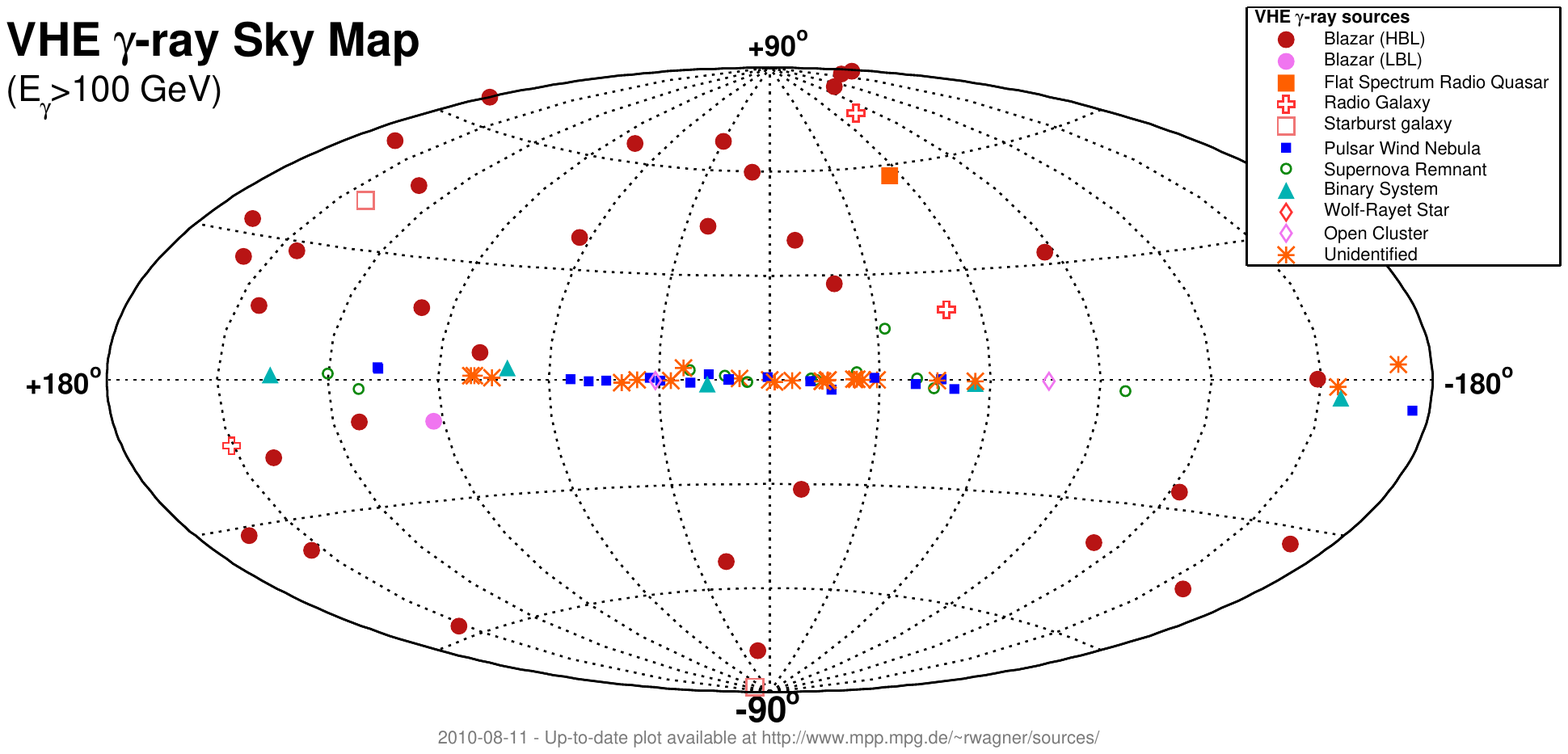}
\caption{\label{VHE-sky}Known sources in the VHE $\gamma$-ray sky as of August 2010. This figure has been adapted from~\cite{Wagner}.}
\end{figure} 
\begin{figure}[t]
\centering
\includegraphics*[width=0.9\textwidth,height=0.5\textheight, angle=0,bb= 124 239 508 514]{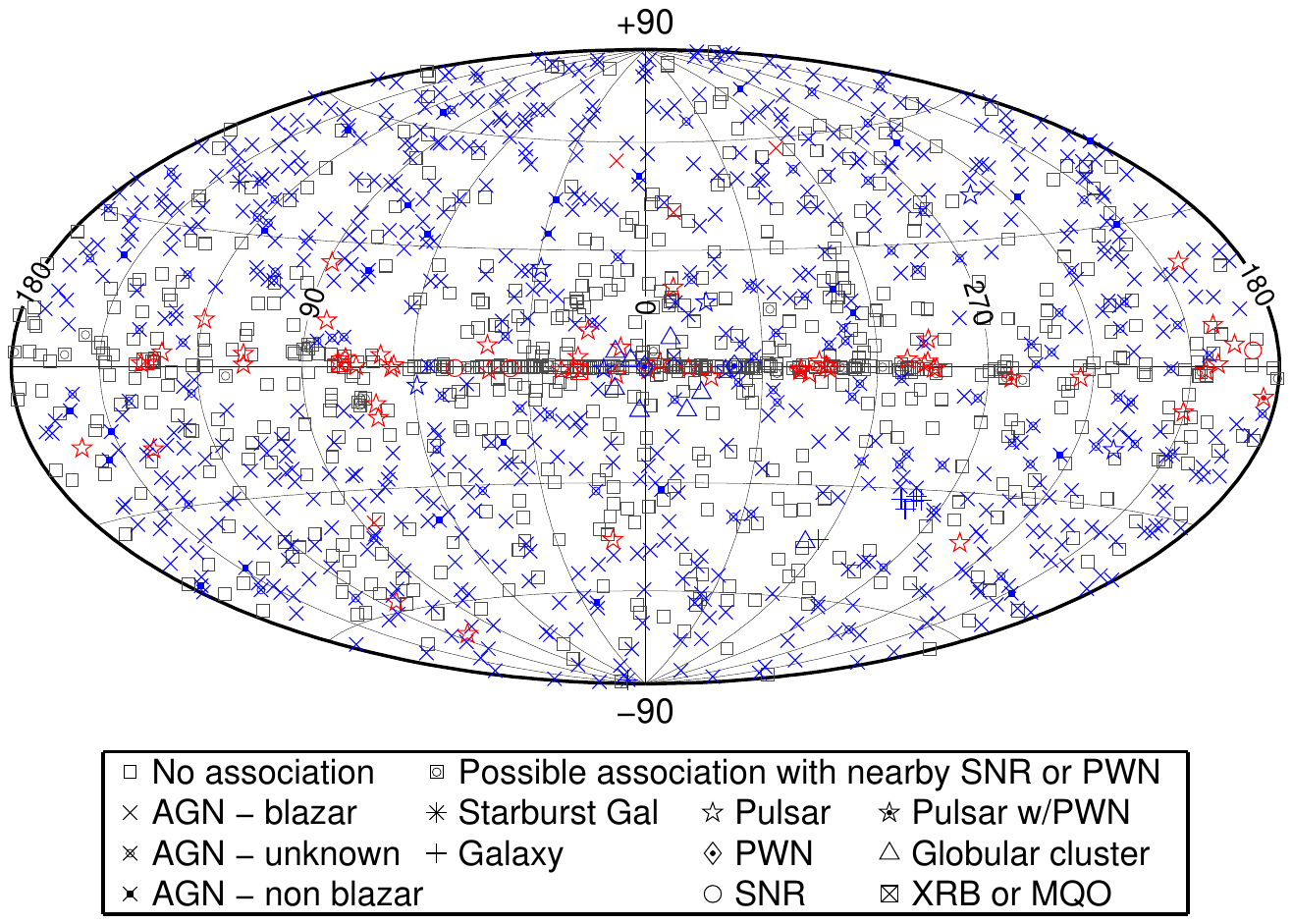}
\caption{\label{Fermi} Fermi LAT first catalog contains 1451 sources detected in the energy range $100\,MeV$ to $100\,GeV$. This figure has been adapted from~\cite{Abdo2010-Fermi-cat}.}
\end{figure} 
\section{Extragalactic background light (EBL) and $\gamma$-ray horizon} Diffuse extragalactic background radiation has been observed from radio to high energy \gams. The faint point sources are the main contributors for this radiation and hence it is a good tool to study global parameters of source populations and universal physics. The optical to infrared diffuse radiation, also called EBL, carries important information about galaxy and star formation history of the universe, and is produced by direct starlight (UV and visible range) and starlight that has been absorbed and reemitted by the interstellar dust (IR to sub-mm range)~\cite{Kneiske2010}. Accordingly, a two peak structure (first peak around $1-2\mu m$ and second in the wavelength range $\sim100-200\,\mu m$) is commonly expected in the  spectral energy distribution (SED) of EBL~\cite{Aharonian-book2004} (Fig.~\ref{ebl-sed}). The SED of EBL is not known accurately, especially in the near and mid infrared range. Upper limits and absolute measurements come from direct observations which might be polluted by foreground emission, while indirect upper limits can also be set by observations of high energy \gam sources. Galaxy number counts integrations of observable galaxies and missing possible faint sources, give strict lower limit~\cite{Kneiske2010}. While traveling from the cosmic sources towards the Earth, \gams can be absorbed in the interstellar or intergalactic space. Attenuation of VHE \gams by matter is negligible, while they are absorbed (Fig.~\ref{abs-illus}) by low energy photons of EBL via pair production (explained in the last section) and leaves an imprint on the measured spectra of different sources. The cross section for this process is described by the Bethe-Heitler formula~\cite{Heitler1960}
\begin{equation}
\sigma(E(z),\epsilon (z), x) = 1.25 \times 10^{-25}(1-\beta^{2})\left[2\beta (\beta^{2}-2) +(3-\beta^{4})\,ln\left(\frac{1+\beta}{1-\beta}\right)\right] \,cm^{2}
\end{equation} 
where\begin{equation}\beta = \left[1-\frac {2(m_{e}\,c^{2})^{2} } {E\,\epsilon\, x(1+z)^{2}}\right]^{1/2}
\end{equation} $m_{e}$ being the mass of electron, $E$ is the energy of a $\gamma$-ray photon, $\epsilon$ is the energy of EBL photon, $z$ is the redshift of the source and $x=(1-cos\theta)$.
The cross section is maximum when $\beta=0.7$ which corresponds to $\epsilon \times E \simeq\,0.5\times 10^{12}V$, where $\epsilon$ and $E$ both are in units of $eV$. Note that while computing the optical depth the energy of a photon scales with redshift $z$ as $(1+z)$. For $E=1\,TeV$, the cross section is maximum at $\epsilon$ $\simeq0.5\,eV$, which means VHE \gams then interact mainly with optical or infrared photons. The cross section with CMBR peaks when the energy of \gam is $\sim1\,PeV$. The attenuation suffered by observed VHE spectra of sources, can be used to derive constraints on the EBL density~\cite{stecker2001}. First limits on the EBL were given by Stecker et al.~\cite{Stecker1992}, while the recent measurements through the detection of distant AGNs can be found in~\cite{Aharonian-2006ebl,Mazin-2007ebl}.
\par
The probability for a photon of observed energy $E$ to survive absorption while traveling from its source at a redshift $z$ is expressed as~\cite{Angelis2008ebl}
\begin{equation}
 e^{-\tau(E,z)}.
\end{equation}
The coefficient $\tau(E,z)$ is called optical depth. The energy dependence of $\tau$ leads to significant deviation of the observed source spectrum as compared to spectrum at the source even for small differences in $\tau$, due to the exponential dependence. Since the optical depth increases with energy, the observed SED becomes steeper than the emitted one.
\par
The \gam horizon~\cite{Blanch2003} also called attenuation edge for a photon of energy $E$ is defined as the distance corresponding to a redshift $z$ for which $\tau(E,z) = 1$, i.e. the path length which provides the attenuation of the photon flux by a factor of 1/e (Fig.~\ref{hor}). Once the opacity of the Universe for \gams is known it is possible to calculate the intrinsic emission spectrum of $TeV$ sources and test the emission models.
\par
Recently detected blazar 3C279 (z=0.536) by the MAGIC collaboration~\cite{Albert2008-ebl} has raised the question of transparency of the Universe to \gams. The redshift region over which the \gam horizon can be constrained by observations has been extended up to z=0.536. The grey area in Fig.~\ref{hor} indicates an optical depth $\tau > 1$, i.e. the flux of \gams is strongly suppressed.
\section{Gamma ray sources}
Almost a century after the discovery of CRs, currently we are in the search for the sources of these energetic radiations. More than one half of the 271 sources detected at energies $>100\,MeV$ in the third EGRET catalog of 1999 (see Fig.~\ref{EGRET}) are still unidentified. Currently, Cherenkov telescopes have made an unprecedented progress in the field of VHE \gam astronomy which has resulted in the detection and identification of a large number of sources (see Fig.~\ref{VHE-sky}). To date more than 100 sources (61 galactic and 42 extragalactic) have been detected in VHE \gams. The Large Area Telescope (LAT) which is the primary science instrument on the \textit{Fermi Gamma-ray Space Telescope} (\textit{Fermi}) started surveying the sky in 2008. After the first 11 months of the science phase of the mission, the first source catalog has been released. This catalog contains 1451 sources (see Fig.~\ref{Fermi}) detected and characterised in the $100\,MeV$ to $100\,GeV$ energy range~\cite{Abdo2010-Fermi-cat}.
\subsection{Galactic sources}
\subsection*{Supernova remnants}When a star reaches its final stage of evolution by burning all its fuel necessary for the fusion reactions that counteract the gravitational pressure, the core collapses followed by ejection of material into outer space. This catastrophic explosion of a star known as a supernova, emits $\sim10^{51}\,ergs$ of energy in a very short time interval~\cite{Weekes2003}. The remains of the destroyed star can be detected for thousands of years over a wide range of energies. The \gams that result from supernova explosions may be detected instantaneously (within few seconds of the explosion) as a GRB, as the steady but periodic emission from pulsars (rotating core) or as the expanding outer shell of the star known as the supernova remnant (SNR).
\subsubsection{Shell type SNR} This is the most common type of SNR, characterized by the shell of interstellar material swept up by the expanding ejected material and shock wave developed along the way, which is clearly visible in X-rays. Particles are accelerated in these shocks by their collisionless interactions with magnetic clouds in the interstellar medium (Fermi acceleration). For decades these sources have been believed to be the principal objects for the production of galactic CRs, based on the kinetic energy available in supernova events. Nearly 10$\%$ of the total mechanical energy released by the supernova explosions in our Galaxy is required to maintain the observed density ($10^{-12}\,ergs\, \,cm^{-3}$) of the galactic CRs~\cite{Aharonian2008}. However it should be noted that some other sources (e.g. pulsars or X-ray binaries) may also meet this energy requirement. The theory of diffusive shock acceleration~\cite{Malkov-Drury2001} provides acceleration efficiency (the fraction of the mechanical energy of the shock transferred to the non-thermal particles) as high as $10\%$ and can explain very hard power law source spectrum ($\alpha \approx$ 2.1) inferred from CR measurements. The high acceleration efficiency and hard energy spectra of protons (up to multi-$TeV$ energies) should lead to production of VHE \gams of hadronic origin (from $\pi^{0}$ decay). Almost all young SNRs are emitters of non-thermal X-rays, presumably originating from the synchrotron radiation by electrons of energy $\gg10\,TeV$. These electrons can also radiate $TeV$ \gams, therefore SNRs are expected to radiate in $TeV$ energy by two competing processes. Recently deep survey of the galactic plane conducted by HESS firmly established the young shell type SNRs as copious $TeV$ \gam emitters. Examples of shell SNRs (emitting $TeV$ \gams) are Cassiopeia A, RX J1713.7-3946, RX J0852.0-4622 etc.

\subsubsection{Pulsar Wind Nebulae (PWN) or Plerions}PWN or plerions are the SNRs which consist of swept up material of supernova explosion and have a rotating magnetized neutron star (pulsar) at their center. The swept up material called nebula has strong non-thermal radio to \gam emission. Generally accepted magnetohydrodynamic (MHD) model presumes that spin down energy of the neutron star is carried away by the stream of particles (mechanism is not completely understood), mostly electron and positron which are produced by the pulsar. The particles in the pulsar wind interact with magnetic field of the nebula and emit radiations from radio to X-ray via synchrotron process. The pulsar wind encounters the interstellar matter at the edges of the nebula which ends in a shock and can accelerate the particles up to $PeV$ energies. These accelerated particles interact with their own synchrotron photons or CMBR and can efficiently emit VHE \gams through IC process. They currently constitute the most populated class of identified galactic VHE sources. As compared to the shell type SNRs the upscattering process in these objects seems to be very efficient which explains their abundance in the Galaxy. The \textbf{Crab Nebula} a prototype object of this class is the brighest known steady VHE \gam emitter (the \textbf{standard candle} source) and was the first source discovered in the VHE regime~\cite{Weekes1989}. 
\subsubsection{Pulsars}Pulsars are fast spining magnetized neutron stars produced in supernova explosions. They are divided into two categories: rotation powered (detectable at radio wavelengths) and accretion powered (detectable at X-ray energies - generally thermal). Another category is known as millisecond pulsars. It is generally believed that rotation-powered pulsars are formed when cores of massive stars ($>$ 8 M$_{\odot}$) collapse~\cite{Lyne-Smith1990}. Pulsars have their masses ranging from 1.4 M$_{\odot}$ to 3 M$_{\odot}$, diameter of about $10\,km$, core density reaching upto $10^{15}\,g/cm^{3}$ and can produce the strongest magnetic field in the universe, of the order of $10^{12}\,G$\footnote{1 G = 10$^{-4}T$}. The rotational energy is given by 
\begin{equation}
 K = \frac{1}{2}\,I\,\omega^{2} = 2\pi^{2}\frac{I}{P^{2}}.
\end{equation}
Here symbols have their usual meanings. For M = 1.4 M$_{\odot}$ and $R=10\,km$, moment of inertia $I=(2/5)\,M\,R^{2}=1.1 \times10^{45}\,g\,cm^{2}$. The rate of rotational energy loss (\textbf{called spin down luminosity or spin down power}) can be written as 
\begin{equation}
 \frac{dK}{dt}=4\, \pi^{2}\,\frac{I}{P^{3}}\,\frac{dP}{dt}. 
\end{equation}
The period $P$ can range from a millisecond to a few seconds. Most of the pulsars show gradual increase in the spin period, for the Crab pulsar, $P = 33.3\,ms$, $dP / dt = 4.21 \times 10^{-13}\, s\, s^{-1}$, and $dK/dt=5\times10^{38}\, ergs \,\,s^{-1}$.
\par
The strong co-rotating magnetic field causes a strong electric field of the order of $6\times10^{10}\,V\, m^{-1}$~\cite{Goldreich-Julian1969} at the poles of the neutron star, which accelerates the particles near the pulsar. The rotation and magnetic axes of pulsars are misaligned, we see the emission only when the beam crosses our line of sight. The emission mechanism is not yet completely understood, however it is generally accepted that the primary radiation mechanism is the synchrotron-curvature radiation taking place near the polar caps of the pulsars. The secondary possible mechanisms are ordinary synchrotron and IC scattering in the outer region of the magnetoshere.
\par
It is not known whether the emission of radiation takes place near the magnetic poles of the rotating neutron star (the polar-cap scenario~\cite{Harding1981}) or farther out in the magnetoshere (slot-gap or outergap scenario~\cite{Arons1979,Cheng1986} and references there in). The high energy part of the \gam spectrum differs substantially between near and far cases. Again the slot gap and the outer gap have different predicted \gam spectrum. The detection of \gams above $10\,GeV$ would allow to differentiate between the different pulsar emission models. The energy spectra and the structure of light curves in the region from several $GeV$ to $30\,GeV$ carry information about the location of \gam production region in the pulsar magnetospere. Generally due to pair production in the strong magnetic field, no $TeV$ \gam emission is expected from the pulsar magnetoshere~\cite{Harding2001,Romani1996,Hirotani2005}. Recently MAGIC collaboration reported the detection of \gams from the Crab pulsar above $25\,GeV$~\cite{Aliu2008} which favors the outer gap model.
\par 
Only six pulsars were firmly established as the sources of high energy \gams by the EGRET on board CGRO launched in 1991~\cite{Weekes2003}. Recent results from the Fermi-LAT collaboration reported the detection of 30 \gam pulsars, out of which 15 were not known earlier at other wavelength~\cite{Abdo-09cat}.
\subsection*{Compact binary systems} Most of the stars in our galaxy are associated with other stars making multiple star systems which are orbiting around each other. Half of the stars in the steller population occur in some kind of binary association~\cite{Weekes2003}. The compact binary system consists of a compact star generally a neutron star or black hole and a normal massive star called the companion star. The compact star accretes mass from the companion star. They are known for the powerfull emission of temporally and spectrally variable X-ray radiation. At the moment four binaries are recognised VHE \gam emitter: PSR B1259-63~\cite{Aharonian2005a}, LS 5039~\cite{Aharonian2005b}, LS I+61 303~\cite{Albert2006} and Cygnus X-1~\cite{Albert2007}. Particle acceleration is possible in these systems either due to the termination of pulsar wind or through the internal shocks in the jets formed in the vicinity of the black hole. The \gams are then produced due to the presence of the dense target material in the form of optical photons or gas provided by the companion star.
\subsubsection{X-ray binary systems}X-ray binaries are a class of binary stars that are luminous in X-rays and are generally treated as thermal sources. The X-rays are produced by matter falling from the companion star to the compact object which effectively transforms the gravitational energy of the compact object into X-ray emission radiated by hot accretion plasma. The most likely emission process is the injection of relativistic electrons into the surrounding medium and further acceleration of these electrons by the shock region created between winds of the two objects. VHE $\gamma$-rays are the result of IC scattering of the electrons with the ambient photons.
\par
Presently, four compact binary systems are found to emit VHE \gams. Three of these binary systems: PSR B1259-63~\cite{Aharonian2005a}, LS 5039~\cite{Aharonian2005b} and LS I+61 303~\cite{Albert2006} are clearly associated to X-ray binaries~\cite{Paredes-Zabalza2010}. These three Binary $TeV$ sources (BTV) have a bright high-mass primary star, which provides intense UV seed photons for IC scattering of particles accelerated around the compact object. All of them have been detected at $TeV$ energies in several parts of their orbits and show variable emission and hard spectrum. The emission is periodic in LS 5039 and LS I +61 303 with periods of 3.9 and 26.8 days respectively~\cite{Aharonian2006,Albert2009} and consistent with their orbital periods~\cite{Aragona2009,Casares2005b,Casares2005a}. These two sources also share the distinction of being the only two known high energy emitting X-ray binaries that are spatially coincident with sources above $100\,MeV$ listed in the Third EGRET catalog~\cite{Hartman1999}. Both these sources have also been recently detected by the Fermi observatory\cite{Abdo2009a,Abdo2009b}. PSR B1259-63  was not detected by EGRET and its periodicity has also not been determined yet, as the orbital period is quite large ($\sim$$3.4\,years$) which requires the monitoring of the source for an extended period (few $years$). 
\par
It is believed that the BTV Cygnus X-1 has a black hole as a compact object where as in the case of PSR B1259-63 it is a neutron star. The nature of compact objects in other two binary systems LS 5039, LS I+61 303 is still not completely understood as there is no strong evidence as yet in support of either a black hole or a neutron star. HESS telescope array discovered a new BTV candiadate HESS J0632+057 which appears to be a point like source within experimental resolution~\cite{Aharonian2007}. It is possibly associated with the massive Be star MWC 148 and has been suggested to resemble known $TeV$ binary systems like LS I +61 303 or LS 5039. No evidence for flux variability from this source was found by the HESS group. The VERITAS collaboration has also observed this source for $31\,hours$ in 2006, 2008 and 2009 with no significant \gam signal above $1\,TeV$ being detected from it~\cite{Acciari2009}. This non-detection excludes with a probability of 99.993$\%$ that the HESS J0632+057 is a steady \gam emitter. Contemporaneous X-ray observations with the $Swift$ X-Ray telescope reveal a factor of 1.8$\pm$0.4 higher flux in the 1-10 $keV$ range than earlier X-ray observations of HESS J0632+057. Future multiwavelength observations combined with results from ground-based and space-based \gam observatories will provide a deeper understanding of the true nature of HESS J0632+057.
\subsubsection{Microquasars}A microquasar (or radio emitting X-ray binary) is a smaller cousin of a quasar. Microquasars are named after quasars, as they have some common characteristics: strong and variable radio emission, often resolvable as a pair of radio jets, and an accretion disk surrounding a compact object. In quasars, the black hole is supermassive (millions of solar masses); in microquasars, the mass of the compact object is only a few solar masses. In microquasars, the accreted mass comes from a normal star, and the accretion disk is very luminous in the optical and X-ray regions. Microquasars are sometimes called radio-jet X-ray binaries to distinguish them from other X-ray binaries. A part of the radio emission comes from relativistic jets, often showing apparent superluminal motion.
\par
Microquasars are very important for the study of relativistic jets. The jets are formed close to the compact object, and timescales near the compact object are proportional to the mass of the compact object. Therefore, ordinary quasars take centuries to go through variations which a microquasar experiences in one day. 
MAGIC observed Cygnus X-1 recently during a short lived flaring episode and reported strong $TeV$ emission~\cite{Albert2007}. This high mass X-ray binary is a microquasar~\cite{Dmazin-thesis}.
\subsection*{Young open star clusters}Star clusters which are groups of stars are of two types. One is the globular clusters which are tight groups of hundreds of thousands of very old gravitationally bound stars, while other is the open clusters, a more loosely clustered group of stars, generally contain less than a few hundred members, and are often very young. The particles can be accelerated to extreme energies by means of shock created by the collision of winds of these massive young stars. HEGRA reported the first evidence of this new class of galactic VHE source in the vicinity of Cygnus OB2 region which was confirmed by MAGIC. HESS detected VHE emission from a massive star called Wolf-Rayet (WR) star (\cite{Dmazin-thesis} and references therein).
\subsection*{Galactic center}The galactic center (\textit{l=0$^\circ$, b=0$^\circ$}), located at a distance of $8.33\pm0.35 \,kpc$ from the Earth~\cite{Gillessen2009}, is a region of interest and it is significant that there is an EGRET source 3EG J1744-3039, positionally coincident with it. Fermi, which is currently surveying the sky in the energy range between $20\, MeV$ to more than $300\,GeV$ with unprecedented sensitivity and resolution  has detected a source at the galactic center with high significance after 3 months of observation, which is listed in the published LAT Bright Source List~\cite{Abdo-09cat} as 0FGL 1746.0-2900. It is believed that there is a Supermassive Black Hole (SMBH) at the center of our Galaxy (as most if not all galaxies are believed to contain SMBHs at their centers). The galactic center emits $TeV$ \gams as reported by the Whipple~\cite{Kosack2004}, CANGAROO~\cite{Tsuchiya2004}, HESS~\cite{Aharonian2004gc,Aharonian2006gc-nature,Aharonian2006gc} and MAGIC~\cite{Albert2006gc} collaborations. The \gam source shows steady emission even during the X-ray flares and behaves like a point source within the angular resolution of the current instruments. Since the region around the galactic center contains many potential \gam sources (the young shell type SNR Sag A East, the newly discovered PWN G 359.95-0.04, the SMBH Sag A$^{*}$), the exact interpretation of the emission process is rather difficult~\cite{Angelis2008ebl}. The \gam production in the vicinity of SMBH is expected to be variable on time scales as short as $1\,hour$. The lack of variations in $TeV$ flux therefore interpretated as \gam production in extended regions however it can not be used as a decisive argument against the black hole origin of $TeV$ emission~\cite{Aharonian2008}. Other possible emissions because of SMBH could be acceleration of electrons at termination shock of its hypothetical wind (similar to PWN but powered by black hole). Plausible radiation mechanism therefore includes IC scattering of energetic electrons, the decay of pions produced in the interactions of energetic hadrons with the interstellar medium or radiation field and curvature radiations of UHE protons close to SMBH. These considerations disfavour dark matter annihilation as the main origin of the detected flux.
\par
The possibility of indirect dark matter detection through its annihilation into VHE \gams has aroused interest to observe the center of our galaxy. Recently HESS and MAGIC observed the galactic center, measuring a steady flux consistent with a differential power law (spectral index of about 2.2) upto energy $\sim20\,TeV$ with no apparent cutoff~\cite{Aharonian2004gc,Albert2006gc}. The detection of $TeV$ \gams from the galactic center (within $\approx$1 arc minute around Sgr A$^{*}$ (a hypothetical SMBH) is the first model-independent evidence of high energy particle acceleration in the central $3\,pc$ region of the galaxy. There are several objects located in this region which can be responsible for VHE emission, in particular the dark matter halo, the young shell type SNR Sag A East, the newly discovered PWN G 359.95-0.04, the SMBH Sag A$^{*}$ itself and the entire central $1\,pc$ diffuse region filled by dense molecular clouds and CRs, are possible counterparts of HESS J1745-290.
\subsection*{Unidentified sources} There exists a population of $TeV$ \gam sources which does not have counterparts at longer wavelength i.e. in the X-ray or radio energy bands which spatially coincide with the location of the observed excess. These sources cluster around the galactic plane, indicating that they are likely of galactic origin. Most of these sources are spatially extended sources, further suggesting that they might be associated with unseen PWNe or SNRs which constitute the largest population of galactic $TeV$ \gam sources. Many previously unidentified TeV gamma-ray sources have subsequently been found to be probably associated with PWNe or SNRs (\cite{Wei2009} and references there in).
\par
The situation for being unidentified is the likely outcome of experimental and physical considerations. As an example of the former, extended sources may have different morphologies at different wavelengths while for the later, sources may emit substantially in the $GeV - TeV$ range and are too dim to be detected at other spectral bands (``dark accelerators'') e.g TeV J2032+4130. The explanation for these unidentified sources may either be exotic where these sources originate from the annihilation of dark matter or conventionally from \textit{p-p} interaction which entails emission mostly in the VHE, contrary to electrons which produce comparable fluxes by synchrotron emission.
\par
TeV J2032+4130 is the first unidentified $TeV$ \gam source and has to date remained as such. This source was first discovered by HEGRA~\cite{Aharonian-02un} with marginal statistical significance initially but latter on confirmed by a number of experiments~\cite{Aharonian-05un, Konopelko-07un, Albert-08un}. The HEGRA results indicated that this source was an extended $TeV$ source and the measured \gam spectrum was hard with a power law index of 1.9. These results have been confirmed by the MAGIC collaboration. The MILAGRO group had discovered three unidentified VHE \gam sources MGRO J2031+41, MGRO J2019+37 and MGRO J1908+06~\cite{Abdo-07una,Abdo-07unb}, among them the last one is still unidentified though the other two have now been identified at other energies (\cite{Angelis2008ebl} and references therein). A catalog of bright \gam sources which has been released recently contains all the sources detected by Fermi LAT with a statistical significance of more than 10$\sigma$~\cite{Abdo-09cat}. The Fermi LAT source 0FGL J2032.2+4122 lies very close to CoG of TeV J2032+4130 (about 8 arc min) and therefore is a promising candidate for being the $GeV$ counterpart of the latter, based on spatial coincidence alone. Interestingly 0FGL J2032.2+4122 is also one of the 29 \gam pulsars detected by the Fermi LAT. This source is named as LAT PSR J2032+41, because it falls in a special category of pulsars which seems to show pulsation only in the \gam band. Being ``dark'' at other wavelengths, these sources could have easily escaped previous pulsar surveys.
\subsection*{Diffuse emission} The diffuse galactic \gam emission is produced by CR protons, nuclei and electrons interacting with the interstellar gas and the radiation fields. It is generally believed (still not experimently proved) that CRs upto the knee point ($\leq 5 \times10^{15}$ eV) are produced in SNRs. The spectrum of these \gams arising from the interaction of the CRs with matter in the galaxy can provide important information to understand the origin and propagation of CRs in our galaxy. The measured differential spectral index of CRs is $\alpha \approx 2.75$~\cite{Asakimori1998}, while shock acceleration theory predicts a spectrum close to $\alpha \approx 2$~\cite{Malkov-Drury2001,Ellison2007}. The difference in the spectral index is attributed to diffuse energy loss of higher energy CRs from our galaxy~\cite{Hillas2005}. Measurements by EGRET indicated an excess of \gam emission at energies $\ge1\,GeV$ relative to diffuse galactic \gam emission models that were consistent with directly measured CR spectra in the Earth's neighbourhood. However, the measurements of galacic diffuse emission with Fermi LAT for energies $100\,MeV$ to $10\,GeV$ do not show the $GeV$ bump~\cite{Torres2009} discovered by EGRET.
\subsection{Extragalactic sources:}
\subsection*{Active Galactic Nuclei}AGN are compact regions at the center of galaxies which have a much higher than normal luminosity over some bands or all of the electromagnetic spectrum. A galaxy hosting an active nucleus is called an active galaxy. AGN host a very compact SMBH with masses between 10$^{6}$ to 10$^{10}$ solar masses. In the standard model of AGN, cold material close to the central black hole forms an accretion disc. Dissipative processes in the accretion disc transport matter inwards and angular momentum outwards, while causing the accretion disc to heat up. At least some accretion discs produce jets, twin highly collimated and fast outflows that emerge in opposite directions from close to the disc and perpendicular to it. The jet production mechanism and indeed the jet composition on very small scales are not known at present, as observations cannot distinguish between the various theoretical models that exist. The jets extend up to several $kpc$ distance scale from the central SMBH and have the most obvious observational effects in the radio waveband. The jets contain excellent conditions for efficient particle acceleration and subsequently highly variable emission of non-thermal radiations which has been measured over all the wavelengths. Currently, the AGN classification schemes are based on the differences due to anisotropic radiation pattern i.e. the pointing directions are more important than the intrinsic physical differences. AGN with relativistic jets close to the line of sight of the observer are classified as blazars. More details about AGN along with their emission mechanism are discussed in the next chapter. 
\subsection*{Gamma Ray Bursts}GRBs are flashes of \gams associated with the extremely energetic explosions in distant galaxies. They are the most luminous electromagnetic emissions occurring in the Universe in any wavelength band and perhaps the brightest phenomenon since the big bang. Bursts can last from milliseconds to nearly an hour, although a typical burst lasts a few seconds and typically (10$^{51}$ - 10$^{54}$) $ergs\,s^{-1}$ are released within this time . The short duration of GRBs hints at very compact progenitors. The initial burst mostly in \gams is usually followed by a longer-lived ``afterglow'' emitting at longer wavelengths. In spite of many successes in the interpretation of the various phases of a GRB~\cite{Zhang2007}, there are still fundamental unanswered questions involving essentially all aspects of the GRB phenomenology. Currently, the most accepted picture of GRBs is the cosmological fireball model~\cite{Zhang2007, Piran1999, Meszaros2006}. In the fireball model the various observed phenomenology from a GRB are due to an ultra-relativistic outflow with initial Lorentz factor between (10$^{2}$ - 10$^{3}$)~\cite{Molinari2007} generated during the final collapse of a high mass star or the coalescence of a binary system made of compact objects. The former case is usually associated with long duration GRBs, while the later to the short duration class. The VHE \gams are expected to be produced in internal shocks due to synchrotron and IC processes. The mean distance of GRBs is about z = 2.8, which essentially means that in most cases the VHE \gams will not be able to reach the Earth because of EBL absorption. In the past several detections of GRBs were made in the $MeV-GeV$ range by EGRET and more recently by AGILE and Fermi. On the contrary, in spite of continuous attempts, so far no confirmed detection at VHE range has been obtained. First upper limit on VHE \gam emission of GRBs has been published by MAGIC collaboration~\cite{Albert-2006GRB}. 
\subsection*{Star-forming galaxies} These young galaxies are characterized with an exceptionally high star formation rate as compared to the normal galaxies. High star formation and supernova rate in starbust galaxies enhances the energy density of energetic nonthermal particles - mostly electrons and protons which are Fermi accelerated in the sites of SNRs. Coulomb, synchrotron and Compton energy losses by the electrons and pions decay (after their production by the interaction of energetic protons with ambient gas) results in emission from radio to high energy \gams. An estimation of VHE emission has also been obtained by analytical calculations of hadronic process. However, only a weak level of emission is expected making the nearby and extremely powerful starburst galaxies like M82, NGC253 and Arp220 the obvious targets for observations. The flux predicted from M82 by Persic et al.~\cite{Persic2008} could be detectable with long exposure by MAGIC II and VERITAS, whereas Cherenkov Telescopes Array (CTA) would detect it in nearly $10\,hours$ of observations. Indeed, VERITAS observed the M82 galaxy during 2007-2009 for about $137\,hours$ and detected the source~\cite{Acciari2009-M82-Nature} at energies $>700\,GeV$. This is one of the weakest VHE sources observed, with a luminosity about $0.9\%$ Crab. Similarly NGC253 has been reported by HESS to be detected in a long campaign during 2005-08 with $119\,hours$ of observation time~\cite{HESS-NGC253}. The source was detected at the $0.3\%$ Crab level, which appears to be lower than what models have predicted~\cite{Domingo2005}. Observations of Arp220 by MAGIC has placed a loose upper limit to its VHE flux~\cite{Albert-2007sbg}.

\chapter{Active Galactic Nuclei}
\section{Introduction}The fundamental building blocks of the Universe are stars which can also be found in clusters. The most common star system is the galaxy which has $10^{8}$ to $10^{12}$ stars. The ``Milky Way'', is a typical Galaxy with $\sim10^{11}$ stars in a disk like structure. Galaxies themselves are found in clusters which are members of so-called superclusters~\cite{Weekes2003}. A Galaxy is called a normal galaxy if it is not ultra-luminous and most of the observed optical radiation seems to be the sum of all the steller emissions. The Andromeda galaxy is our closest normal galaxy at a distance of about $570\,kpc$. On the other hand, active galaxies are those whose core luminosity exceeds that of the normal galaxies and points to the evidence of relativistic particle acceleration~\cite{Weekes2003}. 
\par
In this chapter, we discuss the classification of AGN and models for observed \gam emission in blazars (a particular class of AGN). We also present a list of extragalactic VHE \gam sources detected till date i.e. August 2010. 
\section{Active Galactic Nuclei} AGNs~\cite{Kembhavi1999} unusually have a bright, starlike central core called a nucleus. The nucleus emits more radiation as compared to the rest of the galaxy and the emitted radiation shows high variability of flux. Because of the enormous energies coming from rather small volumes, it is generally accepted that these nuclei have supermassive black holes (usual mass of SMBH is of the order of $10^{6}$ to $10^{10}$ solar masses) at their centers as a source of energy for the extraordinary luminosity. The observed energy spectra of AGN are bright at all wavelengths, indicating the non-thermal nature of emission, unlike stars which show a typical black body radiation. Though the \gam luminosity also varies at all wavelengths spanning from low energy to VHE range, variations at lower energies (radio and optical) are usually weaker and slower than at high energies. These objects are characterized by low emission quiescent state with occasional outbursts referred to as ``flares'', which may last for hours to months.
\subsection{Classification of AGN} 
\begin{figure}[t]
\centering
\includegraphics*[width=0.95\textwidth,height=0.5\textheight,angle=0,bb=0 0 479 308]{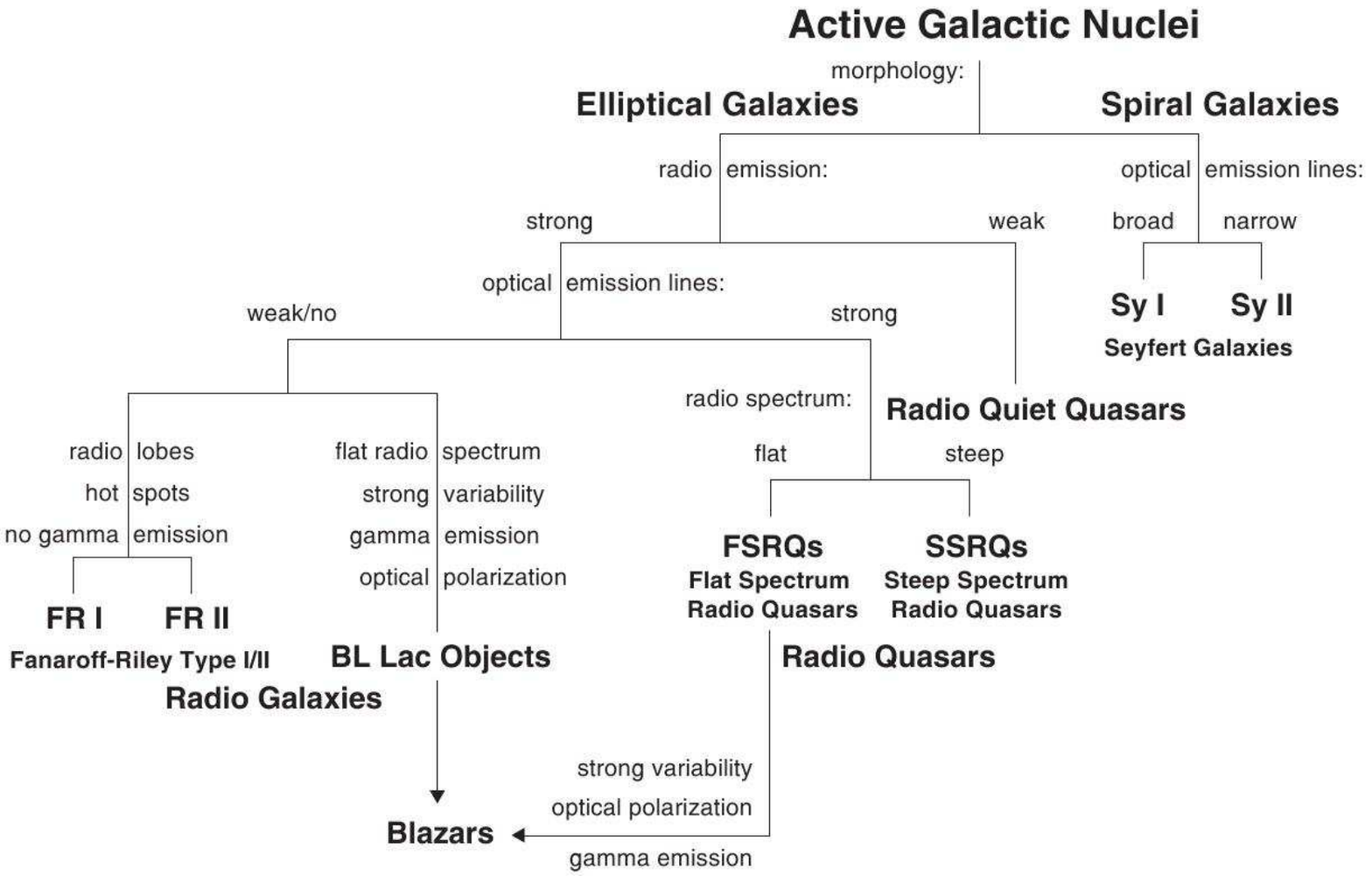}
\caption{\label{AGN-class}Empirical classification scheme of AGNs adapted from~\cite{Wagner2006}.}
\end{figure} 
Detailed description of classes of AGN can be found in~\cite{Urry1995,Padovani1999-1,Padovani1999-2}. Earlier these objects were empirically discerned by their radio emission, the properties of their optical emission lines, morphological considerations, $\gamma$-ray emission and some other properties (Fig.~\ref{AGN-class}). AGN are primarily divided into two groups of spiral and elliptical galaxies. The spiral galaxies are subdivided into Seyfert galaxies (type I and II) based on their optical emission lines. On the other hand, one class of elliptical galaxies with weak radio emission are called Radio Quite Quasars (RQQ), while the other with strong radio emitting objects are further classified on the basis of their optical emission lines. The objects with strong optical emission lines are Radio Quasars (they have two groups: Flat Spectrum Radio Quasars (FSRQs) and  Steep Spectrum Radio Quasars (SSRQs)). The second category in this class i.e. objects having weak or no optical emission lines are further classified as Radio Galaxies having no \gam emission and the BL Lacerate (BL Lac) objects which have \gam emission and also show strong flux variability. The FSRQ which also have \gam emission, strong variability and optical polarization, together with BL Lacs form the class of Blazars. All blazars emit \gams.
\subsection{The unified scheme of AGN}
\begin{figure}[t]
\centering
\includegraphics*[width=0.70\textwidth,height=0.3\textheight, angle=0,bb=0 0 404.949 304]{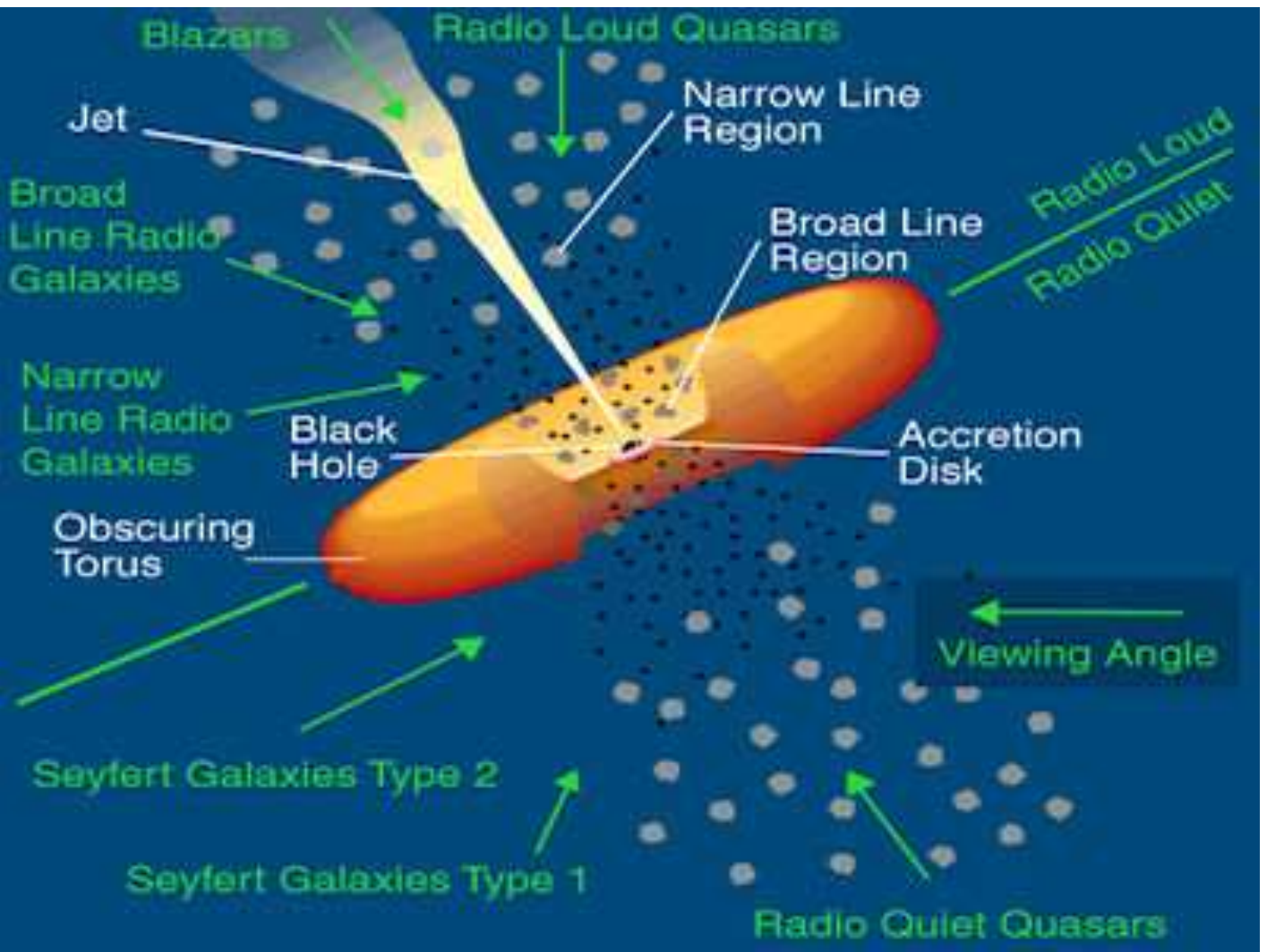}
\caption{\label{AGN-sch}An artistic view of the unified AGN scheme. When the viewing angle is quite narrow with respect to the jet, system is called a blazar. This figure has been adapted from~\cite{AGN}.}
\end{figure} 
\begin{figure}[h]
\centering
\includegraphics*[width=0.7\textwidth,height=0.7\textheight, angle=270, bb=0 0 595 842]{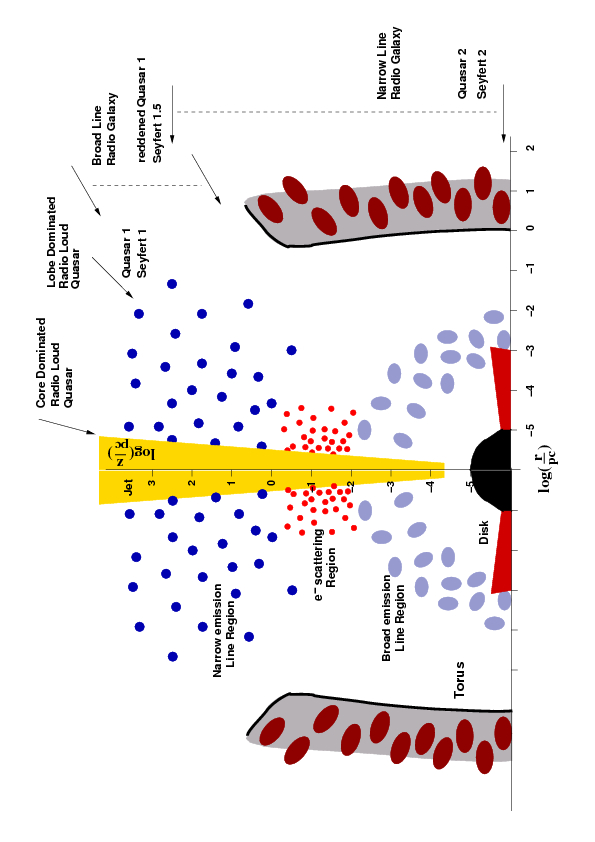}
\caption{\label{AGN-sketch}Another view of the AGN unified scheme wherein the classification has been made on the basis of viewing angle. The figure illustrates the different regions around the central black hole. The image has been adapted from~\cite{Biermann2002}.}
\end{figure} 
A unified theory of all AGN has been proposed~\cite{Urry1995}, which reduces the differences in the observed properties  to mainly the orientation of the sources relative to our line of sight. In this theory the central SMBH~\cite{Lynden1969} accretes surrounding material forming a rotating plasma accretion disk, which is heated up and emits thermal radiations from optical to X-ray wavelength. The resulting spectrum peaks in X-ray band. The gravitational energy of infalling matter is transferred into thermal radiation with a conversion efficiency of about $10 \%$~\cite{Dmazin-thesis}. The rapidly moving gas clouds close to the central engine (SMBH) or above the disk also heat up and form a broad-line emission region, observed by the Doppler-broadened emission lines. This strong optical and ultraviolet radiation is obscured along some lines of sight particularly along the equitorial plane by a dusty torus that surrounds the whole system of AGN. Beyond the torus i.e. farther from the central source, slower moving molecular gas clouds produce narrow line absorption and emission spectra. The broad line region has a particle density of $>10^{8}\, particles\,\,cm^{-3}$ and line widths of a few hundred thousand $km/s$, while the narrow line region has a lesser density of $10^{3}$ to $10^{6}\, particles\,\,cm^{-3}$ and line widths of a few hundred $km/s$~\cite{Sagar-th2007,Wagner2006}. The spin of the black hole induces twisted magnetic field lines as jets at the poles of the rotating source~\cite{Blandford1977}. Particles can be accelerated by strong shock waves in the jets to ultra-relativistic energies. The jets may extend up to $1\,Mpc$ for FR-II type AGN~\cite{Masaaki-th2008}.
\par
For a $10^{8}\,M_\odot$ black hole, the radius is $\sim 3 \times 10^{13}\,cm$, the broad line clouds are located within $\sim 2 - 20 \times 10^{16}\,cm$ of the black hole and the inner radius of dusty torus is perhaps $\sim 10^{17}\,cm$. The narrow line region extends approximately from $10^{18} - 10^{20}\,cm$ and radio jets have been detected on scales from $10^{17}$ to several times $10^{20} \,cm$~\cite{Urry1995}.
\par
A schematic drawing of the unified model can be seen in (Fig.~\ref{AGN-sch} and~\ref{AGN-sketch}). This model suggests that the wide variety of AGN types seen is largely a function of viewing angle , and hence of geometry rather than physics. It is generally believed that the jets are filled with relativistic particles and that much of the observed radio emission is synchrotron radiation from these particles. The thick torus which surrounds the accretion disk may completely obscure it, if the system is viewed edge-on. The radio lobes are then clearly seen and the AGN is called a radio galaxy (with radio emission dominant).
\par
If the line of sight is far away from the jet (almost perpendicular), the optical and radio emission from the core can be seen along with broad emission lines. Such AGN systems are known as \textbf{lobe dominated quasars} (type II quasars and Seyfert II galaxies are generally equitorial views). The core becomes more and more apparent with decreasing viewing angle and the objects are classified as \textbf{core dominated quasars} (type I quasars and Seyfert I galaxies are intermediate views). The synchrotron emission from jets becomes the strongest mechanism and radiation is variable and polarized ~\cite{Weekes2003}. When the viewing angle becomes quite narrow, the jet is the most obvious feature and the system is called a \textbf{blazar}. If the observer is directly looking into the jet (essentially zero viewing angle), it is an extreme blazar. Blazars with no emission lines are classified as BL Lacs, while the ones with emission lines are FSRQ.
\section{Blazars}Currently the blazars constitute the largest class of known $TeV $\gam emitters. They are characterized by a close alignment of the jet axis with the line of sight of the observer. The blazar types have differences in the optical emission, as Optically Violently Variable (OVV) quasars and FSRQs have strong emission lines while BL Lacs show weak / no optical lines. These differences are probably due to the amount of cold molecular clouds encountered along the line of sight. The relativistic jets which are believed to be the key elements of the observed emission of blazars, produce highly variable component of non-thermal emission. The dramatically enhanced fluxes of Doppler-boosted radiation ($\propto$ $\delta^{4}$)~\cite{Aharonian2008}, resulting from the fortuitous orientation of the jets towards us, make these objects ideal laboratories for studying the underlying physics of AGN jets through multiwavelength studies of temporal and spectral characteristics of radiation from radio to VHE \gams. First of all this concerns the BL Lac objects, several of which are already established as $TeV$ \gam emitters. The rapid variability of these objects in brightness and in polarization is explained by relativistic synchrotron jets which indicates the presence of ultrarelativistic electrons. The $TeV$ radiation not only tells us that the particles (electrons and/or protons) in these objects are accelerated to VHE, but also provides the strongest evidence in favor of the commonly accepted paradigm that the non-thermal radiation is produced in relativistic outflows (jets) with Doppler factor $\delta \ge10$~\cite{Aharonian-book2004}.
\par
Indeed, the enormous apparent VHE \gam luminosity of $TeV$ blazars, reaching
\begin{equation}
 L_{app}\, = \, 4\pi d^{2} f_{\gamma}\, \sim \, 10^{45}\,\, ergs/s
\end{equation} 
(where $d\,=\,cz/H_{0}$ is the distance to the source and $f_{\gamma}$ is the integrated \gam energy flux) during the strongest flares of Mrk421 and Mrk501 can be reduced to more reasonable level assuming that the radiation is produced in a source moving with a large Doppler factor,
\begin{equation}
\label{L-int}
 L_{int}\,\approx \, \delta^{-4}L_{app}.
\end{equation}
This assumption also allows a more comfortable (larger) size of the emitter based on the observed variability timescales,
\begin{equation}
\label{size}
 R\, \le \, c\, \Delta t_{var} \, \delta
\end{equation} 
In fact the same arguments apply to other components of non-thermal variable radiation as well, but the uniqueness of \gams comes from their ``fragility'' due to interaction with ambient photon fields, and the assumption about the relativistic bulk motion becomes unavoidable in order to overcome the problem of catastrophic internal $\gamma-\gamma$ absorption.
\par
Observations of multi-$TeV$ \gams from Mrk501, the spectrum of which in the high state extends to $\sim20\,TeV$, provide a very robust constraint on the Doppler factor of the jets. Assuming that the optical/UV flux of this source in a high state $f_{\gamma} \simeq 5 \times 10^{-11} \,erg\,/\,cm^{2} \,s$ ~\cite{Pian1998}, is produced in the jet, the catastrophic absorption of $20\,TeV$ \gams can be avoided only when $\delta \ge10$.
\par
The flux variability on different time scales is a remarkable feature of $TeV$ radiation from BL Lac objects. Since the discovery of $TeV$ $\gamma$ radiation from Mrk421 ($z$ = 0.030) and Mrk501 ($z$ = 0.034) by the Whipple collaboration~\cite{Punch1992,Quinn1996}, these sources have been the subject of intensive studies through multi-wavelength observations. Both sources are variable with typical average $TeV$ fluxes between 10\% to 50\% that of the Crab Nebula, but with rapid variability on time scales of less than $30\,minutes$~\cite{Gaidos1996,Aharonian2002-mrk421}. Recently the MAGIC collaboration detected variability of Mrk 501 on timescales less than a few minutes~\cite{Albert2007-Mrk501}.
\par
IACTs are well suited to search for short duration signals from blazars in the flaring states~\cite{Aharonian2008}. This is a key condition which makes the simultaneous observations of X-ray and \gam wavebands particularly meaningful~\cite{Coppi1999}. This has been demonstrated in many observations of Mrk421, Mrk501, 1ES 1959+650, PKS 2155-304 and some other BL Lacs simultaneously with ground based \gam and satellite based X-ray telescopes. One of the best results in this area has been obtained during the spectacular flares of Mrk421 in 2001. A nice sample of such events, the 19 March 2001 flare detected by RXTE and VERITAS, shows a clear $keV/TeV$ correlation on sub-hour timescales~\cite{Fossati2008}. The correlated flux variability is often interpreted as a strong argument in favor of the synchrotron-Compton jet emission models in which the same population of ultra-relativistic electrons is responsible for production of both X-rays and $TeV$ \gams via synchrotron radiation and IC scattering, respectively. However, the very fact of correlation between X-ray and $TeV$ \gam fluxes does not yet rule out other possibilities, including the so called hadronic models which assumes that the observed \gam emission is initiated by the accelerated protons interacting with the ambient gas, low-frequency radiation and magnetic fields~\cite{Aharonian-book2004}.
\par
Since the intrinsic source \gam spectrum can be modified by both internal and extragalactic absorption of $TeV$ \gams, the identification of radiation processes becomes very difficult. As long as the measurements of CMBR remain highly uncertain at all relevant wavelengths, the information contained in the observed \gam spectrum can not be reliable. On the other hand, the spectral variability of $TeV$ emission as well as the correlation of observed $TeV$ fluxes with other energy bands do not depend on the extragalactic absorption. Therefore, only detailed studies of spectral evolution of $TeV$ \gams and their correlation with X-rays from several X-ray selected BL Lac objects in different states of activity, and located at different distances within $1\,Gpc$, will allow us to follow and resolve simultaneously the fluctuations predicted by different models on timesscales close to the shortest one likely in these objects~\cite{Aharonian-book2004}.
\par
The spectral studies of Mrk501 and Mrk421 performed during quiescent or moderately high states generally show steep energy spectra with photon indices close to $\Gamma = 3$. However the observations of these objects by Whipple, HEGRA and CAT groups during the exceptionally bright and long lasting activity of Mrk501 in 1997 and Mrk421 in 2001 revealed very hard energy spectra~\cite{Aharonian2008}. The high state time-averaged \gam spectra of both sources can be described by a canonical ``power law with an exponential cutoff'' function of the form
\begin{equation}
 dN /dE \,=\, K\, E^{\Gamma}\, e^{(-E/E_{0})}
\end{equation} 
with $\Gamma \sim 2$. The high states of these sources are described by different [$\Gamma , E_{0}$] combinations: $\Gamma = 1.92$ and $E_{0} = 6.2\,TeV$ for Mrk501 1997 flare and $\Gamma = 2.19$ and $E_{0} = 3.6\,TeV$ for Mrk421 2001 flare (\cite{Aharonian2008} and references therein). The exponential cutoff in the spectrum of Mrk421 starts earlier. Since both sources are located at approximately the same distances, the difference in cutoff energies can be interpreted as an indication against the hypothesis that attributes the cutoffs to the pure extragalactic absorption.
\par 
Therefore one may conclude that for unambiguous identification and deep understanding of acceleration and radiation mechanisms in the jets, the correlation of absolute \gam and X-ray fluxes is very important, but not yet sufficient. The key information seems to be contained in the spectral variability in both energy bands on time scales comparable to the characteristic dynamical times of about an hour or less.
\begin{figure}[t]
\centering
\includegraphics*[width=0.85\textwidth,height=0.45\textheight, angle=0,bb=18 144 592 718]{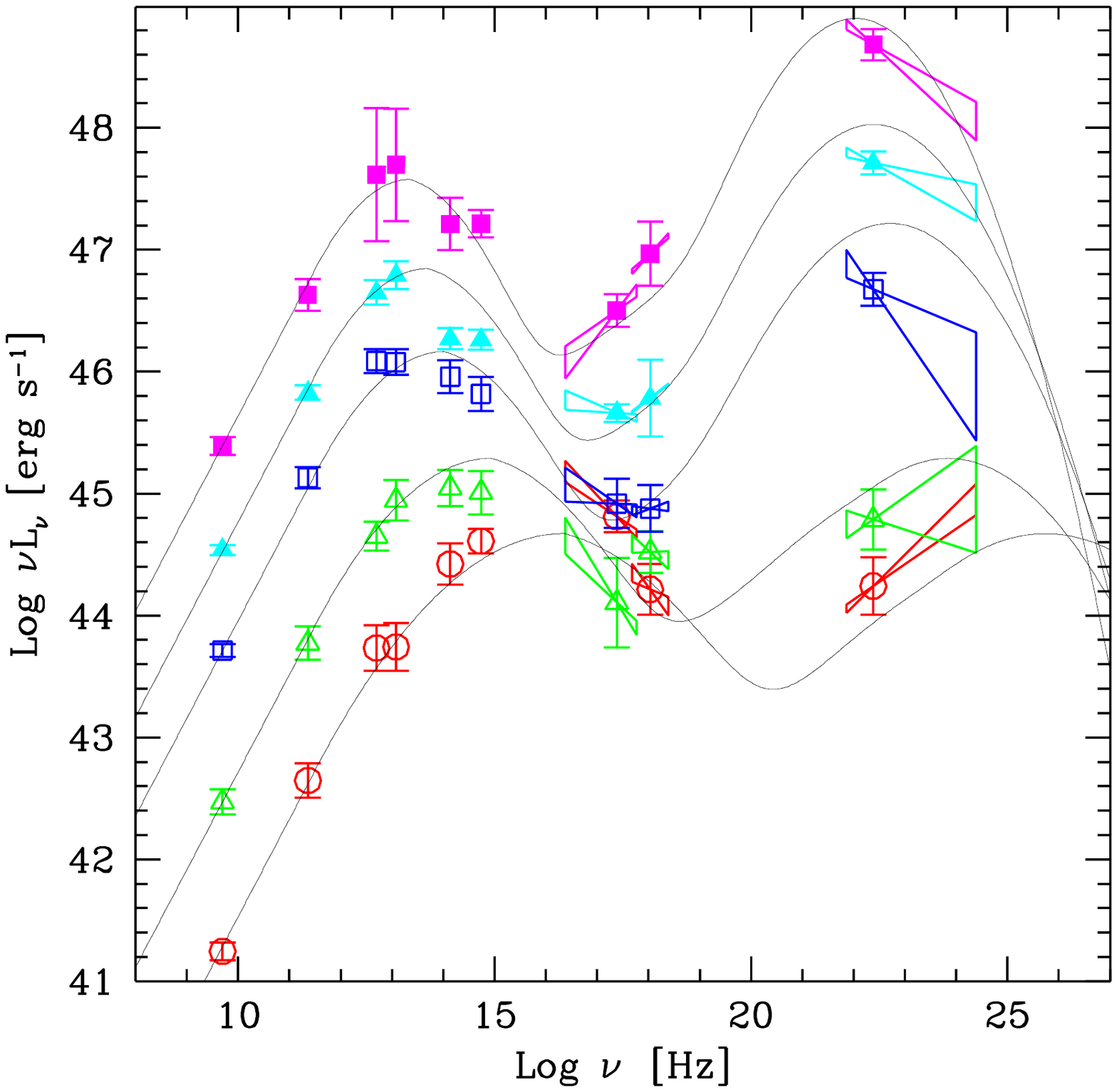}
\caption{\label{sequence}The average SED of the blazars studied by Fossati et al.~\cite{Fossati1998} luminosity, including the average values of the hard X-ray spectra. The thin solid lines are the spectra constructed following the parameterization proposed in~\cite{Donato2001}. This figure has been adapted 
from~\cite{Donato2001}.}
\end{figure} 
\subsection{The blazar sequence}
The observed SED of blazars exhibits a two-bump structure with the first peak called synchrotron peak in the infrared to $keV$ energy range and the second peak or Compton peak in the range of $MeV$ to $GeV$ energies. The statistics of spectral characteristics of the detected blazars allows us to make several phenomenological studies about their properties. A popular scheme to unify the blazar population is the so called blazar sequence proposed by Fossati et al.~\cite{Fossati1998}. The authors analyzed the SEDs of 126 blazars and identifying a remarkable continuity in their sample and drew the following conclusions:
\begin{itemize}
 \item The first peak occurs in different frequency ranges for different samples/luminosity classes, with most luminous sources peaking at lower frequency (i.e. synchrotron peak is anti-correlated with the source luminosity).
\end{itemize}
\begin{itemize}
 \item The X-ray spectrum becomes harder while the \gam spectrum softens with increasing luminosity, indicating that Compton peak also moves to lower frequencies for most luminous sources.
\end{itemize}
\begin{itemize}
 \item The peak frequency of the \gam component correlates with the peak frequency of the lower energy one (the smaller the $\nu_{peak,sync}$, smaller the peak frequency of the high energy component).
\end{itemize}
\begin{itemize}
 \item The luminosity ratio between the high and low frequency components increases with bolometric luminosity.
\end{itemize}
In this blazar sequence a single parameter, related to luminosity, seems to govern the physical properties and radiation mechanisms in the relativistic jets present in the BL Lac objects as well as in FSRQs~\cite{Fossati1998}. In the simplest form of blazar sequence (Fig.~\ref{sequence}), one distinguishes low frequency-peaked BL Lac objects (LBLs) and high frequency-peaked BL Lac objects (HBLs). LBLs are more luminous and have their first peak in the infrared to optical bands and the second one is mostly in $MeV$ region. The energy emitted in the second peak dominates the energy released in the first one. In case of HBLs, the first peak occurs in the energy range of UV to X-ray energies and dominates over the second peak in $GeV-TeV$ energies. A simple one-zone leptonic model~\cite{Ghisellini1998} describes the results satisfactorily by attributing the observed differences between these objects to the maximum energy of the accelerated electrons.
\par
The model given in~\cite{Ghisellini1998} was used to predict sources with detectable VHE emission by~\cite{Costamante2001}. Indeed, after this publication, ten new extragalactic $TeV$ \gam emitters were discovered by 2007 and almost all of them fullfilled the criteria proposed in ~\cite{Costamante2001}, that observe the sources at moderate redshift of highest combination between the radio flux (at $5\,GHz$) and X-ray flux (at $1\,KeV$). This supports the major predictions of the blazar sequence, however there is no proof that there are no VHE \gam emitters outside this selection criteria~\cite{Dmazin-thesis}. It was recently shown that the anti-correlation between the luminosity and the synchrotron peak frequency does not hold and it was an artifact of a selection effect~\cite{Padovani2007}. More recently few intermediate frequency peaked BL Lac objects (IBLs) have been found (see Table \ref{extragalacticlist}) bridging the gap between LBLs and HBLs.
\subsection{Blazars detected in $TeV$ energy}
\begin{figure}[t]
\centering
\includegraphics*[width=1.0\textwidth,height=0.35\textheight, angle=0,bb= 0 0 567 267]{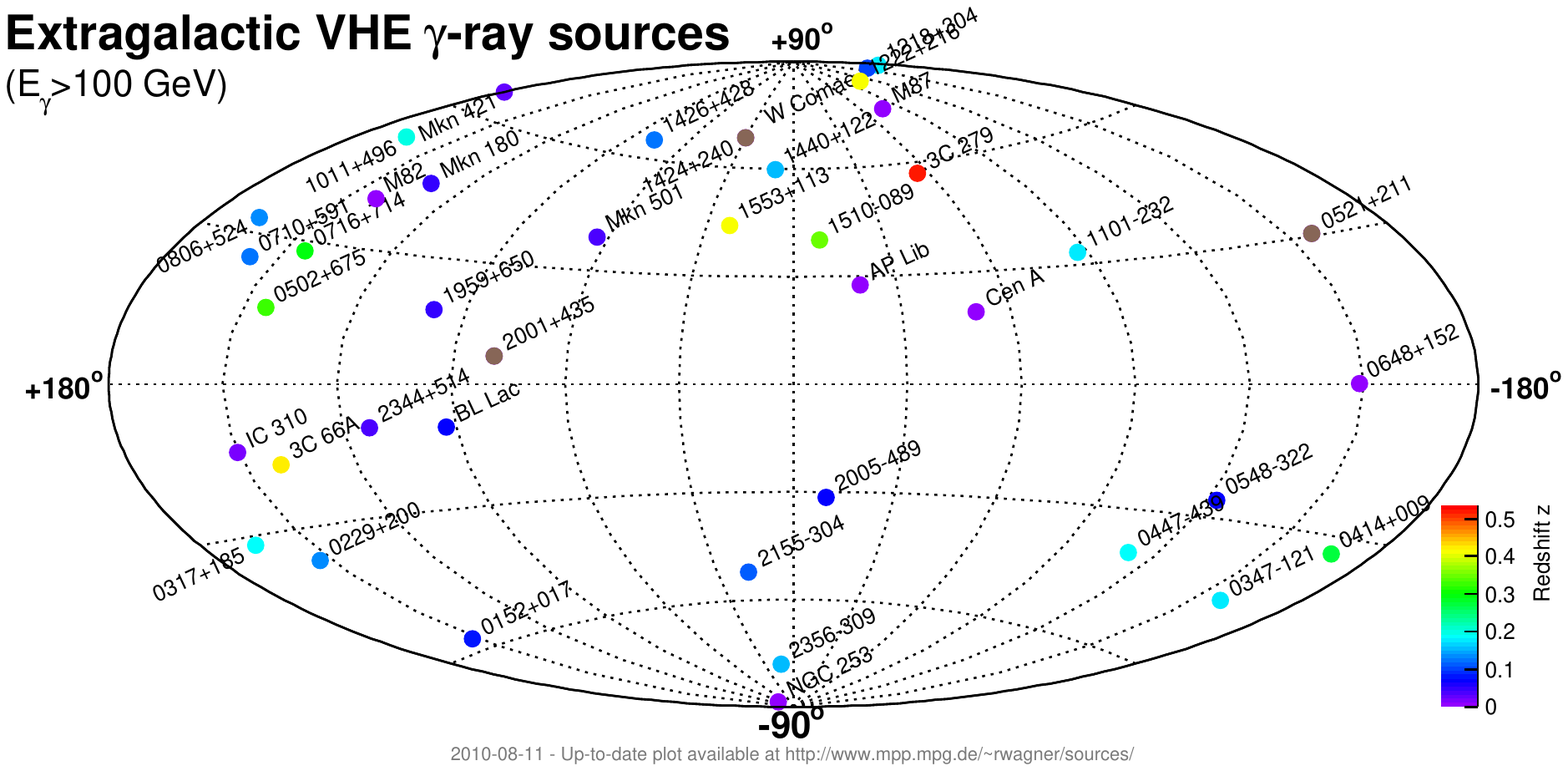}
\caption{\label{extragalactic}Sky map of total 42 extragalactic VHE $\gamma$-ray sources detected till date (August 2010). Redshift of sources indicated in different colors. The image has been adapted from~\cite{Wagner}.}
\end{figure}

All extragalactic sources detected till August 2010 are shown in Fig.~\ref{extragalactic} and also listed in Table \ref{extragalacticlist} along with their discovery references. Most of these detected sources are blazars except two radio galaxies (M87 and Centaurus A), two starbust galaxies (NGC 253 and M82) and three unidentified sources (MAGIC J0223+430, RGB 0648+152 and IC 310).
Majority of the detected blazars as is clear from the Table \ref{extragalacticlist}, belong to HBL type, while as to date only three sources each of LBL and FSRQ have been detected. Four recently detected sources belong to type IBL. A giant elliptical radio galaxy M87 (jet is viewed side on) is in our immediate cosmic vicinity, just 55 million light years away (z=0.0044). While the Whipple group has reported only an upper limit on the $TeV$ \gam flux from this source~\cite{Le-Bohec2004}, the HEGRA collaboration found a weak $TeV$ \gam signal~\cite{Aharonian-2003M87}, which was confirmed by the HESS~\cite{Aharonian-2006M87}, VERITAS~\cite{Copin2007M87} and MAGIC~\cite{Albert-2008M87} observations separately. M87 was also jointly observed by three ground-based VHE \gam telescopes VERITAS, MAGIC and HESS along with radio observations at $43\,GHz$ during 2008. Simultaneous observations at the lowest and highest ends of the electromagnetic spectrum indicate that this AGN accelerates particles to very high energies in the immediate vicinity of its central black hole~\cite{Acciari2009M87}. This source is currently being observed in a coordinated effort by VERITAS and MAGIC and an increasing VHE \gam flux level has been measured  reaching a historic high state of about 20\% of the Crab Nebula flux during the observations on 09 April, 2010~\cite{telegram1}.
\begin{table}[H]
\caption{Extragalactic TeV sources detected so far (August 2010).}
\centering
\begin{tabular}{|c|c|c|c|c|}
\hline 
Sr No & Source	& Type & Redshift & Discovery reference \\
 & & & z & \\
\hline 
1 & Mkn 421 & HBL & 0.030 & Punch et al., Nature 358 (1992) 477 \\
\hline 
2 & Mkn 501 & HBL & 0.034 & Quinn et al., ApJL 456 (1996) L83 \\
\hline 
3 & 1ES 2344+514 & HBL & 0.044 & Catanese et al., ApJ 501 (1998) 616 \\
\hline 
4 & 1ES 1959+650 & HBL & 0.047 & Nishiyama, Proc. 26th ICRC (1999) 3, 370 \\
\hline 
5 & PKS 2155-304 & HBL & 0.116 & Chadwick et al., ApJ 513 (1999) 161 \\
\hline 
6 & H 1426+428 & HBL & 0.129 & Horan et al., ApJ 571 (2002) 753 \\
\hline 
7 & M87 & radio & 0.0044 & Aharonian et al., A\&A 421 (2004) 529 \\
\hline 
8 & PKS 2005-489 & HBL & 0.071 & Aharonian et al., A\&A 436 (2005) L17 \\
\hline 
9 & 1ES 1218+304 & HBL & 0.182 & Albert et al., ApJL 642 (2006) L119 \\
\hline 
10 & H 2356-309 & HBL & 0.165 & Aharonian et al., Nature 440 (2006) 1018 \\
\hline 
11 & 1ES 1101-232 & HBL & 0.186 & Aharonian et al., Nature 440 (2006) 1018 \\
\hline 
12 & PG 1553+113 & HBL & $\>$0.09 & Aharonian et al., A\&A 448 (2006) L19; \\
& & & & Albert et al., ApJL 654 (2007) L119 \\
\hline 
13 & Mkn 180 & HBL & 0.045 & Albert et al., ApJL 648 (2006) L105 \\
\hline 
14 & PKS 0548-322 & HBL & 0.069 & Superina et al., 30th ICRC, Mérida, 2007 \\
\hline 
15 & BL Lacertae & LBL & 0.069 & Albert et al., ApJL 666 (2007) L17 \\
\hline 
16 & 1ES 0229+200 & HBL & 0.140 & Aharonian et al., A\&A 475 (2007) L9 \\
\hline 
17 & 1ES 0347-121 & HBL & 0.185 & Aharonian et al., A\&A 473 (2007) L25 \\
\hline 
18 & 1ES 1011+496 & HBL & 0.212 & Albert et al., ApJL 667 (2007) L21 \\
\hline 
19 & 3C 279  & FSRQ & 0.536 & Albert et al., Science 320 (2008) 1752 \\
\hline 
20 & RGB J0152+017 & HBL & 0.080 & Aharonian et al., A\&A 481 (2008) L103 \\
\hline 
21 & 1ES 0806+524 & HBL & 0.138 & Acciari et al., ApJ 690 (2009) L126 \\
\hline 
22 & W Comae & IBL & 0.102 & Acciari et al., ApJ 685 (2008) L73 \\
\hline 
23 & S5 0716+71 & LBL & 0.31 & Teshima et al., ATel 1500, 1 \\
\hline 
24 & 3C 66A & IBL & 0.444 & Acciari et al., ApJ 693 (2009) L104 \\
\hline 
25 & MAGIC J0223+430 & UNID & & Aliu et al., ApJ 692 (2009) L29 \\
\hline 
26 & Centaurus A & radio &  & Aharonian et al., ApJ 695 (2009) L40 \\
\hline 
27 & RGB J0710+591 & HBL & 0.125 & Ong et al., ATel 1941, 1 \\
\hline 
28 & PKS 1424+240 & IBL &  & Ong et al., ATel 2084, 1 \\
\hline 
29 & NGC 253 & starburst & 0.0008 & Acero et al., Science 326 (2009) 1080 \\
\hline 
30 & M82 & starburst & 0.0007	& Karlsson, N., arXiv:0912.3807  \\
\hline 
31 & VER J0521+211 & others &  & Ong et al., ATel 2260, 1 \\
\hline 
32 & RBS 0413 & HBL & 0.19 & Ong et al., ATel 2272, 1 \\
\hline 
33 & 1ES 0414+009 & HBL & 0.287	& Hofmann et al., ATel 2293, 1 \\
\hline 
34 & 1ES 0502+675 & HBL & 0.341 & Ong et al., ATel 2301, 1 \\
\hline 
35 & PKS 0447-439 & HBL & 0.2	& Raue et al., ATel 2350, 1 \\
\hline 
36 & PKS 1510-089 & FSRQ & 0.36 & Wagner, HEAD 2010 \\
\hline 
37 & RGB 0648+152 & UNID &  & Ong et al., ATel 2486, 1 \\
\hline 
38 & IC 310 & UNID& 0.019 & Neronov et al., arXiv:1003.4615.\\
\hline
\end{tabular} 
\label{extragalacticlist}
\end{table}
\par
The MAGIC collaboration has discovered VHE \gam emission from a new unidentified source MAGIC J0317+413 above $100\,GeV$ in March 2010~\cite{telegram4}. The emission position is consistent with the head-tail radio galaxy IC310 (z=0.0189) located in the outer region of the Perseus cluster of galaxies. Their preliminary analysis indicates emission at the level of $\sim 2.5\%$ of the Crab Nebula flux above $300\,GeV$. Further, the source was also recently detected by the  Fermi LAT satellite based detector at a statistical significance of $\sim6\sigma$ above $30\,GeV$~\cite{Neronov2010}.
\begin{table*}[t]
\centering
\begin{tabular}{|c|c|c|c|c|}
\hline 
39 & 4C +21.35	& FSRQ & 0.432 & Mariotti et al., ATel 2684, 1\\
\hline
40 & AP Lib &	LBL & 0.049	& Hofmann et al., ATel 2743, 1\\
\hline
41 & MAGIC J2001+435 & HBL &	& Mariotti et al., ATel 2753, 1\\
\hline
42 & 1ES 1440+122 & IBL & 0.162 & Ong et al., ATel 2786, 1\\
\hline
\end{tabular} 
\label{extra}
\end{table*}

\par
The VERITAS collaboration has discovered VHE \gam emission from four IBLs (a new class of AGN) W Comae (z=0.102)~\cite{Acciari-2008WComae}, PKS 1424 +240 (unkown distance)~\cite{telegram3}, 3C 66A ($z=0.444$)~\cite{Acciari-20093C66A} and 1ES1440+222 (z=0.162, last source in Table~\ref{extragalacticlist}) above $100\,GeV$. The detection of PKS 1424 +240  was also confirmed by the MAGIC group~\cite{telegram2} and Fermi-LAT~\cite{Abdo-09cat} in VHE and HE respectively. It may be noted that the radio galaxy 3C 66B lies in the same field of view as 3C 66A at a separation of $0.12^\circ$ and is also a plausible source of VHE radiation~\cite{Tavecchio2009}. The MAGIC collaboration has detected VHE \gam emission from 3C 66A/B region~\cite{Aliu2009} favoring 3C 66B as the source. Advanced telescopes with better angular resolution will be able to resolve these sources clearly. In addition, the MAGIC telescope has also detected VHE \gam emission from another interesting object 3C279 which falls in the FSRQ class of AGN  and is the most distant ($z=0.536$) source detected so far~\cite{Albert2008-ebl}. Such detections indeed constrains current theories about the density of the EBL implying that the Universe is possibly more transparent at cosmological distances than ealier believed to be in the \gam band. Further, the addition of new blazar classes namely IBL, HBL and FSRQ to the VHE catalog is expected to play an important role in our understanding of blazar populations, their kinematics and VHE \gam production mechanisms.
\section{Production of \gams in AGN}
Although the SMBH at the center of the AGN, is the putative source of energy for the entire system, it is not intuitively obvious how energy can be extracted from the black hole. Particularly it is not obvious why it should emerge in the form of a relativistic jet. Black holes are generally conceived as energy sinks -a gravitational sink hole that pulls in every thing within the event horizon. While the formation of the hot accretion disk around the black hole is understandable in terms of conservation of angular momentum and hence, the reason that these objects are detectable as bright X-ray sources, it is less clear why beams of relativistic particles should emerge from the vicinity of the black hole~\cite{Weekes2003}. The challenges for understanding the relativistic jets that are the heart of blazars are: what is the source of energy; how these relativistic energies can be achieved; what is the reason that two jets are formed pointing in exactly opposite directions and what is the mechanism by which the jets remain collimated as they emerge through the turbulent surroundings of the black hole~\cite{Weekes2003}.
\par
Though there is no general consensus on the origin of the emission components seen in \gam blazars, it is generally agreed that the low energy component arises from incoherent synchrotron emission by relativistic electrons within the jets. This is supported most strongly by the high-level variable polarization observed in these objects at radio and optical wavelengths. The observation of HE and VHE \gams from blazars provides independent evidence that the production of radiation takes place in the jets and also the jets make a small angle with the line of sight. This conclusion is also drawn from the inferred compact emission region (based on short time variability): given the observed luminosity (Eq. 2.1), it would not be possible for \gams to escape $\gamma-\gamma$ pair production unless there is relativistic beaming.
\par
It is generally agreed that the relativistic jets are caused by bulk relativistic motion, characterized by bulk Lorenz factor $\Gamma$, defined as
\begin{equation}
 \Gamma = (1-\beta^{2})^{-1/2}
\end{equation}  
The Doppler factor of an object moving at $\beta = v/c$, making an angle $\theta$ with the line of sight is
\begin{equation}
 \delta = 1/(\Gamma\, (1\,-\,\beta\, cos\, \theta \,))
\end{equation} 
The introduction of Doppler factor alleviates the explanation of observed properties of blazars in a variety of ways~\cite{Blandford1978}. 
Some of these are explained by equation \ref{L-int} and \ref{size}. In addition the problem of accelerating particles to such high energies is alleviated by $\nu_{obs} = \delta\, \nu_{source}$. In practice, generally $\delta \approx 10 -20$, so that the enhancement in energy can be as much as $10^{3-4}$~\cite{Weekes2003}.
\par
In the following sub-section, we briefly summarize the models generally used to explain \gam emission from blazars. The models are usually aimed not only to explain the observational data but also to constain the jet parameters like the magnetic field, geometry of the emission region and the Lorenz factor of the jet. Since the responsible particles for VHE \gam emission are for example accelerated electrons and protons, accordingly the models are divided into leptonic and hadronic categories. 
\subsection{Leptonic models} The charateristic double peak  shape of SED of blazars is immediately suggestive of the synchrotron Compton model at work in these sources. Electrons are accelerated beyond the velocity of the bulk Lorenz outflow and the probable mechanism of acceleration are the shocks propagating down the jet which are caused by colliding inhomogeneities in the jet i.e blobs of material moving down the jet with different velocities. The electrons radiate synchrotron radiation in the magnetic field associated with the jet and produce the first peak in the SED. The position of synchrotron peak is determined by the efficiency of the shock acceleration and the cooling processes (i.e. synchrotron and Compton scattering)~\cite{Weekes2003}. The VHE \gam emission is explained by IC scattering of optical to X-ray photons off relativistic electrons or positrons in the jet. The following two sub-classes differ in the origin of seed photons for IC. 
\subsubsection{Synchrotron self-Compton (SSC)} In SSC models~\cite{Bloom1996,Ghisellini1998ssc,Maraschi1992,Sikora01,Konigl81} the seed photons are synchrotron photons emitted by the same electron population. Experimental evidence for the SSC mechanism has been provided by the observed correlation of the X-ray and VHE \gam flux levels during large flares of VHE \gam emitting blazars~\cite{Maraschi1999,Takahashi00,Krawczynski2001}. The SSC model is widely accepted to describe VHE \gam emission from the HBL objects.
\subsubsection{External radiation Compton (ERC)} The second class of models involve external ERC~\cite{Sikora1994,Dermer1992,Blandford1995,Dermer97,Ghisellini96,Wagner95}, wherein the seed photons are ambient infrared or optical photons, photons of CMBR or thermal radiation either directly from the accretion disk or first scattered by the surrounding gas and dust clouds. Hence in these models the \gam emission takes place outside the jet. The existence of such sources of soft radiation is deduced from observations of non-blazar AGN, i.e. in which the jet is not seen edge-on. The SED of these sources shows additional thermal bumps in UV, infrared and X-ray energies~\cite{Weekes2003}.
\par
One characteristic feature of BL Lac objects, i.e. lack of strong emission lines, is commonly interpreted as an evidence that ambient photon fields are not as important as synchrotron photons. Therefore SSC models are more likely to explain the data than ERC models. However, often the predicted emission is below the observed emission ( e.g. in FSRQs), making ERC contribution necessary~\cite{Wagner2006}.
\par
Presently, the leptonic models independent of the specific assumption concerning the acceleration and radiation scenarios, represent the preferred concept for $TeV$ blazars. Two important features characterize these models~\cite{Aharonian-book2004}:
\begin{itemize}
 \item Electrons can be readily accelerated to $TeV$ energies e.g. through the shock acceleration mechanism, 
\end{itemize}
\begin{itemize}
 \item They can radiate X-rays and $TeV$ \gams with very high efficiency via synchrotron and IC channels.
\end{itemize}
\subsection{Hadronic models}Although the associated acceleration of protons is expected with at least the same efficiency as that of electrons (for most acceleration mechanisms), the hadronic models require proton acceleration up to energies of $10^{20}\,eV$, otherwise they can not offer efficient \gam production mechanisms in the jets~\cite{Aharonian-book2004}. These models assume that the observed \gam emission is initiated by accelerated protons interacting with;
\begin{itemize}
 \item Ambient matter: the so-called mass-loaded hadronic models~\cite{Pohl2000}.
\end{itemize}
\begin{itemize}
 \item Photon fields: called photo-pion hadronic models~\cite{Mannheim1993,Mannheim1996,Mannheim1998}.
\end{itemize}
\begin{itemize}
 \item Magnetic fields: ``pure'' proton synchrotron model~\cite{Aharonian2000-hadron}.
\end{itemize}
\begin{itemize}
 \item Both magnetic and photon fields~\cite{Mucke2001,Mucke2003}.
\end{itemize}
While the synchrotron and photo-pion hadronic jet models require acceleration of protons to extremely high energies, the mass-loaded hadronic models need protons of relatively modest energies.
\par
Hadron models are attractive as they are characterized by particle acceleration up to energies $10^{20} eV$, and explain CR acceleration to extreme energies. The leptonic models generally lack this virtue. The hadron models have no problem explaining the highest \gam energies but they do have problems accounting for the rapid cooling necessary to account for the short time variations observed. Additionally, they have difficulties in explaining the observed correlations of X-ray and VHE \gam emissions of blazars during flares. This is because proton cooling by synchrotron radiation is less than that by electrons by the factor $(m_e / m_p)^3$. However cooling can also come from collisions with photons or ions~\cite{Weekes2003}. These models also require extreme parameters of the acceleration region, particularly high plasma densities and high magnetic fields $B \gg 10 G$.
\par
A by-product of proton models is the production of energetic neutrinos (from decay of charged pions) which might be detectable with next generation of neutrino telescopes. The detection of such flux would effectively eliminate the lepton models. This possibility is also quoted as the justification for the construction of large neutrino telescopes. Orphan VHE \gam flares i.e. unaccompanied by outbursts in other wavelengths e.g. observed for BL Lac object 1ES 1959+650 in 2002~\cite{Daniel2005} can be interpreted as an indication of hadronic acceleration at work.
\par
The HE and VHE \gam observations strain both the leptonic and hadronic models but do not, at present, rule either out. However, the observations generally seem to favor leptonic models~\cite{Weekes2003}. Note that the leptonic and hadronic acceleration may both be present in blazar jets at the same time e.g. hadronic acceleration (or a leptonic/hadronic admixture) might account for a baseline flux and fast flare could be governed by additional electron acceleration (\cite{Wagner2006} and references therein).
\section{SSC models: Parameters and observables}
SSC models describe the multi-wavelength spectra of HBLs quite successfully, wherein synchrotron and IC emissions are responsible for low and high energy components respectively~\cite{Takahashi00,Tavecchio2001}.
\begin{figure}[t]
\centering
\includegraphics*[width=0.9\textwidth,height=0.29\textheight, angle=0,bb=0 0 415 165]{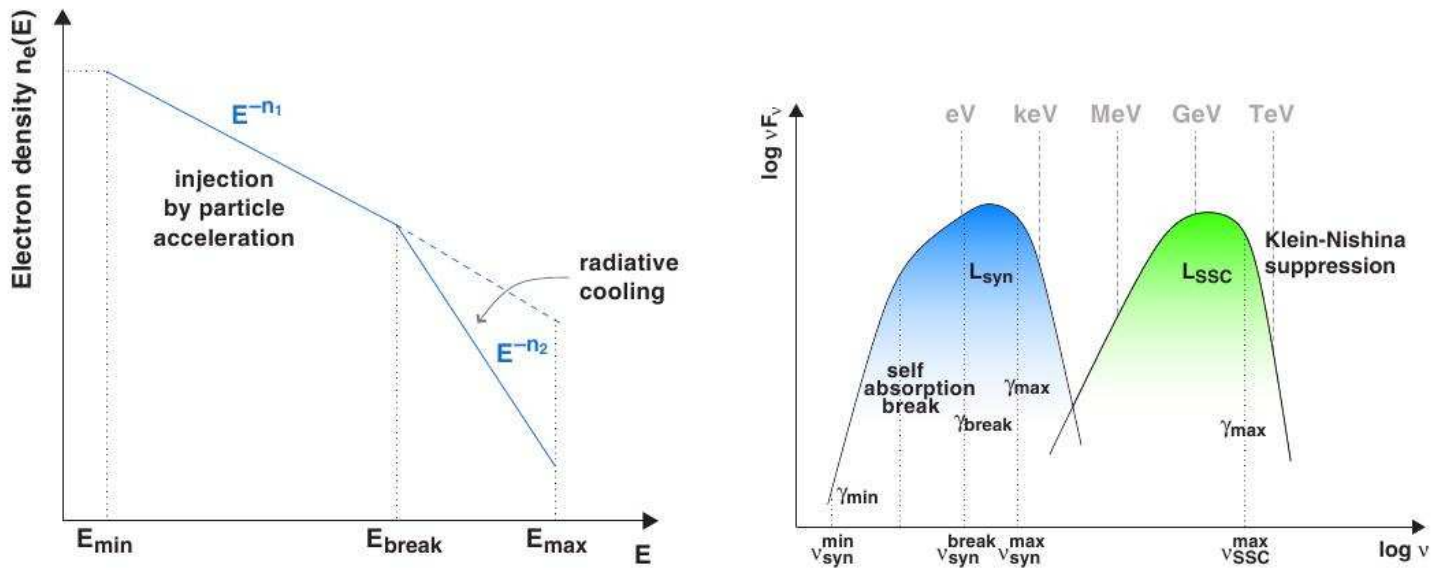}
\caption{\label{ssc}Schematic of the relativistic electron energy spectrum (left) and corresponding SED produced by this electron injection (right). For high Lorentz factors, radiative cooling decreases the number density of electrons and leads to a break in the electron spectrum. On the right side, the breaks stem from radio self absorption of created synchrotron photon, radiative electron cooling and the cutoff in the electron injection spectrum. This figure has been adapted from~\cite{Wagner2006}.}
\end{figure} 
\par
In its simplified one-zone homogeneous version, SSC model assumes a spherical acceleration region (blob of radius R), filled with an isotropic electron population and a random magnetic field (B), moving at relativistic speed $\beta =v/c$ along the jet within a small angle $\theta$ to the line of sight. The observed radiation is strongly affected by the relativistic Doppler beaming factor ($\delta$).
\par
Considering the diffusive shock as the acceleration mechanism, the electron injection spectrum can be taken as a power law $dN / dE \propto E^{-n1}$ with $n1 \simeq 2$. The spectrum is modeled to show a break point where radiative cooling through synchrotron radiation becomes the dominating process. At this energy it steepens to $n2 \simeq 3$ and thus can be described by three parameters, $E_{min}$, $E_{max}$ and $E_{break}$ (left plot of Fig.~\ref{ssc}). as observed in BL Lacs, such an electron spectrum is able to produce the synchrotron photon spectrum in randomly oriented magnetic fields of approximately constant strength~\cite{Kino2002}. Usually nine parameters are required to completely specify the SSC models, namely size of emission region (R), minimum electron energy ($E_{min}$), maximum electron energy ($E_{max}$), break energy ($E_{break}$), spectral index of electron spectrum before break (n1), index after break (n2), normalization factor of electron density (K), magnetic field (B) and the Doppler beaming factor ($\delta$).
\par
The measured SED can be used to drive seven ``observable'' quantities i.e. photon index of synchrotron radiation before break ($\alpha 1$), after break ($\alpha 2$), peak synchrotron frequency ($\nu_{s,b}$), synchrotron frequency at high energy cutoff ($\nu_{s,max}$), peak IC frequency ($\nu_c$), total measured energy flux for synchrotron component ($l_s$) and for the IC component ($l_c$)~\cite{Masaaki-th2008}.
\par
Using the standard formula for radiation one can obtain the relation between the model parameters and observables.
Observed synchrotron frequencies from a single electron of Lorenz factor $\gamma_{break}$ and $\gamma_{max}$ are respectively given by~\cite{Kino2002}
\begin{equation}
 \nu_{s,b} = 1.2 \times 10^6 B \,\gamma_{break}^2 \,\,\frac{\delta}{1+z}
\end{equation} 
\begin{equation}
 \nu_{s,max} = 1.2 \times 10^6 B \,\gamma_{max}^2 \,\,\frac{\delta}{1+z}
\end{equation} 
The maximum value of the observed SSC energy in the Klein-Nishina regime is 
\begin{equation}
 h\nu_c = C_1\, \gamma_{max}\, m_e\, c^2 \,\,\frac{\delta}{1+z}
\end{equation} 
The Constant $C_1 < 1$ represents the uncertainty  of the Klein-Nishina effect.
\par
One more relation can be obtained from the ratio of synchrotron to Compton luminosities (\cite{Masaaki-th2008} and references therein).
\begin{equation}
 u_B=\frac{d_L^2}{R^2c \delta^4} \frac{l_s^2}{l_c}
\end{equation} 
where $d_L$ is the luminosity distance and $u_B (=B^2/8\pi)$ is the energy density of magnetic field. One more observable quantity is the variability time scale which is related to the size of emission region ($R \,\lesssim\, c\,t_{var}\, \delta$).
\par
When the synchrotron photons are upscattered to VHE energies via the IC process by the electron population, the second peak in the SED of blazars is produced (Fig.~\ref{ssc}). The position of the IC peak is determined by $E_{break}$, as the peak synchrotron power is emitted by the electrons with this energy. An approximate mapping between synchrotron and IC photon energies produced by electrons of given energy is represented as follows~\cite{Krawczynski2004} 
\begin{equation}
 E_{IC}/TeV \approx \left(\frac{\delta/10}{B/0.5 G}\right)^{0.05} (E_{sy}/keV)^{0.5} 
\end{equation} 
within the present uncertainties, the left side of both the peaks can be described by the same spectral index~\cite{Tavecchio1998}.
\par
The Doppler boosting factor is an important parameter which describes the effect of time dilation, boosting and jet focusing. Note that the observed source frequency appears as $\nu . \delta$, the emission solid angle is reduced by $\delta^2$, adding to an increase in brightness by $\sim \delta^3$. An additional constant to the model is given by the requirement of source transparency which can be translated into a limit on the optical depth of the source and thus on energy density of the soft radiation (\cite{Wagner2006} and references therein). Time dependent modeling has to be used whenever a flare duration is very short. To fit the observed data consistently from several Mrk501 flares, time dependent modeling and the introduction of the second emission zone is necessary~\cite{Krawczynski2002}. 
\section{Implications of VHE \gam observations}
\begin{itemize}
 \item VHE \gam observations of blazars have helped in resolving the nature of differences between LBLs and HBLs. It had been proposed that these objects are same except that the jets of LBLs are aligned more closely with our line of sight, based on their smaller numbers and higher luminosities. However, the observation of rapid variability and spectra that extends upto $TeV$ energies in HBLs point to the differences between these two sub-classes being more fundamental: the HBLs have higher maximum electron energies and lower intrinsic luminosities.  
\end{itemize}
\begin{itemize}
 \item Simultaneous measurements of the synchrotron and VHE \gam spectra constrain the magnetic field ($B$) and Doppler factor ($\delta$) of the jet. In case of Mrk421, the observed correlation between VHE and optical/UV photons indicates that if both sets of photons are produced in the same region of the jet, $\delta \gtrsim 5$ is required for VHE photons to escape pair-production losses significantly. If SSC mechanism produces the VHE \gams, $\delta = 15-40$ and $B = 0.03-0.09 G$ for Mrk 421 and $\delta \approx 1.5 - 20$ and $B = 0.08-0.2 G$ for Mrk501~\cite{Weekes2003}. Proton models which utilize synchrotron cooling as the primary means for proton energy losses require $B = 30-90 G $ for $\delta \approx 10$, in order to match the timescales of the correlated emission (\cite{Weekes2003} and references therein). The Mrk421 values of $\delta$ and $B$ are extreme for blazars but still within allowable range and are consistent with its extreme variability.
\end{itemize}
\begin{itemize}
 \item VHE observations have constrained the types of models that are likely to produce the \gam emission e.g. the X-ray and VHE flares correlation is consistent with SSC models. The relative absence of flaring at EGRET energies can be explained because the lower energy electrons (responsible for \gam emission at EGRET range) radiate away their energy more slowly.
\end{itemize}
\begin{itemize}
 \item In ERC models, where the target seed photons are the ambient photons, should have energies $< 0.1 eV$ (i.e infrared band) in order to avoid significant attenuation of VHE photons by pair production. 
\end{itemize}
\begin{itemize}
 \item Models in which \gam emission is due to proton proginators through $e^\pm$ cascades originating close to the base of the AGN jet, have difficulty explaining $TeV$ emission observed in Mrk421. The high densities of unbeamed photons near the core, cause high pair opacity to $TeV$ photons. These models predict that the radius at which the optical depth for $\gamma$-$\gamma$ production drops below unity, increases with \gam energy. Hence VHE \gams should vary either later or more slowly than $MeV-GeV$ \gams ~\cite{Blandford1995}. However, this contradicts the Mrk421 observations~\cite{Weekes2003}.
\end{itemize}
\begin{itemize}
 \item Both Mrk421 and Mrk501 are often considered to be very similar, but their \gam observations have shown them to be quit different. Mrk421 always has a higher mean flux, shorter time variation and peak in the synchrotron spectrum ($\nu_{sync}$) always in soft X-ray (does not show dramatic shift) as compared to Mrk501~\cite{Weekes2003}.
\end{itemize}
\section{Flux variability} Most of the blazars are found to be highly variable at all timescales (from minutes to years) with change of flux level by more than an order of magnitude. The first clear detection of flaring activity in the VHE emission of an AGN came in 1994 observations of Mrk421 by Whipple telescope where a 10-fold increase in the flux, from an average level that year of $\approx15\%$  to approximately $150\%$ of the Crab flux, was observed. Many flaring states detected in VHE \gam energies are usually correlated with X-ray and some times also in the optical band~\cite{Catanese1997}. The correlation between X-ray and \gam flux has been seen in Mrk421, Mrk501 and PKS 2155-304. However, a campaign on the $TeV$ blazar 1ES1959+650 has shown an orphan \gam flare without X-ray or optical counterpart. An orphan flare for Mrk421 was also found in X-ray without a $TeV$ counterpart (\cite{Dmazin-thesis} and references therein). Spectral shape variability has been established for the strong $TeV$ blazars like Mrk421, Mrk501, H1426+428~\cite{Aharonian2003-flare} and 1ES1959+650~\cite{Aharonian2003-flare-1ES,Albert2006-1ES1959}. However during a very strong outburst of PKS 2155-304 no spectral change was observed~\cite{Aharonian2007-PKS2155}.
\par
The rapid variability found in $TeV$ blazars provides additional constraints on emission models. The flux doubling time for Mrk421 was found to be $\sim30\,minutes$~\cite{Gaidos1996}, while for Mrk501 and PKS 2155-304 the variability timescales were shorter than  $3\,minutes$~\cite{Albert2007-Mrk501,Aharonian2007-PKS2155}. Very large amplitude flux variability at short timescales implies that $TeV$ emission originates from a small region very close to SMBH and therefore VHE \gam observations provide an unique opportunity to study the close vicinity of the central object~\cite{Dmazin-thesis}. Several mechanisms have been proposed to explain the observed fast flare~\cite{Wagner2006}:
\begin{itemize}
 \item \textbf{Additional population of more energetic electrons or altering the electron injection spectra:} Large variations in the flux are achieved in X-ray and \gam energies if $\gamma_{max}$ is increased by a factor $\sim5$, with no noticable change in flux levels at other energies. Alternatively flare can also be produced by increasing the luminosity of the injected electrons.
\end{itemize}
 \begin{itemize}
  \item \textbf{Modulation in the seed photon field for the IC process:} Assuming a small X-ray hot spot rotates on the surface of the accretion disk whose emission may be collimated by the hot spot geometry by an ordered magnetic field. Although strong absorption does not allow generation of VHE \gams in this region, the energy release of such a region can explain fast flaring e.g. as observed in Mrk421.
 \end{itemize}
\begin{itemize}
 \item \textbf{Sub-shocks inside the jet e.g. unsteady jets, fed by discrete ejecta:} Fast flares can also be explained by assuming unsteady jets, fed by discrete ejecta. For example M87 (often interpreted as misaligned blazar) shows bright knots in the jets in optical region which can be interpreted as internal shocks.
\end{itemize}
\begin{itemize}
 \item \textbf{Change in the acceleration environment:} Fast flaring implies small acceleration regions or high Doppler factors. Considering small blobs with enhanced magnetic field strength inside the acceleration region can explain the flares.
\end{itemize}
The observed correlation of X-ray and VHE \gam emissions can help to discern the models. e.g. the correlation between these two bands points to SSC origin and hence to sub-shock scenarios while orphan VHE flares favor ERC and hence modulation models.
\par
In this thesis we study the blazars Mrk421 and H1426+428 in VHE energy range above $1\,TeV$. The observations were carried out during 2005-2007 and 2004-2007 on Mrk421 and H1426+428 respectively. We present the results of our data analysis for these sources in chapter 6 and 7 of this thesis.

\chapter{Detection technique of VHE gamma-rays}
\section{Introduction} Since \gams encompass a very wide range of energies ($1\, MeV$ to $10^{20}\, eV$), we can not expect that a single type of detector will work at all \gam energies. As already mentioned in chapter 1 that depending on the energy regions the satellite experiments, atmospheric Cherenkov telescopes, air shower arrays and fluorescence detectors are employed to detect \gams. At $MeV-GeV$ energies, \gams have been typically observed with space-based instruments but at VHEs these instruments are completely unusable because of the rapidly falling \gam fluxes. Therefore astronomy at energies more than few tens of $GeV$, can only be done by ground-based instruments using the Earth's atmosphere as the detection medium. With the advent of IACTs in late 1980's, ground-based observations of $TeV$ \gams became possible and since then the number of \gam sources has rapidly grown up to over a hundred. It is worth mentioning here, the recently launched Fermi LAT covers the energy range from about $20\,MeV$ to more than $300\,GeV$.
\par
For most wavelengths, the atmosphere poses significant challenges for astronomy. The quality of optical and infrared astronomy is compromised by observations made through the atmosphere. For UV, X-rays and \gams, direct detection by ground-based instruments is impossible. A cascade of secondary particles called ``\textbf{EAS}'' is initiated by VHE \gams, which propagate down through the atmosphere. The secondary charged particles and photons produced in the EAS can be detected by the ground-level instruments. Thus the possible disadvantage of the atmospheric absorption is converted into an advantage by the ground-based \gam telescopes~\cite{Ong1998}. An IACT detects the ACR emitted in the atmosphere by secondary electrons which are produced in the cascade initiated by a primary \gam photon. CRs (mostly protons, nuclei, electrons) also produce the ACR via hadronic interactions and essentially become background for observations of \gams. These two types of primaries differing in development of EASs, can be distinguished by IACTs.
\par 
In this chapter, we briefly describe the model of the atmosphere which is an integral part of detectors at VHEs, development of an EAS in the atmosphere, production of ACR and the imaging ACT which has been proven to be the most sensitive technique for the detection of VHE \gams.
\begin{figure}[t]
\centering
\includegraphics*[width=0.55\textwidth,height=0.55\textheight, angle=270,bb=66 52 557 766]{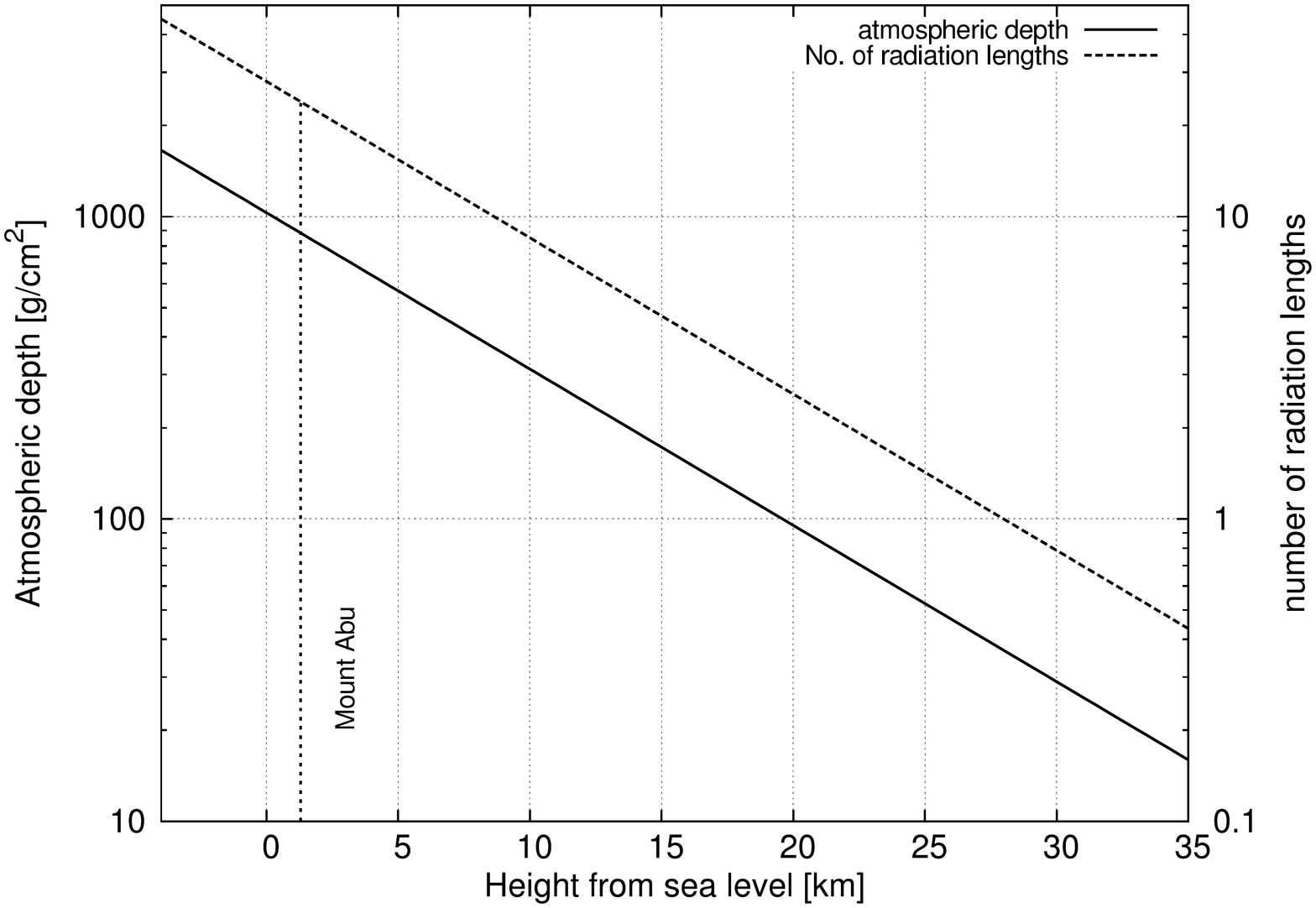}
\caption{\label{atmos}Variation of atmospheric depth or air mass grammage ($g\,cm^{-2}$) with the atmospheric height from sea level. Corresponding variation in number of radiation lengths is also shown. Mount Abu ($1300\,m$ asl) corresponds to atmospheric depth of $\sim890\,g\,cm^{-2}$ and radiation length of $\sim24.2$.} 
\end{figure} 
\section{The atmosphere}\label{atmosphere}The Earth's atmosphere is a layer of gases surrounding the planet that is retained by its gravity. Dry air contains roughly (by volume) $78.09\%$ nitrogen, $20.95\%$ oxygen, $0.93\%$ argon, $0.038\%$ carbon dioxide, and small amounts of other gases. Air also contains a variable amount of water vapor, on an average around $1\%$. The atmosphere does not technically end at any given height, but becomes progressively thinner with altitude.
\par
It is a common practice to remove the effect of the density of the medium in which an EAS develops by defining the pathlengths in units of $g\,cm^{-2}$. The atmospheric depth X [in $g\,cm^{-2}$] is related to the density profile of the atmosphere by (\cite{Masaaki-th2008} and references therein)
\begin{equation}
\label{eq1}
 X = \int_h ^\infty \rho(h')dh' 
\end{equation}
where $\rho(h)$ is the density of the atmosphere at altitude $h$. \\For an isothermal atmosphere, the above equation can be described by means of a constant called \textbf{scale height} $h_0$ as
\begin{equation}
 X = X_g\, exp (-h/h_0)
\end{equation}
where $X_g \approx 1030\, g\,cm^{-2}$ and $h_0 \approx\, 8.5 \,km $.
However, for a more realistic model of the atmosphere the relationship between X and h~\cite{Rao1988} is given by
\begin{equation}
 h = \left(\frac{6740+2.5 X}{1000}\right)ln(1030/X)
\end{equation} 
The change of atmospheric density with altitude is given by~\cite{Aharonian2008} 
\begin{equation}
 \rho = \rho_0\, exp(-h/h_0)
\end{equation}
Where $\rho_0 \approx 1.205 \times 10^{-3}\,g\,cm^{-3}$ is the typical atmospheric density at sea level.
\par
The index of refraction\footnote{\\ $\eta=\sqrt{\epsilon\,\mu}$ where $\epsilon$ is dielectric constant and $\mu$ is the magnetic permeability of the medium.\\ For air $\mu = 1$ and refractive index $\eta=1.00029$} of the atmosphere $n$ is a critical parameter for understanding the production of ACR. If we define $\eta \equiv n-1$ then $\eta$ is proportional to the density and is given by
\begin{equation}
 \eta = \eta_0\,exp(-h/h_0)
\end{equation} 
where $\eta_0 = 2.9 \times 10^{-4}$ and $h_0 = 7.25\,km$, differing slightly from the standard scale height of the atmosphere which is $8.5\,km$.
\par
The atmosphere is far from an ideal detector. Poisson fluctuations in the Night Sky Background (NSB) light limit the energy threshold of an IACT. Atmospheric absorption and scattering significantly modifies the spectrum of the detected ACR, especially in the UV part. The dominant contributions to absorption or scattering of ACR are~\cite{Aharonian2008}:
\begin{enumerate}[(i)]
\item Rayleigh scattering: where photons get scattered by particles of much smaller size than the wavelength of the photon. The absorption length ($\Lambda_R$) depends on the fourth power of photon wavelength i.e. $\Lambda_R \propto \lambda^4$.
\item Mie scattering: where photons are scattered by particles with dimensions comparable to the wavelength (e.g. aerosoles), here $\Lambda_M \propto \lambda$.
\item Absorption by ozone: $O_3 + photon \to O_2 + O$.\\ This strongly absorbs near UV region ( 200 to 300 nm).
\item Scattering by water vapor in clouds or from local humidity.
\end{enumerate}
A rough estimate for the magnitude of these effects under good weather conditions can be obtained by considering the empirical model for the absorption length derived by Hillas~\cite{Hillas1982}
\begin{equation}
 \Lambda = 1385 - 0.3798\,X - 0.000556\,X^2\,\,(g\,cm^{-2}).
\end{equation}
\subsubsection{Radiation length} HE electrons predominantly lose energy in matter by bremsstrahlung while high-energy photons by pair production. The characteristic amount of matter traversed for these related interactions is called the radiation length $X_0$. It is both the mean distance over which a HE electron loses all but 1/e of its energy by bremsstrahlung, and 7/9 of the mean free path for pair production by a high-energy photon. For example an electron with initial energy $E_0$ travels a distance X ($g\,cm^{-2}$) in a dissipative medium, its energy can be written as
\begin{equation}
\label{rad-length}
 E_X=E_0\,exp(-X/X_0).
\end{equation}
The radiation length is given, to good approximation, by the expression (\cite{Tickoo2002} and references therein)
\begin{equation}
 X_0(g\,cm^{-2}) = \left(\frac{716.4\,A}{Z(Z+1)ln(287/\sqrt Z)}\right)
\end{equation}
where $Z$ and $A$ are the effective atomic number and atomic mass of the medium. For air at STP, $Z=7.34$ and $A=14.37$: therefore using above formula, the radiation length for electron $X_0 \simeq \textbf{36.8} \,g\,cm^{-2}$ in air.
Another term called \textbf{shower unit} is defined as $R=X_0 \,ln2$.
\par
Note that total vertical thickness of the atmosphere is $\textbf{1030}\,g\,cm^{-2}$ up to sea level and since the radiation length is $\textbf{36.8} \,g\,cm^{-2}$, this amounts to $\sim$\textbf{28} radiation lengths of atmosphere. The relationship between atmospheric height and the air mass grammage is shown in Fig.~\ref{atmos}. For electrons at lower energies (below few tens of $MeV$s), the energy loss by ionization is predominant. Therefore one more parameter called critical energy is defined which characterizes the effectiveness of a medium as a radiator.
\subsubsection{Critical energy} Critical energy ($E_c$) is the energy at which the rate of energy loss by ionization becomes comparable with the rate of energy loss by radiation (e.g. Bremsstrahlung in case of electron). The $E_c$ for electrons (positrons) in gases can be computed by the following expression (\cite{Tickoo2002} and references therein):
\begin{equation}
 E_c \,(MeV) = \left(\frac{710}{Z+0.92}\right).
\end{equation} 
The electron has a critical energy $\sim$\textbf{\textit{86} $MeV$} in the atmosphere (air).
\section{Extensive air shower} When VHE \gam photons or other CR particles enter the atmosphere, they interact with nuclei of atmosphere molecules typically at an altitude of about $20-25\,km$ (the point of first interaction) and generate the so-called EAS. The interaction of electrons, muons and \gams with nuclei is electromagnetic i.e. they generate secondary particles by pair production and Bremsstrahlung, while protons and ionized nuclei interact via hadronic interactions i.e. they produce pions, muons and kaons as secondary particles. As shown in Fig.~\ref{air-showers} the secondary particles further interact with atmospheric nuclei starting a cascade of particles i.e. an EAS. The number of secondary particles grow up rapidly, while their energy decreases continuously in each interaction. Since the energy of the primary particle is distributed over all the secondary particles, at some point of shower development the energy of the shower particles is too small to produce further particles i.e. it falls below $E_c$. At this point shower reaches to its maximum size called ``\textbf{shower maximum}'' beyond which particles suffer energy loss due to ionization, as a result the number of shower particles decay exponentially and finally the shower dies out. Depending on the type of primary particle we deal with two types of air showers: 
\begin{figure}[t]
\centering
\includegraphics*[width=0.7\textwidth,height=0.25\textheight, angle=0,bb=0 0 1128 566]{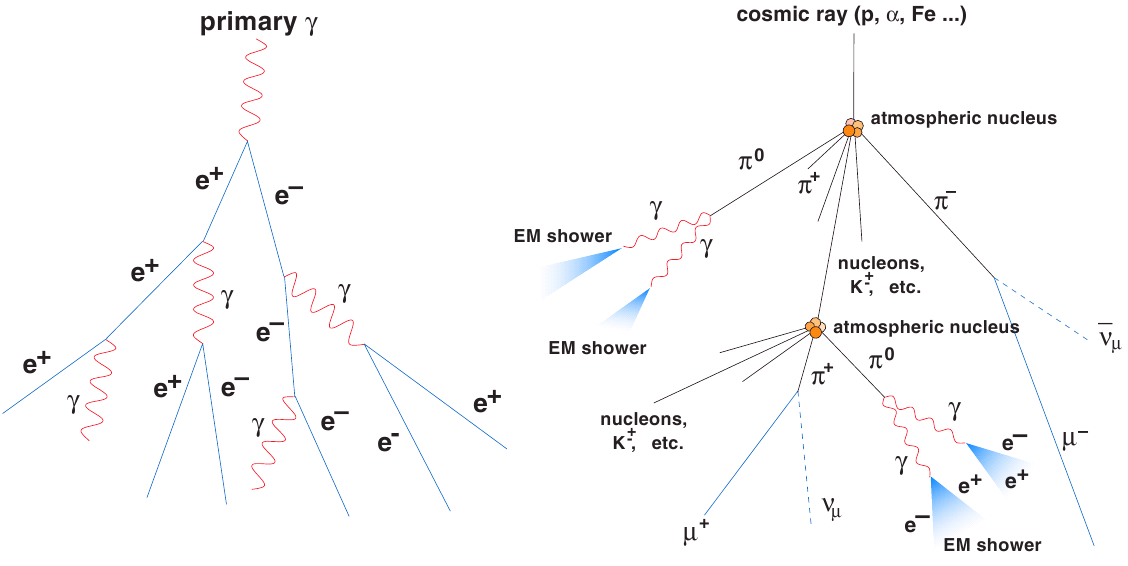}
\caption{\label{air-showers}Schematic description of the extensive air showers. Left sketch shows the development of an electromagnetic shower while right one is for the hadronic shower. This figure has been adapted from~\cite{Wagner2006}.}
\end{figure}
\subsection{Electromagnetic EAS} 
The Earth's atmosphere is opaque to high energy \gams. At sufficiently high energies (more than twice the electron rest mass) the predominant interaction of \gams with atmospheric nuclei is pair production. For a \gam of energy $>10\,GeV$, the photon is effectively replaced by two charged particles ($e^\pm$) sharing its energy and traveling in almost the same direction as that of the \gam photon. Basically two processes mainly contribute to the shower development: pair production and Bremsstrahlung. Pair produced electrons and positrons further produce photons via Bremsstrahlung; such photons in turn produce $e^\pm$ pair, and so on, giving rise to a shower of charged particles (mainly $e^\pm$) and photons called an electromagnetic EAS, as shown in the left of Fig.~\ref{air-showers}. The process is described in~\cite{Rossi1941,Nishimura1952}.
\par
At shower maximum the $e^\pm$ energy is $\approx 300\, MeV$. The energy loss of an electron due to Bremsstrahlung is proportional to its energy:
\begin{equation}
 -dE_x/dx = E_e/X_0
\end{equation} 
Once the energy of $e^\pm$ reaches $E_c=86\,MeV$, shower gets absorbed. As already mentioned the radiation length of $e^\pm$ for Bremsstrahlung is $36.8 \,g\,cm^{-2}$ and the \textbf{interaction length} or \textbf{mean free path} of photons for pair production is $9/7 X_0$ i.e. $\textbf{47.3} \,g\,cm^{-2}$. Therefore the same order of interaction lengths in both the processes, indicates a quite simple and symmetric structure of an electromagnetic shower with respect to the original \gam trajectory (also called \textbf{shower axis}). The average opening angle of emission in all these processes is $1/\gamma$, where $\gamma$ is the Lorenz factor of electron. The resulting electromagnetic cascade is remarkably tightly bunched along the shower axis. However, tansverse broadening (lateral development) of the shower takes place which is predominantly due to multiple coulomb scattering process of electrons traversing in the air. Although small, the lateral spread of secondary particles also occurs due to the Earth's magnetic field. Bremsstrahlung and pair production also contribute to some extent in the transverse development of the shower. The shower particles spread out laterally to tens or even hundreds of meters, moving downwards with relativistic speeds.
\par
The distribution of Coulomb scattering is described by the theory of Moliere~\cite{Bethe1953}. It has a roughly Gaussian shape for small scattering angles, while at larger angles it approximates Rutherford scattering with longer tail than a Gaussian distribution. The lateral spread of electromagnetic showers in different materials scales quite accurately with \textbf{Moliere radius}\footnote{The Moliere radius is a characteristic constant of a material giving the scale of the transverse dimension of the fully contained electromagnetic showers initiated by an incident high energy electron or photon. By definition, it is the radius of a cylinder containing on average 90\% of the shower's energy deposition. It is related to the radiation length $X_0$ by the following approximate relation:
\begin{equation*}
R_M=0.0265X_0(Z+1.2)
\end{equation*}
where $Z$ is the atomic number.} ($R_M$), which in turn depends on radiation length and critical energy in a given medium. Accounting that the distance corresponding to $X_0$ varies with density of atmosphere (equation~(\ref{eq1})) the expression for $R_M$ is (\cite{Fuste2007} and references therein): 
\begin{equation}
 R_M = \frac{9.6 (g\,cm^{-2})}{\rho_{air}}. 
\end{equation} 
At sea level $R_M$ is approximately $80\,m$.
\begin{figure}[t]
\centering
\includegraphics*[width=0.5\textwidth,height=0.3\textheight, angle=0,bb=0 0 204 170]{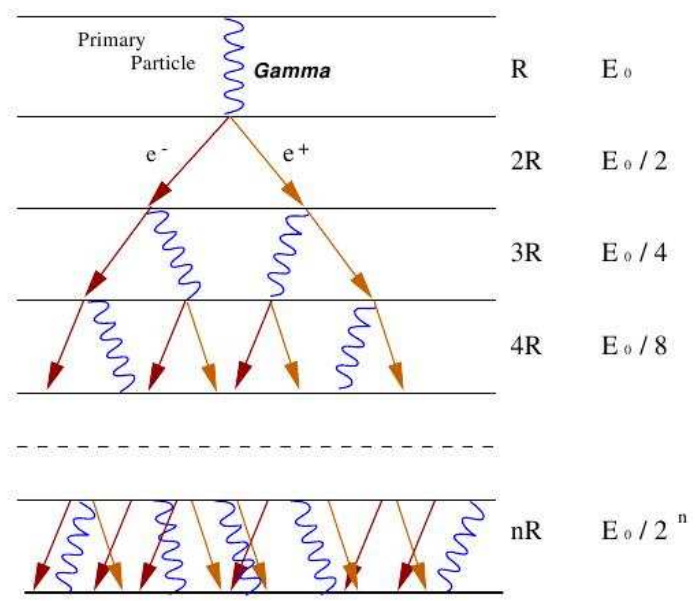}
\caption{\label{model}A simple model of an electromagnetic shower development proposed by Heitler. This figure has been adapted from ~\cite{Polarization}.}
\end{figure}
\par
The basic features of the longitudinal development of an electromagnetic shower, illustrated by a toy model in Fig.~\ref{model} has been proposed by Heitler~\cite{Heitler1944}. The model considers only pair production and Bremsstrahlung process with assumption that both radiation length for electron and interaction length for photon have same value i.e. $X_0$. Instead of $X_0$ the quantity defined earlier called shower unit ($R=X_0\,ln2$) is used so that the probability of an electron or photon to interact along a distance $R$ is therefore, $exp(-R/X_0)=1/2$. It is also assumed that in each interaction the energy of the interacting particle is equally divided between the products of the interactions. 
\par
As shown in Fig.~\ref{model}, a \gam photon of initial energy ($E_0$) which enters the atmosphere, undergoes pair production after traveling a distance $R$. The resulting $e^\pm$ each has on an average half of the initial energy i.e. $E_0/2$. After another distance $R$, each particle produces a photon of an average energy $E_0/4$ via Bremsstrahlung, while retaining $E_0/4$ energy with itself. At a distance $nR$ from the first interaction, there are $2^n$ particles each with an average energy of $E_0/2^n$. The particle multiplication continues till the average energy of particles drops below critical energy ($E_c=86\,MeV$ for air) and thus halts the cascade.
\par
While for \gams, the critical energy corresponds to a point where cross section for Compton scattering is approximately equal to pair production cross section ($ \sigma_{Compton}\approx \sigma_{pair}$), for electron it is the energy at which energy loss due to ionization is comparable to that by Bremsstrahlung ($(dE/dX)_{radiation} \approx  (dE/dX)_{ionization}$). 
At shower maximum total number of particles ($e^\pm$ and photons) are 
\begin{equation}
 N_{max}=E_0/E_c = 2^n.
\end{equation}
Assuming that the shower stops abruptly at shower maximum, the maximum penetration depth of the shower can be expressed as
\begin{equation}
 X_{max}=nR=\frac{lnE_0/E_c}{ln2}R=X_0\,ln(E_0/E_c).
\end{equation}  
A more realistic approximation for the number of particles ($e^{\pm}$) produced in an electromagnetic shower as a function of atmospheric depth $t(=X/X_0)$ expressed in units of radiation lengths and primary energy $E_0$ is given by Greisen~\cite{Greisen1960}
\begin{equation}
\label{eq14}
 N_e(t,E_0)=\frac{0.31}{\sqrt{ln(E_0/E_c)}}\, exp\,[t\,(1-1.5\,ln\,s)]
\end{equation} 
Where $s$ is a dimensionless quantity called \textbf{shower age} and defined as
\begin{equation}
 s=\frac{3t}{t+2\,ln(E_0/E_c)}
\end{equation} 
The parameter $s$, basically indicates the degree of development of the shower whose meaning is given by the derivative $dN_e(t,E_0)/ds$,
\begin{equation}
 \frac{dN_e(t,E)}{ds} = N_e(t,E)\times \frac{3\,ln(E/E_c)(s-3\,ln\,s -1)}{(3-s)^2}
\end{equation} 
As one can see from the above derivative $\frac{dN_e}{ds} > 0$ in the range $0 <s < 1$, i.e. number of $e^\pm$ grows as $s$ increases. At $s=1$, $\frac{dN_e}{ds} = 0$, i.e. $N_e$ becomes maximum, while in the range $s > 1$, $\frac{dN_e}{ds} < 0$, shower dies out. Therefore the parameter related to the shower such that~\cite{Rossi1941}:
\begin{itemize}
 \item $s=0 \to$ first interaction point,
\end{itemize}
\begin{itemize}
 \item $s=1 \to$ the maximum development of the shower, and
\end{itemize}
\begin{itemize}
 \item $s=2 \to$ the point at which shower begins to extinguish.
\end{itemize}
\begin{figure}[t]
\centering
\includegraphics*[width=0.6\textwidth,height=0.6\textheight, angle=270,bb=66 56 557 758]{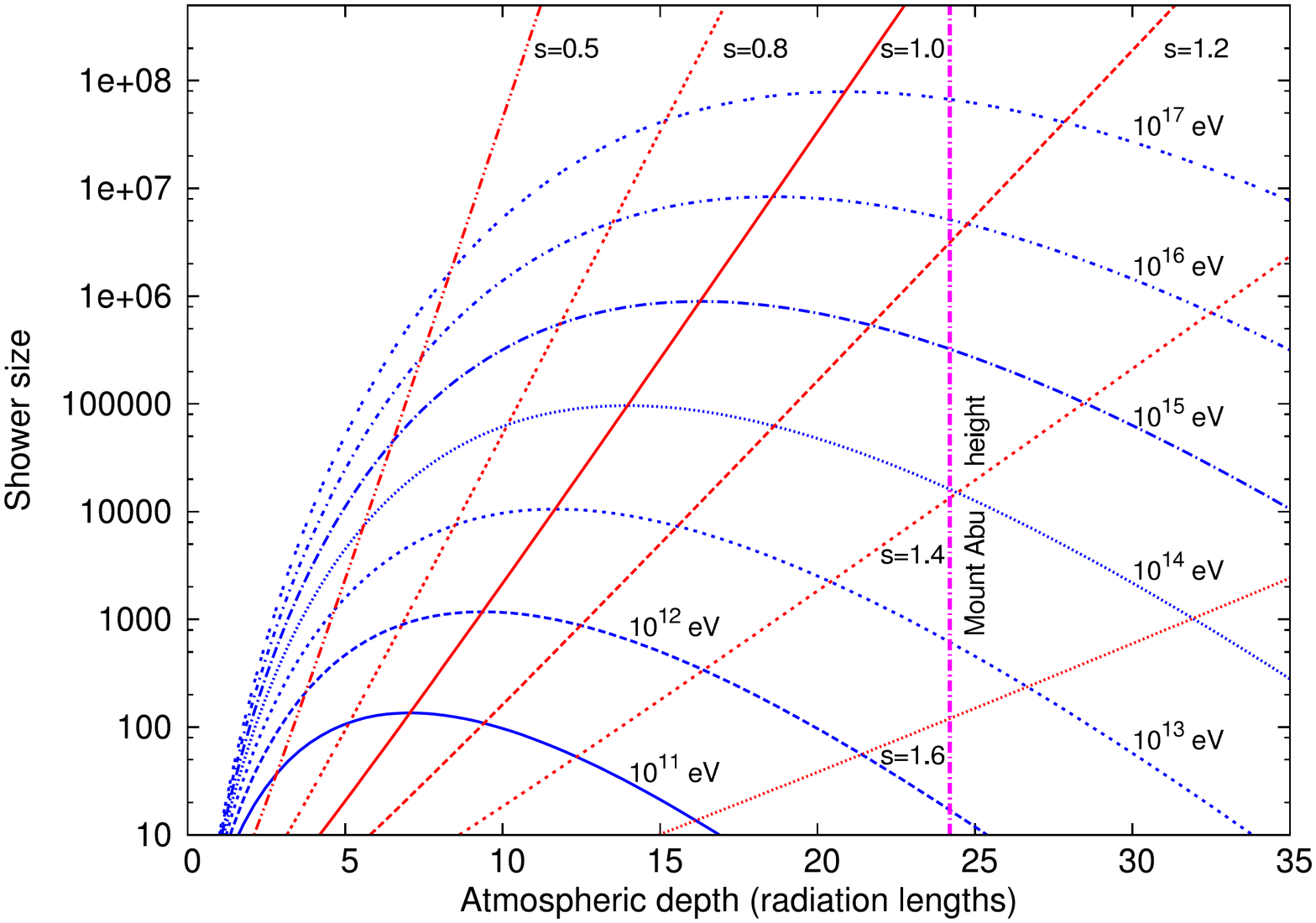}
\caption{\label{shower-age}Longitudinal development of an electromagnetic shower for different primary energies using Greisen approximation. Different blue curves represent different energies ($E_0$) of primary photons. The green lines characterize the shower development by the shower age $s$. $s=1$ represents the maximum development of the shower. The altitude of the TACTIC telescope ($1300\,m$ asl) is also indicated.}
\end{figure}
\par
By using this definition of shower age, the depth of shower maximum can be computed in terms of number of radiation lengths as $t_{max} = ln(E_0/E_c)$ i.e. logarithmic dependence on initial energy, same as obtained with the simple model of Heitler. Fig.~\ref{shower-age} shows the longitudinal development of the shower as described by equation~(\ref{eq14}), for different primary energies. The higher the energy of the primary, the deeper the air shower penetrates into the atmosphere. The shower maximum is at about $13$ to $7\, km$ above sea level for primary \gam energies between $50\,GeV$ to $10\,TeV$.
\par
The lateral distribution of electrons is modeled by the NKG-formula (Nishimura-Kamata-Greisen)~\cite{Greisen1960, Kamata1958}, which describes the electron density as a function of the distance $r$ from the shower axis
\begin{equation}
 \rho(r,t,E)=\frac{N_e(t,E)}{R_M^2}\left(\frac{r}{R_M}\right)^{s-2} \left(1+\frac{r}{R_M}\right)^{s-4.5}\frac{\Gamma(4.5-s)}{2\pi \Gamma(s)\Gamma(4.5-2s)}
\end{equation}  
where $\Gamma$ is the Gamma function and $R_M=21.2\, MeV.(X_0/E_c)\,\approx 80\, m$ in air at sea level. Nearly $99\%$ of shower energy is contained within $3.5R_M$.
\par
Air showers induced by $e^\pm$ primaries, pose an irreducible background for \gam showers, as they develop in the same way as \gam showers do. Due to the steep electron spectrum of index $\sim3.3$, these showers can normally be ignored above energies $500\,GeV$ (\cite{Wagner2006} and references therein).
\par
Roughly $10\%$ of the energy of primary \gams reaches an observational level of $5200\,m$ above sea level in the form of electromagnetic particles, while only $5\%$ of primary particle's energy reaches to an altitude $4300\,m$. The fluctuations in the development of an EAS are due to fluctuations in the depth of the first interaction and the probability that the primary particle will survive to an atmospheric depth of $N$ radiation lengths is given by~\cite{Aharonian2008}
\begin{equation}
 P(N)= exp(-\frac{9}{7} N)
\end{equation} 
The shower fluctuations excluding the point of first interaction, are described by~\cite{Wagner2006}
\begin{equation}
 \triangle N_e(s)\simeq \frac{9}{14}(s-1-3\,ln\,s)\, N_e(s)
\end{equation} 
Due to their statistical nature, the fluctuations from shower to shower are quite large and have important consequences particularly for the energy estimation of the primary particles. 
\subsection{Hadronic EAS}When $CR$ particles, most of which are protons enter the atmosphere, the inelastic scattering  with atmospheric nuclei causes nuclear disintegration and production of mesons e.g pions and kaons as well as nucleons. The secondary particles in turn themselves produce disintegrations thus starting a cascade of particles called hadronic EAS, as shown at the right of Fig.~\ref{air-showers}. This interaction can be represented by the following equation
\begin{equation*}
 CR + Atmospheric Nuclei (AN) \Rightarrow CR' + AN' + m\pi^\pm + n \pi^0 + \,other\, mesons
\end{equation*}
$CR'$, which is a fragment of the primary $CR$ can keep repeating the above interaction or even reach the ground if the original $CR$ has sufficient energy.
\par
Nearly $90\%$ of the secondary particles in a hadonic shower are pions and remaining $10\%$ are kaons and antiprotons. One third of the total pions are the neutral pions($\pi^0$).  
\begin{table}[t]
\caption{Mesons produced in hadronic EAS.}
\centering
\begin{tabular}{|c|c|c|c|c|}
\hline 
particle & rest mass ($MeV/c^2$) & mean life (s) & decay mode & probability \\
\hline 
$\pi^0$ &  135 & $8.4 \times 10^{-17}$ & $\gamma$ + $\gamma$ & 0.988 \\
 &   &  & $\gamma$ + $e^-+e^+$ & 0.012 \\
\hline
$\pi^\pm$ &  139.6 & $2.6 \times 10^{-8}$ & $\pi^+ \to \mu^+ \,+\, \nu_{\mu}$ & 0.999 \\
& & &  $\pi^- \to \mu^- \,+\, \nu_{\mu}^-$ & 0.999\\
& & &  $\pi^+ \to e^+ \,+\, \nu_e$ & 0.0001\\
& & &  $\pi^- \to e^- \,+\, \nu_e^-$ & 0.0001\\
\hline
$K^\pm$ &  493.7 & $1.24 \times 10^{-8}$ & $K^\pm \to \mu^\pm \,+\, \nu_{\mu}$ & 0.635\\ 
& & & $K^\pm \to \pi^\pm \,+\, \pi^0$ & 0.212\\
\hline
$\mu^\pm$ &  105.7 & $2.2 \times 10^{-6}$ & $\mu^+ \to e^+ \,+\, \nu_e\,+\,{\nu_{\mu}}^-$ & --\\ 
& & & $\mu^- \to e^- \,+\, \nu_e^-\,+\,\nu_{\mu}$ & --\\
\hline
\end{tabular} 
\label{mesons}
\end{table}
The decay modes of mesons along with their rest mass, mean life and corresponding probability of decay are summarized in Table~\ref{mesons}. It is clear from the Table that $\pi^0$ has very short life and decays into two photons before it undergoes any hadonic interactions. Even a $\pi^0$ with $100\,GeV$ energy will decay in the laboratory frame in $6\times10^{-12}\, s$. Photons from $\pi^0$ decay initiate the electromagnetic EAS.
\par
In the early phase of shower development, high Lorenz factor of mesons results in hadronic interactions while at a later stage their decay becomes more important. On the other hand, due to their longer lifetime and smaller cross section most of the produced muons are able to retain their energy and reach the ground. Due to high energy, their life is further extended by a factor $E_{\mu}/m_{\mu}c^2$, where $E_{\mu}$ and $m_{\mu}$ are energy and rest mass of the muon. The major energy loss of muon's energy is through ionization, while the decay of low energy muons can start the electromagnetic sub-shower. Muons and neutrinos are the main source of energy loss in hadronic EAS~\cite{Fuste2007}.
\par
The interaction length or mean free path ($\lambda$) for a proton is more than two times larger than the radiation length in an electromagnetic shower. The interaction lengths of protons and mesons in air at $1\,TeV$ energy are~\cite{Gaisser1990}\\
$\lambda_{p}=83\,g\,cm^{-2}$,\\
$\lambda_{\pi}=107\,g\,cm^{-2}$,\\
$\lambda_{K}=138\,g\,cm^{-2}$.\\
As a result, hadronic showers penetrate deeper into the atmosphere (i.e. depth of shower maximum is larger) as compared to a pure electromagnetic EAS of similar primary energy. A hadronic shower grows until the energy per nucleon is below the pion production threshold of about $1\,GeV$. A hadron-induced EAS therefore consists mainly of three components:
\begin{enumerate}[(i)]
 \item A hadronic core, containing strongly interacting high energy nucleons,
 \item Mesons constantly feeding electromagnetic sub-showers (mainly from $\pi^0$ decay), and
 \item Non-interacting muons and neutrinos, carrying most of the primary energy.
\end{enumerate}
Study of hadonic-induced EAS can be performed by detailed Monte Carlo simulations. However incomplete understanding of hadronic interactions at VHE leads to rather large uncertainties in the results.
\subsection{Differences between photon and hadron initiated showers} \gams are observed in the unavoidable huge background (orders of magnitude more) created by hadronic CRs, making it necessary to search for background rejection methods. These methods rely on the characteristic differences between the two shower types, which are as follows:
\begin{itemize}
 \item As already mentioned, because of the smaller cross sections of interaction of proton with air, its interaction length is much larger ($\sim$$83\,g\,cm^{-2}$) compared to radiation length of \gam photon ($\sim$$37\,g\,cm^{-2}$). The longitudinal development of an EAS is governed by the interaction length of primary particle (also on secondary particles), the hadronic shower develops deep into the atmosphere as compared to that of an electromagnetic EAS.
\end{itemize}
\begin{itemize}
 \item The strong interactions are different from the electromagnetic ones and as an effect the mean transverse momentum carried by pions and kaons is  higher than electrons and positrons. The hadronic showers therefore have more lateral extent.
\end{itemize}
\begin{figure}[H]
\begin{center}
\centering
\subfigure[$1\,TeV$ \gam photon, overhead view]{\label{gam-o}\includegraphics*[width=0.4\textwidth,height=0.4\textheight, angle=0,bb=0 0 128 128]{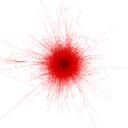}}
\subfigure[$1\,TeV$ proton, overhead view]{\label{pro-o}\includegraphics*[width=0.4\textwidth,height=0.4\textheight,angle=0,bb=0 0 128 128]{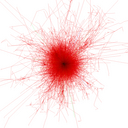}}
\subfigure[$1\,TeV$ \gam photon, lateral view]{\label{gam-l}\includegraphics*[width=0.4\textwidth,height=0.5\textheight, angle=0,bb=0 0 128 256]{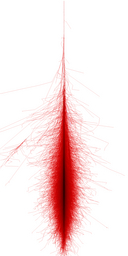}}
\subfigure[$1\,TeV$ proton, lateral view]{\label{pro-l}\includegraphics*[width=0.4\textwidth,height=0.5\textheight,angle=0,bb=0 0 128 256]{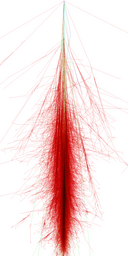}}
\caption{\label {EAS} Monte-Carlo simulated extensive air showers of vertically incident $1\,TeV$ \gam photon and proton. Courtesy F. Schmidt, ``CORSIKA Shower Images''~\cite{Schmidt}.}
\end{center}
\end{figure}
 \begin{itemize}
  \item Generally, hadronic showers have several sub-cascades, resulting in much more wider and irregular shower than electromagnetic showers. This feature is used to separate 
\gam induced showers from hadronic ones. However if only the sub-cascade of the hadronic EAS is detected, it can mimic \gam induced EAS and becomes an irreducible background.
\end{itemize}
\begin{itemize}
 \item In addition to these intrinsic differences arising during shower development, one characteristic difference comes from the nature of primary particle. The \gam showers point back to the source of origin of the primary \gam, while the arrival direction of proton-induced-showers is isotropized due to the deflection in galactic and extragalactic magnetic fields. This feature is the back bone for the background rejection in IACTs particularly stand alone systems like TACTIC.
\end{itemize}
\begin{itemize}
 \item In general, the hadron induced showers produce less ACR (discussed in the following section) than the \gam induced showers.
\end{itemize}
\begin{itemize}
 \item The proton showers are subjected to larger fluctuations in the shower development, both within the shower and between individual showers. These fluctuations are caused by the larger interaction length, larger transverse momentum of secondary particles and variety of secondary particles.
\end{itemize}
\begin{itemize}
 \item Photoproduction or photomeson is another process occurring in both electromagnetic and hadronic showers. Here $\gamma$ + nucleus $\to$ hadrons: Hadronic states, mostly pions are produced~\cite{DanielK}.
The cross section for this process increases with increasing photon energy, but has a small value of $1-2\,mb$ for $0.1-20\,TeV$ photons~\cite{Krys1991}. This is much smaller than the cross section for pair production which is $520\,mb$. By measuring the muon component, it is possible to distinguish between \gam and proton induced air showers. As these muons originate from decay of pions, their rate is at least an order of magnitude less in \gam showers. This procedure of $\gamma$ and hadron separation is limited by the smaller number of muons at lower energies i.e. $<10\,TeV$ and by the increased occurrence of photoproduction processes at energy $>1\,PeV$ (\cite{DanielK} and references therein).
\end{itemize}
\par
Fig.~\ref{EAS} illustrates the simulated EAS for \gam and proton primaries of same energy ($1\,TeV$).
As an EAS develops through the atmosphere, the charged particles mainly electrons produce ACR by polarizing the medium and fluorescent radiations by exciting the nitrogen molecules of the atmosphere. The atmospheric Cherenkov radiation is the main topic of concern here.
\section{Atmospheric Cherenkov Radiation}
\begin{figure}[t]
\begin{center}
\centering
\subfigure[Particle moves with low velocity ($v < c/n$)]{\label{slow-3}\includegraphics*[width=0.22\textwidth,height=0.25\textheight, angle=0,bb=0 0 101 157]{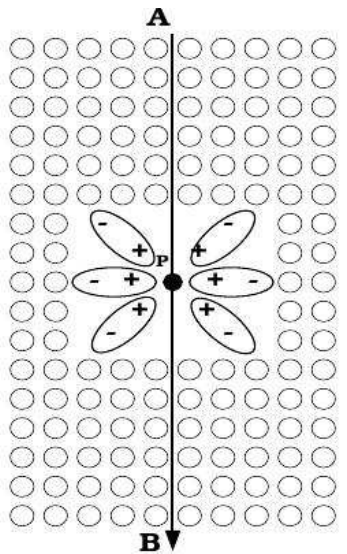}}
\subfigure[Particle moves with high velocity ($v > c/n$)]{\label{fast-3}\includegraphics*[width=0.22\textwidth,height=0.25\textheight,angle=0,bb=0 0 100 158]{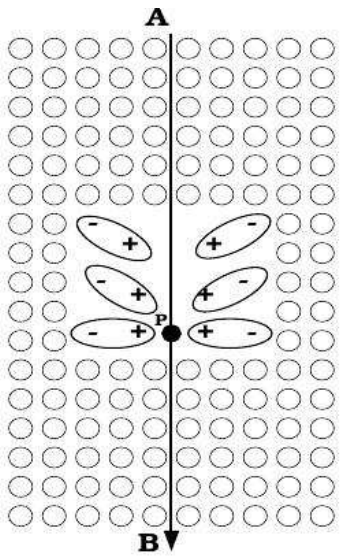}}
\subfigure[Propagation of ACR in the                   medium]{\label{cheren}\includegraphics*[width=0.33\textwidth,height=0.28\textheight,angle=0,bb= 0 0 513 466]{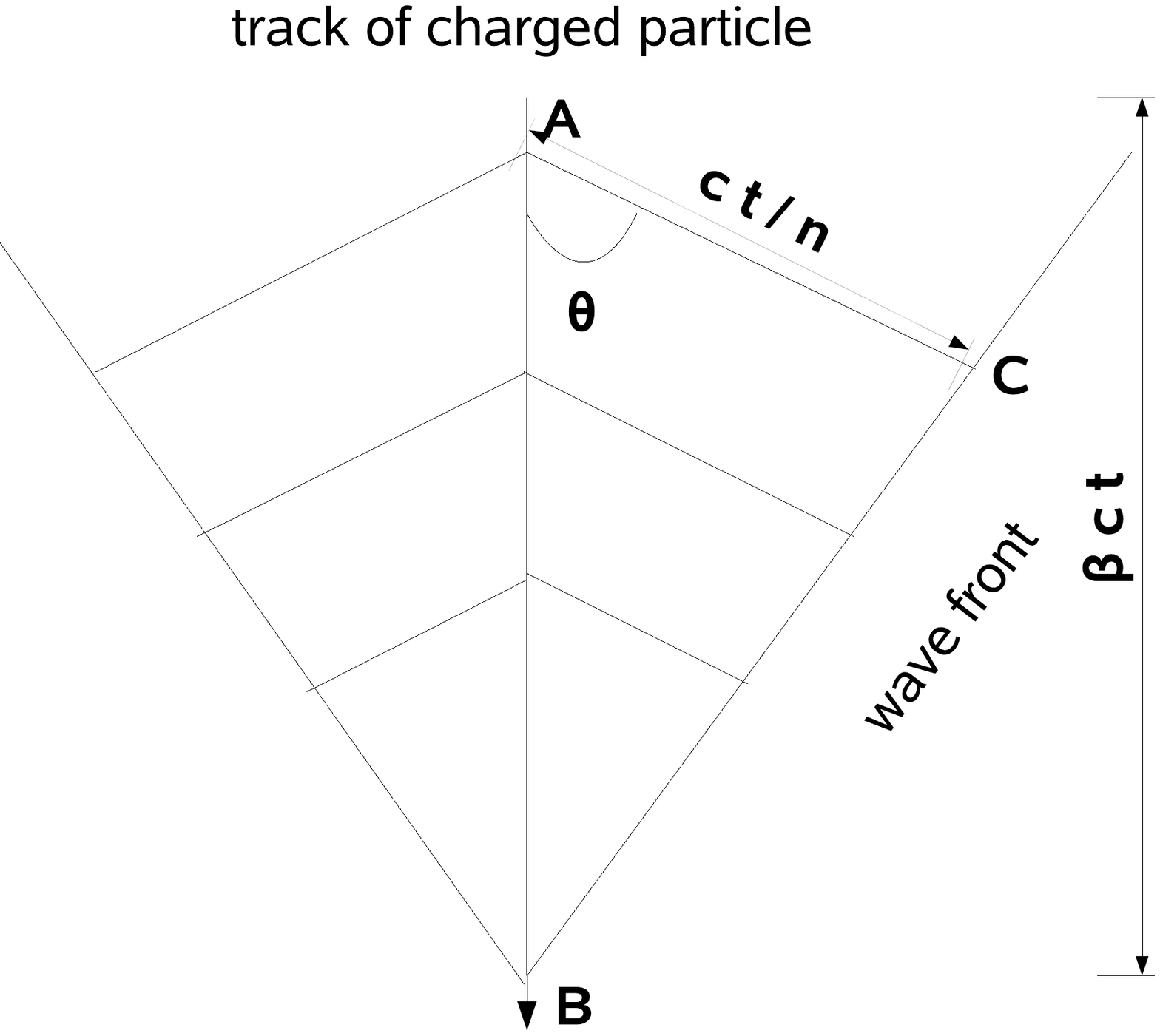}}
\caption{\label {Pol}Illustration of the polarization produced in a dielectric medium by a charged particle as it passes through it.}
\end{center}
\end{figure}

When a charged particle moves through a medium with the relativistic speed which exceeds the local speed of light in that medium, the emitted radiation is called Cherenkov radiation\footnote{In this thesis, we call it Atmospheric Cherenkov Radiation (ACR) as in the case of the VHE \gam detection the Earth's atmosphere acts as a detector wherein the Cherenkov radiation is produced whenever a VHE photon or proton impinges at the top of the atmosphere.}. This radiation was discovered by P.A. Cherenkov in 1934~\cite{Cherenkov1934}, while the theoretical explanation for the origin of ACR was given by Frank and Tamm in 1937~\cite{Tamm1937}.
Considering the case of an electron moving in a transparent medium (i.e. zero absorption coefficient thus refractive index is real). As shown in Fig~\ref{slow-3}, electron is moving slower than the speed of light. The atoms of the medium are represented by circles near the track AB of the electron. At a position P on the track, the electric field of the particle distorts the atoms such that positive charges of nuclei are displaced towards the moving electron. The medium is thus polarized and radiates electromagnetic pulses, as atoms return to their normal position once the particle moves away. Because of the complete symmetry of polarization, the resultant field at a large distance is zero hence no emission is observed. However, when the electron moves faster than the velocity of light, the symmetry in the plane perpendicular to the track of particle still exists but there is a resultant dipole field along the particle track (Fig.~\ref{fast-3}). The waves from each small track interfere constructively and the radiation is observed. When $v < c/n$, the electromagnetic waves travel ahead of the particle, setting instant dipoles in the medium resulting in no net polarization. In the other case, no instant dipoles are set ahead of particle and net polarization exists along the track of the particle~\cite{Polarization}.
\par
Fig.~\ref{cheren} illustrates the geometrical interpretation of the ACR. The emission is observed only at an angle with respect to the particle track and is called \textbf{Cherenkov angle} ($\theta$). This angle represents the direction in which waves from arbitrary points on a particle track are coherent and forming a plane wavefront. The coherence takes place when time taken by the particle to move from A to B is same as the time taken by the wave from A to C i.e.
\begin{equation}
\label{cherenkov-cond}
 cos\, \theta = \frac{c\,t\,/\,n }{\beta\,c\,t} = \frac{1}{\beta\,n}
\end{equation} 
where $v=\beta\,c$ is the particle's velocity and $n$ is the refractive index of the medium. It follows from equation (\ref{cherenkov-cond}) that for a given medium there is minimum velocity for a particle called \textbf{threshold} or \textbf{critical velocity} 
\begin{equation}
\label{threshold-velocity}
\beta_{min} = \frac{1}{n}
\end{equation}
below which no radiation takes place. Therefore the minimum velocity required for emission of ACR with angular frequency $\omega$ depends on the refractive index of the material for the frequency $\omega$. At $\beta_{min}$ , $\theta=0^\circ$ i.e. radiations are emitted along the particle track. On the other hand, if particle moves slower than the radiation ($v\,<\,c/n$), the equation (\ref{cherenkov-cond}) has no solution and hence no radiation takes place. Moreover, angle of emission is maximum when $\beta=1$ i.e. particle is ultra relativistic, therefore 
\begin{equation}
\label{cherenkov-max}
\theta_{max}=cos^{-1}\left(\frac{1}{n}\right)
\end{equation}
The emitted radiation is in the visible and near visible wavelength band for which $n(\omega)\,>\,1$. A real medium is always dispersive, so radiations are restricted to those frequencies for which $n(\omega)\,>\,\frac{1}{\beta}$. For X-ray region $n(\omega)\,<\,1$, i.e. radiation is forbidden since equation (\ref{cherenkov-max}) can not be satisfied~\cite{Polarization}.
\par
There are two more conditions, which need to be fulfilled to achieve coherence~\cite{Polarization}
\begin{enumerate}[(i)]
\item The length of the track of the particle in the medium should be larger than the wavelength of the radiation, otherwise diffraction would be dominant.
\item The velocity of the particle should be constant during its passage through the medium i.e differences in the time for particle to travel successive distances $\lambda$ should be less than the period ($\lambda$/c) of the emitted radiation. 
\end{enumerate}
The quantum theory of Cherenkov radiation includes the effects of emitted radiation on the motion of the particle. These effects are rather small as the energy of the radiated quanta is much smaller than the energy of the particle. Quantum treatment modifies the equation (\ref{cherenkov-cond}) to the following expression (\cite{Tickoo2002} and references therein)
\begin{equation}
 cos\, \theta = \frac{1}{n\beta}+\frac{\lambda_0}{\lambda}\frac{(n^2-1)}{(2n^2)}\frac{\sqrt{(1-\beta^2)}}{(\beta)}
\end{equation}
where $\lambda_0=0.0024\,nm$ is the Compton wavelength.
\subsection{Spectral distribution of Cherenkov radiation}Considering a particle of charge $ze$, the number of Cherenkov photons emitted per unit path length and per unit wavelength interval are given by~\cite{Polarization}
\begin{equation}
 \frac{d^2N}{dxd\lambda} = \frac{2\pi\alpha z^2}{\lambda^2} \times \left(1-\frac{1}{\beta^2\,n^2(\lambda)}\right)
\end{equation}
where $\alpha=e^2/\hbar\,c = 1/137$ is the structure constant. For electron moving along a track of length $l$, the number of emitted Cherenkov photon in the wavelength range $\lambda_1$ and $\lambda_2$ is then given by~\cite{Jelley1958}
\begin{equation}
 N = 2\pi\alpha l \times \left(\frac{1}{\lambda_2} -\frac{1}{\lambda_1}\right)\times \left(1-\frac{1}{\beta^2\,n^2(\lambda)}\right)
\end{equation} 
\begin{figure}[t]
\centering
\includegraphics*[width=0.6\textwidth,height=0.3\textheight, angle=0,bb=0 0 506 367]{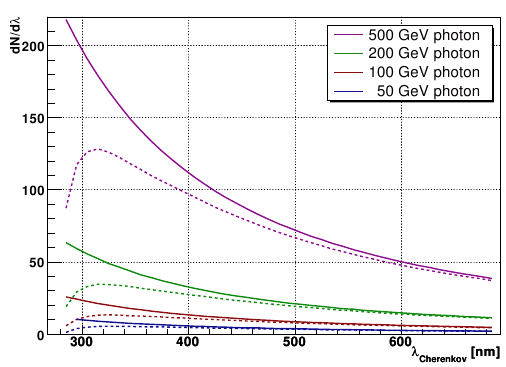}
\caption{\label{ch-spect}Spectra of Cherenkov light emitted by vertical \gam showers of different primary energy at $10\,km$ height is represented by solid curves while the corresponding detected spectra (effect of absorption by ozone and scattering) at observational level is represented by dashed curves. This figure has been adapted from~\cite{Wagner2006}.}
\end{figure}
Most of the photons are therefore produced in a short wavelength range i.e. in UV and blue waveband. No ACR is produced below $180\, nm$. Considering the frequency $\nu$ of Cherenkov photons, the total energy radiated is $W=N\,h\,\nu$. The spectral distribution may be represented in terms of energy per unit path length per unit frequency interval i.e. $(d^2W/dl\,d\omega) \propto \omega$ or energy per unit path length per unit wavelength interval i.e.$(d^2W/dl\,d\lambda) \propto 1/\lambda^3$ or number of photons per unit path length per unit frequency interval i.e.$(d^2N/dl\,d\omega) \propto const $ or number of photons per unit path length per unit wavelength interval i.e.$(d^2N/dl\,d\lambda) \propto 1/\lambda^2$~\cite{Jelley1958}. The differential photon spectrum of Cherenkov photons is shown in Fig.~\ref{ch-spect}, for photons emitted at $10\,km$ and observed at $2\,km$ altitude. The observed Cherenkov photon spectrum differs considerably from the produced spectrum due to the transmission characteristics of the atmosphere, explained in section~\ref{atmosphere}.
\par
The energy carried by ACR is given by the Frank-Tamm formula which yields the amount of ACR emitted as a charged particle moves through a medium at superluminal velocity. The energy emitted per unit length per unit frequency is~\cite{Tamm1937}
\begin{equation}
\frac{dE}{ dx d\omega} = \frac{\mu(\omega) q^2} {4 \pi} \omega {\left(1 - \frac{1} {\beta^2 n^2(\omega)}\right)}
\end{equation}
where $\mu$ is the permeability of the medium, $q$ is the electric charge of the particle. ACR does not have characteristic spectral peaks, as typical for fluorescence or emission spectra. Higher frequencies are more intense in ACR. This is why visible ACR is observed to be brilliant blue. In fact, most ACR is in the ultraviolet spectrum. The total amount of energy radiated per unit length in air by an electron is given by
\begin{equation}
\frac{dE}{dx} = \frac{e^2}{4 \pi} \int_{v>c/n(\omega)} \omega {\left(1 - \frac{1} {\beta/^2 n^2(\omega)}\right)} d\omega.
\end{equation}
This integral is done over the frequencies ω for which the particle's speed $v$ is greater than speed of light in the media $\frac{c}{n(\omega)}$. The integral is non-divergent because at high frequencies the refractive index becomes less than unity.
\subsection{Cherenkov radiation in EAS}
\begin{figure}[t]
\centering
\includegraphics*[width=0.5\textwidth,height=0.5\textheight, angle=270,bb=66 56 557 758]{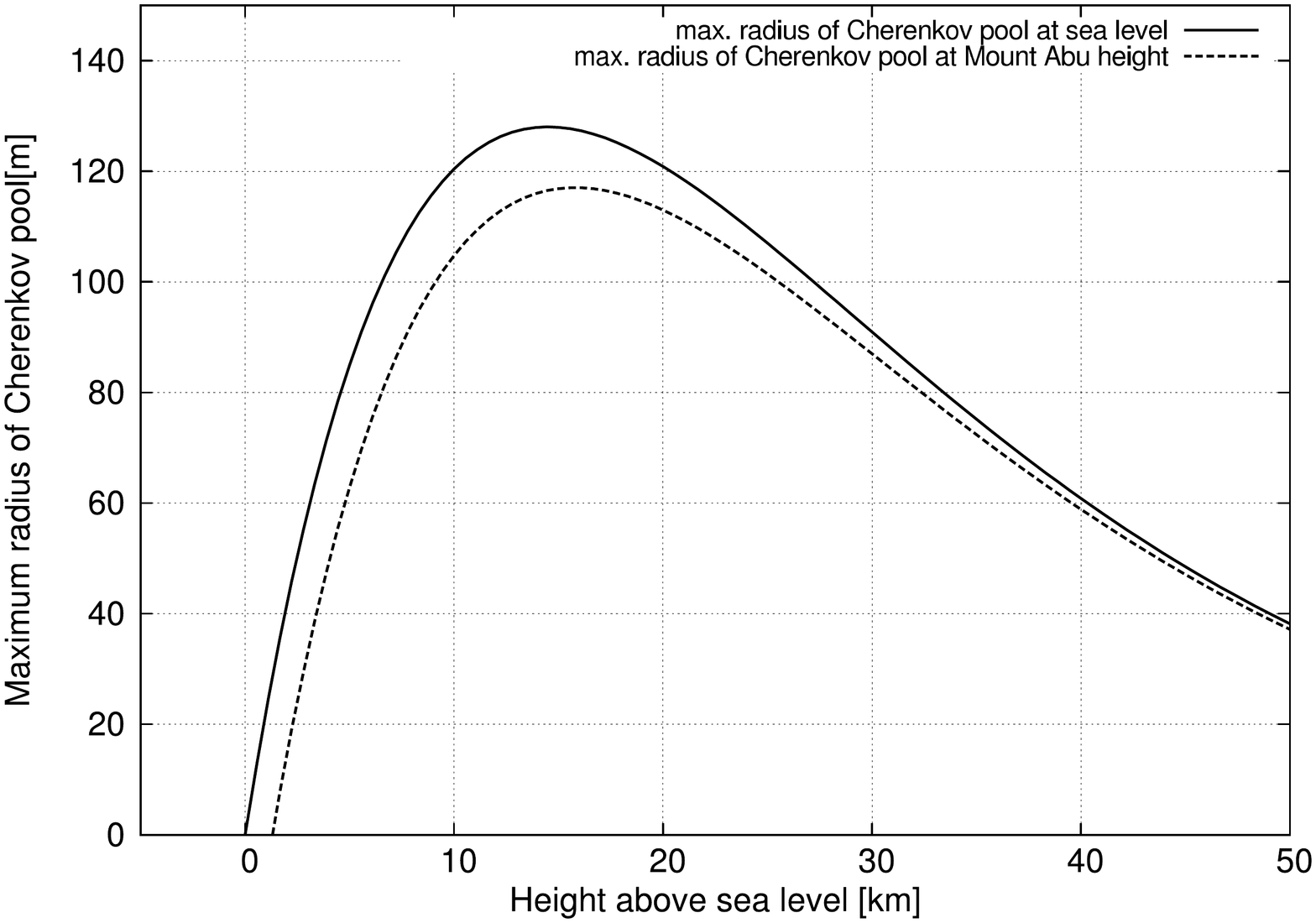}
\caption{\label{core} Radius of Cherenkov light cone as a function of height of emission. This is shown for sea level and Mount Abu altitude ($1300\,m$ asl).}
\end{figure}

If the energy of a primary particle is less than a few $TeV$, the shower gets absorbed in the atmosphere, while the ACR produced by the secondary charged particles can be detected at the ground level. It was Blackett~\cite{Blackett1948}, who suggested that there should be a small contribution of the order of $10^{-4}$ from the ACR produced by CR particles in the atmosphere, to the Light Of Night Sky (LONS). Galbraith and Jelley~\cite{Galbraith-Jelley1953,Galbraith1955} detected the ACR pulses associated with EAS which are subjected to large fluctuations in the point of first interaction and development of cascade subsequently. The ACR carries information about the point of origin (direction) and energy of the primary \gam photon. There are some important characteristics of ACR emitted in an EAS:\\
 \begin{figure}[t]
\centering
\includegraphics*[width=0.5\textwidth,height=0.5\textheight, angle=270,bb=59 56 557 766]{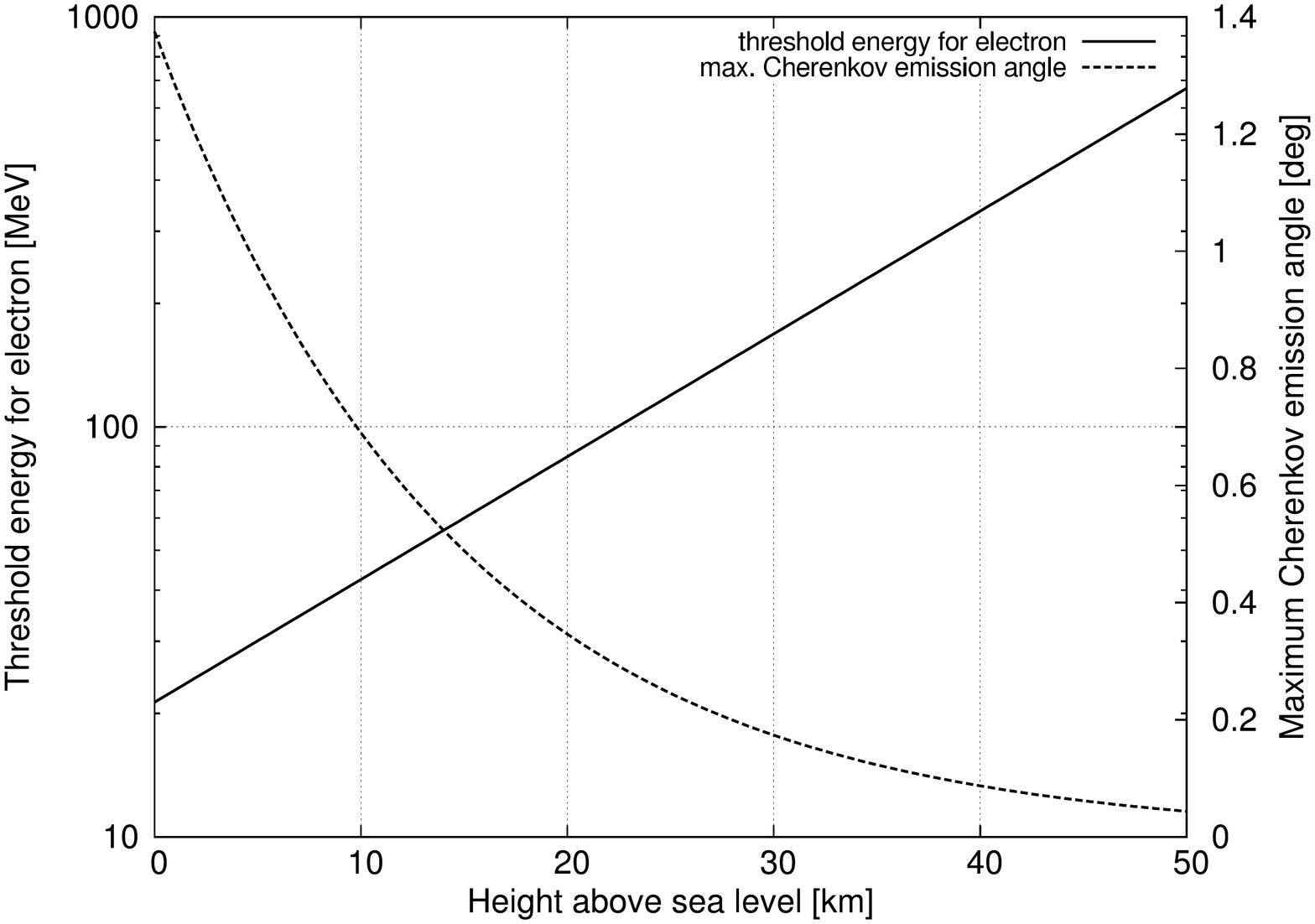}
\caption{\label{energy-height-angle}Variation of electron's threshold energy for Cherenkov emission in the Earth's atmosphere and maximum angle of emission as a function of altitude.}
\end{figure}
\begin{figure}[t]
\centering
\includegraphics*[width=0.6\textwidth,height=0.4\textheight, angle=0,bb=0 0 470 693]{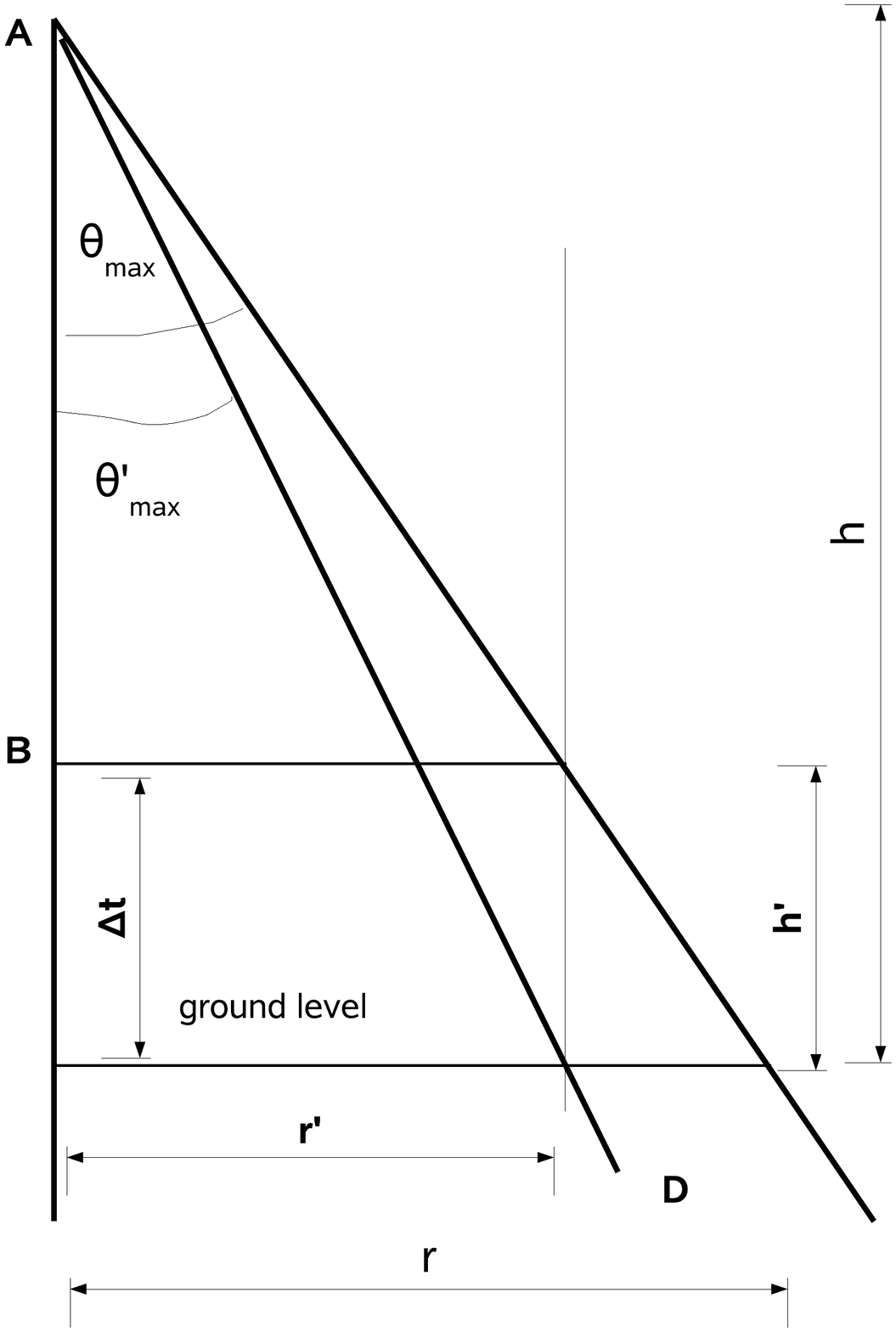}
\caption{\label{geo}Radius of ACR pool and Cherenkov pulse duration.}
\end{figure}
\textbf{(1) Threshold energy for Cherenkov emission:}\, Equation (\ref{threshold-velocity}), can be re-written in terms of threshold energy ($E_{min}$), a particle must have in order to emit ACR
\begin{equation}
\label{threshold-energy}
 E_{min} = \gamma_{min}\,m_0c^2 = \frac{m_0c^2}{\sqrt{1-\beta^2_{min}}} = \frac{m_0c^2}{\sqrt{1-1/n^2}}
\end{equation} 
where $m_0$ is the rest mass of the particle and $n=0.00029$ at sea level. Equation (\ref{threshold-energy}) implies that the minimum energies required to produce ACR at sea level are $21.3\, MeV$, $4.4\,GeV$ and $39.1\,GeV$ for electrons, muons and protons respectively. Thus the light particles like electrons are the most efficient in producing the ACR in an EAS. Since the refractive index decreases and hence the $\beta_{min}$ increases with height, the \textbf{threshold energy for the production of ACR also increases with height}. For electron $E_{min}$ is $35\,MeV$ at an altitude of $7.5\,km$. However, the increase in thereshold energy with altitude does not practically affect the fraction of electrons emitting ACR due to the fact that in the early phase of shower development the mean energy of electrons is significantly higher than in the shower tail. The smaller number of electrons at high altitudes produce only a small amount of ACR, while most of this is produced in the region of shower maximum~\cite{DanielK}.\\
\textbf{(2) Maximum Cherenkov emission angle:}\,\, The maximum Cherenkov emission angle given by equation (\ref{cherenkov-max}) also depends on height of emission in the following manner
\begin{equation}
 cos\,\theta_{max}= \frac{1}{1+\eta_0 \,exp(-h/h_0)} = 1-\eta_0\, exp(-h/h_0) 
\end{equation} 
On expanding $cos\,{\theta_{max}}$ ($1-\frac{\theta^2_{max}}{2!}+ \frac{\theta^4_{max}}{4!}- \frac{\theta^6_{max}}{6!}+ - - - $) we get
\begin{equation}
\label{thetamax}
 \theta_{max} = \sqrt{2\eta_0}\, exp(-h/2h_0).
\end{equation} 
Thus the maximum emission angle (or emission angle in general for $\beta \approx 1$) for ACR with respect to shower angle depends on refractive index (and height of emission). \textbf{The angle of emission increases with decreasing height}. Note that the average scattering angle $\theta_{col}$ due to multiple Coulomb scattering which is described by (\cite{Masaaki-th2008} and references therein)
\begin{equation}
\langle \theta^2_{col}\rangle \approx \left(\frac{E_s}{E}\right)\frac{X}{X_0}: \,\, E_s = \sqrt{\frac{4\pi}{\alpha}}.m_e\,c^2 = 21\,MeV
\end{equation}  
also increases with shower age. Both these effects lead to somewhat increased Cherenkov emission angle with decreasing height. Fig.~\ref{energy-height-angle} shows the variation of $E_{min}$ for electron and maximum angle for ACR as a function of atmospheric height above sea level.\\
\textbf{(3) Cherenkov light pool radius:}\,\, Let us consider a case wherein an energetic particle emits ACR from a height $h$ as shown in Fig.~\ref{geo}. The emission is in a narrow cone with maximum opening angle $\theta_{max}$ around its trajectory resulting in a nearly circular ring (also called \textbf{Cherenkov ring}) of radius $r$ which can be determined by using the equation (\ref{thetamax})
\begin{equation}
 r(h)=h\,tan\,\theta_{max}= \sqrt{2\eta_0}\,h\,\, exp(-h/2h_0).
\end{equation}  
The value of $r(h)$ is maximum when $h=2h_0$ which is $r_{max}\simeq128\, m$. Fig.~\ref{core} shows the dependency of the radius of Cherenkov ring on the height of emission at two observational levels (at sea level and at Mount Abu altitude). It is clear purely from geometrical consideration that the radius of Cherenkov ring shrinks with increasing observational height for fixed height of emission. This radius for an observational height $h_{obs}$ can simply be written as
 \begin{equation}
 r(h)=(h-h_{obs})\,tan\,\theta_{max}= \sqrt{2\eta_0}\,(h-h_{obs})\,\, exp(-h/2h_0).
\end{equation} 
It gives the maximum radius of the Cherenkov ring for Mount Abu altitude as $r_{max} \simeq 116.5 \,m$.
The ACR distribution of an EAS on the ground is the superposition of ACR cones from individual shower particles (mostly electrons). In case of an electromagnetic shower, ACR illuminates typically a circle called \textbf{ACR pool} with an average maximum Cherenkov radius of $\simeq120\,m$ at moderate altitudes like Mount Abu. Note that the circle is a special case for a vertically incident shower, while for an inclined EAS, the overlapping ACR forms an ellipse. Such ACR pools are typically produced by emissions taking place between $10$ to $20\,km$ above sea level. 
\begin{figure}[t]
\begin{center}
\centering
\subfigure[]{\label{mod}\includegraphics*[width=0.5\textwidth,height=0.4\textheight,angle=0,bb=0 0 542 519
]{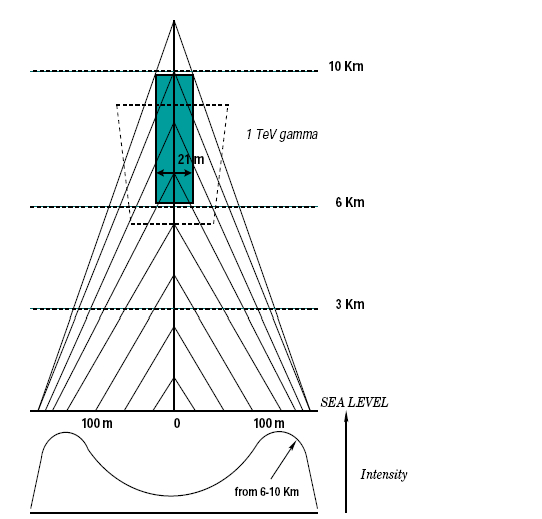}}
\subfigure[]{\label{gam-pro}\includegraphics*[width=0.45\textwidth,height=0.3\textheight, angle=0,bb= 0 0 546 493  ]{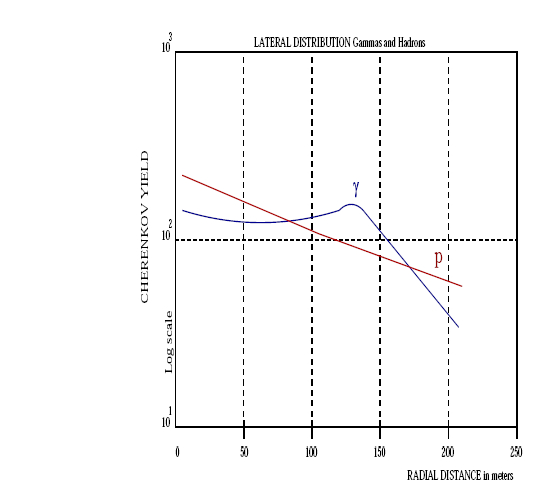}}
\caption{\label {}(a) Geometrical model for emission of Cherenkov radiation from $\gamma$ and hadron induced showers. The stripped region is the main region for the emission of Cherenkov light from \gam showers. (b) Lateral Cherenkov photon distribution for both \gam photon ($1\,TeV$) and proton ($2\,TeV$) showers. The \gam induced shower shows a relatively flat lateral distribution of Cherenkov light upto nearly $125\,m$ radial distance from the shower core. These figures have been adapted from~\cite{Polarization}.}
\end{center}
\end{figure}


\textbf{(4) Geometrical Model:}\,\, The ACR as seen at detector level, can be considered to come from three portions of a typical $1\,TeV$ shower shown schematically in Fig.~\ref{mod}. The first portion which contains $\sim25\%$ of total light comes from shower particles at elevations between the height of first interaction down to an elevation of $10\,km$. The mean value of the Cherenkov emission angle is about $1^\circ$ in air. It increases from $0.66^\circ$ at $10\,km$ to $0.74^\circ$ at $8\,km$ and finally becomes close to $1.4^\circ$ at sea level. This results in focusing of light on the ground into a blurry \textbf{Cherenkov ring} with radius of $r\approx 10\,km\, \times\,0.012 = 120\,m$ for a typical $1\,TeV$ \gam shower. Multiple Coulomb scattering gives rise to an exponential distribution of scattering angles with respect to the shower axis $\sim e^{-\theta/\theta_0}$ where the characteristic scattering angle is about $\theta_0 = 0.83\,E^{-0.67}_{min}$ giving a typical value of $\sim5^\circ$ (\cite{Aharonian2008} and references therein). While this angle is much larger than the Cherenkov angle, the exponential distribution is still strongly peaked in the forward direction and the Cherenkov ring, while blurred by scattering, is still discernable. The Cherenkov ring is most prominent when the shower dies out far above the observational level, as in low energy \gam showers. The light from the highest altitude arrives at the same time as that from the lower altitudes as the particles in the shower travel close to velocity of light and compensate for the greater geometrical path travelled because light from higher altitudes goes at slower velocity in air. The result is that the light in the annulus is strongly bunched in time with a spread of $\sim1\,ns$~\cite{Weekes2003}.
\par
The bulk of the light ($\sim$$50\%$) comes from a cylinder (see Fig.~\ref{mod}) of length $\sim 4\,km$ (nearly between $6$ to $10\,km$ height) and radius $21\,m$ centered on the shower core; this cylinder contains the shower maximum and hence bulk of the emitted light. The Cherenkov photons on reaching ground level form a ring like structure which is also called the \textbf{hump} of the Cherenkov pool. The light from this region is a good measure of the total energy and is best measured at a distance of nearly $100\, m$ from the shower axis on the ground. The angular spread of this light will have a half-width of $\sim 0.2^\circ$~\cite{Weekes2003}. For a proton cascade of same primary energy the similar column is represented by dashed line. The differences in the depth of shower maximum for a \gam photon and proton is due to the difference in their interaction lengths.
\par
The last $\sim$$25\%$ of the light originates from the few surviving particles of the shower, radiating below an elevation of $6\,km$ which is subjected to large fluctuations, and generally falls close to the shower axis.
\par
The intensity of ACR on the ground as a function of distance from the shower core is shown in Fig.~\ref{mod}. In this lateral distribution, any particular radial distance on the ground from the core represents a particular height in the atmosphere. By comparing the lateral distribution of ACR for different energies it has been shown by Rao and Sinha~\cite{Rao1988} that the hump feature is due to high energy electrons above $1\,GeV$. Electrons of lower energies are more scattered and do not contribute to the formation of the Cherenkov ring. The intensity near the shower axis rises when the tail of the shower of a high energy particle penetrates close to the observational level. This is because the light is emitted closer to the the ground and has a shorter distance to travel. The lateral distribution of ACR yield on the ground for a typical $1\,TeV$ \gam shower is shown in Fig.~\ref{gam-pro}. The \gam induced shower shows a relatively flat distribution upto the hump, in contrast to a proton shower of a similar energy.\\
\textbf{(5) Cherenkov pulse duration:}\,\, In a non-dispersive medium, the wavefront is infinitely thin and therefore the duration of ACR pulse is infinitely short. However, in a dispersive medium, the Cherenkov angle is different for different wavelengths and the wave trains will spread out. The observed duration along the particle track at a distance $r'$ as shown in Fig.~\ref{geo}~\cite{Polarization} is
\begin{equation}
\triangle t = \frac{r'}{\beta \cdot c} \cdot (tan\,\theta'_{max} - tan\,\theta_{max}).
\end{equation}
Another effect that increases the pulse duration is the different heights of emission of the Cherenkov photons. Again referring to Fig.~\ref{geo}, emissions of Cherenkov photons take place at two points A and B, assuming that photons are emitted along the track of the particle, the measured time difference between the detection of photons can be written as\begin{equation}
\delta t = \frac{\eta_0 \, h_0}{c}(exp^{-h'/h_o} - exp^{-h/h_0}).                                                                                                              \end{equation}  
Taking $h'=6\,km$ and $h=10\,km$, we get $\delta t~\sim 1.3\,ns$. The total pulse duration will be the sum of these two effects. When the development of a $1\,TeV$ \gam photon induced EAS shower is simulated through the Monte Carlo procedure, the value of Cherenkov pulse duration turns out to be $\sim 5\,ns$~\cite{Tickoo2002}.\\
\begin{figure}[t]
\centering
\includegraphics*[width=0.95\textwidth,height=0.35\textheight, angle=0,bb=0 0 610 296]{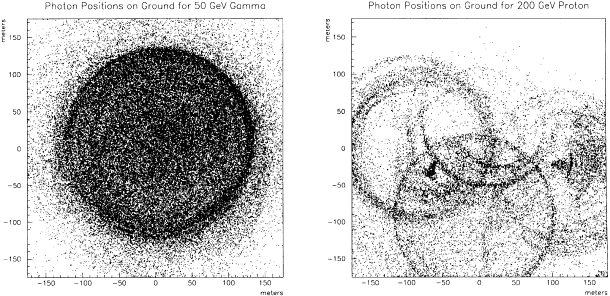}
\caption{\label{lat-gam-prot}Comparision of the lateral distributions of ACR for vertically incident \gam and proton showers. The left shows the arrival positions on the ground of all Cherenkov photons produced in an air shower created by a single $50\,GeV$ \gam. The right shows the same information for an air shower created by a single $200\,GeV$ proton. This figure has been adapted from~\cite{Ong1998}.}
\end{figure}
\textbf{(6) Differences in ACR distribution of \gam and proton induced showers:}
\begin{enumerate}[(i)]
 \item Because of the longer interaction length, the particles producing ACR in proton showers reach closer to the ground level than \gam showers, which gets reflected in Fig. \ref{gam-pro} as the intensity is more near the shower axis for protons. The Cherenkov ring also gets smeared out in proton showers.
 \item The ACR pool created by \gams is more regular in comparison to the protons since there are more sources for the Cherenkov emission other than electrons and positrons (e.g. muons) contributing to the light pool. Muons produce intense rings in ACR distribution on the ground.
 \item For the same primary energy, proton showers have a ACR pool less intense than the \gam showers, as only 1/3 of the primary energy (to the first approximation) in proton shower contributes in the development of electromagnetic cascade.
 \item The Root Mean Square (RMS) of Cherenkov photon density measured over a $10\,m^2$ area  is calculated to be $16\%$ for \gam showers and $36\%$ for the proton showers at around $1 \,TeV$ energy and at $100\, m$ from the shower core~\cite{Polarization}. 
\end{enumerate}
These differences are demonstrated in Fig.~\ref{lat-gam-prot}.
\section{Detection techniques of extensive air showers}The basic purpose of detection of an EAS experimentally, is to determine the type (though difficult) and properties of primary particle, such as direction, energy and chemical composition in case of CRs. The detection methods are mainly based on the energy region of interest. In a \gam induced EAS, the particles are beamed to the ground and their trajectories retain the directionality. The shower front is narrow with a thickness of $\sim 1\,m$ at the center. Tight longitudinal structure (i.e. fast timing $\sim$ few $ns$) can be used to measure the shower orientation and hence the primary direction. Fast timing allows to separate the signal produced due to the shower front from random background processes. Finally the lateral distribution of air showers allows detector spread out on the ground to be sensitive to particles whose initial trajectory would not have intersected a detector in the absence of the atmosphere. Thus the effective area of an EAS detector is much larger than its physical area.
\par
At energies more than few $TeV$, sufficient number of secondary particles are able to reach the ground, which can be detected with a suitable detector e.g. array of detectors based on scintillation counters. Such a detector is normally called an air shower array. The fast timing between detector elements is used to estimate the direction of the primary. The shower core is approximated by the location of the highest particle density and the lateral distribution is reconstructed from measurements of the particle density at different radii from the core~\cite{Ong1998}. The advantage of this technique is the $100\%$ duty cycle, while angular resolution is poor and has only modest possibility of $\gamma$-hadron separation.
\par
A small fraction of the primary energy is emitted as isotropically distributed fluorescent light (yield is $\simeq 10^{-5}$ of the total energy~\cite{DanielK}) from the excitation of nitrogen molecules\footnote{during de-excitation, a distinct well known line spectrum $337\, nm\,N_2$-line is emitted near UV.}. Operating at rather high threshold energies ($\ge 10^{15}\,eV$), fluorescent detectors can be operated only during dark periods. Detection of neutrinos is also possible, but due to their weakly interacting nature with matter, these detectors are not very successful.
\par
On the other hand, the ACR produced by an EAS can be measured very efficiently above energies of more than few tens of $GeV$. The Cherenkov photon spectrum peaks  near UV region ($1/\lambda^2$ dependence) and the sky is also transparent to light between 300 to 600 $nm$, most of the produced Cherenkov photons reach the ground under good atmospheric conditions\footnote{For example: $1\,TeV$ \gam shower produces on an average $\sim 3\times10^6$ Cherenkov photons at ground.}. Preserving the directional information contained in the shower particles, the Cherenkov photons are strongly beamed in the forward direction and arrive at the ground with a tight time structure ($\sim 5\,ns$). This permits the detection of ACR by fast photosensitive detectors amidst the LONS. This technique is discussed in detail below.
\subsection{Atmospheric Cherenkov technique}This technique has been used for more than forty years to search for point sources of VHE radiation. The details of the technical aspects of the ACT can be found in~\cite{Cawley1995,Weekes1988,Ong1998}. The first generation systems consisted of just a single light detector in the focal plane of a search-light mirror coupled to fast pulse counting electronics. These telescopes are characterized by the mirror collection area ($A$), the reflectivity ($R$), the solid angle ($\Omega$) and the signal integration time ($\tau$). Even with a simple detector like $A=2\,m^2$, $R=85\%$, $\Omega=10^{-3}$ and $\tau=10\,ns$, it is possible to detect the ACR signal from EAS (of few $TeV$ energy) with high efficiency.
\subsection{Trigger threshold of the telescope}An atmospheric Cherenkov telescope generates a trigger whenever an ACR signal hits the detector with a sufficient strength from both \gam photon or proton induced EAS amidst the noise of NSB photons. Characteristic features of an ACR pulse e.g. narrow pulse width ($\sim$$5\,ns$), limited angular size ($<1^\circ$) on the ground and nature of its photon spectrum\footnote{Cherenkov light peaks at short wavelength (blue/UV) whereas LONS peaks at longer wavelength} facilitate its detection. Within the ACR pool (circle of approximately $125\,m$ radius), the number of Cherenkov photons is proportional to the energy of the primary \gam. Defining a scaling factor for Cherenkov photon yield ($y_\gamma$) as
\begin{equation}
 y_\gamma \equiv \rho_\gamma/E
\end{equation} 
where $\rho_\gamma$ is photon density of Cherenkov photons on the ground and $E$ is energy of the primary particle. The value of $y_\gamma$ depends on energy e.g it is $0.033\,photons/m^2/GeV$ at $25\,GeV$ for wavelength between $300$ and $550\,nm$ and increases to $0.065\,photons/m^2/GeV$ at $1\,TeV$~\cite{Ong1998}. It also increases with height because of lesser atmospheric attenuation.
\par
The flux of night sky photons is highly location dependent and has a typical value of $2\times10^{12}\,photons/m^2/s/sr$ for a good quality site in the wavelength region of $300$ to $600\,nm$ (\cite{Ong1998} and references therein). A telescope with $1^\circ$ field of view and $10\,ns$ trigger formation time would receive $\sim 5 \,photons/m^2$ from the LONS while a much higher expected value of $\sim 65 \,photons/m^2$ from a $1\,TeV$ \gam shower. Thus, a $1\,TeV$ shower should be detectable above NSB with a telescope having $1\,m^2$ mirror area.
\par
The signal produced by Cherenkov photons in terms of a number of photoelectrons in a photosensitive instrument, which is usually a photomultiplier tube (PMT) can be written as
\begin{equation}
 S=\rho_\gamma\,A\,R\,\eta_{pmt} = y_\gamma \,E\,A\,R\,\eta_{pmt}
\end{equation}
where $\eta_{pmt}$ is the quantum efficiency of the PMT. The ACR must be detected above the fluctuations in the NSB in the integration time $\tau$. The noise level can be expressed in terms of fluctuations as
\begin{equation}
 N = \sqrt{\phi_{LONS}\,\Omega\,A\,R\,\eta_{pmt}\,\tau}
\end{equation}  
where $\phi_{LONS}$ is the NSB photon flux and $\Omega$ is the solid angle subtended by a PMT. Note that, in actual case the wavelength dependence of both Cherenkov and LONS production and collection should also we considered.
The energy threshold of an atmospheric Cherenkov telescope can be defined in a number of ways. At the moment the most appropriate definition is the minimum \gam energy for which the signal to noise ratio ($S/N$) is sufficient to adequately trigger the telescope. The smallest detectable light pulse is therefore inversely proportional to $S/N$, i.e.
\begin{equation}
\label{thre-eq}
 E_{th} \propto (1/y_\gamma) \sqrt{\frac{\phi_{LONS}\,\Omega\,\tau}{A\,R\,\eta_{pmt}}}.
\end{equation} 
As is clear from equation (\ref{thre-eq}), the energy threshold of a Cherenkov telescope can be reduced by working at darker sites, reducing field of view and integration time and at the same time increasing the mirror collection area, its reflectivity and quantum efficiency of the PMT.
\subsection{Flux sensitivity of the telescope}The most troublesome factor in \gam astronomy is the huge background of CRs. The ACR distribution from these hadronic showers is similar to that from \gam showers so that first-generation Cherenkov telescopes were unable to distinguish between the two. In the field of view of a simple telescope, whose solid angle is optimized for \gam detection, the background CR events are $10^3$ times as numerous as the strongest steady \gam discrete source thus far detected. Since the arrival directions of CRs are isotropic due to interstellar magnetic fields, a discrete source of \gams can stand only as an anisotropy in an otherwise isotropic distribution of air showers. However, there are very distinct differences in the cascade development in electromagnetic and hadronic showers, as already discussed and can also be found in~\cite{Hillas1996,Jelley-Porter1963}, that make the study of sources of VHE \gams with Cherenkov telescopes possible.
\par
The flux sensitivity basically defines the performance of a Cherenkov telescope and is determined by its ability to detect a \gam signal over the CR background. Defining the $\eta_\gamma$ and $\eta_{cr}$ as the \gam and CR retention factors, after using some discriminating tools for two types of shower events, a ``figure of merit'' called \textbf{quality factor} ($Q$) of a telescope can be defined as:
\begin{equation}
 Q = \frac{\eta_\gamma}{\sqrt{\eta_{cr}}}.
\end{equation} 
The CR background has a power law type spectrum, which can be expressed in the form of integral flux as 
\begin{equation}
 \phi_{cr}(>E) \equiv C_{cr}\,E^{-\alpha _{cr}}.
\end{equation}  
In the range of interest $\alpha_{cr} = 1.7$. Similarly, the \gam source energy distribution can be assumed to have the form
\begin{equation}
 \phi_\gamma(>E) \equiv C_\gamma\,E^{-\alpha_\gamma}
\end{equation}
$\alpha_\gamma$ can have values from 1 to 3 and is generally assumed to increase with energy i.e. the spectrum steepens~\cite{Weekes2003}.
\par
If $S$ is equal to the number of \gams detected from a given source in observation time $T$, and $A_\gamma$ is the collection area for \gam detection, then the \gam signal for some threshold energy $E$ can be written as
\begin{equation}
 S = \phi_\gamma(E)\,A_\gamma(E)\,T.
\end{equation}
The background CRs recorded during the same time is given by
\begin{equation}
 B = \phi_{cr}(E)\,A_{cr}\,\Omega \,T
\end{equation} 
where $A_{cr}$ is the collection area for the detection of CRs of energy $E$. Defining $N_\sigma$ as the minimum number of standard deviations at which a source must be detected to be believable\footnote{$N_\sigma$ for a reliable source detection is generally taken as 5}, we have~\cite{Ong1998}
\begin{equation}
\label{sensitivity}
 N_\sigma \propto \frac{S}{\sqrt{B}}=\frac{C_\gamma}{\sqrt{C_{cr}}}\,E^{(0.85-\alpha_\gamma)}\,\frac{A_\gamma(E)}{\sqrt{A_{cr}(E)\Omega}}\,T^{1/2}
\end{equation}
As is evident from the equation (\ref{sensitivity}), to maximize $N_\sigma$ i.e. optimizing the sensitivity, the product $A_{cr}\Omega$ should be minimized, while at the same time maintaining a reasonable collection area for \gams, which neccessitates the requirement of an efficient technique for background rejection. The source related information can also be used to optimize sensitivity, e.g. steady sources like Crab Nebula should be observed for longer period. For typical sources, the telescope can be operated at lowest possible energy threshold, which can also help the detection of sources having cutoffs at higher energies. However, sources with extremely hard spectra ($\alpha_\gamma < 0.85$, though rare), require observations at higher energies. 
\subsection{Methods of rejecting cosmic ray background}Since more than $99\%$ of the air showers produced in the atmosphere are caused by the isotropic CRs, the study of \gams requires an efficient technique recognising and rejecting the hadronic showers and accepting only \gam showers. The characteristics of the ACR produced in air showers induced by \gams and hadrons are similar to a large extent. However, as already pointed out Monte Carlo simulations show that there are perceptible differences between the two types of primaries, which makes their discrimination possible. Many techniques  to achieve a good $\gamma$-hadron separation exist but all of them are based on geometrical considerations and involvement of different processes in the way the two kind of showers develop in the Earth's atmosphere. Here are some of the characteristics of the ACR that can be used for their segregation:\\\\
\textbf{Lateral distribution:} The ACR is sampled by an array of small telescopes. The arrival direction of the shower is determined through the arrival time information of the shower front at different detectors. The photon density is also sampled by digitizing the Cherenkov pulse. A \gam induced shower is expected to have a flatter distribution of photons as compared to hadrons. Examples of such telescopes are given in the next section. The flatness parameter is used for background rejection with a quality factor of $\sim 1.8$~\cite{Bhat2001}.\\\\
\textbf{Pulse duration:} The shape of the Cherenkov pulse is related to the development of the shower cascade, e.g. the rise time of the pulse reflects the longitudinal development and decay time is a measure of attenuation of the cascade below shower maxima~\cite{Tickoo2002}. Width of the Cherenkov pulse represents the production profile of Cherenkov photons. Some of these properties are exploited by the Pachmarhi group~\cite{Chitnis2001}.\\\\ 
\textbf{Spectral content:} Hadron showers are comparatively richer in muon contents, which produce ACR closer to the observational level. The UV part of ACR from a \gam shower gets almost absorbed in the atmosphere, thus making it possible to differentiate between \gam and hadron based on their UV to visible light ratio. This approach was used by the Crimean group~\cite{Zyskin1987}.\\\\
\textbf{Shower image parameters:} In addition to other details (discussed in next section) the Cherenkov image also contains the important information about the nature of the primary particle e.g. the image that results from a typical \gam shower is elliptical and compact with an orientation that points towards the center of the field of view. In contrast, Cherenkov images from hadrons are less regular and randomly oriented in the focal plane~\cite{Hillas1985}. This is the imaging technique explained in next section and has been used in the work being presented in this thesis. This technique provides very high quality factor with the rejection of $\sim99.7\%$ hadrons, while retaining $\sim50\%$ \gams.
\section{The imaging technique}
The atmospheric Cherenkov telescopes fall into two broad classes: Wavefront sampling atmospheric Cherenkov telescopes and IACTs. The former records the integrated while the
\begin{figure}[H]
\centering
\includegraphics*[width=0.9\textwidth,height=0.8\textheight, angle=0,bb=0 0 585 770]{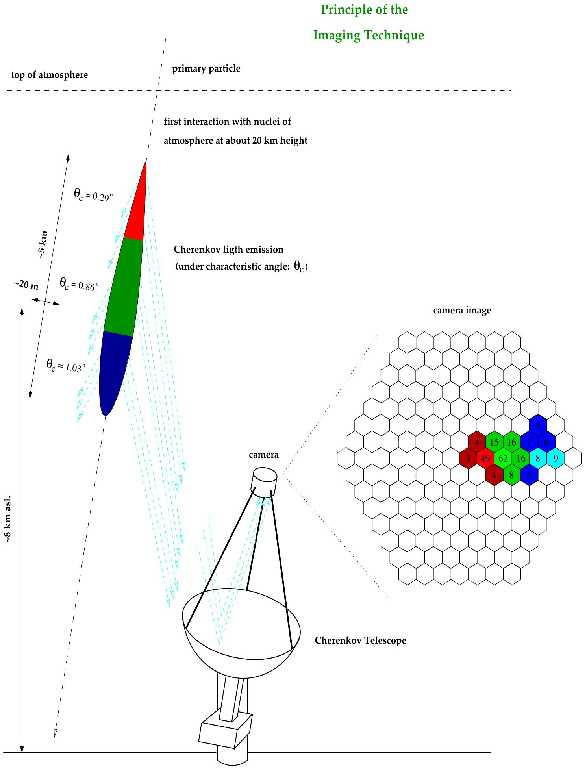}
\caption{\label{IACT}Basic principle of the imaging technique, illustrating the detection of an EAS. This figure has been adapted from~\cite{DanielK}.}
\end{figure}
later records the differential information of the ACR. Wavefront sampling instruments make use of a large array of mirrors that cover a large fraction of ACR pool and record the light intensity at a number of points on the ground. Examples of these instruments are THEMISTOCLE experiment~\cite{Baillon1993}, AGSAT~\cite{Goret1993}, Solar-II / CACTUS~\cite{Tumer2002}, STACEE~\cite{Hanna2002}, CELESTE~\cite{Par2002}, GRAAL~\cite{Arqueros2002}, PACT~\cite{Bhat1997} and HAGAR~\cite{hagar}.
\par
On the other hand, the development of atmospheric Cherenkov imaging technique gave the first effective discrimination between \gam and background hadron showers (\cite{Weekes2003} and references therein). In the following, we explain the working principle of the imaging technique. An array of PMTs in the focal plane of a large optical reflector constitutes a camera to record ACR image for each triggered shower. The camera is triggered when a preset number of PMTs detect a light level above a set threshold within a short integration time. The light level in all the pixels is then recorded digitally and the image is analyzed offline to determine whether it has the expected characteristics of a \gam shower with a point of origin at the center of the field of view. Discrimination against the background CR showers is based on geometry of the image e.g. \gams arriving parallel to the optic axis of the telescope have roughly elliptical images which appear to originate from the center of the camera.
\begin{figure}[t]
\centering
\includegraphics*[width=0.8\textwidth,height=0.35\textheight, angle=0,bb=0 0 551 369]{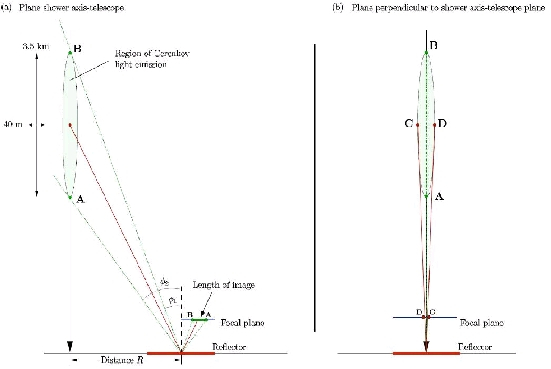}
\caption{\label{shower-image}Sketch of a shower and its image. (a) Longitudinal distribution of the shower related to the length of the image. (b) Lateral distribution of the shower and determination of the width of the image. This figure has been adapted from~\cite{Fuste2007}.}
\end{figure}
\par
As depicted in Fig.~\ref{IACT}, due to the mirror and camera geometry, Cherenkov photons viewed within the opening angle of a pixel are imaged onto the same point in the camera, and therefore the camera coordinates are usually defined in angular coordinate system. Since the height of emission of photons, governs the angle of emission, the different emitting regions of the shower are mapped at different positions in the camera plane. The image formation of different parts of the shower in the camera is shown in different colors in Fig.~\ref{IACT}. The upper part of the shower is mapped onto the camera closer to its center, whereas the lower part of the image is mapped further away. The shower axis which defines its orientation is mapped onto the camera plane as a straight line called \textbf{image axis} pointing towards the shower impact position on the ground. The image of the shower thus recorded in the camera contains information about the development of the shower and the properties of emitted ACR. It is possible therefore to infer the information of the primary particle, such as its energy, direction of incidence and also its nature. The energy of the primary particle is largely determined from the light content of the image, while the direction is determined through the orientation and shape of the image. Thus the procedure of inferring the properties of the primary particles based on the information from the image formed in the camera is called the imaging technique.
\par
Fig.\ref{shower-image} demonstrates, how the geometry of an EAS is related to the shape of the light distribution imaged in the camera. Fig.\ref{shower-image}(a) represents the view of the plane containing the telescope and the shower axis, wherein length of the camera image determines the angle ($\phi_2 - \phi_1$) under which the shower is seen. If the shower is viewed over longer heights (points A and B are separated farther away) or shower is viewed from larger distance (large core distance), the camera image results in an elongation. Fig.~\ref{shower-image}(b) demonstrates the perpendicular view i.e. the width of the camera image determines the lateral extent of the shower.
\par
Based on the geometrical relations, explained above A. M. Hillas in 1985~\cite{Hillas1985} constructed a set of image parameters called \textbf{Hillas parameters}, which are very efficient tools for reconstruction of an EAS and improved the capabilities of imaging technique to distinguish between \gam and hadron initiated showers. The Hillas parameters are still providing the best results in the field of VHE \gam astronomy. We will discuss the Hillas parameters in Chapter 5.

\chapter{TACTIC telescope}
\section{Introduction}
\begin{figure}[t]
\centering
\includegraphics*[width=0.4\textwidth,height=0.4\textheight,angle=0,bb=0 0 389 639]{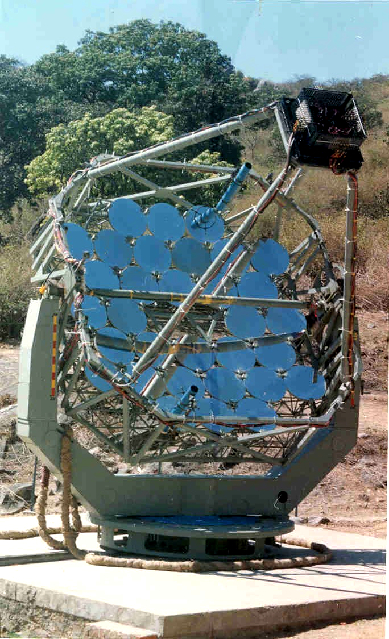}
\caption{\label{tactic}Photograph of the TACTIC  imaging  element.}
\end{figure}

In this chapter, we describe the instrumentation aspects of the TACTIC (TeV Atmospheric Cherenkov Telescope with Imaging  Camera)~\cite{Bhat1996} imaging element. We discuss the main features of its various subsystems and its overall  performance with regard to (\textit{i}) tracking accuracy of its 2-axes drive system, (\textit{ii}) spot size of the light collector, (\textit{iii}) back-end  signal processing electronics and topological trigger generation scheme, (\textit{iv}) data acquisition and control system and  (\textit{v}) relative and absolute gain calibration methodology. The TACTIC  has been in operation at Mount Abu ($24.6^\circ$ N,  $72.7^\circ$ E, $1300\,m$ asl), a hill resort in Western Rajasthan, India, for the last several years to study $TeV$ \gam emission from celestial sources. The mean NSB brightness levels  at Mount Abu is found to be $\sim(1.34 \pm 0.50)\times10^{12}\, photons\, m^{-2}\,s^{-1}\,sr^{-1}$ (pre-monsoon) and $\sim(1.54 \pm 0.55)\times10^{12}\, photons\, m^{-2}\,s^{-1}\,sr^{-1}$ (post-monsoon). The telescope uses a tessellated light-collector of area $\sim 9.5\,m^2$ and deploys a 349-pixel imaging camera, with a uniform pixel resolution of $\sim 0.3^\circ$ and a total field-of-view $\sim 6^\circ \times 6^\circ$. It records the images of the EAS via ACR which is produced by the secondary relativistic particles of the EAS. Using a trigger field of view of $\sim 3.4^\circ \times3.4^\circ$ (corresponding matrix of $11\times11$ pixels) the telescope records a CR event rate of $\sim2.5 \,Hz$ at a typical zenith angle of $15^\circ$. The photograph of the TACTIC imaging telescope is shown in Fig.~\ref{tactic}.
\section{Mechanical structure of the telescope}
The main sub-assemblies of the TACTIC  telescope are : mirror basket, mirror fixing/adjusting frames, zenithal and azimuthal gear  assembly, encoders/motors for its two axes, camera support boom assembly and the PMT-based imaging camera. A three dimensional truss type structure has been used to support the mirror frame.  The mirror basket is a 3 layer welded mildsteel  tubular  grid structure   fabricated in 3 parts  which are bolted together. The basket is provided with a central tie rod  which extends  into   short stub  shafts  at the two ends  which are housed in bearings. The individual  mirror facets are supported on three levelling studs, so that desired inclination of the specific mirror with respect to the telescope axis can be achieved. The weight of the moving part of the telescope is around $6.5\,ton$. The complete basket assembly is held by the fork of the telescope which transmits the load through the central  roller thrust bearing to the telescope foundation. The horizontal part of the fork frame is clamped on to a large gear of the azimuthal drive rotary table of the telescope. The thrust bearing  using  hardened rollers was specially  designed and  fabricated  in-house.   A central pin and a taper roller bearing, housed in the large gear is used to ensure proper rotation about its vertical axis. The base structure has been anchored to the ground by 4 foundation bolts. The  zenithal  motion  to the telescope is given from only one end of the basket. A common shaft assembly fixed in self aligning ball bearing having a large gear fixed on it, provides the zenithal rotation of the mirror basket. Five stage gear box providing a total speed reduction of 5189.14 has been used in the zenithal drive of the telescope.  Likewise, a four stage gear box  leading to a speed reduction of 6348.94  has been used in the azimuth drive.  A large capacity circular cable drag chain has been provided for easy and free movement of the nearly 700  signal and high voltage cables.
\par
Deflection analysis using FEM software package has also been performed at various zenith angles to arrive at the shift in the focal point position due to structural  deformations of the mirror basket frame, mirror holding attachment and support booms. The results of this study indicate that a shift of $<3\,mm$ in the focal point position can be produced over the zenith angle  range of  $0^\circ$ to $70^\circ$.
\section{Drive control system}
The large light collector aperture ($\sim3.5\,m$) coupled with the attendant large telescope weight ($\sim6.5\,ton$) have led to the choice of an altitude-azimuth (alt-azm) mounting for the TACTIC telescope, as against the comparatively  simpler equatorial mounting. The main advantage of the alt-azm mount is that the telescope weight is supported uniformly on a horizontally-placed central thrust-bearing. For an alt-azm mount, the immobile axis is set towards the zenith and the telescope guidance is provided by the azimuth and zenith angles. As the source co-ordinates (right ascension, $\alpha$ and declination, $\delta$) are known in the equatorial co-ordinate system, a
co-ordinate transformation is required to find the position and speed of the celestial source along either axis of the telescope. The general equations for such a transformation are given by:
\begin{equation}
tan \: A = \frac{cos\delta \: sinh}{sin\phi \: cosh \: cos\delta - cos\phi \: sin\delta}
\end{equation}
\begin{equation}
cos z = sin\phi \: \; sin\delta + cos\delta \: \; cos\phi \:\: cosh
\end{equation}
\begin{equation}
dz/dh = cos\phi \:\:sinA
\end{equation}
\begin{equation}
dA/dh = sin\phi \: + \: \; cos\phi \: \; cosA \: \; cotz
\label{equation:dA_dh}
\end{equation}
where $\phi$ is the latitude of the observatory (for Mount Abu $\phi=24.63^{\circ}$), $z$ is the zenith angle measured from local vertical, $A$ is the azimuth angle measured from the geographical south with positive sign for the angle towards the west and $h$ is the source hour angle. The hour angle of the source is related to the local sidereal time (LST) by the equation 
\begin{equation}
h = LST - \alpha.
\end{equation}
The design of the TACTIC  drive control system is  based on the CAMAC standard. 
The telescope uses two $100\,N\,cm$ hybrid stepper motors (Pacific Scientific make, Model H 31NREB, NEMA  Size 34)  for driving   its  azimuthal and zenithal axes through  multistage gear-trains.  The maximum azimuthal speed of the TACTIC telescope is $100\,rad/rad$ (equivalent to $1500^0\,hour^{-1}$), and the resulting blind spot has a diameter $\sim$1.2$^0$. A GPS based CAMAC-compatible  digital  clock  (Hytec Electronics Ltd. make; GPS92)  with a resolution of   $\sim$$10\,ns$ and absolute  time accuracy of $\sim$$100\,ns$, is used to compute  the source co-ordinates  in real time.  The new co-ordinates of the source  are calculated  after  every  second   while  tracking  a candidate source. More details regarding various  hardware  components and related software of the telescope drive system  are discussed in~\cite{Tickoo1999,Tickoo2002,Koulr2007} .
\par 
The tracking accuracy  of the telescope  is also checked  on a regular basis  with so called ``point runs'', where a reasonably bright star, having a declination close  to that of the  candidate \gam source is  tracked continuously  for about $5\,hours$. The point run calibration data  (corrected zenith and azimuth angle  of the telescope  when the star image is centered)  are  then incorporated in the telescope drive system software so that appropriate corrections can be  applied directly  in real time  while tracking  a candidate \gam source. 
\section{Light collector design}
\begin{figure}[h]
\centering
\includegraphics*[width=1.0\textwidth,angle=0]{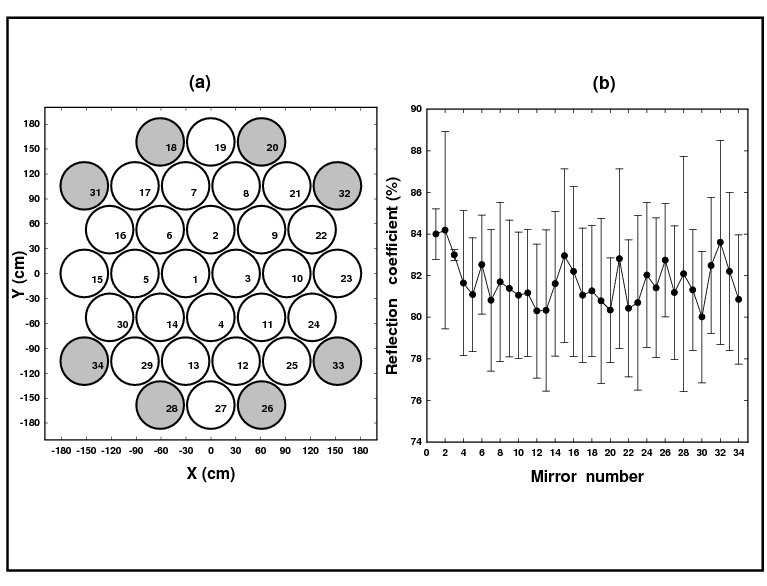}
\caption{\label{mirror} (a) Projection of the 34 mirror facets of the tessellated light collector of the TACTIC  telescope. The shaded circles represent  the peripheral mirrors that have been  raised along the z-axis of the light collector for obtaining  a better point spread function (b) Reflection coefficients for all the 34 mirrors used in the telescope. Seven randomly selected  locations on the mirror surface  have been used to quantify  the  mean reflection coefficient  of the mirror. The error  bars represent standard dispersion ($1\sigma$ value) of the 7 reflection coefficient values.}
\end{figure}

The  TACTIC light collector uses 34 front-face aluminium-coated, glass spherical mirrors of $60\,cm$ diameter  each  with the following characteristics (\textit{i}) focal length $\sim400\,cm$, (\textit{ii}) surface figure $\sim$ few $\lambda$  (\textit{iii}) reflection coefficient $>$80$\%$ at a wavelength of $\sim400\,nm$ and (\textit{iv})  thickness $20\,mm$ to $40\,mm$. Fig.~\ref{mirror}(a) shows the projection of the mirror layout  onto  a plane transverse to the symmetry axis of the basket. The  shorter focal length  facets are deployed close to the principal axis of the basket  while  the longer focal length facets are deployed around the periphery. The peripheral mirrors have the effect of increasing  the overall spot size as they  function in an off-axis mode.  In order to  make the design as  close to the Davies-Cotton design as possible, we  used  longer studs on the mirror frame structure to raise the pole positions of the peripheral mirrors. Fig.~\ref{mirror}(b)  shows the measured reflection coefficients of individual mirrors. The  reflection coefficient  measurements  of the mirrors were made  using a reflectometer (Dyn-Optics make; Model 262). 
The alignment procedure of the various mirror facets is discussed in~\cite{Tickoo-raytrace,Koulr2007}. After mirror aligment, a common focus with the minimum possible image spread was obtained at a focal plane distance of $386\,cm$ instead of at $400\,cm$, as would have been expected for a standard paraboloid or a Davies-Cotton design of the reflector. This value of focal plane distance is chosen on the basis of the simulation results~\cite{Tickoo-raytrace}.
The alignment of the mirror facets is further confirmed by observing a bright star image at the focal plane. Gross misalignment in any of the facets is easily identified  as it results in  multiple images being seen on the focal plane.
\begin{figure}[h]
\centering
\includegraphics*[width=1.0\textwidth,angle=0]{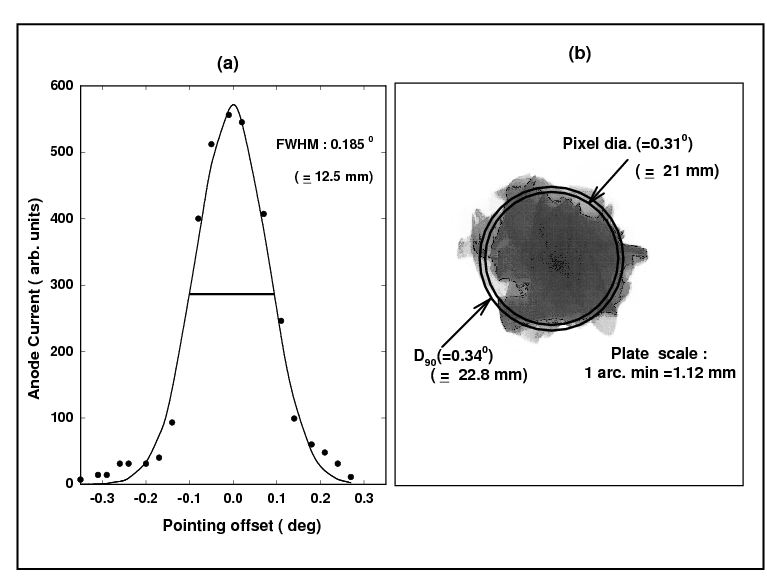}
\caption{\label{spotsize} (a) Measured point spread function of the TACTIC telescope  light collector. (b) Photograph of the image produced  by Sirius. Circles  superimposed on the image  have diameters 
$\sim 0.31^\circ$ and  $\sim 0.34^\circ$ and represent  the diameter of the camera pixels and  $D_{90}$, respectively.}
\end{figure}
\par
In order to evaluate the optical quality of the  light collector experimentally, the telescope was pointed towards the bright star  $\zeta$-Tauri and its image recorded by monitoring  the anode current of  the central pixel of the imaging  camera. The anode current versus angular offset plot  is  shown in Fig.~\ref{spotsize}(a). 
The point-spread function shown has a FWHM of $\sim 0.185^\circ$ ($\equiv12.5\,mm$)  and  $D_{90}\sim 0.34^\circ$ ($\equiv22.8\,mm$). Here, $D_{90}$ is defined as the diameter of  circle, concentric with the centroid of the image, within which $90\%$ of reflected rays lie. An image of the star Sirius recorded at the focal plane of the telescope  has also been shown in Fig.~\ref{spotsize}(b) and it has superimposed on it two circles which correspond to the diameter  of the pixel and measured $D_{90}$ value calculated on the basis of Fig.~\ref{spotsize}(a). The value of  $D_{90}\sim0.29^\circ$ ($\equiv19.3\,mm$), predicted on the basis of the simulation for an incidence angle of $0^\circ$, matches reasonably well with the measured  value mentioned above. Other details regarding the ray-tracing simulation procedure  and comparison of the measured point-spread function of the TACTIC light collector with the simulated performance of ideal Davies-Cotton and paraboloid designs  are discussed in~\cite{Tickoo-raytrace}.
\section{The imaging camera}
\label{6}
The imaging camera with an angular resolution of $0.3^\circ$ has been used in the TACTIC focal plane. The camera frame is made up of two $5\,mm$ thick  aluminium  plates  in which  $19\,mm$  diameter  holes  are drilled at a pitch of $22\,mm$  corresponding  to the locations of the 349-pixels. The PMT is held in place by a metallic  collar  fixed to its  socket  which  in turn  is held  to the rear plate by a specially designed  fastner. The central pixel of the camera which is on the  principal  axis of the light collector is used  for checking the  alignment of the mirror facets and the tracking/pointing  accuracy of the telescope.  The pixels  are numbered sequentially  clockwise  from the central pixel  which is designated  as pixel number 1.
\par
The camera uses $19\,mm$ diameter PMT (ETL-9083 UVB), with bialkali photocathode having maximum quantum efficiency of $\sim$$27\%$ at $340\,nm$ and UV glass for the window to enhance its sensitivity in the $280-300\,nm$ wavelength  band as well. The 10 stage  linear focussed PMT has a rise time of $\sim$$1.8\,ns$ which is compatible with  the time profile of the Cherenkov pulse. A low current zener diode-based voltage divider network (VDN) is used with the PMT. This VDN design~\cite{Bhat1996-vdn} has the advantage of ensuring stable  voltages at the last two dynodes with VDN current of only about $240\,\mu A$  which is a factor of 5  less than the minimum current  recommended  for a resistive VDN. The VDN uses  negative voltage and the photocathode is at a high voltage of $1000-1400\,V$ while the anode is at the ground potential. The VDN of the PMT is permanently soldered to its socket  and two  RG174 coaxial cables from each VDN circuit board are terminated with coaxial connectors on the connector panels fixed to 4 sides of the camera.
\par 
The Compound Parabolic Concentrator (CPC)  shape  was chosen for the light guides (made of SS-304) to ensure better  light collection efficiency and reduction in the NSB light falling on the PMTs. Some of the important geometrical parameters of the CPCs  used in the TACTIC telescope camera are the following: entry aperture $\sim21.0\, mm$, exit aperture  $\sim15.0\,mm$, acceptance angle $\sim45.58^\circ$ and height $\sim17.6\, mm$. The light collection efficiency of the CPC, which includes both the geometrical collection efficiency and the reflectivity of the surface was experimentally measured to be $\sim65\%$.
\section{Backend signal processing electronics and trigger generation }
\begin{figure}[h]
\centering
\includegraphics*[width=0.7\textwidth,angle=0,bb=0 0 435 292]{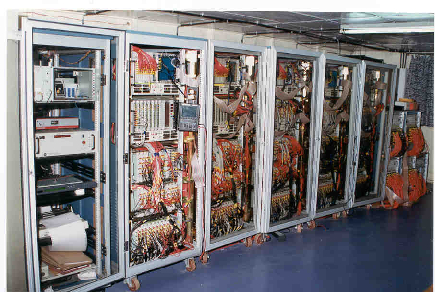}
\caption{\label{control}Photograph of TACTIC control room, showing signal processing electronics.}
\end{figure}
\subsection{Signal processing electronics}
\begin{figure}[h]
\centering
\includegraphics*[width=1.0\textwidth,angle=0]{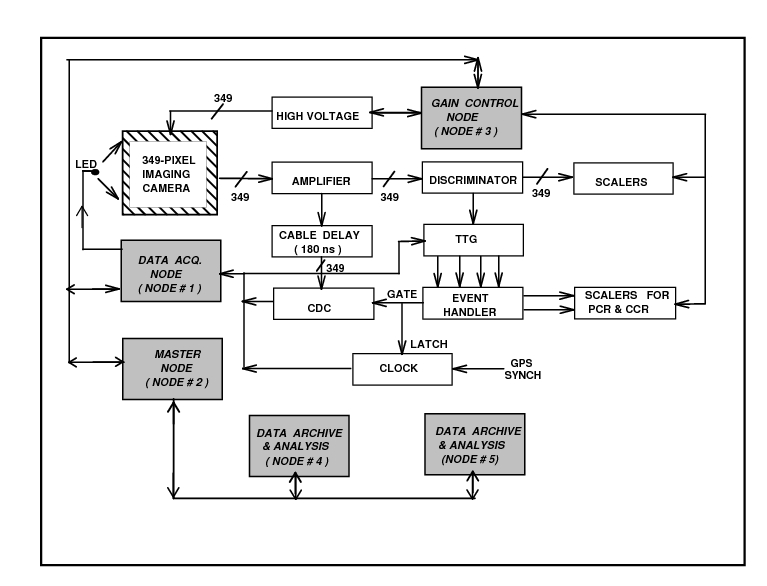}
\caption{\label{signal-elect} Block diagram of the back-end  signal processing  electronics  used in the TACTIC  imaging element; TTG - TACTIC  Trigger Generator; PCR-Prompt Coincidence Rate ; CCR- Chance Coincidence Rate.}
\end{figure}

The image of a typical Atmospheric Cherenkov Event (ACE) is registered in the form of varying amplitude pulses of $\sim3-60\,mV$ produced by a group of $5-20$ pixels of the camera. These voltage pulses are brought to the control room, using $55\,m$ long high quality RG 58 coaxial cables. In-house developed fast NIM Hex amplifier modules with a user selectable gain range of $2-50$ and amplitude discriminator modules of $50-500\,mV$ range are used for amplification and threshold selection of the PMT signals. 
While the multichannel fast NIM-based amplifier and fixed threshold discriminator modules have front panel adjustments for gain and threshold respectively, the charge content and the scaler rates are directly read off the CAMAC bus. The complete back end instrumentation shown in Fig.~\ref{signal-elect} is based on inhouse developed medium channel density modules and is housed in seven $19$ inch racks of 36 U (1U=1.75 inch) height (Fig.~\ref{control}).
One of the two outputs of a discriminator channel is used for monitoring the Single Channel Rate (SCR) with a CAMAC scaler while the other output  is connected to the  TACTIC Trigger Generator (TTG). The outputs from each of the independently operating TTG modules are then collated in an Event Handler (EH) which generates the $22\,ns$  duration gate pulse  and interrupts the data acquisition for reading the charge ADC data  from all the 349 pixels.  The  final trigger  pulse  is also  used  for   latching  the   GPS-based clock. The scalers and charge-to-digital converters (CDC) for the 349 channels use 5 CAMAC crates each.  Each of these crates is controlled using an in-house developed multi-crate CAMAC controller and  five such controllers are daisy chained and connected to a data acquisition Personal Computer (PC).  A  similar  strategy  following a custom built standard has been used for the computer-programmable High Voltage (HV) units. 
\subsection{Trigger generation }
\begin{figure}[h]
\centering
\includegraphics*[width=1.0\textwidth,angle=0]{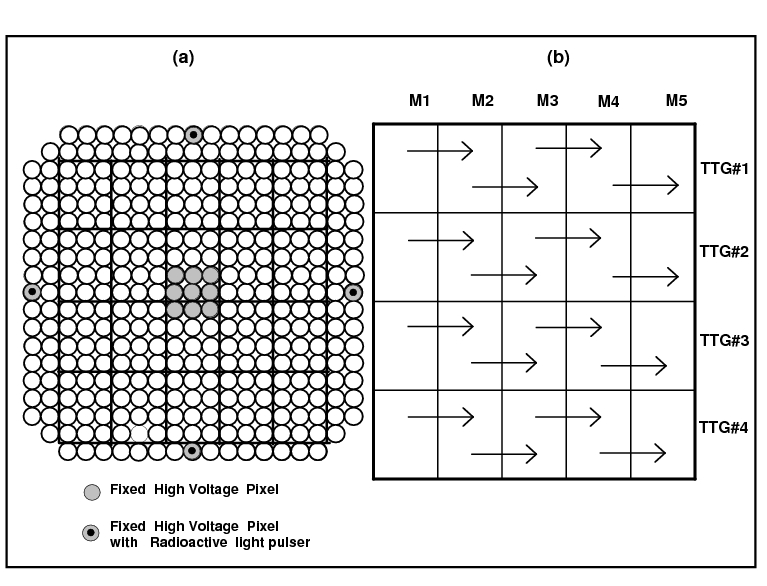}
\caption{\label{trigger} (a) Layout of the  349-pixel imaging camera of the TACTIC  telescope. Shaded pixels in the center and the same  filled  with circles at the periphery of the camera  represent  PMTs for which single channel rate is not stabilized. (b) Pictorial representation of the horizontal cascading followed  within each  trigger generator module. Each trigger generator module  can handle 60 channels  by using 5  memory ICs (M1-M5)  of   type TC55B417.} 
\end{figure}

The imaging camera uses a programmable topological trigger~\cite{Bhat1994-trigger} which can pick up events with a variety of trigger configurations. As  the trigger  scheme is not hard wired, a number of coincidence trigger options e.g Nearest Neighbour Pairs (NNP),  Nearest Neighbour  Non-collinear Triplets (3NCT) and Nearest Neighbour  Non-Collinear  Quadruplets (3NCQ) can be generated under software control.  The  layout of the 349-pixel imaging camera, which can  use  a maximum  of 240  inner pixels for trigger generation is  depicted in Fig.~\ref{trigger}(a).
The trigger criteria have been implemented  by  dividing  these inner  240 pixels into 20 groups  of 12 (3 $\times$4) pixels. A section of 5 such groups is connected to a TTG module and a total of 4 TTGs are required  for a maximum of 15$\times$16 matrix of trigger pixels. The trigger scheme has been designed around  16k x 4 bit  fast static RAM 
(Toshiba make TC55B417; access time of $<8\,ns$). Each  of the 4 TTG  modules  uses 5  memory ICs (indicated by  M1-M5  in Fig.~\ref{trigger}(b)) and  has  horizontal cascading built into it. A pictorial representation of the horizontal cascading  followed within each TTG module  is  shown in Fig.~\ref{trigger}(b).   Vertical cascading has not been provided in the TTG due to non-availability of more than 3 outputs from the discriminator. The loss of events   as a result  of the absence of vertical  cascading has been estimated to be about $12\%$ through Monte-Carlo simulations and actual on site experiment~\cite{Kaulsr2003}. The memory address lines are connected to the CAMAC address lines and the front panel receptacles (for connection to the  discriminator outputs) through two sets of tri-state buffers. The TTG operation starts with writing of the data, as per a user defined topology, from a disk file into each of its memories under CAMAC control. Once programmed, the TTG outputs follow the event topology as described earlier. Apart from generating the prompt trigger, the TTG has a provision for producing a chance coincidence output based on $^{12}C_{2}$ combinations from various groups of closely spaced 12 channels. This chance coincidence output is used as a system monitor for evaluating its overall functioning during an observation run. Monitoring of the Chance Coincidence Rate (CCR) has also helped  in keeping a close check on the operation of the telescope and the quality of the collected data. Other details  regarding the  design, implementation and performance evaluation  of the programmable TTG for the TACTIC  telescope  are discussed in~\cite{Kaulsr2003}.
\subsection{Stabilization of single channel rates}
A cost-effective  method  for operating  the  imaging camera of the TACTIC \gam telescope at stable SCR and safe anode current values  is being  used  despite  variations in the LONS experienced by the individual  pixels from time to time~\cite{Bhattn2001}. The camera  operates  9 central pixels with fixed high voltages and the remaining pixels at different high voltages to ensure  their operation within a pre-determined SCR range. The purpose  behind using the central 9 pixels of the camera at fixed high voltages is to facilitate the gain normalization (flat fielding) of the remaining 336 camera pixels, so that the  event sizes (sum of CDC counts in the clean, flat-fielded image) recorded during a night's observations can be directly compared to one another. Operation of the pixels in a narrow SCR band  has the advantage of ensuring  a stable CCR which can be used  as a system diagnostic parameter. An elaborate  algorithm~\cite{Bhattn2001} has been developed to monitor the SCR rates of all pixels using the CAMAC front ends and ensure their operation within a narrow range despite changes  in the NSB light level incident on them due to changes in the  sky brightness and star-field  rotation. The algorithm  also ensures that all the pixels of  the camera operate within  safe anode current ranges. The feedback loop of the algorithm  changes the HV to the various pixels using  a multi-channel HV unit which has a resolution of $1\,V$. The decision of operating  a pixel under enhanced light levels is solely based on the comparison of the SCR and applied high voltage with reference data generated under controlled light level conditions. The PMTs of TACTIC telescope are operated with a safe anode current limit of $\le 30\,\mu A$. A detailed description of the SCR stabilization scheme can be found in~\cite{Bhattn2001}.
\section{Data acquisition and control system}
The data acquisition and control (DAC) system of the telescope has been designed around a network of  PCs running on the QNX (version 4.25~\cite{QNX}) Real-Time Operating System (RTOS). The software is designed for the real time acquisition of event and calibration data and on-line display of telescope status in terms of Prompt Coincidence Rates (PCR), CCR and the functional status of each of the 349 pixels of the camera.
\subsection{System architecture}
\begin{figure}[t]
\centering
\includegraphics*[width=1.0\textwidth,angle=0]{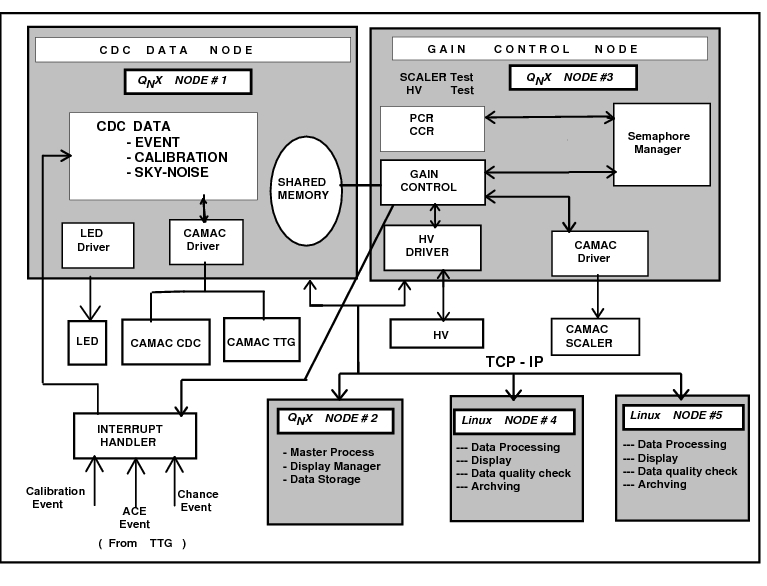}
\caption{\label{daq} Block diagram  of the multi-node  PC-based  data acquisition and control system.}
\end{figure}

As the number of jobs to be handled by the DAC system is fairly large in addition to the high throughput of scaler and CDC data from the 349 channels, we have used a network of 3 PCs equipped with QNX RTOS to perform the various functions of the data acquisition and control. The QNX RTOS was chosen for its multi-tasking, priority-driven scheduling and fast context switching capabilities.  In addition, the QNS RTOS also provides a  powerful set of Inter Process Communication (IPC) capabilities via messages, proxies  and signals. The DAC of the TACTIC is handled by a network of three  PCs as shown in Fig.~\ref{daq}. 
While one PC  is used to monitor the SCRs and control the HV to the PMTs, the other PC handles the acquisition of the event and calibration data and the programming of the TTGs.  These two front-end PCs,  referred to as the rate stabilization node and the data acquisition node respectively,  along with a master node form the multinode DAC network of the TACTIC Imaging telescope. All executable routines stored on the master node are spawned on to the other two front-end nodes as and when required.
The same network is extended to two more LINUX-based PCs which are used for on-line data analysis and archiving.
An EH module described in the next section, controls the whole process of data acquisition.
\subsection{Event handler: Interface between CAMAC hardware and the application software}
\begin{figure}[t]
\centering
\includegraphics*[width=0.8\textwidth,angle=0]{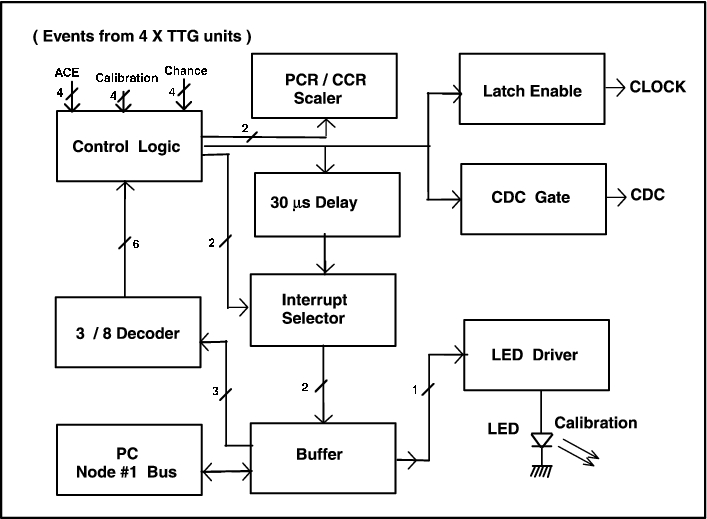}
\caption{\label{ev-handler}Block diagram of the event handler module which provides the interfacing between the CAMAC hardware and application software of the data acquisition and control system of the TACTIC Imaging Element.}
\end{figure}

The EH module shown in a block diagram form in Fig.~\ref{ev-handler} provides the link between the TACTIC hardware and the application software. The EH accepts the ACEs, calibration and chance trigger outputs from various TTG modules and interrupts the front end data acquisition node. The interrupt driven system uses two interrupts at levels 3 and 4 for better data segregation of the various event types of the telescope as two different interrupt service routines are used for the data acquisition. The system acquires the 349 channel CDC data for the trigger selected ACEs, relative calibration flashes generated by the calibration LED and sky pedestal events, in addition to CDC data for the 4 absolute calibration channels.  The HV and scaler data are also logged continuously, though at a much lower frequency.
\par
The acquisition of CDC data for an ACE is handled at interrupt level 3 while the calibration and sky pedestal data are acquired using interrupt level 4. The switch over of these interrupt levels is managed by the EH hardware under the control of the data acquisition software and works in the following manner. The rate stabilization node detects the stabilization of the camera and informs the data acquisition node accordingly which in turn sets a flag in the EH resulting in the change in the interrupt level from 3 to 4. A logic pulse is also sent to the LED driver which produces a $20\,ns$ duration light flash using the LED (Nichia, Japan make type SPB 500) for the purpose of Relative Gain Calibration (RGC) of the 349 pixels of the camera. This flash which is sensed by all the operating pixels of the camera produces a ‘calibration output’ from the TTG modules and interrupts the DAC system at the interrupt level 4 selected earlier. At event occurrence the EH also generates a TTL output for latching the system clock and a $20\,ns$ wide NIM pulse for gating the CDC modules. The hardware interrupt to the data acquisition node is generated after a delay of $\sim30\,\mu s$ to account for the CDC conversion time. After every event the EH goes into a blocking mode until the CDC data of all the channels have been read out. After the generation of the preset number of calibration pulses, the EH sends out user programmed number of ‘false triggers’ to acquire the sky pedestal data from all the channels. Even though the RGC data and the sky pedestal data are acquired using the same interrupt level, their segregation is possible because of the fixed number of these events. After the acquisition of the preset number of sky pedestal events the control software resets the EH flag leading to the selection of interrupt level 3 for the acquisition of the ACE data. This process of interrupt level switching continues during the course of observation.
\subsection{Data acquisition and control software}
The multinode DAC software is designed for the real-time acquisition of event and calibration data and on-line display of telescope status in terms of PCR, CCR and the status of each of the 349 pixels of the camera. As each specific job is handled by an independent process, the software comprises of about 6000 lines of C-code distributed over 15 routines and device drivers. Ten concurrent processes running across the network of 3 PCs under the QNX RTOS handle the various aspects of DAC. All executable routines stored on the master node have priorities (0 to 31) assigned to them and are spawned on to the other nodes as and when required. Some of the main processes of the data acquisition system are listed in Table~\ref{processes}. The software is also provided with all necessary safeguard measures to ensure trouble-free operation of the telescope. Furthermore, the system also provides single-point monitoring and control from a designated node with elaborate GUI facilities for monitoring the functions of various sub-systems. Quasi on-online analysis of the data recorded by the telescope is also performed to evaluate the data quality and to look for the presence of a \gam signal in the data.  More details regarding  hardware and software features of the DAC system of the telescope are discussed in~\cite{Yadav2004-daq}.\\ 
\begin{table}[t]
\caption{List of processes involved in the TACTIC data aquisition and control system.}
\centering
\begin{tabular}{|c|c|c|}
\hline 
Process name & Description & Priority\\ 
& & (node)\\
\hline
$st\_tactic$ & Spawns and controls the execution of all other processes & 23\\
& & (2)\\
\hline
$gain\_ctl$ & Monitors the scaler rates and controls the high & 20\\
& voltage of all the channels& (3)\\
\hline
$da\_ctl$ & Controls the execution of cal\_rel, cal\_sky and da\_acq pro- & 20\\
& cesses and keeps the time information in the shared memory & (1)\\ 
\hline
$cal\_rel$ & Acquires CDC data for all the channels corresponding & 21\\
& to relative calibration & (1)\\
\hline
$cal\_sky$ & Acquires CDC data for all the channels corresponding& 21\\
& to sky pedestal runs & (1)\\
\hline
$da\_acq$ & Acquires CDC data for all the channels corresponding& 21\\
& to a Cherenkov event & (1)\\
\hline
$semaphore$ & Regulates the access to a shared hardware & 17\\
& amongst various peocesses & (3)\\
\hline
$abs\_calib$ & Acquires the absolute calibration data from 4 & 21\\
& specified channels and is part $da\_acq$ process& (1)\\
\hline
$pcr\_ccr$ & Acquires and displays the prompt and chance & 11\\
& coincidence rates & (3)\\
\hline
$prog\_mcc$ & Programs the RAM of the trigger generator modules& 10\\
& as per user selected trigger configuration & (1)\\
\hline
$cdc\_ped$ & Acquires the CDC pedestal data for all the channels & 10\\
& & (1)\\
\hline
$sync\_clk$ & Synchronises the clocks of all the PCs on the network & 10\\
& with the master node & (2)\\
\hline
\end{tabular} 
\label{processes}
\end{table}
\\
\textbf{Master process:} \\The software is activated by the main process $st\_tactic$ which runs on the master node with the highest priority and controls the execution of all other processes. On activation this process checks the network integrity and synchronizes the clocks of the other two PCs with the master node clock. Clock synchronization of the three nodes is important because after being spawned from the master node the various processes use the local PC clock for reference. Network integrity is checked by establishing a virtual circuit between processes on the local and remote nodes. The clock synchronization is followed by the spawning of a few other basic processes like $Semaphore$, $gain\_ctl$ and $da\_acq$ on their respective nodes. A menubar of 16 icons each of which is associated with a specific task is also displayed on the master node. This master process communicates with other processes running on remote nodes to coordinate their execution by using messages and signals. The master process has a number of sequential interlocks built into it to ensure safe operation of the telescope.
\par
A $semaphore$ has been developed under QNX to coordinate the functioning of multiple processes in the TACTIC software which are running on the same node and accessing the same CAMAC bus. This semaphore has been implemented using the IPC features of QNX. All processes accessing shared resources need to go through the semaphore to ensure that no other process is currently using the resource. In case of simultaneous access a semaphore list is prepared in which the highest priority process gets preference in execution.\\
\begin{figure}[t]
\centering
\includegraphics*[width=0.8\textwidth,angle=0,bb=0 0 358 239]{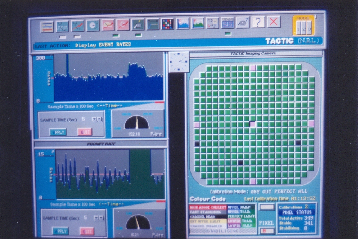}
\caption{\label{node}The pictorial representation of the 349-pixel imaging camera as created on the master node by the Gain Control Process. scrolling plots for PCR and CCR are also shown at the left of the pannel.}
\end{figure}
\\\\\\
\textbf{Gain control process:} \\This process running on the Gain Control node controls the gain of the PMTs by changing the HVs applied to them on the basis of their SCR resulting from the ambient light-induced shot noise fluctuations~\cite{Bhattn2001}. The process adjusts the HV of each of the PMTs to ensure that its SCR is within the predefined range. The information about channel number, HV and SCR is sent to the remote master node for display. The pictorial representation of the 349-pixel imaging camera as depicted in Fig.~\ref{node} is also created by this process on the master node. This colour coded pictorial representation makes it convenient for the user to identify and switch off/on any malfunctioning pixel. An alarm is sounded whenever a pixel operates beyond its safe limits and the HV to the camera is switched off whenever more than $30\%$ of the pixels operate in an unsafe range.
\par
The gain control process apart from ensuring the operation of the PMT within safe anode current limits is also responsible for maintaining a stable value of SCR of individual pixels despite being invariably exposed to different NSB light levels which can also include a bright star sometimes. These changes in the LONS background are further enhanced by the phenomenon of star field rotation~\cite{Tickoo1999} which is inherent to altitude-azimuth mounted telescopes, like the TACTIC. This causes the image of an off-axis bright star in the field of view of the telescope to trace a circular path in the focal plane of the telescope thereby giving rise to large variations in the anode current and the SCR of concerned pixels over a time scale of tens of minutes. Since TACTIC imaging element hardware does not use any anode current monitoring ADCs, the same is measured in an indirect manner by keeping track of the SCR. If the SCR of a pixel under actual conditions is higher than the predesignated SCR range, the first attempt of the gain control process is to operate it at a lower HV to ensure that the PMT still remains ON, so that minimal information content is lost when the telescope gets triggered by an ACE. But if the recorded SCR is found to be more, indicating that the anode current of the PMT is still more than $30\,\mu A$, the pixel is switched off. \\\\
\textbf{Data acquisition and control process:}\\ This process which runs on the data acquisition node controls the two processes on the local node which acquire event and calibration data. These two processes running with the same priority record the event data and the Relative Gain Calibration (RGC) data from all the 349 channels along with time information on the occurrence of an event. A proxy is attached to each of these processes which is triggered by the interrupt handler. Operating in a mutually exclusive mode the $da\_acq$ process records the ACE data while the $cal\_rel$ and $cal\_sky$ processes record the RGC and sky noise data, respectively. The Absolute Gain Calibration (AGC) data of the 4 designated channels is acquired by the $abs\_calib$ process which is also controlled by the $da\_acq$ process. This process also makes available the real-time event information to all the processes which need it by using the shared memory.\\\\
\textbf{Prompt and chance coincidence rates:} \\The PCR and CCR processes running on the gain control node generate scrolling displays of the rates on the master node as shown in Fig.~\ref{node}. Any malfunctioning of the system is reflected immediately on these displays in terms of variations in these rates. This process provides facilities for changing online the integration time and the full scale count of the rate displays. These processes access the same CAMAC hardware as the gain control process through the Semaphore. The number of calibration runs taken by the system and the effective observation duration is also displayed on the console.
\par
Apart from the above listed main processes a number of stand alone routines for testing the various subsystems of the telescope are also available in the software. These routines are operated from the menu bar as and when required.\\\\
\textbf{Data filing system and data quality evaluation:} \\In view of the large data volume recorded in an observation spell, the TACTIC data are stored on an hourly basis in a binary format on the master node. The system generates 15 files at the start up and adds 5 new files to it every hour. These 5 files which are referred to as ‘cdc’, ‘cal’, ‘sky’, ‘scr’ and ‘abs’ contain information regarding ACEs, RGC of the pixels, sky pedestal values, SCR values and AGC data, respectively. Apart from the AGC file which contains data for only 4 channels, all other files mentioned above have information about all the 349 pixels used in the camera. These files are embedded with absolute time information with a resolution of $1\,\mu s$ so that the data can be collated properly at the analysis stage. In addition to these files, several other files common to the whole observation spell are also generated by the system. These contain details regarding HV applied to the pixels, list of disabled pixels and pixels operating at fixed HV, PCR, CCR and log of other system parameters. With each event consuming a space of $\sim708\,bytes$, the amount of data collected during a typical observation spell lasting for $5\,hours$ (assuming an event rate of $\sim2.5\,Hz$ with 7 calibration runs) turns out to be about $60\,MB$.
\par
Quasi on-line analysis of the data recorded by the telescope is also performed to evaluate the quality of the data recorded by the system and to identify hardware-related problems. The procedure involves transferring the hourly data files from the Master Node to Node4/Node5 by using standard FTP protocol. Some of the checks which are routinely followed include analyzing the RGC and sky pedestal files so that pixels with either low or high gain can be identified and adjusted during the course of an observation spell. Checks like this have been found to be very useful for ensuring recording of good-quality data with minimum loss of observation time. The analysis code written in C++ runs under the Linux operating system and is equipped with elaborate data visualization tools using the interactive data analysis system ROOT~\cite{ROOT,ROOT-NIM}. Apart from providing a graphical display of image parameter distributions, the software also contains suitable routines for extracting \gam signal from the recorded data by applying various event selection cuts.
\par 
The data acquisition and control system of the TACTIC telescope has been working quite satisfactorily. The dead time of the system has been experimentally measured to be $\sim2.5\,ms$ by collecting the RGC data along with absolute time information at a trigger rate of $\sim400\,Hz$. After performing a time series analysis of the timing data it was found that any further increase in the event rate resulted in loss of some events by the data acquisition system. However, the loss of actual Cherenkov events corresponding to an event rate of $\sim2.5\,events\,s^{-1}$, by considering a non-paralizable dead time model turns out to be only $0.63\%$ which is safely acceptable.
\par
The gain control algorithm of the data acquisition and control system has been upgraded recently so that minimum possible time is spent on stabilizing the SCR of camera pixels. The modified approach being used now requires at least 3 pixels from the ‘trigger zone’ to cross the calibration limit for initiating a fresh calibration call. This has resulted in a substantial saving of the PMT gain control stabilization time and has allowed us to use about $80\%$ of the actual observation time for collecting data on a candidate \gam source.
\section{Gain calibration scheme}
There are two schemes employed for calibrating the camera PMTs of the TACTIC telescope and are known as RGC and AGC. Currently the procedure followed by us for AGC, uses the same data which have been collected for the purpose of RGC.
\subsection{Relative gain calibration}
RGC scheme uses a high intensity  blue LED placed at a distance of $\sim2\,m$ from the camera surface to determine the relative gain of the camera pixels. The LED has been enclosed with a light-diffusing medium to ensure the uniformity of its photon field  onto the imaging camera within $\sim \pm 6\%$. The mean light intensity from the pulsed LED recorded by each pixel, in response to 2000 light flashes is subsequently used for off-line RGC of the camera PMTs.
\subsection{Absolute gain calibration}
In order to determine the AGC of PMTs, for the TACTIC telescope,  two different procedures have been used:\\  
(\textit{i}) The first procedure for determining AGC of the camera PMTs  involves  monitoring the absolute gain of a  set of 4 gain calibrated  pixels placed at the periphery of the camera as shown in  Fig.~\ref{trigger}(a). Since measurement of the absolute gains of these PMTs  by determining  their  single  photoelectron peaks  a number of times  during  an observation run is rather time consuming, we have instead  used a relatively simpler method  of  measuring the light pulser yield of a calibrated  source  for  the  in-situ  determination  of the absolute gain of these calibration channels. The calibrated  light  sources used is the Am$^{241}$-based light pulsers (Scionix make) which produces fast optical flashes at an average rate of $\sim20\,Hz$ with maximum emission at a wavelength of $\sim370\,nm$. The underlying principle for converting the charge content of an uncalibrated pixel from CDC counts to  photoelectrons, uses the fact that the calibration pixels are also exposed to the light flashes from the LED during the relative calibration run and hence it becomes possible to obtain the conversion factors for all the remaining 345 pixels of the camera~\cite{Tickoo2002abscalib}.\\
(\textit{ii}) The conversion for image size in CDC counts to number of photoelectrons  has also been  performed  independently by using the excess noise factor method. Here we use the RGC data to estimate the conversion factor. After compensating for the gain normalisation between PMTs, we calculate the variance in CDC counts between tubes for each RGC event and relate to the average number of photoelectron ($pe$). This relationship involves a larger variance than would be inferred from Poisson counting statistics alone due to fluctuations in the electron multiplication process~\cite{Biller1995}. Considering $N_{pe}$ and $\sigma_{pe}$ are the average number of $pe$ and correponding variance respectively at the photo cathode while $N_{cdc}$ and $\sigma_{cdc}$ are measured average number of digital counts and measured variance in this count respectively. The following relations can be written between these quantities:
\begin{equation}
 \sigma_{pe}=\sqrt{N_{pe}}
\end{equation} 
\begin{equation}
 N_{cdc}=g\,\sigma_{pe}
\end{equation}
where g is the gain in units of digital counts per pe. The measured variance in digital counts can be written as
\begin{equation}
 \sigma_{cdc}=g\,\sqrt{\alpha}\,\sqrt{\sigma_{pe}}
\end{equation}
where $\alpha$ is a correction factor to take into account the additional fluctuations (caused due to electron multiplication process at the dynodes of the PMT). Using these relations the expression for the gain in terms of the measured quantities can be expressed as
\begin{equation}
 g=\frac{1}{\alpha}\frac{\sigma^2_{cdc}}{N_{cdc}}
\end{equation}  
The analysis of relative calibration data yields a value of $1pe\cong$ ($6.5\pm1.2$) CDC for this conversion factor ($g$) when an average value of $\alpha \sim1.7$ is used for excess noise factor of the PMTs. Further work to determine the excess noise factor values more precisely for TACTIC PMTs is  still underway. 
\chapter{Data analysis procedure}
\section{Introduction}IACTs do not directly detect the radiation emitted by the astrophysical sources being observed, but indirectly by monitoring the Cherenkov photons emitted by the secondary particles (mostly electrons) produced in the EAS initiated by the incident \gam photon. The resulting ACR pulse is digitized by using the camera PMTs and associated electronics. The ACR images recorded in the camera are analyzed offline by subjecting them to various standard analysis steps like noise cleaning, gain normalisation of PMTs and their parameterization. The statistical significance of the source detection is then evaluated by applying ``gamma domain cuts'' to the parameterized images (also called events). The energy estimation of a primary \gam photon and subsequent determination of the observed source spectrum are the important steps in the data analysis chain. The data analysis software used to obtain results presented in this thesis has been developed on the LINUX plateform using C code integrated with the ROOT-library which is an object oriented data analysis framework developed at CERN~\cite{ROOT}.
\par
Monte Carlo simulations have been performed to optimize gamma domain cuts, determination of effective collection area and the quality factor of the ACT used for signal processing. In this chapter, we shall briefly discuss the simulation studies carried out for the TACTIC imaging element relevant to the present work and the detailed analysis procedure (starting with raw image to the spectrum determination), which have been followed. Details of the simulation studies for the TACTIC telescope like its energy threshold, trigger efficiency, optimization of imaging cuts and their energy dependence, estimation of the sensitivity can be found in~\cite{Koulmk1999,Koulmk2002,Sapru2002,Sathyabama1999,Koulmk2003,Koulmk2005}. We begin with the discussion of Hillas parameters of an ACR image by describing its analytical formalism and physical meaning.
\section{Parameterization of ACR images}
\subsection{Hillas parameters}
\begin{figure}[t]
\centering
\includegraphics*[width=0.75\textwidth,height=0.55\textheight, angle=0,bb=0 0 555 735]{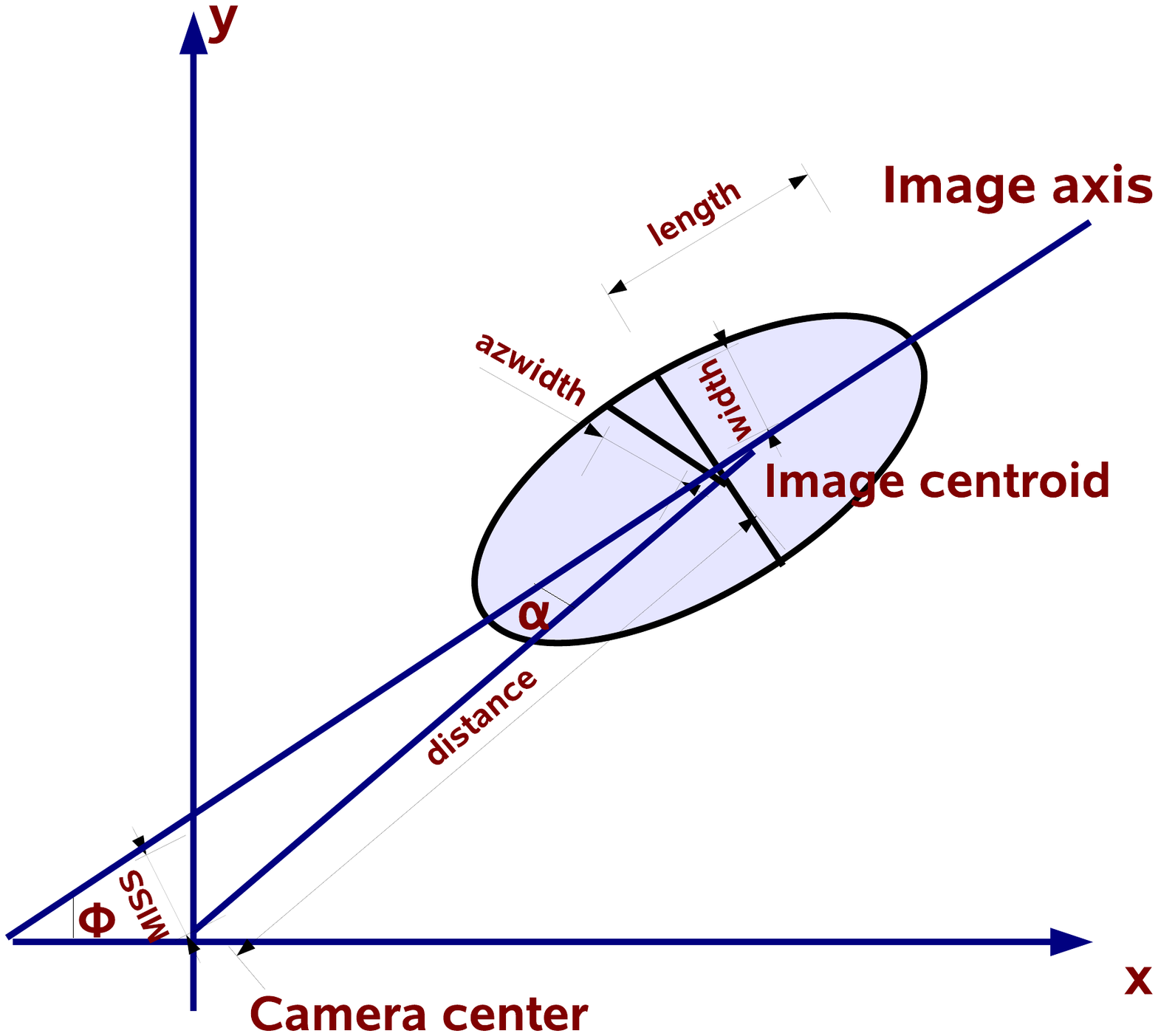}
\caption{\label{hillas} Illustration of image parameters.}
\end{figure}

The images of ACR observed by an individual IACT are typically analyzed by using moment analysis of the images to derive a set of quantities that characterize their roughly elliptical shape. These are referred to as the Hillas parameters after M. Hillas, who first proposed these parameters~\cite{Hillas1985} (see Fig.\ref{hillas} for their illustration). Moments are based on the CDC counts recorded in each pixel, together with its coordinates. Hence image parameters are reconstructed from the information contained in the pixels. Here we summarize the important image characteristics used in IACT image analysis
\begin{itemize}
 \item \textbf{Size (S):} The overall light content of the image is parameterized by the total number of $pe$ contained in it. To the first approximation, the size is proportional to the energy of the primary particle. The size energy relation also depends on the impact parameter and and zenith angle of the observation. 
\end{itemize}
\begin{itemize}
 \item \textbf{Length (L):} The RMS angular size along the major axis of the ellipse and related to the longitudinal development of the shower. This parameter gives a measure of the parallax angle to the shower maximum and increases with the impact parameter.
\end{itemize}
\begin{itemize}
 \item \textbf{Width (W):} The RMS angular size along the minor axis of the ellipse and related to the lateral development of the shower.
\end{itemize}
\begin{itemize}
 \item \textbf{Distance (D):} The distance from the centroid of the image to the center of the field of view of the camera. This parameter gives information about the impact point of the shower i.e. a crude measure of core distance for a \gam shower. It can also be viewed as the angle between the shower axis and the line joining the shower maximum and the telescope. In case of a stand alone telescope, this parameter is very important for energy estimation.
\end{itemize}
\begin{itemize}
\item \textbf{Miss (M):} The perpendicular distance of the major axis of the image from the center of field of view of the camera. It is a measure of the shower orientation.
\end{itemize}
\begin{itemize}
\item \textbf{Azwidth:} The RMS angular size along a line, which is perpendicular to the line joining the image centroid to the center of field of view. It is a measure of both the shape and the orientation of the image.
\end{itemize}
\begin{itemize}
\item \textbf{Concentration} or \textbf{Frac2 (F2):} Ratio of the two largest pixel signals to the sum of all signals. It represents the degree of light concentration and hence is a measure of the compactness of the image.
\end{itemize}
\begin{itemize}
 \item \textbf{Alpha ($\alpha$):} The angle between the image axis ( i.e. shower axis) and the line joining the image centroid and the camera center (source position). It is basically the angle between shower axis and the optical axis of the telescope and hence is an orientation parameter. As events with small $alpha$ point towards the source position, it is one of the most powerful parameters for the $\gamma$-hadron separation. The uncertainty in $alpha$ is mainly due to the finite pixel size and the signal fluctuations ($\propto\,1/\sqrt{size}$)~\cite{Wagner2006}.
\end{itemize}
\begin{itemize}
 \item \textbf{Asymmetry:} It is a measure of the symmetric shape of the image. The \gam images should have tails which preferentially point away from the source position. 
\end{itemize}
\subsection{Estimation of image parameters}
\label{imageparameters}
Estimation of the image parameters can be done on the basis of moment analysis. The zero-order moment is the size, i.e. the sum of CDC signals in all the PMTs (also called pixels) which are non-zero in a clean image. The first-order moments describe the position of the image in the focal plane of the camera and the second-order moments describe the extent of the image. Generally first and second order moments are sufficient for parameterization of the images. In this section, we summarize the formulae used to calculate the Hillas image parameters from the calibrated PMT signals~\cite{Weekes1989}. To calculate the image parameters, we first need to determine the image axis, which can be expressed as the equation of a straight line with slope $m$ and intercept $c$ 
\begin{equation}
y = mx +c
\end{equation} 
The constants $m$ and $c$ are estimated in such a way that the resulting line minimizes the signal weighted sum of squares of perpendicular angular distances of the pixels i.e $\chi^2$ minimization. Considering x$_i$ and y$_i$ denote the pixel coordinates of the $i^{th}$ pixel in degree with origin being the center of the camera and $s_i$ denotes the calibrated signal in the PMT, we can write the expression for $\chi^2$ as  
\begin{equation}
\label{chi}
\chi^{2} = \sum_{i=1}^n \frac{s_{i}\,(y_{i}-m\,x_{i}-c)^{2}}{1+m^{2}}
\end{equation} 
where $n$ is the total number of pixels. Assuming that both x$_i$ and y$_i$ have equal errors $m$ and $c$ can be obtained by differentiating the equation (\ref{chi})  
The following quantities can be defined according to their standard definitions, 

\begin{equation}
<x> = \frac{\sum_{i=1}^n (s_i\,x_i)} {\sum_{i=1}^n s_i} \qquad\qquad <y> = \frac{\sum_{i=1}^n (s_{i}\,y_{i})} {\sum_{i=1}^n s_i}
\end{equation} 

\begin{equation}
<x^2> = \frac{\sum_{i=1}^n (s_i\,x_i^{2})} {\sum_{i=1}^n s_i} \qquad\qquad  <y^2> = \frac{\sum_{i=1}^n (s_i\,y_i^2)} {\sum_{i=1}^n s_i}
\end{equation} 

\begin{equation}
<x^3> = \frac{\sum_{i=1}^n (s_i\,x_i^3)} {\sum_{i=1}^n s_i}  \qquad\qquad  <y^3> = \frac{\sum_{i=1}^n (s_i\,y_i^3)} {\sum_{i=1}^n s_i}
\end{equation} 

\begin{equation}
<xy> = \frac{\sum_{i=1}^n (s_i\,x_i\,y_i)} {\sum_{i=1}^n s_i}
\end{equation} 

\begin{equation}
<x^2y> = \frac{\sum_{i=1}^n (s_i\,x_i^2\,y_i)} {\sum_{i=1}^n s_i} \qquad\qquad  <xy^2> = \frac{\sum_{i=1}^n (s_i\,x_i\,y_i^2)} {\sum_{i=1}^n s_i}.
\end{equation} 
Spread of the images in different directions can then be defined in terms of moments as
\begin{equation}
\sigma_{x^2}= <x^2> - <x>^2   \qquad\qquad \sigma_{y^2}= <y^2> - <y>^2 
\end{equation} 

\begin{equation}
\sigma_{xy}= <xy> - <x><y>. 
\end{equation} 
Some other higher order moments are
\begin{equation}
\sigma_{x^3}= <x^3> -3<x^2><x> + 2<x>^3 
\end{equation}

\begin{equation}
\sigma_{y^3}= <y^3> -3<y^2><y> + 2<y>^3  
\end{equation}

\begin{equation}
\sigma_{x^2y}= <x^2y> -2<xy><x> - <x^2><y> + 2<x>^{2}<y>  
\end{equation}

\begin{equation}
\sigma_{xy^2}= <xy^2> -2<xy><x> - <y^2><x> + 2<y>^2<x>.  
\end{equation}
Defining
\begin{equation}
d = \sigma_{y^2} - \sigma_{x^2} 
\end{equation}

\begin{equation}
z = \sqrt{d^2+ 4(\sigma_{xy})^2}.
\end{equation}

The constants of the image axis are then written as follows:

\begin{equation}
m = \frac{d+z}{2\sigma_{xy}}
\end{equation}

\begin{equation}
c = <y> - m<x>.
\end{equation}

The image parameters can be calculated from these moments and can be written as
\begin{equation}
length, \,\,\,L = \sqrt{\frac{\sigma_{x^2} + \sigma_{y^2} + z}{2}}
\end{equation}

\begin{equation}
width, \,\,\,W = \sqrt{\frac{\sigma_{x^2} + \sigma_{y^2} - z}{2}}.
\end{equation}
Since $<x>$ and $<y>$ are the coordinates of the image centroid, therefore
\begin{equation}
distance, \,\,\,D = \sqrt{<x>^2 + <y>^2}
\end{equation}

\begin{equation}
miss, \,\,\,M = \sqrt{\frac{1}{2}((1 + \frac{d}{z}) <x>^2 + (1 - \frac{d}{z})<y>^2) - \frac{2\sigma_{xy}<x><y>}{z}}
\end{equation}

\begin{equation}
alpha, \,\,\,\alpha = sin^{-1} \left(\frac{M}{D}\right)
\end{equation}

\begin{equation}
azwidth, \,\,\, Az = \sqrt{(L^2\,sin^2\alpha + W^2\,cos^2\alpha)}
\end{equation}

\begin{equation}
asymmetry = \frac{(\sigma_{x^3}\,cos^3\phi + \sigma_{y^3}\,sin^3\phi + 3\sigma_{x^2\,y}\,cos^2\phi\,sin\phi + 3\sigma_{x\,y^2}\,cos\phi\,sin^2\phi)^{1/3}}{L} 
\end{equation}
\begin{equation}
 concentration \,\,\,F2=\frac{s_{Ist\,max}+s_{IInd\,max}}{\sum_{i=1}^n s_i}.
\end{equation} 
Where $s_{Ist\,max}$ and $s_{IInd\,max}$ are maximum and second maximum signal respectively contained in pixels of an image.
\par
The image parameters described here are used for $\gamma$-hadron separation. Even after applying \gam domain cuts (discussed later in this chapter) some fraction ($\sim 0.2\%$) of CR images is still classified as \gams due to the similarity of processes between shower development initiated by \gam and CR primaries. These CR events essentially become background for the \gam images. We have described here the $\alpha$ parameter generally used to extract the \gam signal from the background of CR images. This method of $\gamma$-hadron separation called ``$\alpha$ approach'' is source dependent i.e. it uses the information of the source position in the camera. Another source independent procedure called ``$\theta^2$ approach'' can also be used to extract the \gam signal. In this procedure source independent parameters e.g. $length$, $width$, $size$ etc. are used. While the ``$\alpha$ approach'' is useful for point like \gam sources, the analysis of many extended galactic sources or on purpose off-source observation can be better performed by ``$\theta^2$ approach''. The later is often used for data taken in the Wobble mode (discussed later) of observation and uses the method called $Disp$ analysis discussed here. However, in this thesis, we follow the ``$\alpha$ approach''.
\begin{figure}[t]
\centering
\includegraphics*[width=0.55\textwidth,height=0.3\textheight, angle=0,bb=0 0 293 242]{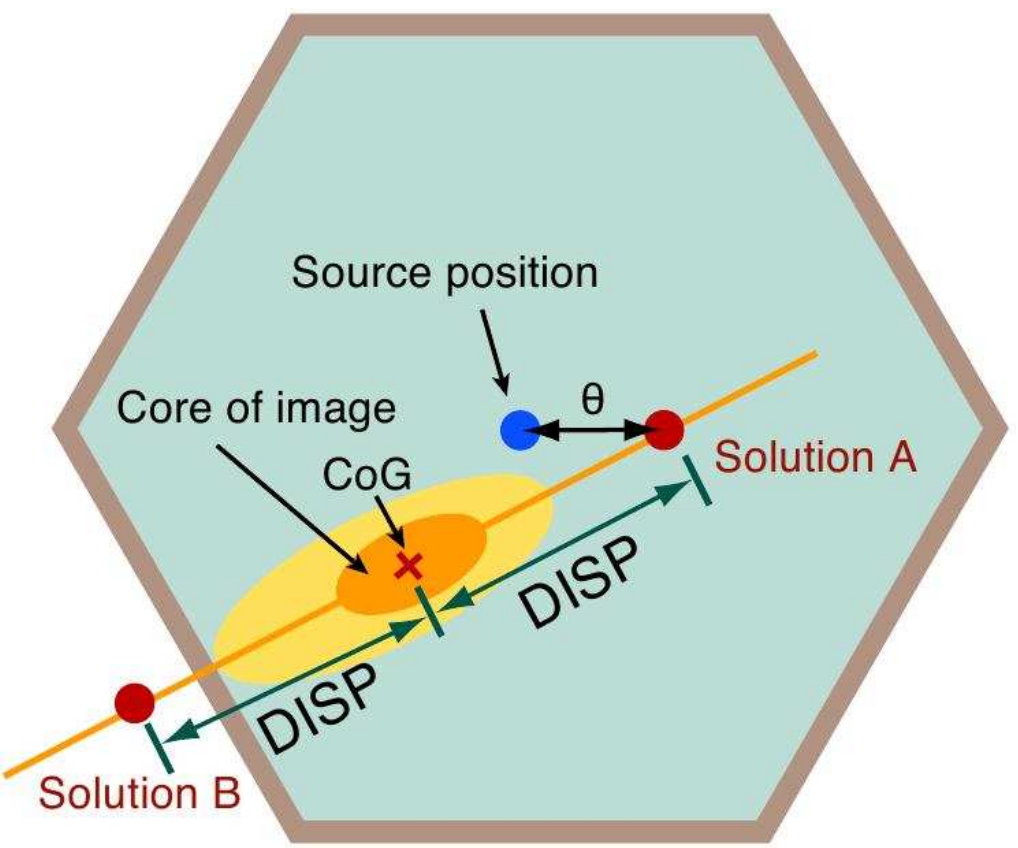}
\caption{\label{disp} Sketch demonstrating the $Disp$ and $\theta$ parameters. The $Disp$ parameter leads to two estimated sky directions of an event (represented by solution A and B). The measurement of the asymmetry in the image along the major axis i.e. determining the head and tail of the image can provide the correct solution. Here the source is in the camera center and therefore solution A is the correct one. $\theta$ is the angle between the estimated and real source positions. This figure has been adapted from~\cite{Dmazin-thesis}.}
\end{figure}
\subsection{$Disp$ procedure}This procedure was first put forward by Fomin~\cite{Fomin1994} and exploits the fact that the image elongation $\left(\frac{Width}{Length}\right)$ depends on the distance of the image centroid or Center of Gravity (CoG) of a \gam image from the source position in the camera plane. As depicted in Fig.~\ref{disp},  the angular distance between the real and estimated source positions called $\theta$ is obtained using the $Disp$ method~\cite{Lessard2001,Domingo-Santamaria2005}. The $Disp$ parameter is defined as the angular distance between the expected source position and the CoG of the shower image along the major axis. In the $Disp$ method the arrival direction of an EAS is estimated from the image shape parameters on an event-by-event basis assuming that the source (incoming direction of the primary particle) lies on the major axis of the image. This is plausible because the major axis is a projection of the shower axis onto the camera. Ideally the image axis coincides with the trajectory of a \gam photon~\cite{Dmazin-thesis}.
\par
The elongation of the shower image and the value of the $Disp$ parameter depend upon the impact parameter of the shower on the ground. For small impact parameter, the image should have a form close to that of a circle and be positioned very near the source position in the focal plane~\cite{Fomin1994}. For increasing impact parmeter, it should become elongated and have a form close to that of an ellipse and be positioned farther from the source position in the focal plane. A simple form for the relationship between the elongation of an image and the $Disp$ is~\cite{Lessard2001} 
\begin{equation}
 Disp = \xi \left( 1- \frac{Width}{Length}\right)
\end{equation} 
where $\xi$ is a scaling parameter. Generally, $\xi$ depends on the height of the shower in the atmosphere, the zenith angle of observation, parameters of the model of the atmosphere and the energy of the primary particle. Experimentally, $\xi$ is determined from the data so that the calculated shower arrival directions line up with a known source position.
\par
There is an important degeneracy in the $Disp$ method: it provides two source positions along the major axis of the image (see Fig.~\ref{disp}) i.e. it does not provide information on which side of the shower image the arrival direction lies. Therefore a method to select the correct source position is needed. Images in the camera plane carry information about longitudinal development of the shower. The ``asymmetry'' charge distribution in the image contains the ``head and tail'' information of the recorded shower i.e. which edge of the image is closer to the source position in the camera plane.  Cherenkov photons from the upper part of the shower produce a narrow section of the image with higher photon density (head) than the photons arriving from the shower tail which normally produce spread out end of the image~\cite{Domingo-Santamaria2005}. 
\par
The $\theta^2$ distribution is expected to be flat for the background CR events and has an exponential shape for a \gam signal peaking at $\theta^2=0^\circ$~\cite{Dmazin-thesis}. By Using $Disp$ method and a proper camera acceptance model a sky map of the arrival directions of the \gam events can be produced~\cite{Wagner2006}.
\section{Monte Carlo Simulations for TACTIC}
The development of an EAS in the atmosphere is a stochastic random process whose details can not be described by parametric functions or analytic expressions. Therefore Monte Carlo simulation are required to optimize the design and performance of an IACT. Further Monte Carlo simulations are helpful in understanding the reconstruction of \gam initiated showers and their segregation from much more abundant showers of hadronic origin. In order to determine the effective collection area of the telescope, it is necessary to simulate its trigger efficiency. The energy reconstruction procedure also requires simulated \gam showers to estimate the energy of the primary particle.  While \gam simulations agree quite well with the simple theory forming regular elliptical images, hadronic showers are dominated strongly by fluctuations in the initial nuclear interactions.
\subsection{CORSIKA}While a number of Monte Carlo simulation codes e.g. CORSIKA (COsmic Ray SImulations for KAscade)~\cite{corsika}, MOCCA, KASCADE, ALTAI and GrISU (\cite{Aharonian2008} and references therein) are being used in VHE \gam astronomy, we have used the CORSIKA (version 5.6211) air shower simulation code~\cite{corsika} for predicting  and optimizing  the performance of the TACTIC imaging  telescope. Originally developed to perform simulations for KASCADE experiment~\cite{Doll1990}, CORSIKA is a detailed Monte Carlo code to study the evolution of EAS in the atmosphere initiated by protons, nuclei or any other particle. Its applications range from Cherenkov telescope experiments ($\sim10^{12}$\,eV) up to the highest energies ($\sim10^{20}$\,eV) observed in EAS. The development of CORSIKA is guided by the idea to predict not only the correct average values of the observables but also to reproduce the correct fluctuations around the average value. The simulations of the electromagnetic part in an air shower in CORSIKA are treated with the code called EGS4, while hadronic interactions are simulated by models, GHEISHA for low energies ($<10\,GeV$) and VENUS, QGSJET, DPMJET, SIBYLL, etc. for higher energies ($>10\,GeV$) (\cite{corsika} and references therein).
\subsection{TACTIC simulation chain}
\label{simulationchain}
Generally, the complete execution of Monte Carlo simulations for an IACT is sub-divided into two parts. The first part comprises generation of air showers induced by different primaries and recording the relevant raw Cherenkov data (data-base generation), while folding in the light collector characteristics and PMT detector response is performed in the second part. In order to carry out the simulation study for the TACTIC telescope, we have used the altitude of the Mount Abu \gam observatory i.e.$1300\,m$, corresponding to air mass grammage $\sim890\,g\,cm^{-2}$ and the horizontal and vertical magnetic field component values of $35.9\,\mu\,Tesla$ and $26.6\,\mu\,Tesla$ respectively~\cite{Tickoo2002}. The Cherenkov photon wavelength band of $300\,- 450\,nm$ have been chosen for generating the data-base. The simulated data-base for \gam showers uses about $34000$ showers in the energy range $0.2-20 \,TeV$ with an impact parameter of $5-250\,m$. These showers have been generated at 5 different zenith angles ($\theta= 5^\circ$, $15^\circ$, $25^\circ$, $35^\circ$ and  $45^\circ$). A data-base of about $39000$ proton initiated  showers in the energy range $0.4-40\, TeV$, have been used for studying the $\gamma$-hadron separation capability of the telescope. The representative CR integral energy spectrum of the form $E^{-1.7}$ has been used in the data-base generation. The incidence angle of the proton showers is simulated by randomizing the shower directions in a field of view of $6^\circ\times6^\circ$  around  the pointing direction of the telescope. A photon bunch received by a detector at the observatory level is characterized using the CORSIKA code by seven parameters, namely bunch size, x and y coordinates, two direction cosines, height of production and arrival time at the ground. The Cherenkov photons are ray-traced~\cite{Tickoo-raytrace} to the detector focal plane and the number of \textit{pe} likely to be registered in a PMT pixel are inferred after folding in the atmospheric absorption of the photons, optical characteristics of the mirrors, the metallic compound-paraboloid light concentrators at the entrance window of the pixels and the photocathode spectral response. The Cherenkov  photon data-base, consisting  of number of \textit{pe} registered by each pixel, is then subjected to noise injection, trigger condition check and image cleaning.
\par
The Cherenkov pulses are detected in the presence of LONS in actual experimental conditions. To estimate this contribution, measurements on a few PMTs are carried out in the laboratory with the help of a steady light source of a similar intensity as that of LONS at Mount Abu. The resulting shot noise is then superimposed on the ACR generated \textit{pe} content of each pixel. To revert it back i.e. to exclude the shot noise contribution generated by LONS, a standard two level ``tail cut'' procedure has been used to obtain the clean image~\cite{Fegan1996}. The aim of this procedure (i.e. image cleaning) is to retain maximum pixels containing Cherenkov light while suppresing the pixels with signal from LONS alone. The procedure involves the following steps:
\begin{enumerate}[(i)]
 \item Pixels with less than $3.0\sigma$ signal (so called ``\textbf{boundary threshold}'') are set to zero.
 \item Pixels with more than $6.5\sigma$ signal (known as ``\textbf{picture threshold}'') are retained.
 \item Pixels with signal in between $3.0\sigma$ and $6.5\sigma$ are retained provided they are adjacent to a pixel satisfying condition (2).
\end{enumerate}
The resulting data-bases, consisting of \textit{pe} distribution in the imaging camera at various core distances and zenith angles are then used for estimating trigger efficiency, effective detection area, optimum ranges of Cherenkov image parameters for discriminating between \gam and CR events, differential count rate and effective threshold energy of the telescope for \gam and CR proton events and finally developing the energy reconstruction procedure for \gam events. The clean Cherenkov images are characterized by calculating their standard image parameters like $length$, $width$, $distance$, $alpha$, $size$ and $Frac2$ as described in section~\ref{imageparameters}. The standard Dynamic Supercuts procedure~\cite{Mohanty1998} is used to separate \gam like images from the background CR. 
\subsection{Procedure for $\gamma$-hadron separation}ACT is quite efficient in $\gamma$-hadron separation above $100\,GeV$. The technique exploits the characteristic differences between the shower development processes initiated by these two primary particles and consequently the Cherenkov image generated at the telescope camera. Based on these differences several strategies have been developed for $\gamma$-hadron separation, which include supercut procedure, extended or dynamic supercut procedure and cluster analysis~\cite{Mohanty1998}. The sensitivity of an IACT strongly depends on the rejection capabilities of CR background events and can be improved by considering the energy and zenith angle dependence of image parameters. A comparative study has been carried out for the TACTIC telescope to evaluate the performance of supercuts, dynamic supercuts and zenith angle dependent dynamic supercuts. The details of this study can be found in~\cite{Koulmk2005}.
\par
The performance of different \gam selection methodologies is quantified by the quality factor ($Q$) for the telescope, which is basically the ratio of fractions of \gam and CR retained events after the selection procedure is applied. In the following we summarize the supercut procedure since it is the basic methodology and then the dynamic supercut procedure which accounts for the energy dependence of image parameters.\\\\
\textbf{Supercut procedure:} \gam images have $length$, $width$ and $alpha$ parameters, which are restricted to a narrow range of values as compared to proton images. This can be used to optimize the range of parameters to maximize the significance of the source detection. We have optimized the ranges of image parameters for the simulated data-base that results in maximum Q value i.e. the maximum acceptance of \gams with minimum contamination of background CR proton events. The range of these values for \gam domain are listed in Table~\ref{supercuts}.
\begin{table}[h]
\caption{\gam domain supercuts.}
\centering
\begin{tabular}{|c|c|}
\hline 
size & $S\ge50$ pe \\
\hline 
distance & $0.5^\circ \le D \le 1.2^\circ$\\
\hline
length & $0.11^\circ \le L \le 0.33^\circ$\\
\hline
width & $0.05^\circ \le W \le 0.17^\circ$\\
\hline
alpha & $\alpha \le 18^\circ$\\
\hline
\end{tabular} 
\label{supercuts}
\end{table}
Note that the size cut is applied to eliminate events dominated by noise, while $distance$ parameter (rather poor discriminator) is restricted to maintain the energy resolution for the telescope~\cite{Tickoo2002}.
\par
Apart from resulting in comparatively lower quality factor, there are other disadvantages in using this approach directly for the extraction of spectra because $length$ and $width$ of the images increases with the energy of the primary VHE \gams, and this exactly is taken care of in dynamic supercut methodology.\\\\
\textbf{Dynamic supercut procedure:} A straight forward method to incorporate the energy dependence in selection procedure is to scale the upper and lower values for each parameter with the energy of the primary \gam photon, which of course is unknown. However the image size is a fairly good estimate of energy and $length$ and $width$ are also well correlated with the $log(size)$. The first order polynomial fit of these correlations ($length$ v/s $log(size)$ and $width$ v/s $log(size)$) can be used to obtain the dynamic supercut ranges of parameters. The upper values of these parameters for the TACTIC telescope have been optimized for higher \gam acceptance compared to the supercuts, while lower values have been kept fixed because of insignificant effect. The ranges of dynamic supercuts for VHE \gams are given in Table~\ref{dynamicsupercuts}.
\begin{table}[h]
\caption{Range of \gam domain dynamic supercuts.}
\centering
\begin{tabular}{|c|c|}
\hline 
size & $S\ge50$ pe \\
\hline 
distance & $0.5^\circ \le D \le 1.2^\circ$\\
\hline
length & $0.11^\circ \le L \le (0.27527 + 0.023\,ln(S))$\\
\hline
width & $0.05^\circ \le W \le  (0.06356 + 0.026\,ln(S))$\\
\hline
alpha & $\alpha \le 18^\circ$\\
\hline
\end{tabular} 
\label{dynamicsupercuts}
\end{table}
 \begin{figure}[t]
\centering
\includegraphics*[width=0.5\textwidth,height=0.6\textheight, angle=270,bb=59 56 557 747]{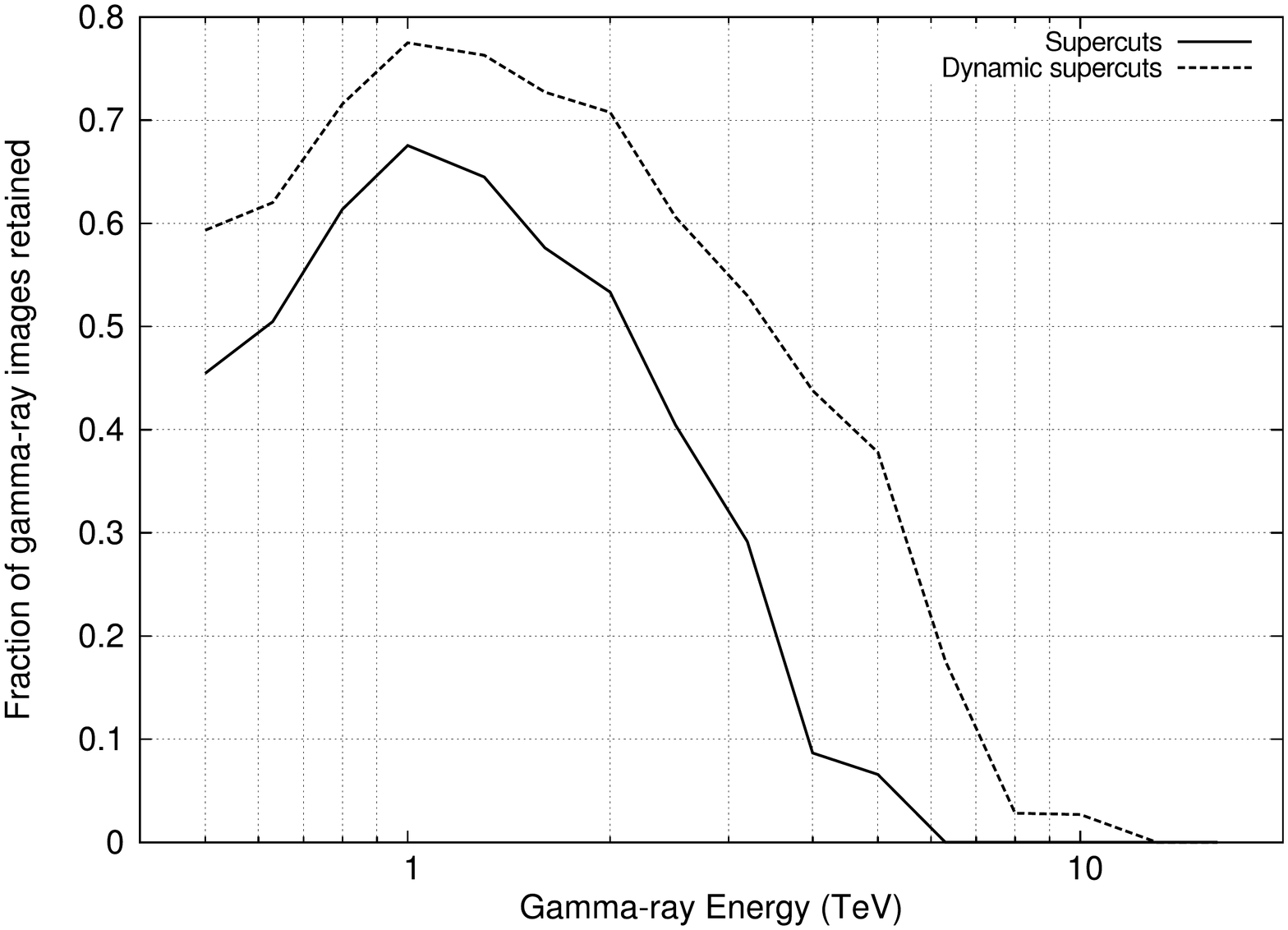}
\caption{\label{gam-ret}\gam retention factor as a function of primary energy.}
\end{figure}
\par
The \gam retention factor as a function of \gam energy has been compared in Fig.\ref{gam-ret} for supercut and dynamic supercut procedures. Comparatively better \gam retention for dynamic supercuts is evident. We have used the dynamic supercut procedure in the work presented in this thesis. 
\subsection{Comparison with real data}
\begin{figure}[t]
\centering
\includegraphics*[width=0.9\textwidth,height=0.4\textheight, angle=0]{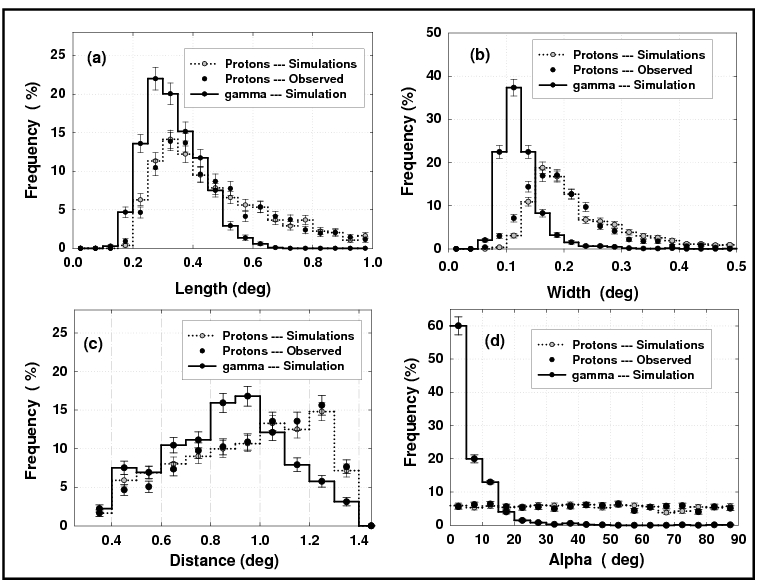}
\caption{\label{para-com}Comparison of observed and simulated image parameter distributions.}
\end{figure}

The agreement between the expected and actual performance of the telescope can be checked by comparing the expected and observed image parameter distributions. Fig.\ref{para-com} shows the distributions of the image parameters $length$, $width$, $distance$ and $alpha$~\cite{Hillas1985,Weekes1989} for simulated protons and for the actual Cherenkov images recorded by the telescope. The simulated distributions of these image parameters for \gams have also been shown in the figure for comparison. The observed image parameter distributions are found to closely match the distributions obtained from the simulations for proton-initiated showers, testifying to the fact that the event triggers are dominated by background CRs. 
\section{Evaluation of telescope parameters}In order to obtain the observed \gam flux from the detected number of \gams, we need to use several characteristic parameters of the telescope e.g trigger efficiency, effective area, threshold energy and CR detection rate, etc. These characteristic parameters in turn depend on design features of the telescope i.e. size of the light collector, pixel size, field of view of the camera, trigger generation scheme, single pixel threshold level and the image parameter cuts. We describe here some of these important parameters.
\subsection{Single pixel threshold and topological trigger}
\begin{figure}[h]
\centering
\includegraphics*[width=0.6\textwidth,height=0.6\textheight, angle=270]{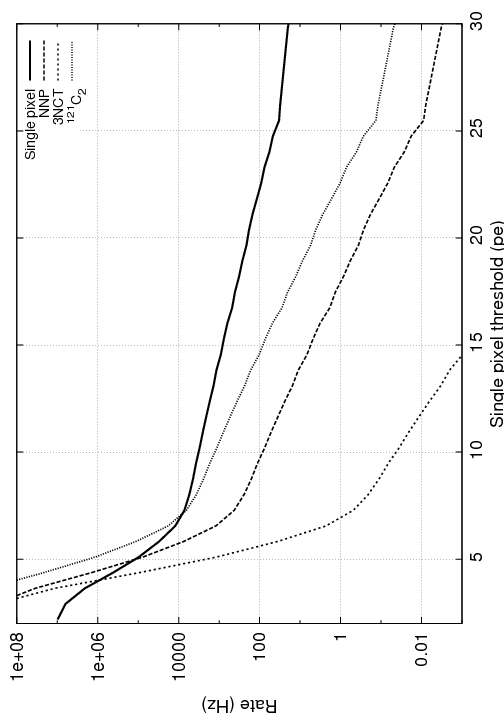}
\caption{\label{pixel-th}Chance coincidence rates for different topological trigger schemes as a function of single pixel threshold. The discrimination level is converted into $pe$ after obtaining the single photoelectron spectrum for the PMT.}
\end{figure}
The NSB strongly limits the energy threshold of an IACT since the trigger threshold is normally set to the LONS fluctuations. However, LONS contribution can be reduced by employing the smaller pixel size and limiting the charge integration time. The variation of SCR as a function of its trigger threshold can be used to select the appropriate trigger scheme, which also governs the trigger threshold energy of the telescope. Further, the variation of SCR with threshold level can be different from a pure Possionian  distribution because of the afterpulsing in PMTs~\cite{Mirzoyan1997}. The measured variation of SCR with trigger threshold for a PMT is shown in Fig.\ref{pixel-th}. Given the large number of pixels forming the trigger region, the telescope trigger rate turns out to be extremely high. In order to suppress these unwanted random triggers, we use some topological schemes for trigger generation. Several topological trigger configurations e.g. NNP, 3NCT and $^{121}C_2$ have been worked out and resulting CCR have been plotted for comparison in Fig.\ref{pixel-th}.
\par
By fixing up the tolerable CCR (say $10^{-2}\, Hz$ as shown in the figure) one can compare the single pixel thresholds for different trigger schemes. For example, in NNP trigger configuration the single pixel threshold can be set at $\sim25\,pe$ while 3NCT configuration allows us to go down to $\sim8\,pe$ level for the same value of CCR. It is important to mention here that lowering the single pixel threshold can in principle be compensated by increasing the trigger multiplicity. However, the optimum value of single pixel threshold which finally decides the threshold energy of the telescope is strongly dependent on the intensity of Cherenkov photons in the image plane and the pixel size used~\cite{Tickoo2002}. Both NNP and 3NCT trigger schemes have been used for the TACTIC telescope from time to time, here we shall use NNP trigger scheme for the evaluation of other telescope parameters.
\subsection{Trigger efficiency}Trigger efficiency $\eta(r,E)$, is the probability that the telescope gets triggered by a \gam or proton shower of primary energy $E$, arriving at a distance r from the telescope. In simulations, it is defined as
\begin{equation*}
\eta(r,E) = \frac{number \,of\, showers\, triggering \,the\, telescope}{total\, number\, of \,showers\, generated}. 
\end{equation*}
Apart from core distance and energy of the primary particle, trigger efficiency also depends on the nature of primary particle and hence on the Cherenkov photon density, trigger field of view of the camera, the single pixel threshold level and trigger generation scheme. The trigger efficiencies for the TACTIC imaging telescope have been evaluated for 3NCT and NNP trigger schemes at $8\,pe$ and $25\,pe$ threshold levels respectively, both for \gam and proton generated Cherenkov events.
\subsection{Effective collection area}
\begin{figure}[t]
\centering
\includegraphics*[width=0.9\textwidth,height=0.45\textheight, angle=0]{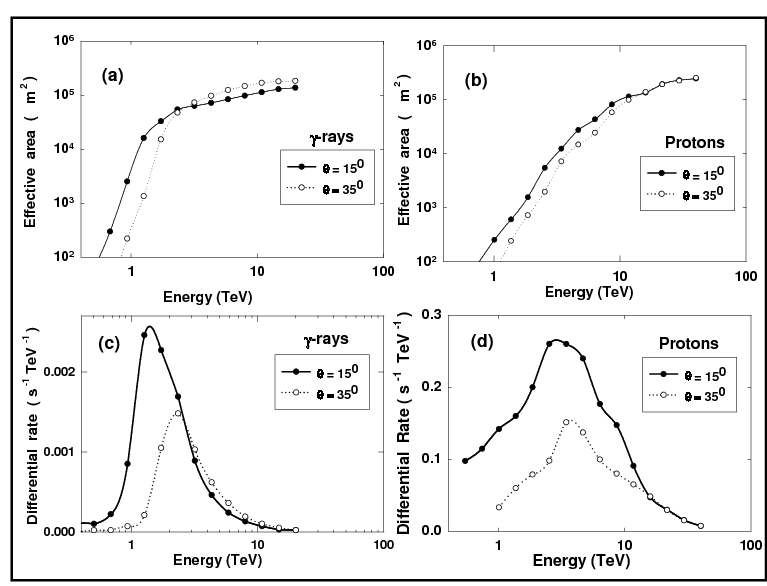}
\caption{\label{eff-area-rate}Effective collection area of the telescope for \gam (a) and protons (b) as a function of the primary energy at zenith angles of $15^\circ$ and  $35^\circ$. Differential trigger rates for \gam (c) and proton events (d) as a function of the primary energy. These results have been obtained by using the NNP trigger scheme with $11\times11$ pixels trigger field and a single pixel threshold of $\geq25 pe$.}
\end{figure}

Cherenkov telescopes have an effective collection area determined by the area of the ACR pool at the ground which has a typical radius of $\sim120\,m$. Such a huge effective collection area makes the Cherenkov telescopes highly efficient for the detection of VHE \gams where flux becomes significantly low due to the power law nature of most of the emissions from astrophysical sources. The energy dependent effective collection area $A_{eff}(E)$ can be written in a general form as
\begin{equation}
 A_{eff}(E) = 2\pi \int_0^\infty r \times \eta(r,E)\,dr.
\end{equation}
This integration is normally discretized in simulations and for the TACTIC imaging telescope the effective collection area corresponding to each primary energy value is evaluated from the following equation
\begin{equation}
 A_{eff}(E) = \pi \sum_{i=1}^n (r_{i+1}^2-r_i^2)\frac{\eta(r_{i+1})+\eta(r_i)}{2}
\end{equation}
where $n=20$, $r_1=5\,m$ and $r_{21}=205\,m$. The evaluated effective collection area for the TACTIC imaging telescope has been shown in Fig.\ref{eff-area-rate} (a) and (b) for \gam and proton induced ACEs respectively, at two representative zenith angle values of $15^\circ$ and $35^\circ$. These results were obtained by using the NNP trigger scheme with $11\times11$ pixels trigger field and a single pixel threshold of $\ge25\,pe$. While significant variation can arise in the calculation of effective area at higher primary energies, the contribution of such events is small due to the power law nature of \gam spectrum of the VHE sources.
\par
The effective collection area for the isotropic CR background can be evaluated from the following equation
\begin{equation}
  A_{eff}(E) = 2\pi \int_0^\infty\int_0^\infty r \times \eta(r,E,\Omega)\,dr\,d\Omega
\end{equation}
where $d\Omega$ is the solid angle subtended by a detector element. 
\subsection{Telescope trigger rates and threshold energies}The differential trigger rate ($\frac{dR}{dE}$) of a telescope is defined as the number of particles with energies between $E$ and $E\,+\,dE$, triggering the telescope per unit time. In simulations, the differential trigger rate for a particle with known differential spectrum ($\frac{dN}{dE}$), is obtained by directly multiplying the differential spectrum with the energy dependent effective collection area ($A_{eff}$). Therefore the differential trigger rate can be written as 
\begin{equation}
\label{diff-rate}
 \frac{dR}{dE}= A_{eff} \times \frac{dN}{dE}.
\end{equation} 
The peak of the differential trigger rate curve as a function of energy determines the threshold energy of the telescope for a given type of primary particle. The integral trigger rate is then obtained by simply integrating equation (\ref{diff-rate})
\begin{equation}
 R(E > E_{th})=\int_{E_{th}}^{E_{max}} \frac{dR}{dE}\,dE
\end{equation} 
where $E_{th}$ is the energy threshold of the telescope for a given type of particle and $E_{max}$ is the maximum energy of the primary particle that can be detected by the telescope. The differential energy spectra used in the TACTIC simulations in the energy range $0.3-50\,TeV$ has the following form for \gam~\cite{Aharonian00-simul} and CR protons~\cite{Aharonian1997}
\begin{equation}
\label{gamma-flux}
\frac{dN_\gamma}{dE}= 2.79\times10^{-7}E^{-2.59}\qquad photons\,\,m^{-2}\,\,s^{-1}\,\, TeV^{-1}
\end{equation}
  \begin{equation}
\label{proton-flux}
\frac{dN_p}{dE}= 1.10\times10^{-1}E^{-2.75}\qquad particles\,\,m^{-2}\,\,sr^{-1}\,\,s^{-1}\,\, TeV^{-1}
\end{equation}
respectively. Fig.\ref{eff-area-rate} (c) and (d) show the differential event rates as a function  of the primary energy  for \gam and proton events, respectively. It is evident from these plots that the \gam trigger threshold energy of the telescope is $\sim1\,TeV$. The corresponding trigger threshold  energy  of the telescope for protons turns out to be $\sim2.5\,TeV$. The reason for this difference in threshold energies is the higher Cherenkov photon density (almost twice) produced by \gam photon EAS as compared to that of proton initiated EAS. Additionally, proton images are more diffuse, and are less likely to reach the trigger threshold. At higher energies ($>20\,TeV$), the photon density is sufficiently high for both \gams and protons to efficiently trigger the telescope and the effective area is mainly governed by the radius of the ACR pool. The effective area for protons, at still higher energies ($>50\,TeV$) increases as compared to \gams, (see Fig.\ref{eff-area-rate} (b)), due to the triggering of the telescope by proton showers, whose directions fall even outside the trigger field of view of the camera.
\par
Showers arriving at large zenith angles develop their shower maximum at relatively greater heights with respect to the telescope. This dependence goes approximately as $Height$ $\propto1/cos(\theta)$, where $\theta$ is the zenith angle of the telescope. This allows the showers arriving at larger core distances to remain within the trigger field of view of the camera. Thereby geometrically increasing the effective collection area at large zenith angles. Again, because of the geometry, the projection of ACR pool on the ground for a shower arriving at large zenith angle will be larger. The dependence of radius of ACR pool also goes approximately as $\propto 1/cos(\theta)$ and consequently the Cherenkov photon density goes down as nearly $\propto 1/cos^2(\theta)$. At higher energies, the effective collection area increases with increasing zenith angle purely due to the geometrical effect as long as the photon density is sufficient to trigger the telescope. At lower energies, the trigger efficiency decreases nearly the same way as the Cherenkov photon density and results in decrease in  effective collection area with increasing zenith angle. Correspondingly, the threshold energy of the telescope also increases at higher zenith angles. These effects have been taken into account while obtaining the energy spectrum of a \gam source.
\par
The agreement between the predictions from Monte Carlo simulations and the actual performance of the telescope have been checked by comparing the observed trigger rate of the telescope with the predicted value. The expected prompt coincidence rate at a zenith angle of $15^\circ$  turns out to be $\sim2.5\, Hz$ for the NNP trigger mode.  This value  has been  obtained  on the basis of integrating the differential rate  curve for protons Fig.\ref{eff-area-rate} (d).  Reasonably good  matching of this with the experimentally  observed value of $\sim 2-3\, Hz$ suggests that the response of the telescope is very close to that predicted by simulations. 
\subsection{Sensitivity}As already discussed in chapter 3, the sensitivity of a \gam telescope is its ability to detect a \gam signal over the CR background. The sensitivity of an IACT can be expressed in two ways:
\begin{enumerate}[(i)]
 \item If the flux of the source is known (e.g. Crab Nebula like sources), we estimate the sensitivity in terms of the \textbf{minimum time required to detect the \gam signal}, which is statistically significant (usually taken as $5\sigma$).
 \item When the flux of the source is unknown, the source is observed for a fixed time period (usually $50\,hours$) and sensitivity is expressed as \textbf{the minimum \gam flux required to detect the signal} at $5\sigma$ statistical significance level.
\end{enumerate}
We shall follow the first approach here. The number of \gam events detected by an IACT, which survive \gam domain parameter cuts can be written as
\begin{equation}
 N_\gamma = \int_{E_{min}}^{E_{max}} \frac{dN_\gamma}{dE}\,A_\gamma(E)\,\eta_\gamma(E)\,T\,dE
\end{equation}  
where $\eta_\gamma(E)$ is \gam retention factor and the integration is over the dynamic energy range of the TACTIC telescope i.e. $\sim1.0\,TeV$ to $\sim20.0\,TeV$. The lower limit of the dynamic energy range signifies the threshold energy while the upper limit corresponds to the energy at which the charge content of the ACE is sufficiently high to saturate the CDC counts. The number of detected events, generated by protons and again surviving after \gam domain parameter cuts, can similarly be written as
\begin{equation}
 N_p = \int_{E_{min}}^{E_{max}} \frac{dN_p}{dE}\,A_p(E)\,\eta_p(E)\,T\,\Omega\,dE
\end{equation}
where $\eta_p(E)$ is proton retention factor, $\Omega$ is the solid angle of the telescope and $T$ is the observation time. The statistical significance ($N_\sigma$) of a \gam signal detection is defined as
\begin{equation}
 N_\sigma = \frac{N_{on}-N_{off}}{\sqrt{N_{on}+N_{off}}} = \frac{N_\gamma}{\sqrt{N_\gamma + 2N_p}}
\end{equation}
where $N_{on}=N_\gamma + N_p$ and $N_{off}= N_p$ are the number of events detected in on-source and off-source regions. The above equations can be solved for the minimum time required to obtain $N_\sigma$ significance, which can be expressed as
\begin{equation}
 T_{min} = N^2_\sigma\left( \frac{1}{\int_{E_{min}}^{E_{max}} \frac{dN_\gamma}{dE}\,A_\gamma(E)\,\eta_\gamma(E)\,dE}+2\Omega \frac{\int_{E_{min}}^{E_{max}} \frac{dN_p}{dE}\,A_p(E)\,\eta_p(E)\,dE}{(\int_{E_{min}}^{E_{max}} \frac{dN_\gamma}{dE}\,A_\gamma(E)\,\eta_\gamma(E)\,dE)^2}\right).
\end{equation}
After substituting the appropriate values for $\eta_\gamma$, $\eta_p$, $A_\gamma$, $A_p$, $\Omega$ and flux values for \gam and protons from equations (\ref{gamma-flux}) and (\ref{proton-flux}) respectively, the minimum time required to detect a \gam signal from the Crab Nebula (flux is given by equation (\ref{gamma-flux})), at a statistical significance of $5\sigma$ turns out to be $\sim25\,hours$.
\par
This sensitivity figure has been  confirmed repeatedly by  analyzing the data collected on the Crab Nebula. Data collected on the Crab Nebula for $\sim$$101.44\,hours$ between 10 November 2005 - 30 January 2006 have yielded  an excess of $\sim$(839$\pm$89) \gam events with a statistical significance of $\sim$9.64$\sigma$.
\subsection{Energy reconstruction procedure for \gam events}
Given the inherent power of Artificial Neural Network (ANN) to effectively handle the multivariate data fitting, an ANN-based energy estimation procedure has been developed for determining the energy spectrum of a candidate \gam source. An ANN is a non-linear statistical data modeling tool which can be used to model complex relationships between inputs and outputs. The tool can be applied to problems like function approximation or regression analysis (similar to  what is being attempted here), classification (pattern recognition) etc. Although not for energy estimation, the idea of applying ANN to imaging telescope data was attempted for the first time by Reynolds and Fegan \cite{Reynolds1995}. While a detailed description of ANN methodology can be found in \cite{Reynolds1995,Lang1998}, we present here only the main steps which have been followed to ensure reliability of end results.
\par 
The \gam energy reconstruction with a single imaging telescope, in general, is a function of image $size$, $distance$ and $zenith\, angle$. The procedure followed  by us uses a 3:30:1 (i.e.3 nodes in the input layer, 30 nodes in hidden layer and 1 node in the output layer) configuration of the ANN with resilient back propagation  training algorithm \cite{Reidmiller1994} to estimate the energy of a \gam like event on the basis of its $image size$, $distance$ and $zenith\, angle$. The training and testing of the ANN  was done in accordance with the standard procedure of dividing the data base into two parts so that one part could be used for the training and the remaining for testing.  We used about 10,000 events out of a total of 34000 parameterized \gam images for training. The 3 nodes in the input layer correspond to $zenith\, angle$, $size$ and $distance$, while the 1 node in the output layer  represents  the expected  energy (in $TeV$) of the event. In order to make ANN training easier and to smoothen inherent event to event fluctuations, we first calculated the $<size>$ and $<distance>$ of the training data sample of 10000 images by clubbing together showers of a particular energy in various core distance bins  with  each bin having a size of $40\,m$. Once satisfactory training of the ANN was achieved, the corresponding ANN generated weight-file was then used as a part of the main data analysis program so that the energy of a \gam like  event could be predicted without using the ANN software package. Rigorous checks  were  also performed to ensure that the network did not become ``over-trained'' and the configuration used was properly optimized  with respect to number of iterations and number of nodes in the hidden layer.  Optimization of the network was done  by  monitoring the RMS error while training the ANN. The optimized configuration (3:20:1 with 5000 iterations) yielded a final RMS error of $\sim0.027$ which  reduced only marginally when the number of nodes in the hidden layer or the number of iterations were increased further.
\par
\begin{figure}[t]
\centering
\includegraphics*[width=0.95\textwidth,height=0.5\textheight,angle=0]{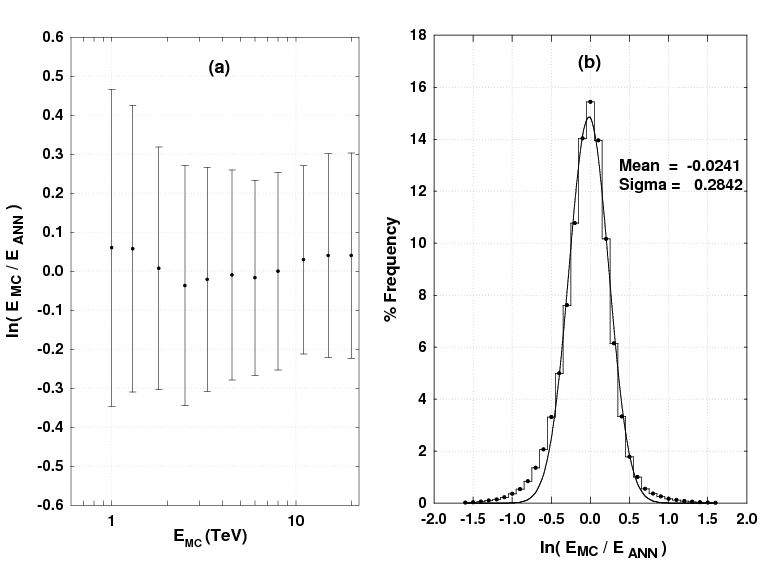}
\caption {\label{ann}(a) Error in the energy resolution function ln($E_{MC}$/$E_{ANN}$) as a function of the  energy of the \gams for the simulated  showers generated  within $5^\circ-45^\circ$  zenith angle range. (b) Frequency distribution of ln($E_{MC}$/$E_{ANN}$) for all events shown in Fig.~\ref{ann}(a) along with a Gaussian fit.}
\end{figure}

A plot of energy reconstruction  error obtained  for a test data sample of 24000 parameterized images  is shown in Fig.~\ref{ann}(a). This plot has been obtained  by testing the ANN  with individual parameterized images at different energies within $5^\circ-45^\circ$ zenith angle range. Denoting the estimated energy by $E_{ANN}$ and the actual energy used during  Monte Carlo simulations by $E_{MC}$, Fig.~\ref{ann}(b) shows the frequency distribution of ln(E$_{MC}$/E$_{ANN}$) for all events shown in Fig.~\ref{ann}(a) alongwith a Gaussian fit. 
Having a  value of $\sigma$(ln E) $\sim28.4\%$  Fig.~\ref{ann}(b),  directly implies that  the  procedure should allow  us to retain  $\sim84\%$  \gams in most of the energy bins from $1-20\,TeV$, if 6 bins per energy decade i.e. $\sigma$(ln E) of$\sim40\%$ are used for determining the energy spectrum of a candidate \gam source. Note that the energy resolution of $\sim40\%$ has been used here to avoid spill over of events in between the neighbouring energy bins. The performance of the ANN-based energy reconstruction procedure was also compared with the results obtained from the linear least square fitting method. This method yielded a $\sigma$(ln E) of $\sim35.4\%$. It is also worth  mentioning here that the new ANN-based energy reconstruction method used here, apart from yielding a lower  $\sigma$(ln E) of $\sim$ 28.4$\%$ as compared to  $\sigma$(ln E) of $\sim$ 36$\%$ reported by the Whipple group \cite{Mohanty1998}, has the added advantage that it considers zenith angle dependence  of $size$ and $distance$ parameters as well. The procedure thus allows data collection over  a much wider zenith angle range as against a coverage of upto $35^\circ$ in case the zenith angle dependence is to be ignored.
\begin{figure}[h]
\centering
\includegraphics*[width=1.0\textwidth,height=0.55\textheight, angle=0,bb=0 0 756 555]{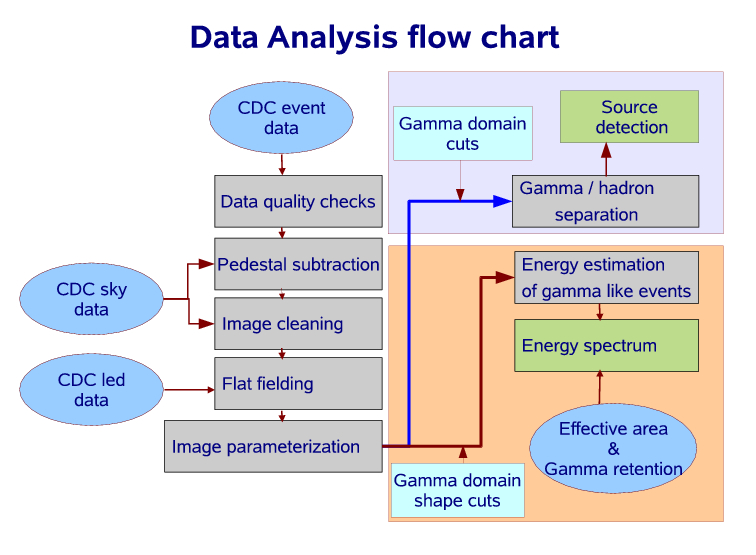}
\caption{\label{data-ana}Flow chart of data analysis procedure followed for \gam signal extraction and energy spectrum determination.}
\end{figure}
\section{TACTIC data analysis}
Presently, there are two primary goals in analyzing the data recorded with the TACTIC telescope:
\begin{enumerate}[(i)]
\item $\gamma$-hadron separation that allows to detect a \gam source.
\item Estimation of primary \gam energy of the $\gamma$-like events, to derive an energy spectrum of a detected \gam source.
\end{enumerate}
Detailed analysis of the telescope data involves several steps. Before applying the \gam domain parameter cuts, the raw Cherenkov images recorded by the telescope need to be processed through the data quality checks, pedestal subtraction, noise cleaning and PMT gain normalization (commonly called flat fielding). All the main steps involved in the analysis have been shown by means of a flow chart in Fig.~\ref{data-ana}. Here, we describe the detailed analysis procedure which has been used in analyzing the data recorded with the TACTIC telescope on different $TeV$ \gam sources.
\begin{figure}[t]
\centering
\includegraphics*[width=0.95\textwidth,height=0.5\textheight, angle=0,bb=0 0 735 555]{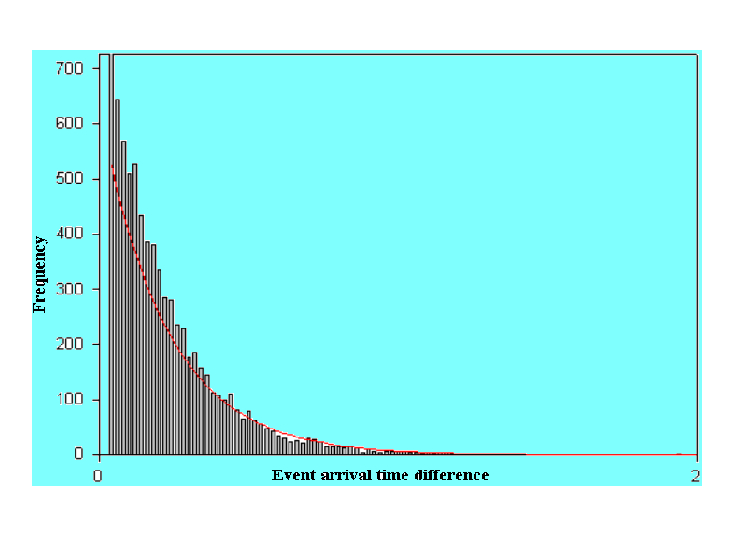}
\caption{\label{arrivaltime}Distribution of arrival time differences of successive events, recorded with the TACTIC telescope. Red curve shows an exponential fit to the experimental data yielding a mean rate of $\sim2.5\,Hz$. Here data used have been recorded in the zenith angle range of $10^\circ$ to $35^\circ$ from the Crab Nebula direction.}
\end{figure}
\subsection{Data quality checks}
\label{dataquality}
The quality of data recorded with the TACTIC telescope depends on a number of parameters like sky conditions, dust levels, instrumentation malfunctions etc. In order to ensure that the quality of recorded data is of a high order, number of data quality checks have been used which includes the compatibility of event rate with the Poissonian statistics and its expected dependence on zenith angle. Some of these are described here:
\begin{itemize}
\item A visual log of \textbf{sky conditions} is recorded through out the observations and referred to while handling the data. 
\end{itemize}
\begin{itemize}
\item CRs reaching the Earth isotropically are random both in direction and timing. It is therefore expected that the \textbf{frequency distribution of the arrival time differences} of these events should be exponential with the exponent determining the mean event rate detected by the telescope. This frequency distribution for one spell of recorded data is shown in Fig.~\ref{arrivaltime} and is indeed exponential. The exponent obtained by fitting the data is in agreement with the observed rate of $\sim2-3\,Hz$ for good spells, as confirmed with visual inspection of sky conditions. 
\end{itemize}
\begin{figure}[t]
\begin{center}
\centering
\subfigure[An example of a good data spell]{\label{goodspell}\includegraphics*[width=0.8\textwidth,height=0.27\textheight, angle=0,bb=0 0 510 250 ]{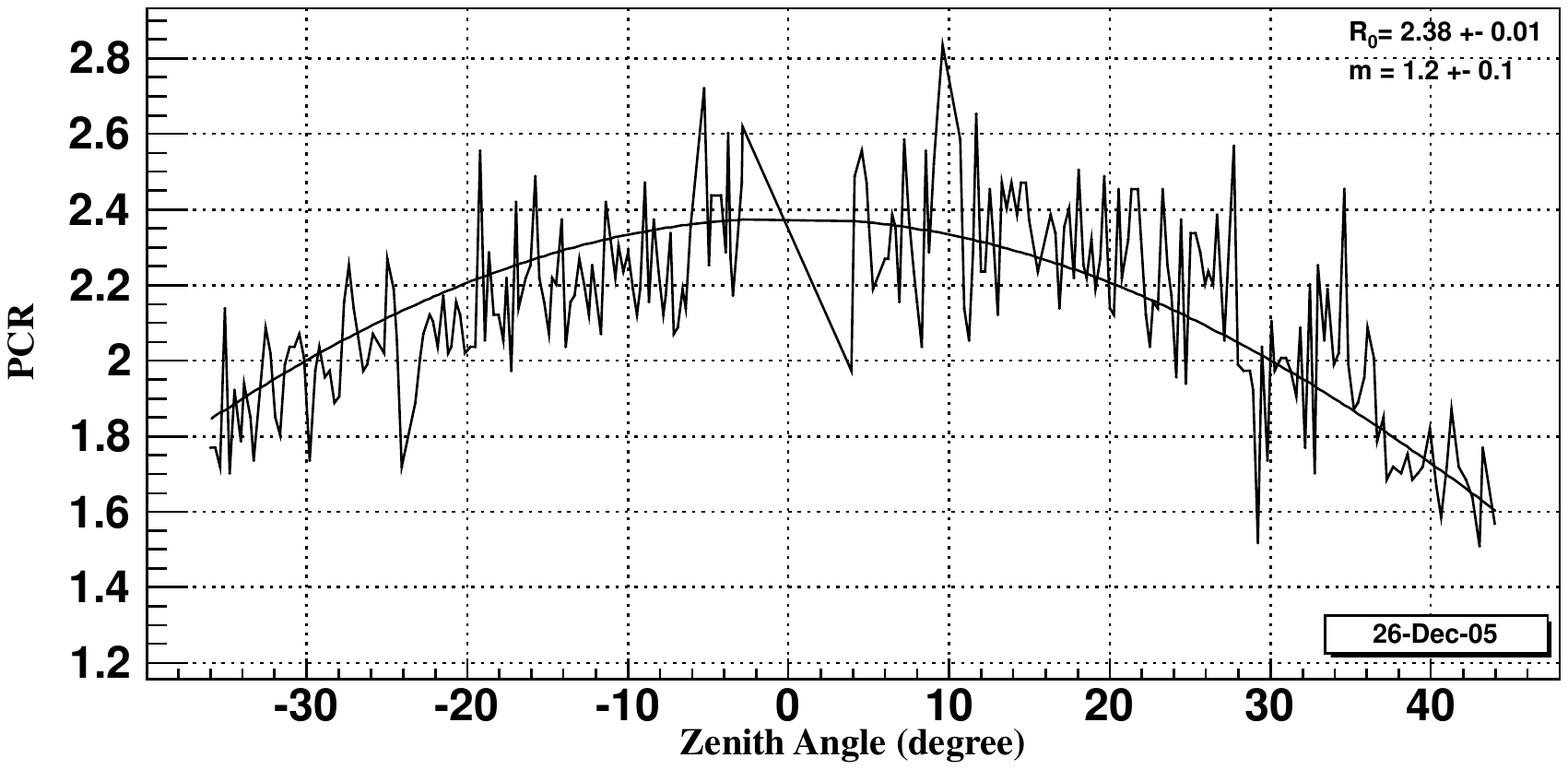}}
\subfigure[An example of a bad data spell]{\label{badspell}\includegraphics*[width=0.8\textwidth,height=0.27\textheight,angle=0,bb= 0 0 510 250 ]{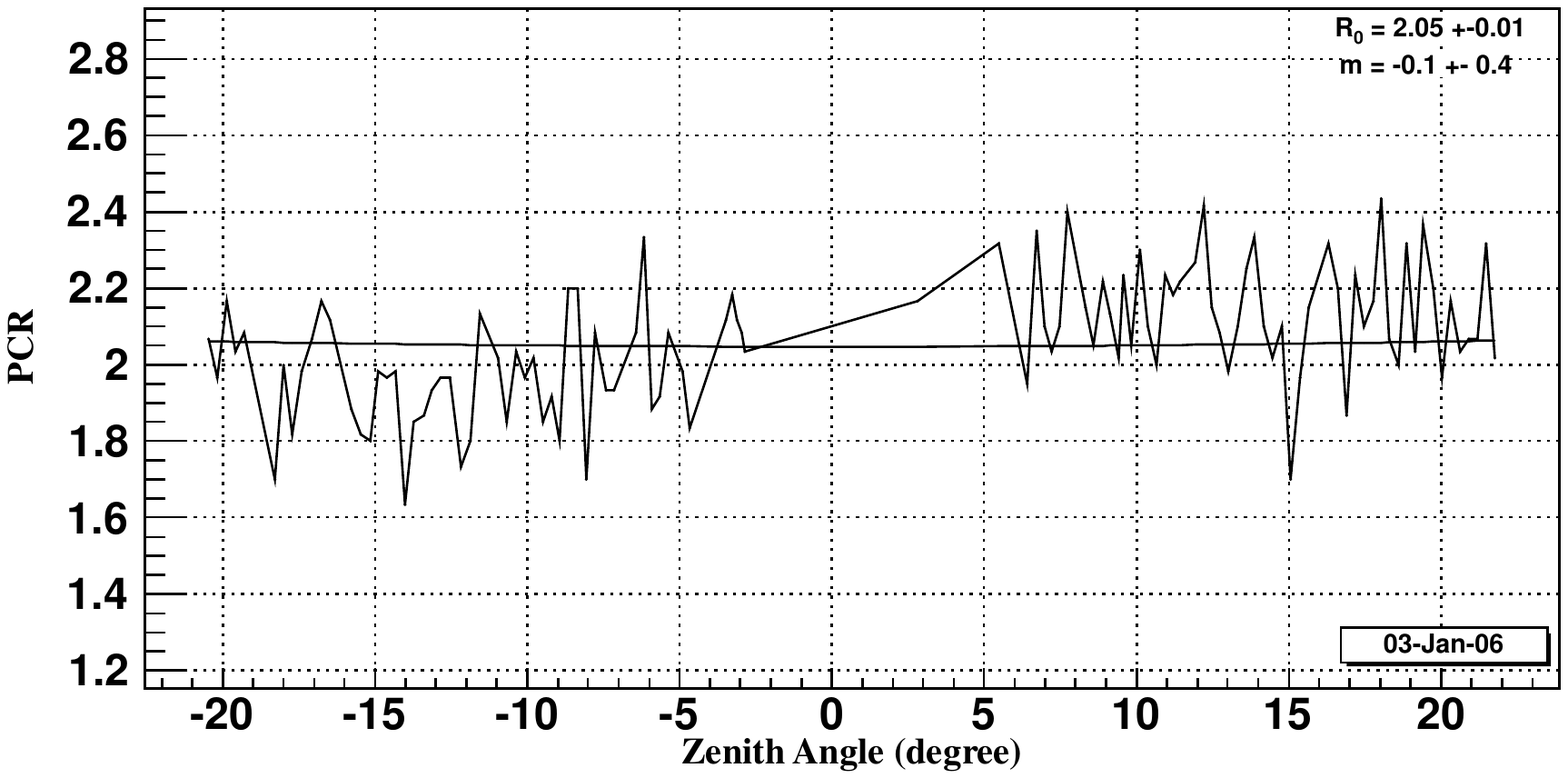}}
\caption{\label{pcr-zen}Variation of prompt coincidence rate as a function of zenith angle of observation. A fit to the data identifies (a) the good data run and (b) the bad data run.}
\end{center}
\end{figure}
\begin{itemize}
\item The main data quality check, which indicates the possibility of problems during observations is the event trigger rate. As discussed earlier, the effective area increases with increasing zenith angle while the photon density decreases and as a net effect the \textbf{telescope trigger rate decreases with increasing zenith angle}, following the relationship
\begin{equation}
\label{pcrzen}
 R_{\theta} = R_0\, (cos\,\theta)^m
\end{equation}
where $\theta$ is the zenith angle of observation, $R_\theta$ is the event rate at zenith angle $\theta$ and $R_0$ is at $\theta=0^\circ$. The value of $m$ depends upon the sky transparency and we have used $m$ between 1-3 for accepting good quality data.
\end{itemize}
\begin{figure}[t]
\centering
\includegraphics*[width=0.5\textwidth,height=0.55\textheight, angle=270,bb=59 56 557 763]{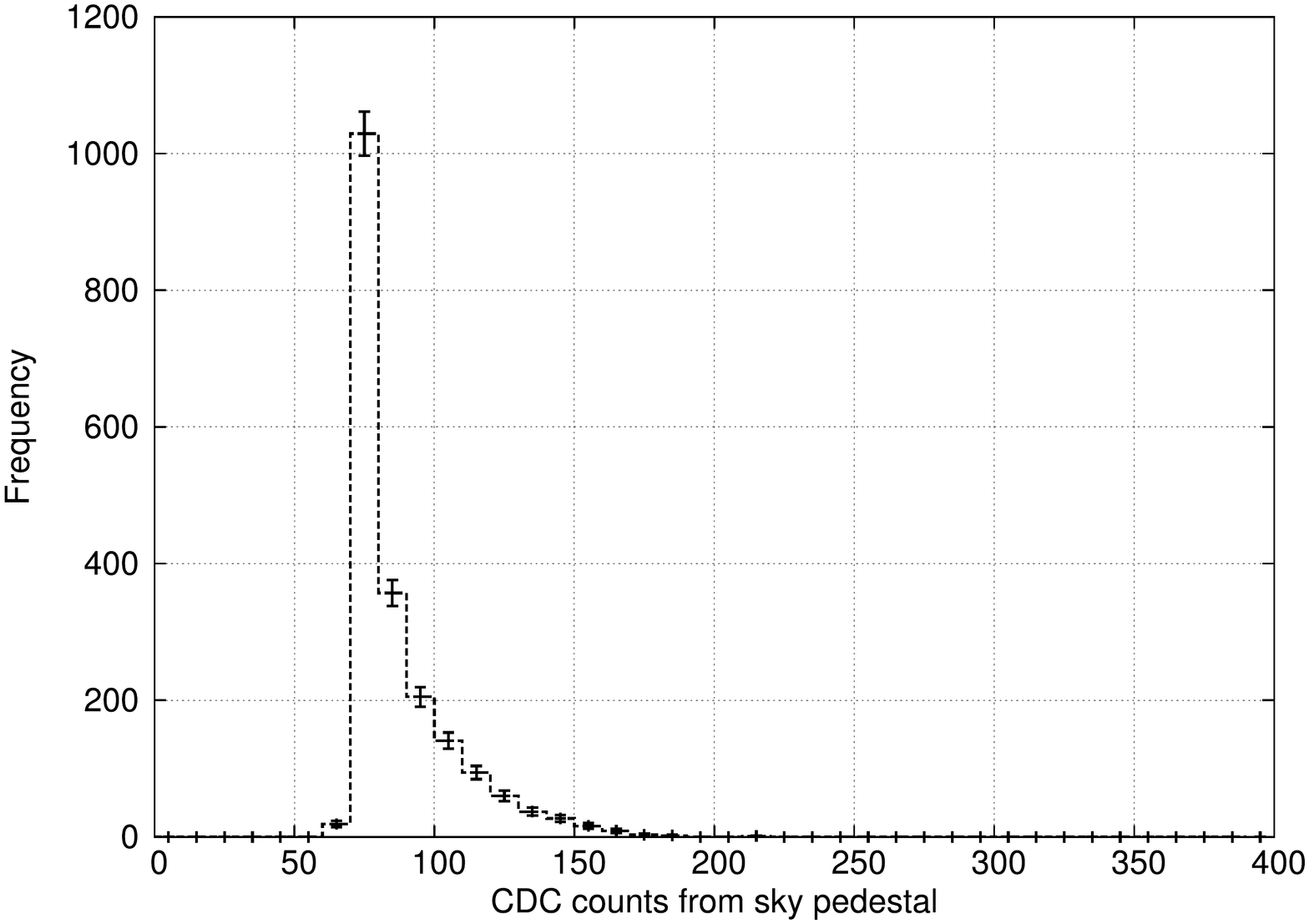}
\caption{\label{sky60}Distribution of sky counts for pixel no. 60.}
\end{figure}
\begin{figure}[h]
\centering
\includegraphics*[width=0.5\textwidth,height=0.55\textheight, angle=270,bb=59 56 554 749]{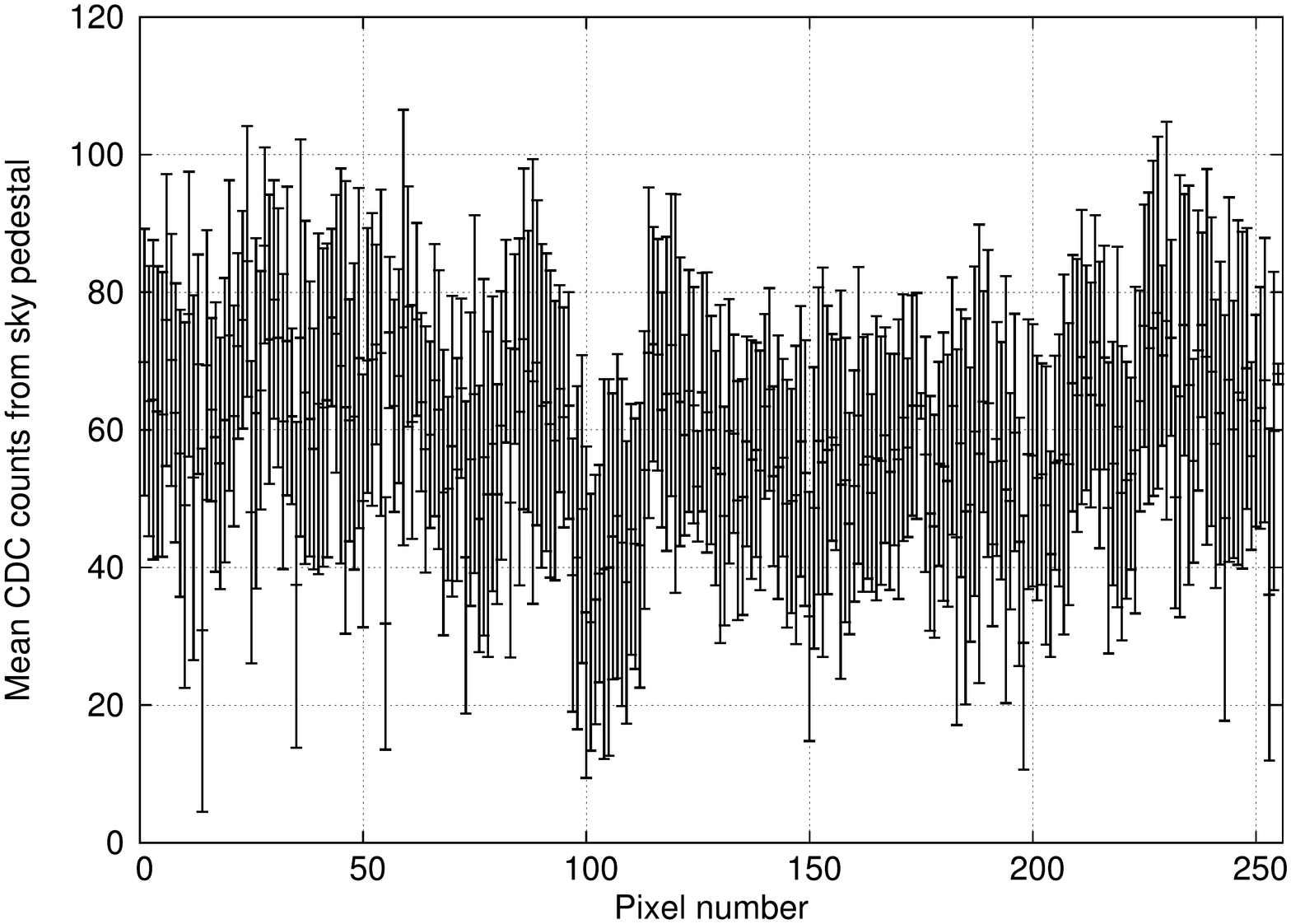}
\caption{\label{skyall}Sky pedestal run, taken during an actual observation run. Plot shows the mean sky counts for 256 pixels with error bars of $1\,\sigma$ statistical fluctuation from 2000 samples.}
\end{figure}
A significantly lower trigger rate than the expected one, generally indicates bad atmospheric conditions or hardware malfunctioning, while higher rates that exceed the allowed statistical fluctuations points to the presence of bright stars in the field of view or even could again be a hardware related problem, like the instability in some of the discriminator channels or malfunctioning of trigger generator module. A representative example of selecting good data spells based on this procedure has been shown in Fig.~\ref{pcr-zen}, wherein a fit to the observed data based on equation (\ref{pcrzen}) qualifies the spell shown in Fig.~\ref{goodspell} as good quality data, while Fig.~\ref{badspell} shows a bad spell, where rate increases with increasing zenith angle and the value of $m$ obtained is negative and falls beyond its allowed values for good sky.
\par
Online tests have also been performed, which demonstrate the over all status of the channels including PMTs and associated electronics. These tests are carried out by analyzing the calibration data collected several times during a typical data taking run, immediately after their acquisition. While results of these tests are demonstrated in Figures~\ref{sky60},~\ref{skyall},~\ref{led60} and \ref{ledall}, we discuss them in the following sections.
\begin{figure}[t]
\begin{center}
\centering
\subfigure[Raw image]{\label{raw}\includegraphics*[width=0.47\textwidth,height=0.33\textheight, angle=270,bb=50 50 554 770]{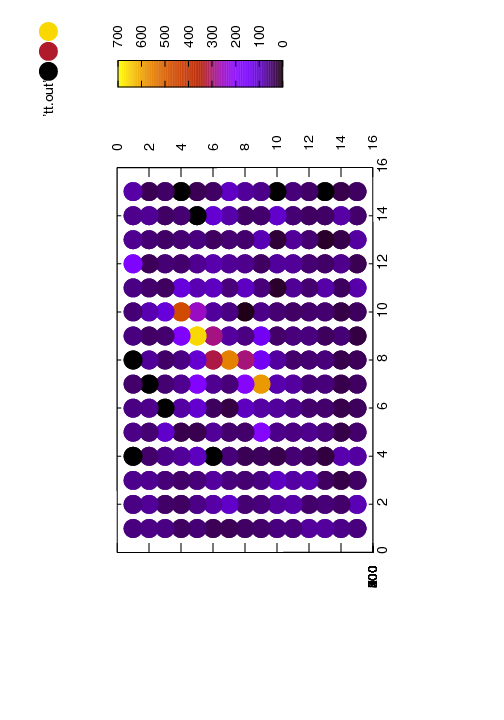}}
\subfigure[Clean image]{\label{clean}\includegraphics*[width=0.47\textwidth,height=0.33\textheight,angle=270,bb=50 50 554 770]{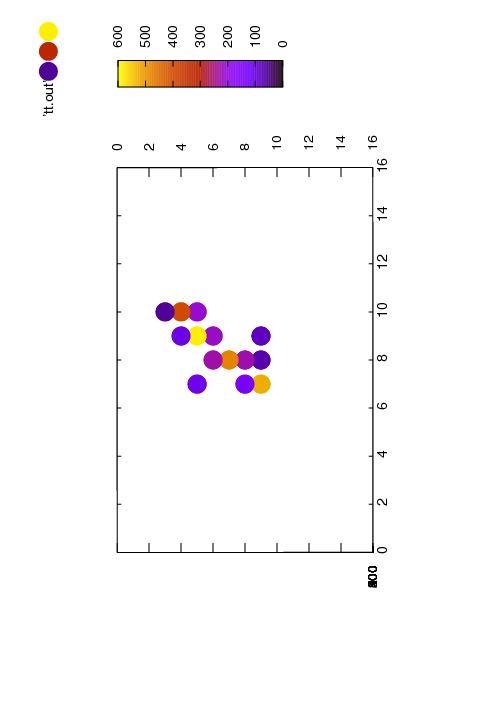}}
\caption{\label {image} Example of an image of a typical shower recorded with the TACTIC telescope. Color code is an indication of the charge content of the pixel. (a) the raw image i.e. prior to the processing procedure, and (b) the same image after pedestal subtraction and application of picture and boundary thresholds.}
\end{center}
\end{figure}
\subsection{Pedestal subtraction} 
The pedestal of a CDC channel is the value of it's charge content converted into CDC counts (1CDC count $\equiv\,0.125\,pc$) which it outputs in the absence of any input signal. This is normally set around 100 CDC counts so that small positive polarity fluctuations on the signal line due to LONS induced shot noise fluctuations, do not generate erroneous value of CDC counts. The pedestal value for each PMT channel is obtained by artificially triggering the camera and recording CDC counts. The trigger rate is kept quite high ($\sim 400\,Hz$), so that the probability of recording an ACE while recording pedestal data is very low. Acquiring the calibration data in this artificial trigger mode is called the sky pedestal run. A typical sky pedestal run consists of 2000 events. The mean value of CDC counts and its variance over 2000 events is then calculated and used for pedestal subtraction and noise cleaning respectively. Fig.~\ref{sky60} shows the distribution of CDC counts for pixel number 60, while the mean value and associated standard deviation for all the pixels have been shown in Fig.~\ref{skyall}. The pedestal subtraction process involves the subtraction of the mean values of sky pedestal CDC counts for all the pixels from the total counts of corresponding pixels recorded in presence of the ACR pulses. Clearly, the aim is to subtract the mean value of LONS, which contributed in the signal while digitizing the ACR events.
\subsection{Image cleaning}Image cleaning is one of the most crucial step in the analysis chain to be followed for an IACT, as the subsequent parameterization of the images uses only the signal contents of the pixels, without its corresponding errors. Most of the pixels in the camera contain only fluctuations due to LONS as indicated in the previous section, while the information about the air shower which triggered the readout is confined to smaller number of pixels. Therefore, in order to reconstruct the image of an EAS, the contribution of the pixels containing the LONS fluctuations alone have to be suppressed. Moreover, the image is discarded if any pixel in the image is saturated either due to presence of a bright star or even due to genuine Cherenkov signal, since the corresponding energy of the primary particle can not be estimated properly.
\par
The purpose and procedure of image cleaning have already been discussed in section~\ref{simulationchain}, we summarize the standard image cleaning procedure again here. A pixel is selected as part of the image if it has a signal above a certain threshold called \textbf{picture threshold} or is beside such a pixel and has signal above a lower threshold called \textbf{boundary threshold}. The picture and boundary thresholds are multiples of the RMS sky pedestal deviation which PMT's signal must exceed to be considered part of the picture or boundary, respectively~\cite{Lessard2002}. The RMS deviations of the pixels are evaluated by using a sky pedestal run and shown in Fig.~\ref{skyall}. The picture and boundary pixels together make up the image, while all others are set to zero. This image cleaning procedure is depicted in Fig.~\ref{image}. A picture threshold of $6.5\,\sigma$ and a boundary threshold of $3.0\,\sigma$ have been used to select the largest number of pixels while at the same time limiting the inclusion of pixels with noise alone.
\begin{figure}[t]
\centering
\includegraphics*[width=0.5\textwidth,height=0.55\textheight, angle=270,bb=59 56 554 768]{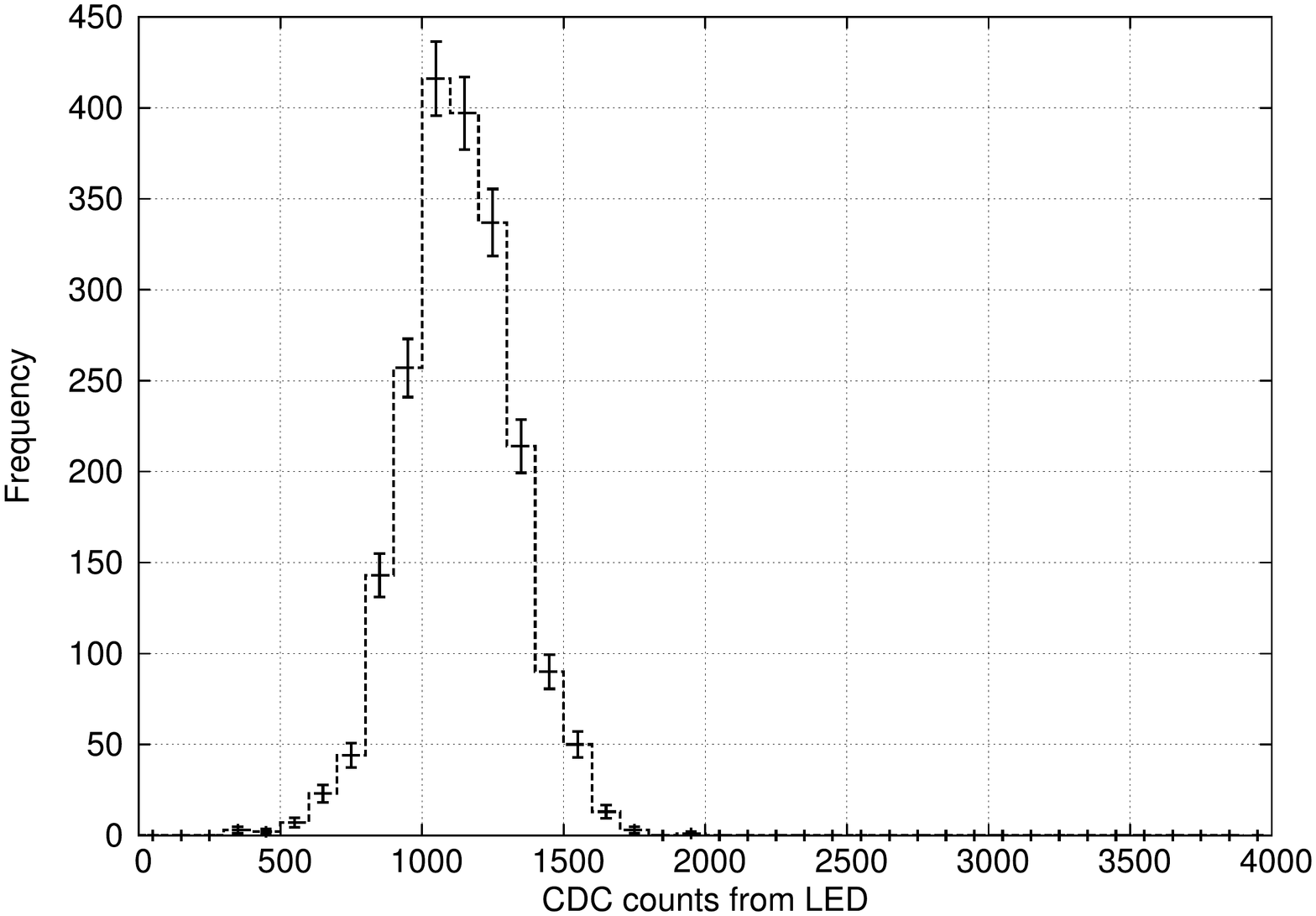}
\caption{\label{led60}Distribution of LED counts for pixel no. 60.}
\end{figure}
\begin{figure}[h]
\centering
\includegraphics*[width=0.5\textwidth,height=0.55\textheight, angle=270,bb=59 56 554 749]{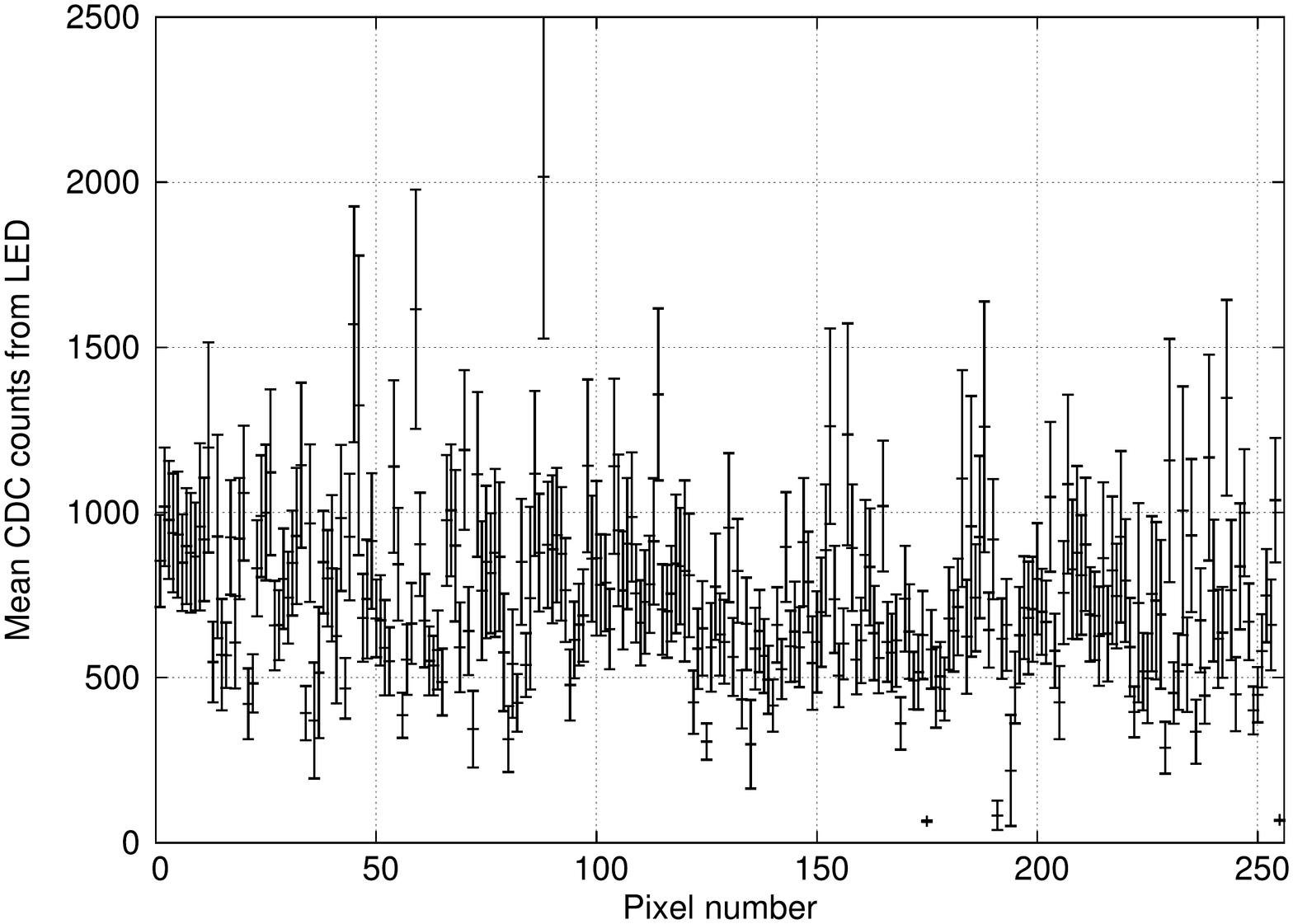}
\caption{\label{ledall}Relative calibration run, taken during an actual observation run. Plot shows the mean LED counts for 256 pixels with error bars of $1\,\sigma$ statistical fluctuation from 2000 samples.}
\end{figure}
\subsection{Gain normalization} This process is also called ``\textbf{flat fielding}''. In this process, we account for the differences in the relative gains of the PMTs. Operating in the pulsed mode, a blue LED based light pulser illuminates the camera uniformally ($\pm 5\%$) from a distance of $\sim2\,m$ from the camera surface. In response to the LED pulse, the readout is triggered and corresponding CDC counts are recorded for all the pixels. A total number of 2000 such calibration pulses are recorded, by triggering the camera at a rate of $\sim400\,Hz$. The calibration data collected in this mode is called the relative calibration run. The mean and standard deviation is then calculated for each pixel on the basis of these 2000 events. The frequency distribution of CDC counts, for pixel number 60 is shown in Fig.~\ref{led60} for a typical relative calibration run. The mean values and associated variances for 256 pixels are depicted in Fig.~\ref{ledall}. The calibration data collected are used to determine the gain normalisation factors of each PMT by comparing their mean CDC counts to the reference pixel. Nine pixels at the center of the camera are kept at fixed HV during the course of observation and any one of them can be used as a reference pixel.
\par
A sky pedestal run and a relative calibration run are together referred  to as a calibration run. In order to achieve the minimum possible trigger threshold, we allow the control system of TACTIC telescope to optimize the HV and therefore the gain of the PMTs. The values of HV are changed only for those PMTs whose SCRs have changed beyond the allowed SCR range in responce to the NSB light. This also ensures the safe anode current operation of PMTs. We need to recalibrate the camera after each such change in the HV values. By examining the mean values and standard deviation of the charge content of the pixels, the possible problems in the hardware can be identified since these were set in an allowed range for a particular setting of the LED pulse. Also by comparing them from one calibration run to another, the online tests for overall channel health can be performed as mentioned in section~\ref{dataquality}.

\subsection{$\gamma$-hadron separation}
\begin{figure}[h]
\begin{center}
\centering
\subfigure[a \gam like event]{\label{slow}\includegraphics*[width=0.5\textwidth,height=0.33\textheight, angle=270,bb=50 50 554 770]{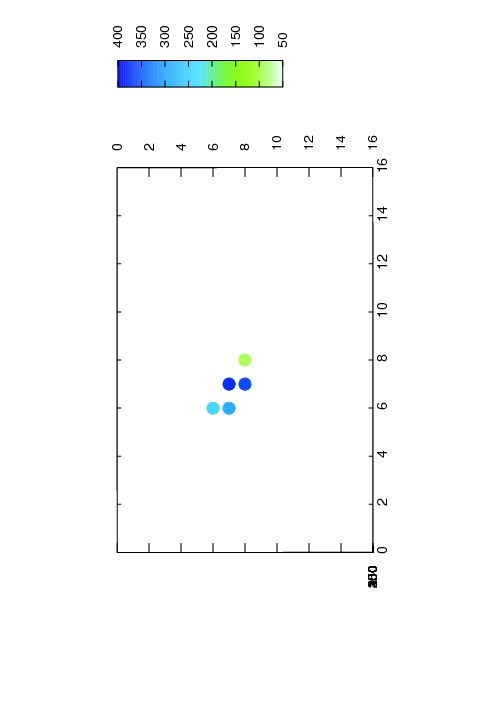}}
\subfigure[a proton event]{\label{fast}\includegraphics*[width=0.5\textwidth,height=0.33\textheight,angle=270,bb=50 50 554 770]{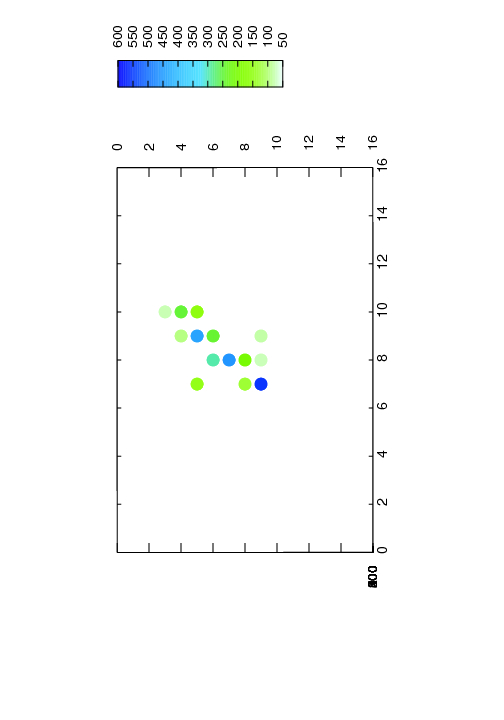}}
\caption{\label {gam-prot}Representative example of a \gam like and a proton shower image, recorded with the TACTIC telescope. Color code is an indication of the charge content of the pixel.}
\end{center}
\end{figure}
The $\gamma$-hadron separation is based on image parameters obtained using the Hillas image parameterization methodology~\cite{Hillas1985} of the cleaned images which has been discussed in section~\ref{imageparameters}. The derived image parameters ($length$, $width$, $size$, $distance$, $alpha$, $frac2$ etc.) are then used (except $\alpha$, which is used separately) to segregate \gam events from the overwhelming background of hadron events by applying the $\gamma$-domain cuts, which are obtained from the Monte Carlo simulations. The representative examples of a \gam like image and a proton image as recorded with the TACTIC telescope have been shown in Fig.~\ref{gam-prot}. The shape selected images (after applying $length$, $width$, $size$, $distance$, and $frac2$) are used to plot the $\alpha$-distribution, the example of which has been depicted in Fig.~\ref{crab-alpha}. The $\alpha$ is also known as image orientation parameter and its value is expected to be smaller for \gam showers having directions close to the optical axis of the telescope, while a nearly flat distribution is expected for an isotropic background of CR protons. For the TACTIC telescope, the excess of \gam events from a point \gam source is expected to be within an $\alpha$ range of $0^\circ-18^\circ$. 
\subsection{Statistical significance of source detection}In the ACT, the observations on a candidate \gam source are normally carried out in two modes; (i) \textbf{ON-OFF mode:} in this mode an on-source run is taken with the \gam source at the center of the camera, while off-source run is taken separately covering the same region of the sky with no probable \gam source in the field of view. (ii) \textbf{Tracking mode:} the tracking mode is also known as the discovery mode of observations, wherein the candidate \gam source is continuously tracked, without taking any off-source run. This mode of observation is preferred to maximize the on-source observation time and to increase the possibility of recording the flaring activity from the source. However, the systematic errors are relatively more as compared to the ON-OFF mode of data taking. Currently a third method called \textbf{Wobble mode} is also being used, wherein the source is kept at off-center position and a few other off-center positions are marked to carry out OFF-source observations simultaneously.
\par
Certain percentage of CR events pass through all the $\gamma$ domain cuts and essentially become background for \gam events, which need to be determined in order to calculate the statistical significance of the detected excess of \gams. While the off-source run provides an estimate of the background events in ON-OFF mode of observation, the tracking mode of data taking utilizes the on-source data itself for the calculation of background events. To estimate the expected background for the tracking mode, we use the events which have passed the dynamic supercut selection criteria except the $\alpha$ parameter. The distribution of $\alpha$ parameter is expected to be nearly flat between $0^\circ$ and $90^\circ$ for the background CR proton events, we therefore use the events with $\alpha$ value between $27^\circ$ to $81^\circ$, for estimation of the background level. While the total range of $\alpha$ values i.e. $0^\circ$ to $90^\circ$ is divided into 10 bins with bin size of $9^\circ$, the on-source region is defined by the first two bins ($0^\circ$ to $18^\circ$). The third and last bins are not included for estimation of the background events, due to possible spill over of on-source events and comparatively less stability of events (possibly due to image truncation) from run to run, respectively.
\par
The significance of the signal can be estimated using either of the following two formulae proposed by Li and Ma~\cite{Li1983} (equation 5 and 17 of the paper)
\begin{equation}
\label{eq5}
 S = \frac{N_{on} - \eta\, N_{off}}{\sqrt{N_{on}+\eta^2 N_{off}}} 
\end{equation} 
or
\begin{equation}
\label{eq17}
 S = \sqrt{2} \left\lbrace N_{on}\,ln\left[ \frac{1+\eta}{\eta}\left( \frac{N_{on}}{N_{on}+N_{off}}\right) \right]  + N_{off}\,ln\left[\left( 1+\eta\right)\left( \frac{N_{off}}{N_{on}+N_{off}}\right)  \right] \right\rbrace ^{1/2}
\end{equation} 
where $N_{on}$ is the number of events in the on-source region of $\alpha$ distribution (events in $0^\circ$ to $18^\circ$), $N_{off}$ is the number of background events (events in $27^\circ$ to $81^\circ$) and $\eta$ is the ratio of number of bins in on-source to number of bins in off-source region. In our case $\eta=3$.\\
Equation (\ref{eq5}) is derived by estimating the standard deviation of the observed signal and it simply follows from the Poisson law of the counts i.e. $N_{on}$ and $N_{off}$. For the case $\eta \sim 1$ ($0.5 \lesssim \eta \lesssim 1.5$), the equation (\ref{eq5}) can be used to evaluate the significance. The other equation (\ref{eq17}), is derived by applying the method of statistical hypothesis test and can be applied in a general case of $\eta\not\approx1$ and also with fewer observed counts.
\par
While evaluating the significance of a \gam signal in our data analysis, we have used both the equations and found that equation (\ref{eq17}) provides only marginal improvement in significance compared to evaluation by equation (\ref{eq5}). 
\subsection{Energy spectrum estimation} As already discussed, we use an ANN based technique to estimate the energy of the primary \gams detected with the TACTIC telescope. The energy spectrum of the \gam source is then determined by binning these \gam photons in terms of energy. The ANN is trained using a Monte Carlo \gam sample and generates a weight file after achieving a satisfactory training. The ANN generated weight file is used in the analysis program, for determining the energy of the primary \gam photons, based on the $size$, $distance$ and $zenith\, angle$ parameters. The energy is determined only for those images which pass through all the \gam domain parameter cuts except $\alpha$. Having an estimate of energy resolution of the telescope and also some idea about its threshold energy, the energy bins i.e. the energy bin width and first energy are selected, accordingly. The energy bin width is defined as
\begin{equation}
 \Delta E = ln\left( \frac{E_{m_{i+1}}}{E_{m_i}}\right) 
\end{equation} 
where $E_{m_i}$ is the middle energy of $i^{th}$ energy bin. The middle energy of a particular bin is related to lower and upper bounds of the same bin and to the middle energy of the other energy bins in the following manner, which follows from the above equation;
\begin{equation}
 E_{l_i}=E_{m_i}\cdot e^{\frac{-\Delta E}{2}} \qquad;\qquad  E_{h_i}=E_{m_i}\cdot e^{\frac{\Delta E}{2}}
\end{equation} 
\begin{equation}
 E_{m_{i+1}}= E_{m_i}\cdot e^{\Delta E} \qquad;\qquad E_{m_{i+1}}= E_{m_1}\cdot i\cdot e^{\Delta E}. 
\end{equation}
After determining the energy, events are put in corresponding energy bins. The number of excess events for each energy bin are then determined by performing the frequency distribution of their $\alpha$ parameter.\\
The differential photon flux  per energy bin has been computed using the formula
\begin{equation}
\frac{d\Phi}{dE}(E_i)=\frac {\delta N_i}{\delta E_i \sum \limits_{j=1}^5 A_{ij}\, \eta_{ij}\, T_j}
\end{equation}
where $\delta N_i$ and $d\Phi(E_i)/dE$ are the number of events and the differential flux at energy $E_i$, measured in the $i^{th}$ energy bin of width $\delta E_i$ and over the zenith angle range of $0^\circ-45^\circ$, respectively. $T_j$ is the observation time in the $j^{th}$ zenith angle bin with corresponding energy-dependent effective area ($A_{ij}$) and $\gamma$-ray acceptance ($\eta_{ij}$). The 5 zenith angle bins ($j=1-5$) used are $0^\circ-10^\circ$, $10^\circ-20^\circ$, $20^\circ-30^\circ$, $30^\circ-40^\circ$  and $40^\circ-50^\circ$ with  simulation data  available at $5^\circ$, $15^\circ$, $25^\circ$, $35^\circ$ and $45^\circ$. The number of \gam events  ($\delta N_i$) in a particular energy bin is calculated by subtracting the expected number of background events, from the \gam domain events.
\begin{figure}[t]
\centering
\includegraphics*[width=0.95\textwidth,height=0.35\textheight, angle=0,bb=0 0 510 329]{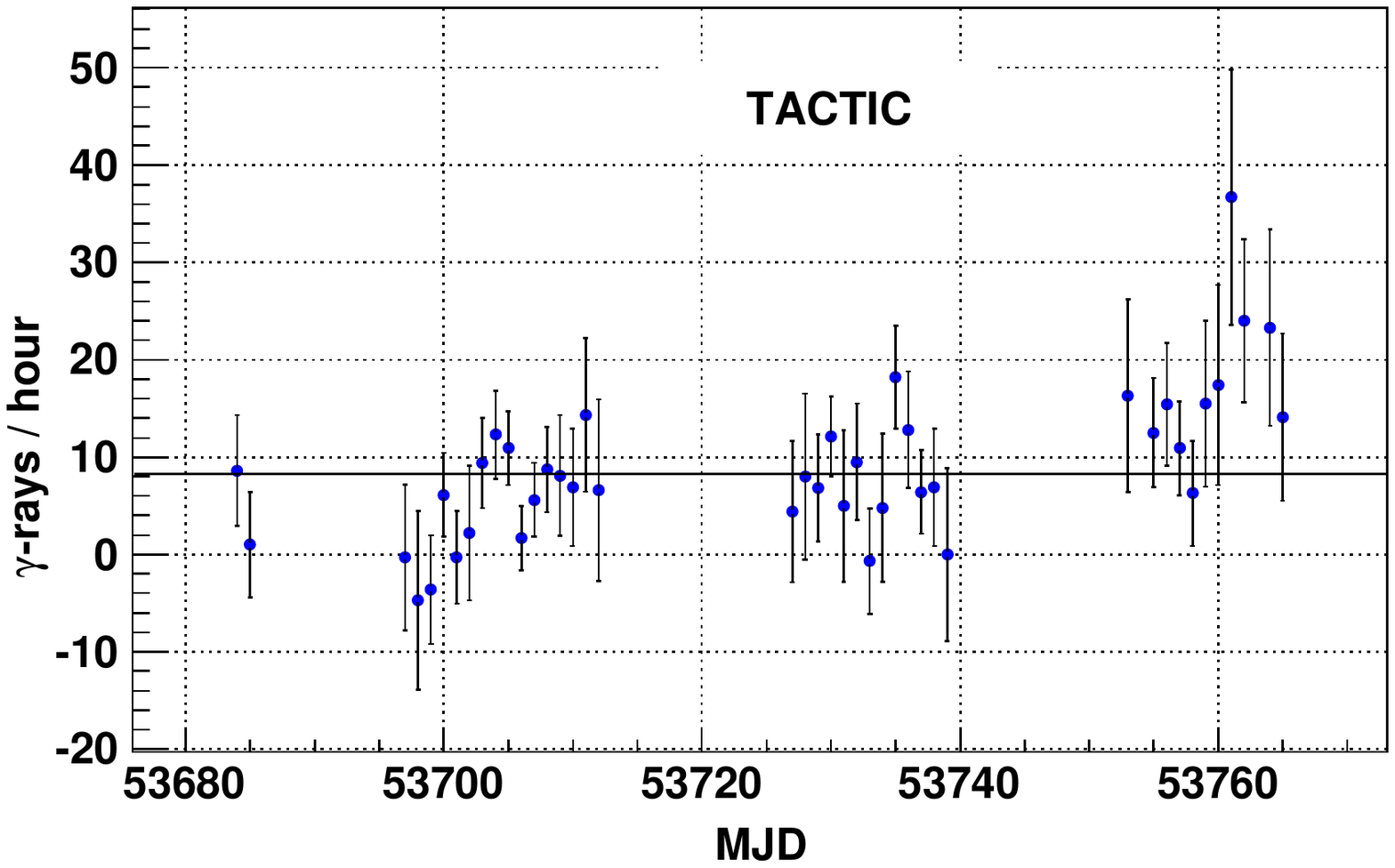}
\caption{\label{crab-lc}Light curve for Crab Nebula as recorded by the TACTIC imaging telescope during 2005-06 observation. Error bars shown  are for statistical errors only.}
\end{figure}
\subsection{Light curve} When the flux is determined with respect to time bins, the so-called light curve is generated. Conceptually, its determination is similar to that of the flux. Instead of determining the number of excess events in the bins of energy, it is determined in the bins of time and energy. If the statistics allow, this can be done fully differentially, otherwise integral flux can be calculated. Generally, in IACT astronomy, the time bin covers the whole observation night. In case of TACTIC, we calculate the number of \gam like events from the source direction for the whole observation night and generate a light curve in terms of $photons/hour$ as a function of Modified Julian Day (MJD). Fig.~\ref{crab-lc} shows the light curve for the Crab Nebula as recorded by the TACTIC telescope during 2005-06 observations. The horizontal line in the figure shows the average \gam rate.
\section{Validation of analysis procedure}
\begin{figure}[t]
\centering
\includegraphics*[width=0.95\textwidth,height=0.35\textheight, angle=0,bb=0 0 567 325]{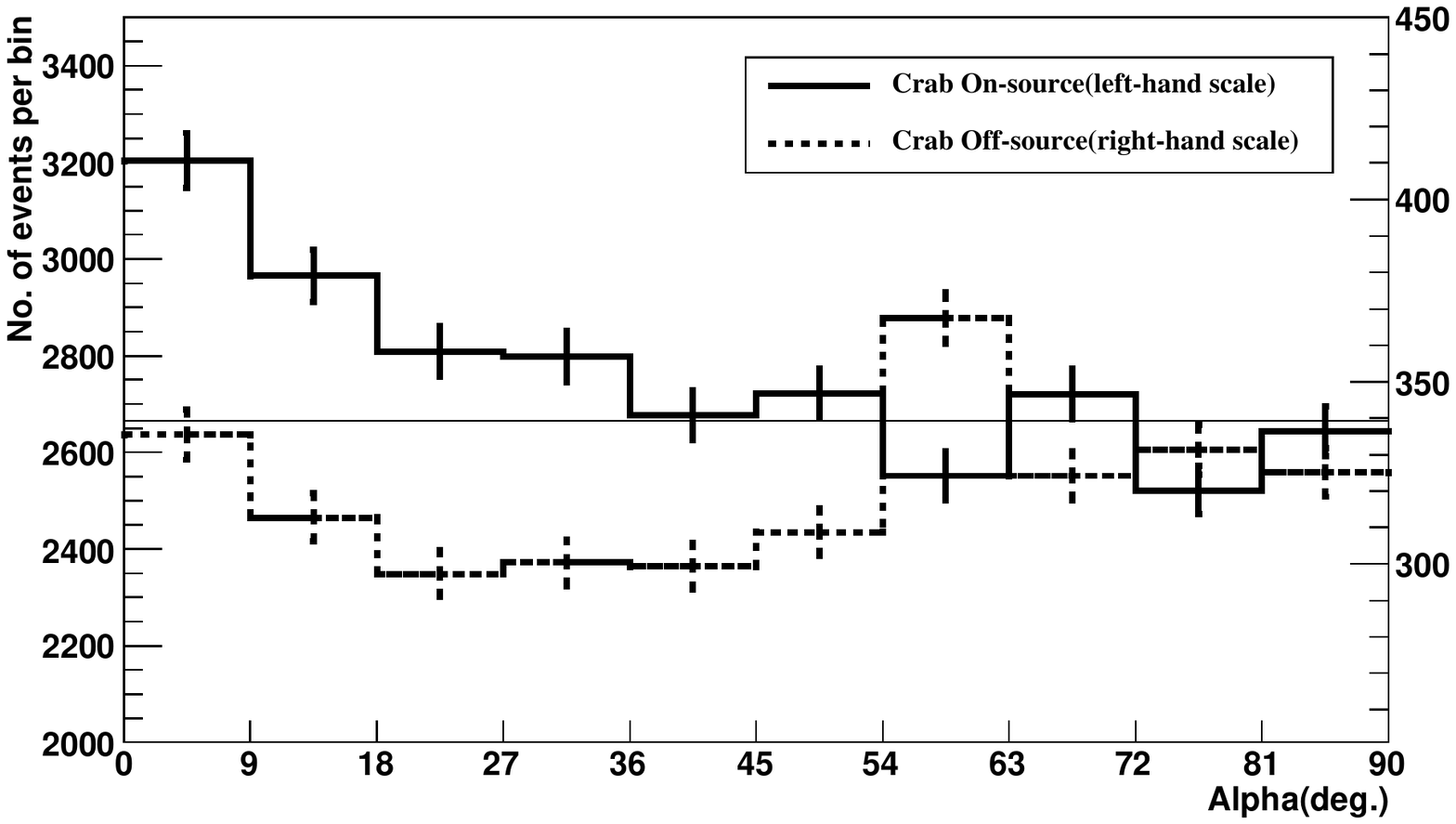}
\caption{\label{crab-alpha}Distribution of $\alpha$ parameter from the Crab Nebula direction (solid line and left hand scale) observed for $\sim$$101.44\,hours$ during 10 November 2005 - 30 January 2006. Horizontal line represents the background level per $9^{\circ}$ bin derived using  reasonably flat $\alpha$ region  from  $27^{\circ}$ $\leq$ $\alpha$ $\leq$ $81^{\circ}$.  Distribution of $\alpha$ parameter is also shown (dotted line and right-hand scale) from the Crab Nebula off-source  direction observed  for $\sim$$12.5\,hours$ during 2006. Error bars shown  are for statistical errors only.}
\end{figure}
The complete data analysis chain has been validated by analyzing the data, collected on the standard candle Crab Nebula, with the TACTIC imaging telescope, for $\sim101.44\,hours$ during 10 November 2005 - 30 January 2006. The excess number of \gam like events selected after using the Dynamic Supercuts procedure was determined to be $\sim$($839\pm89$) events with a statistical significance of $\sim9.64\sigma$. Fig.~\ref{crab-alpha} shows the frequency distribution of the $\alpha$-parameter. The corresponding average \gam rate turns out to be $\sim$($8.27\pm0.88$) / $hour$. In order to generate the Crab Unit (CU) of flux in terms of the detected \gam like events, we analyzed the same data sample of Crab Nebula again by restricting the zenith angle of the observations from $15^\circ$ to $45^\circ$. This zenith angle range is normally covered by Mrk421, during its observations with TACTIC telescope. The resulting \gam rate from Mrk421 (discussed in next chapter) can then be converted into reference CU, while interpreting it's data. This re-analysis yielded an excess of $\sim$($598\pm69$) \gam events in an observation time of $\sim$$63.33\,hours$ with a corresponding \gam rate of
$\sim$($9.44\pm1.09$) / $hour$, thus leading to the conversion: $1CU\equiv$ ($9.44\pm1.09$) / $hour$. While one would have expected this rate to decrease because of increase in threshold energy, the reason behind this increase in \gam rate is the superior \gam acceptance of Dynamic Supercuts at higher zenith angles which over-compensates the decrease in the rate due to the increase of the threshold energy of the telescope.
\begin{figure}[t]
\centering
\includegraphics*[width=0.5\textwidth,height=0.5\textheight, angle=270,bb=59 52 557 747]{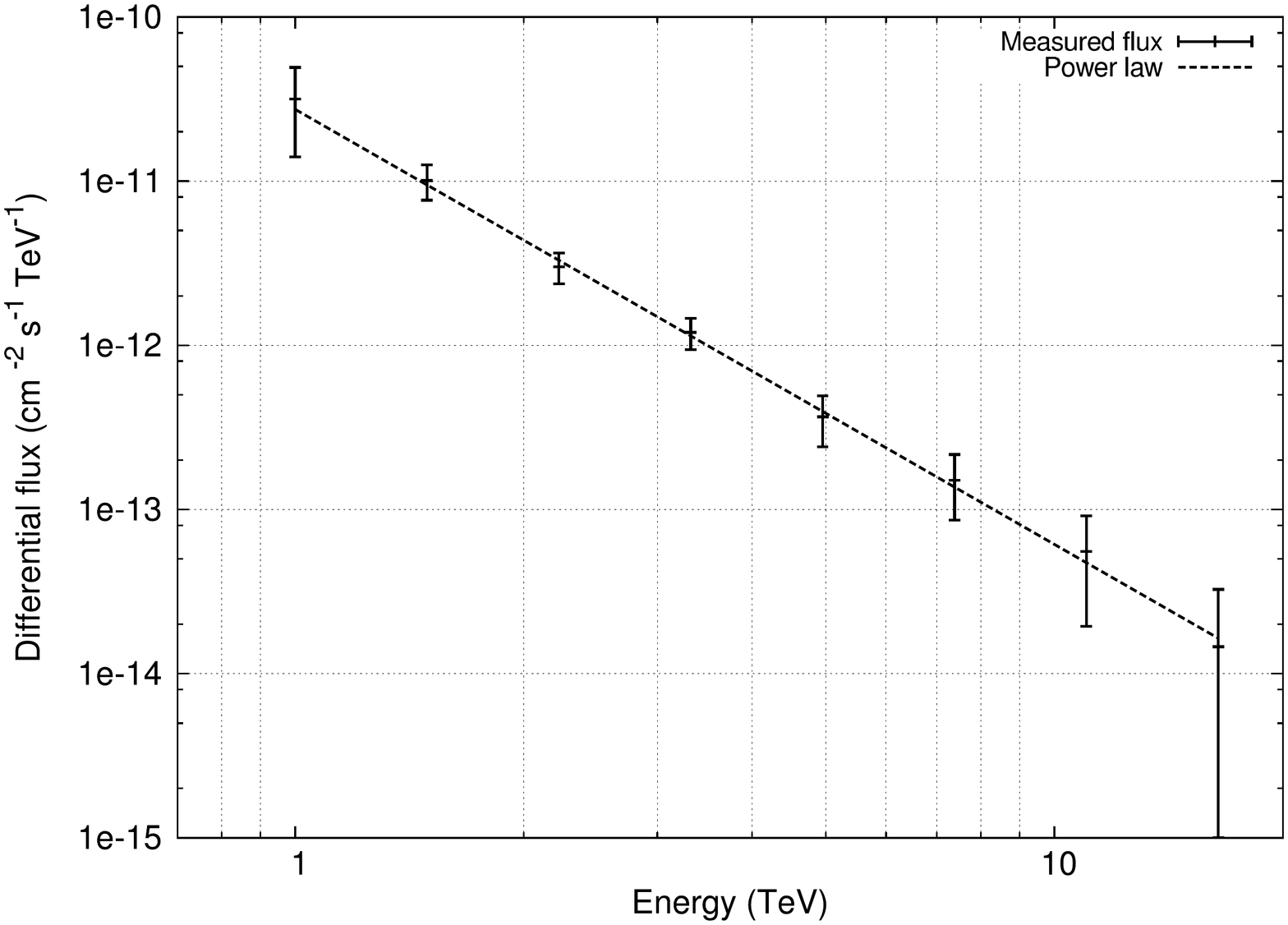}
\caption{\label{crab-spectrum}Spectrum of Crab Nebula as measured by the TACTIC telescope during 2005-06. Error bars shown  are for statistical errors only.}
\end{figure}
\par
In order to test the validity of the energy estimation procedure, we have used the entire data i.e. for $\sim$$101.44\,hours$, collected on Crab Nebula. The \gam differential spectrum obtained after applying the Dynamic Supercuts and appropriate values of effective collection area and \gam acceptance  efficiency (along with their  energy and zenith angle dependence) is shown in Fig.~\ref{crab-spectrum}. While  determining the energy spectrum, we used the excess noise factor method~\cite{Biller1995} for converting the image size in CDC counts to number of photoelectrons. The analysis of RGC data yields a value of $1pe\cong\,(6.5\pm1.2)\,CDC$ when an average value of $\sim1.7$ is used for excess noise factor of the photomultiplier tubes. The differential energy spectrum of the Crab Nebula shown in Fig.~\ref{crab-spectrum} is a power law fit  $(d\Phi/dE=f_0 E^{-\Gamma})$  with $f_0=(2.74\pm0.19)\times 10^{-11} cm^{-2}s^{-1}TeV^{-1}$  and $\Gamma=2.65\pm0.06$. The fit has a $\chi^2/dof=0.53/6$ (here and afterwards the denominator represents  the value for the degrees of freedom ) with a corresponding probability of 0.997. The errors in the flux constant and the spectral index are standard errors. Excellent matching of this spectrum with that obtained  by the Whipple and  HEGRA  groups \cite{Hillas98,Aharonian04} reassures that the procedure followed by us for obtaining the energy spectrum  of a \gam source is quite reliable.
\chapter{TeV observations of Markarian 421}
\section{Introduction}
Mrk421 is a nearby blazar (z=0.030) and one of the brightest BL Lac objects located at a distance of $\sim134.1\,Mpc$ ($H_0=71\,km\,s^{-1}\,Mpc^{-1}$, $\Omega_m=0.27$, $\Omega_\lambda=0.73$)~\cite{don09}. Blazars are the most extreme and powerful variable sources of photons with energies ranging from the radio to the \gam regimes. These objects are a subclass of AGN and are believed to have their jets more aligned toward the Earth as compared to any other class of radio loud AGN. They display high luminosity, irregular amplitude variability in all accessible spectral bands and have a core-dominated radio morphology with flat radio spectra, which join smoothly to the infra-red (IR), optical and ultra-violet spectra. In all these bands the flux exhibits high and variable polarization. With more observations, variety of features of such objects are becoming evident and many different models can explain their observed properties~\cite{boettcher07,sambruna07}. Their high variability and  broadband emission make long-term observations of blazars very important for  understanding their emission mechanisms and other related properties~\cite{horan09}.
\par
The blazar SED has a double-peaked structure in the $\nu$F$_{\nu}$ versus frequency plot. Both peaks are found to vary, often both in intensity and  peak frequency as the activity level of the blazar changes. The first peak is usually referred to as the synchrotron peak, both in leptonic and hadronic models for blazar emission. Further, it is  generally believed to be the result of incoherent synchrotron emission from relativistic electrons and positrons, which are conjectured to be present in the magnetic fields of the jet. The origin of the second peak, usually referred to as the IC peak, is less well determined.
\par
In SSC models~\cite{Bloom1996,Konigl81,Maraschi1992,Sikora01,Ghisellini1998ssc} it is assumed that the synchrotron photons are up-scattered to higher energies by the electrons while in EC models~\cite{Blandford1995,Dermer97,Ghisellini96,Wagner95,Sikora1994,Dermer1992}, these seed photons can come from the accretion disk, the broad-line region, the torus, the local infrared background, the cosmic microwave background, the ambient photons from the central accretion flow or some combination of these sources. Hadronic models have also been invoked to explain the broadband spectra of blazars~\cite{Aharonian2000-hadron,Mannheim1993,Mucke2003}. It is  proposed that the X-ray to \gam emission is synchrotron radiation from protons accelerated in highly magnetized compact regions of the jet. Other authors propose that proton-proton collisions, either within the jet itself or between the jet and ambient clouds, give rise to neutral pions which then decay to \gams~\cite{Beall99,DarLaor97,Pohl2000}.
\par
Mrk421 was the first extragalactic source detected at $TeV$ energies in 1992 by Whipple observatory~\cite{Punch1992} and subsequently by HEGRA group~\cite{Petry96}, using IACTs. The observations which led to the detection of Mrk421 at $TeV$ energies were initiated in response to the detection of AGNs by EGRET. The initial detection indicated a $6\sigma$ excess and the flux above $500\,GeV$ was approximately $30\%$ of the Crab Nebula at these energies. The source has been regularly monitored by different groups since then \cite{Zweerink97,Piron01,Aharonian03,Boone02,Aharonian05,Albert07-m421,Smith06,Rannot2005,Yadav07}. It has been seen that the $TeV$ \gam emission from Mrk421 is highly variable with variations of more than one order of magnitude and ocassional flaring doubling time of as short as $15\,minutes$ \cite{Gaidos1996,Aharonian2002-mrk421}. The first clear detection of the flaring activity in the VHE emission of an AGN came in 1994 observations of Mrk421 by the Whipple telescope where a 10-fold increase in the flux, from an average level that year of approximately $15\%$ of the Crab flux to approximately $150\%$ of the Crab flux, was observed. The observations of Mrk421 in 1995~\cite{Buckley96} revealed several distinct episodes of flaring activity as in previous observations. Perhaps more importantly, they indicated that the VHE emission from Mrk421 was best characterized  by a succession of day scale or shorter flares with a baseline emission level below the sensitivity limit of the Whipple detector.
\par
The hypothesis that the VHE emission from Mrk421 could flare on sub-day time scales was borne out in a spectacular fashion in 1996, with the observation of two short flares~\cite{Gaidos1996}. In the first flare which was observed on 07 May 1996, the flux increased monotonically during the course of $\sim2\,hours$ of observations.
The doubling time of the flare was $\sim1\,hour$. The next night the flux had dropped to a flux level of $\sim30\%$ of the Crab Nebula flux, implying a decay time scale of $< 1$ day. The second flare, observed on 15 May 1996, although weaker, was remarkable for its very short duration: the entire flare lasted approximatey $30\,minutes$ with a doubling and decay time of less than $15\,minutes$. Further in 2001, Mrk421 underwent an extraordinary period of activity during which it was consistently brighter than the Crab Nebula over a three-month period~\cite{Weekes2003}.
\par
Since its detection in the $TeV$ energy range, Mrk421 has also been the target of several multiwavelength observation campaigns \cite{ale10,don09,wag09,Buckley96,Takahashi96,Takahashi00,Blazejowski05}. One of the earliest multi-wavelength compaigns was organized in 1995 to measure the multi-wavelength properties of Mrk421. This compaign revealed, for the first time, correlations between VHE \gams and X-rays. The source also exhibited a large amplitude flare in VHE \gams which was also clearly seen in the ASCA (X-ray) and Extreme Ultaviolet Explorer (EUVE) obsevations. The X-rays and \gams appear to vary together, limited by the one-day resolution of the VHE observations. Observations in April 1998 at $TeV$ energies with the Whipple telescope and at X-ray wavelengths with the BeppoSAX satellite, established an hour scale correlations between the X-rays and \gams in a blazar~\cite{Weekes2003}.
\par
Measurements of the VHE spectra are important for a variety of reasons. While the shape of the high energy spectrum is a key input parameter of AGN emission models, the way in which the spectrum varies with flux compared to lower energy observations provide further tests  for the models. Several groups have determined the energy spectrum  of Mrk421, both at low average flux levels of $<1\,CU$ and from intense flares of $>2\,CU$. The recent results of these studies \cite{Piron01,Aharonian03,Aharonian05,Krennrich02} suggest that  the spectrum is compatible with a power law combined with an exponential cutoff of the form $dN/dE=K\,E^{-\Gamma} \,exp(-E/E_0)$. It has also been reported  that the spectrum hardens as the flux increases \cite{Aharonian03,Krennrich02}, either because of an increase in the cutoff energy or a change in the spectral index itself. Differences in the energy spectrum of Mrk421 and Mrk501 have also been addressed to understand the  \gam production mechanisms of these objects and  absorption effects at the source or in the intergalactic medium due to interaction of VHE \gams with the EBL photons \cite{Dwek05}. The energy spectra of Mrk501 and Mrk421 measured by HEGRA telescope system during their high states in 1997 and 2001, respectively, are described by $\Gamma=1.92$ and $E_0=6.2\,TeV$ for Mrk501 and $\Gamma=2.19$ and $E_0=3.6\,TeV$ for Mrk421. The exponential cutoff in the spectrum of Mrk421 starts earlier. Since both the sources are located at approximatey the same distance, the differences in the cutoff energies can be interpreted as an indication against the hypothesis that attributes the cutoffs to the pure intergalactic absorption effect. The earlier flux measurement of Mrk421 with the TACTIC telescope during 2004 observations have revealed a power law spectrum with a photon index of $2.4\pm0.2$ in the energy range of $2-9\,TeV$~\cite{Rannot2005}.
\par
In this chapter, we present the TACTIC results obtained during 2005-2007 observation period. In addition, we also compare the TACTIC $TeV$ light curves with those obtained with the RXTE/ASM~\cite{ASM} and $Swift$/BAT~\cite{swift} in the X-ray energies for the contemporary periods.  
\begin{table}[t]
\caption{Details of Mrk421 ON-source observations during the period 2005-06.}
\label{obs2005-06}
\centering
\begin{tabular}{|c|c|c|c|}
\hline 
Spell & Observing period & No. of nights & Observation time(hours)\\
\hline
I & 07 Dec 2005 - 11 Dec 2005 & 5 & 09.24 \\
\hline
II & 27 Dec 2005 - 09 Jan 2006 & 12 & 35.71 \\
\hline
III & 23 Jan 2006 - 07 Feb 2006 & 16 & 61.53 \\
\hline
IV & 19 Feb 2006 - 02 Mar 2006 & 10 & 34.54 \\
\hline
V & 19 Mar 2006 - 30 Mar 2006 & 11 & 31.14 \\
\hline
VI & 18 Apr 2006 - 30 Apr 2006 & 12 & 29.55 \\
\hline
VII & 16 May 2006 - 21 May 2006 & 05 & 06.21 \\
\hline
II+III & 27 Dec 2005 - 07 Feb 2006 & 28 & 97.24 \\
\hline
Total  & 07 Dec 2005 - 21 May 2006 & 71 & 207.92\\
\hline
\end{tabular}
\end{table}   
\begin{table}[h]
\caption{Details of Mrk421 ON-source observations during the period 2006-07.}
\label{obs2006-07}
\begin{center}
\begin{tabular}{|c|c|c|c|c|}
\hline 
Spell & Observing period  & No. of nights & Observation time (hours) \\
\hline
I & 18 Dec 2006 - 29 Dec 2006 & 6 & 8.96\\
\hline
II & 17 Jan 2007 - 27 Jan 2007 & 11 &26.54\\ 
\hline
III & 11 Feb 2007 - 22 Feb 2007 & 11 &16.19\\ 
\hline
IV  & 09 Mar 2007 - 22 Mar 2007 & 12 &26.76\\ 
\hline
V  & 06 Apr 2007 - 18 Apr 2007 & 7 &4.61\\ 
\hline
Total & 18 Dec 2006 - 18 Apr 2007 & 57 & 83.06 \\
\hline
\end{tabular} 
\end{center}
\end{table} 
\section{TACTIC observations of Mrk421 (2005-07)}
We have used the imaging element of the TACTIC telescope at Mount Abu to observe Mrk421 during the period 07 December 2005 to 18 April 2007. These data were recorded using two trigger configurations NNP and 3NCT for the observation periods from 07 December 2005 to 21 May 2006 and 18 December 2006 to 18 April 2007 respectively. The single pixel threshold was set to $\geq$ 8/25 $pe$  for 3NCT/NNP topological trigger logic. The telescope was operated at an individual SCR of $3\,kHz$, ensuring the safe operation of PMTs with typical anode currents of $\leq3\mu\,A$. Mrk421 was observed for $\sim208\,hours$  between 07 December 2005 to 21 May 2006 and for $\sim123\,hours$ between 18 December 2006 to 18 April 2007. However, $\sim202\,hours$ and $\sim83\,hours$ of good quality data collected during 2005-06 and 2006-07 respectively were used for further analysis. The observations were carried out in the tracking mode, where the source is tracked continuously without taking  off- source data \cite{Quinn1996}. This  mode  improves the chances of recording possible flaring activity from the candidate source direction.  Nearly $30\,hours$ of the off-source data were recorded during 2005-06, in order to cross-check the background estimation. The total on-source data for 2005-07 have been divided into different spells, where each spell corresponds to one  lunation period. The zenith angle of the observations was $\leq$45$^\circ$. Details of the these observations are given in Table~\ref{obs2005-06}.and \ref{obs2006-07} for 2005-06 and 2006-07 respectively.
\section{Data analysis and results}
The general quality of the recorded data was checked by referring to the sky condition log and compatibility of the PCR and CCR with Poissonian statistics, as already explained in this thesis while discussing the data analysis procedure. The imaging data recorded by the telescope were corrected for inter-pixel gain variation and then subjected to the standard two-level ``image cleaning'' procedure \cite{Konopelko96} with picture and boundary thresholds of 6.5$\sigma$ and 3.0$\sigma$, respectively. The image cleaning threshold levels were first optimized on the Crab data and then applied to the Mrk421 data. The clean Cherenkov images were characterized by calculating their standard image parameters like $length$, $width$, $distance$, $alpha$, $size$ and $frac2$~\cite{Hillas1985,Weekes1989,Konopelko96}. The standard dynamic supercuts \cite{Mohanty1998} procedure was then used to segregate \gam like images from the huge background of CRs. The \gam selection criteria (Table~\ref{M421cuts}) used in the analysis have been obtained
\begin{table}[t]
\caption{ Dynamic Supercuts  selection  criteria used for analyzing the TACTIC data.}
\label{M421cuts}
\centering
\begin{tabular}{|c|c|}
\hline 
Parameter  & Cut Values\\
\hline
$length$ (L) & $0.11^\circ\leq L \leq(0.235+0.0265 \times \ln S)^\circ$\\
\hline
$width$ (W) & $0.06^\circ \leq W \leq (0.085+0.0120 \times \ln S)^\circ$\\
\hline
$distance$ (D) & $0.52^\circ\leq D \leq 1.27^\circ cos^{0.88}\theta$ ;($\theta$$\equiv$zenith ang.)\\
\hline
$size$ (S)  & $S \geq 450 d.c$ ;(6.5 digital counts$\equiv$1.0 pe )\\
\hline
$alpha$ ($\alpha$) &  $\alpha \leq 18^\circ$\\
\hline
$frac2$ (F2) &  $F2 \geq 0.35$ \\
\hline
\end{tabular}
\end{table}  
on the basis of  dedicated  Monte Carlo simulations carried out for  the TACTIC telescope. The cuts have also been validated further by applying them to the TACTIC Crab Nebula data  analysis.
\begin{table}[h]
\caption{ Detailed spell wise analysis of 2005-06 data.}
\label{result0506}
\centering
\begin{tabular}{|c|c|c|c|c|c|c|}
\hline 
Spell & Obs. time  &$\gamma$-ray  & $\gamma$-ray   &  Signifi-         & $\chi^2$ /dof  &   Prob. \\ 
      &  (hours)  & events      & rate (hour$^{-1}$) &  cance($\sigma$)  & (27$^\circ$ $\leq$$\alpha$$\leq$81$^\circ$)               &         \\
\hline
 I    &    9.24   & 9 $\pm$25    & 0.97 $\pm$2.67    & 0.37              & 2.97 /5        & 0.705  \\
\hline 
 II   &   35.71   & 275$\pm$49   & 7.70 $\pm$1.37    & 5.79              & 5.77 /5        & 0.329  \\
\hline
 III   &   61.53   &676 $\pm$66  & 10.99 $\pm$1.07  & 10.64              & 8.57 /5        & 0.127  \\
\hline
 IV   &    34.54   & 91 $\pm$47  & 2.64$\pm$1.37     & 1.94              & 5.10 /5        & 0.404  \\ 
\hline
 V     &   31.14   & 61$\pm$38   & 1.96 $\pm$1.23    & 1.61              & 2.44 /5        & 0.785  \\
\hline
 VI    &   29.55   & 123$\pm$33  & 4.16 $\pm$1.11   & 3.86              & 3.01 /5        &  0.698  \\
\hline
\hline
 All data &  201.72 & 1236$\pm$110  & 6.13 $\pm$ 0.55  & 11.49           & 4.57 /5        & 0.471  \\
\hline
 II +III  &  97.24  & 951$\pm$82   & 9.78 $\pm$0.84   & 12.00            & 5.00 /5        &  0.416  \\
\hline
\end{tabular}
\end{table}   

\par
A well established procedure to extract the $\gamma$-ray signal from the CR background using a stand alone imaging telescope is to plot the frequency distribution of the $\alpha$ parameter of shape and $distance$ selected events. This distribution is expected to be flat for the isotropic background of CR events. For \gams, coming from a point source, the distribution is expected to show a peak at smaller $\alpha$ values. Defining $0^\circ\leq\alpha\leq18^\circ$ as the \gam domain and $27^\circ\leq\alpha\leq81^\circ$ as the background region, the number of \gam events is then calculated  by subtracting the expected number of background events (calculated on the basis of background region) from the \gam domain events. Estimating the expected background level in the $\gamma$-domain by following this approach is well known \cite{Catanese98} and has been used quite extensively by other groups when  equal amount of off-source data is not available. However, we have also  validated  this  method  for the TACTIC telescope by using separate off-source data  on a regular basis and the $\alpha$ distribution of this data in range $\alpha\leq81^\circ$ is in good agreement with the expected flat distribution. The significance of the excess events has  been finally calculated by using the maximum likelihood ratio method of Li $\&$ Ma \cite{Li1983}.
\begin{figure}[t]
\centering
\includegraphics*[width=0.95\textwidth,angle=0,bb=13 13 780 599 ]{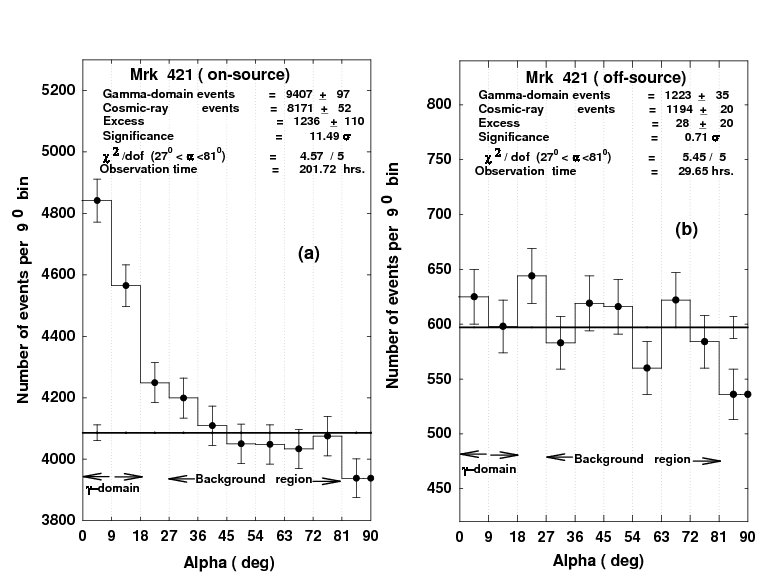}
\caption {\label{alpha0506}(a) On-source $\alpha$ plot for Mrk421  when all the data  collected  between 07 December 2005 and  30 April 2006  for  $\sim201.72\,hours$  are analysed. (b) Off-source $\alpha$ plot for $\sim29.65\,hours$ of observation time. The horizontal lines in these figures indicate the expected background in the $\gamma$-domain obtained by using the background region ($27^\circ\leq\alpha\leq81^\circ$).}
\end{figure} 
\subsection{Results of 2005-06 data analysis}
For the data recorded during 2005-06, we have first analyzed the data for each spell separately. Furthermore, it has also been ensured that these data are subjected to exactly similar analysis procedures as we did in the case of Crab Nebula to avoid any source dependent bias while determining the energy spectrum of Mrk421. The results of the spell wise analysis are presented in Table~\ref{result0506}. An examination of this table clearly indicates that Mrk421 was most active  during  spell II and spell III observations. Fig.~\ref{alpha0506}(a) gives the $\alpha$-distribution when all the data collected for $\sim$$201.72\,hours$ (07 December 2005 to 30 April 2006, spell I to spell VI) are analysed together. The total data  yields an excess of $\sim$(1236$\pm$110) $\gamma$-ray like events with a statistical significance of $\sim$11.49$\sigma$. Fig.~\ref{alpha0506}(b) gives the corresponding $\alpha$-distribution for $29.65\,hours$ of off-source data.  The procedure for estimating the expected background in the $\gamma$-domain by using the background region is also consistent with the results from the off-source  alpha plot (Fig.~\ref{alpha0506}(b)), when  its $\gamma$-ray domain events ($0^\circ\leq\alpha\leq18^\circ$ events) are appropriately  scaled up (to account for the difference in the on-source and off-source observation time) and compared with the $\gamma$-ray domain events of Fig.~\ref{alpha0506}(a). The results of this calculation  yield a value of $\sim$(8320$\pm$238) events for the background level which is in close agreement with the value of $\sim$(8171$\pm$52) events obtained when background region of Fig.~\ref{alpha0506}(a) is used itself. It is worth mentioning  that $\chi^2$/dof of the background region (indicated by column 6 of Table~\ref{result0506}) is also consistent with the assumption that the background  region is flat and thus can be reliably used for estimating the background level in the  $\gamma$-domain.
\begin{figure}[h]
\begin{center}
\mbox{\hspace{0cm}\includegraphics[scale=0.85, bb=0 0 510 329]{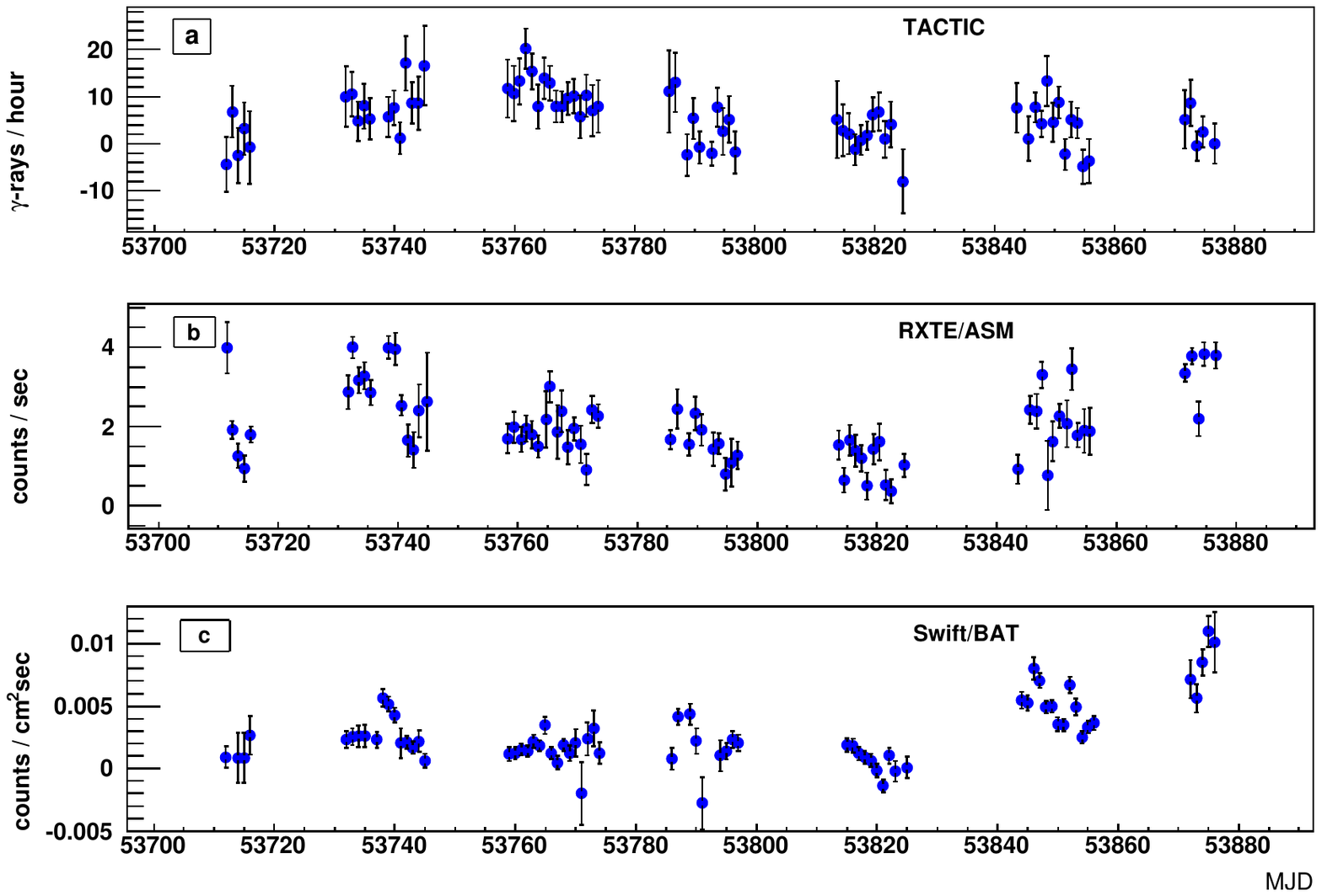}}
\caption{(a) Light curve for Mrk421 as recorded by the TACTIC imaging telescope between 07 December 2005 and 30 April 2006 for a total observation time of $\sim201.72\,hours$.  (b) Daily plot of the average count rate in the $2-10\,keV$ energy range as determined by RXTE/ASM  X-ray detector. (c) Daily plot of the average $counts/cm^2/s$ in the $15-50\,keV$ energy range as determined by \textit{Swift}/BAT detector. Error bars shown are for statistical errors only.}
\label{lc0506}
\end{center}
\end{figure}
\par
It is clear that the source was in a high state during 27 December 2005 to 09 January 2006  and 23 January 2006 to 07 February 2006  at a combined  average flux of $\sim$(1.04$\pm$0.14)CU. It is worth mentioning here that the results of the Whipple group \cite{Veritas,horan09} also indicate that the source was in a high state during the period of their observations from 27 December 2005 to 07 January 2006 and 23 January 2006 to 03 February 2006. Furthermore, our results  for spell 5 (19 March 2006 to 30 March 2006) are also consistent with the Whipple observations  from 23 March 2006 to 04 April 2006  when the source was seen to flare at a lower flux level. These observations clearly  indicate  that, despite  differences in the observation  times  between the two telescope systems, Mrk421  was  observed to be in  a high state by both the systems for a prolonged duration, somewhat similar to the flaring episodes during January-February 2001 and April-May 2004.

\subsection{Light curves}
\subsection*{TeV light curve}
We have examined day-to-day flux variability in the $TeV$ light curve of Mrk421 for spell II and spell III data. The results indicate that we cannot claim any statistically significant flux variation on a nightly basis because of rather large error bars. The light curve of Mrk421 as recorded by the TACTIC imaging telescope is shown in Fig.~\ref{lc0506}(a), where the average hourly \gam rates on daily basis have been plotted as a function of the modified Julian date. The $TeV$ light curve is characterised with a reduced $\chi^2$ value of $115.1/70$ with respect to the mean level of $4.97\pm0.49$ \gam photon events, with corresponding probability of $0.000554$ which is consistent with the variability hypothesis.
\subsection*{ASM X-ray light curve}
The Rossi X-ray Timing Explorer (RXTE) satellite observes the time structure of astromonical X-ray sources in the energy range $2-200\,keV$. It is designed for studying known sources, detecting transient events, X-ray bursts, and periodic fluctuations in X-ray emissions. The RXTE has three instruments~\cite{RXTE}:\\
(\textit{i}) All Sky Monitor (\textbf{ASM}): $2-10\,keV$\\
(\textit{ii}) Proportional Counter Array (PCA): $2-60\,keV$ and\\
(\textit{iii}) High-Energy X-ray Timing Experiment (HEXTE): $15-200\,keV$\\
The ASM was built by the Centre for Space Research (CSR) at the Massachusetts Institute of Technology (MIT). It rotates in such a way as to scan most of the sky ($\sim80\%$) every $1.5\,hours$ monitoring the long-term behavior of a number of the brightest X-ray sources, and giving observers an opportunity to spot any new phenomenon quickly. The ASM consists of three wide-angle shadow cameras equipped with position-sensitive Xenon proportional counters with a total collecting area of $90\,cm^2$. The detector has a sensitivity of $\sim20\,mCrab$~\cite{ASM-1}.
\par
The web page of RXTE~\cite{ASM} makes available all ASM data products, including light curves data in ASCII format. Each raw data point represents the fitted source flux from one $90\,second$ dwell and the Crab nebula flux is about 75 ASM $counts/sec$. Each ``one-day average'' data point represents the one-day average of the fitted source fluxes from a number (typically 5-10) of individual ASM dwells. The error value used in plotting the one-day averages is the quadrature average of the estimated errors on the individual dwells from that day~\cite{ASM}. Corresponding to the TACTIC $TeV$ light curve discussed earlier, RXTE/ASM \cite{ASM} light curve is shown in Fig.~\ref{lc0506}(b). This light curve is characterised with a reduced $\chi^2$ value of $574.4/70$ with respect to the mean level of $2.12\pm0.04$ counts/sec. The probability obtained is very low which is consistent with the variability hypothesis.
\subsection*{\textit{Swift}/BAT light curve}
\textit{Swift} is a multi-wavelength observatory which deploys three instruments~\cite{Swift-1}\\
(\textit{i})  UV/Optical Telescope (UVOT): $170-600 nm$\\
(\textit{ii}) X-ray Telescope (XRT): $0.3-10 keV$ and\\
(\textit{iii}) Burst Alert Telescope (\textbf{BAT}): $15-150\,keV$\\
BAT is a large field of view ($1.4\,steradian$) X-ray telescope with imaging capabilities. It has a coded aperture mask with $0.52\,m^2$ CdZnTe detectors. The BAT typically observes $50\%$ to $80\%$ of the sky each day and has accumulated light curves of many non-GRB sources.
In order to compare the TACTIC light curve of the source obtained during the observation period with that at $15-50\,keV$ we have used \textit{Swift}/BAT~\cite{swift} archived data of Mrk421 to plot its light curve which is shown in Fig.~\ref{lc0506}(c). It is characterised with a reduced $\chi^2$ value of 661/69 (very low probability, consistent with the  variability hypothesis) with respect to the mean level of $0.0025\pm0.000077$ $counts\,cm^{-2}\,s^{-1}$.
\begin{figure}[t]
\centering
\includegraphics*[width=0.71\textwidth,height=0.6\textheight,angle=270,bb= 59 52 557 747]{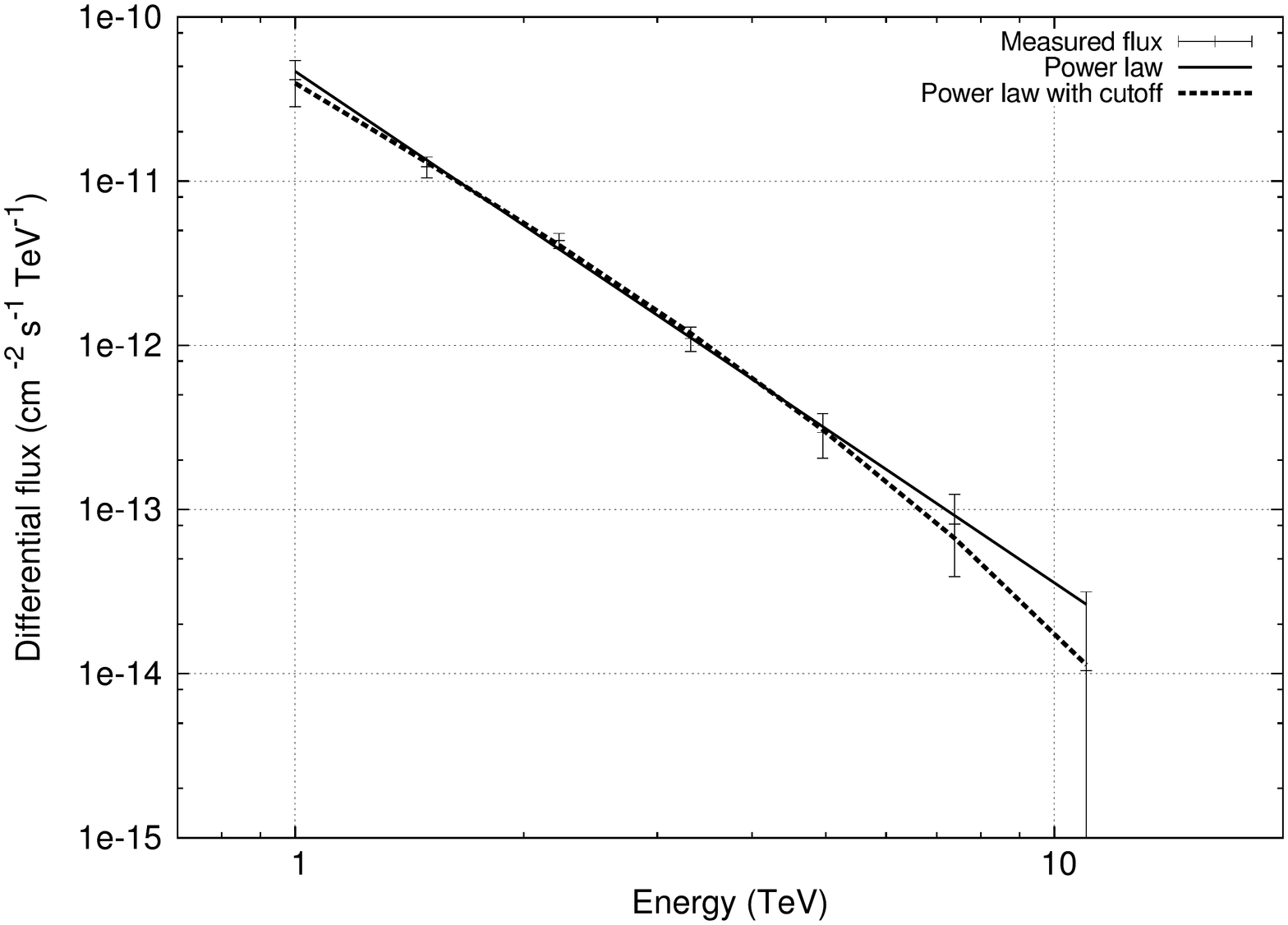}
\caption{\label{m421spectrum}Differential energy spectrum of Mrk421 for the data collected between 27 December 2005 - 07 February 2006 when the source was in a high state.}
\end{figure} 
\begin{table}[t]
\caption{Differential energy spectrum of Mrk421 derived using 2005-06 observations with the TACTIC telescope.}
\begin{center}
\begin{tabular}{|c|c|c|c|}
\hline 
Energy & Energy bin width &  Differential flux & Error in flux\\
($TeV$) & ($TeV$) & ($photons\,cm^{-2}\,s^{-1}\,TeV^{-1}$) & ($photons\,cm^{-2}\,s^{-1}\,TeV^{-1}$)\\ 
\hline
1.0000&   0.403&	4.1399$\times10^{-11}$&   1.2975$\times10^{-11}$\\  
\hline
1.4918&   0.601&	1.2261$\times10^{-11}$&   1.7846$\times10^{-12}$\\ 
\hline
2.2255&   0.896&	4.3461$\times10^{-12}$&   4.5502$\times10^{-13}$\\ 
\hline
3.3201&   1.337&	1.1012$\times10^{-12}$&   1.8538$\times10^{-13}$\\ 
\hline
4.9530&   1.994&	2.9385$\times10^{-13}$&   8.9087$\times10^{-14}$\\  
\hline
7.3891&   2.975&	8.1301$\times10^{-14}$&   4.2371$\times10^{-14}$\\ 
\hline
11.0232&   4.439&	1.0425$\times10^{-14}$&   2.1035$\times10^{-14}$\\  
\hline
\end{tabular} 
\label{flux-value}
\end{center}
\end{table}
\begin{figure}[t]
\centering
\includegraphics*[width=1.0\textwidth,height=0.5\textheight,angle=0,bb=0 0 567 401]{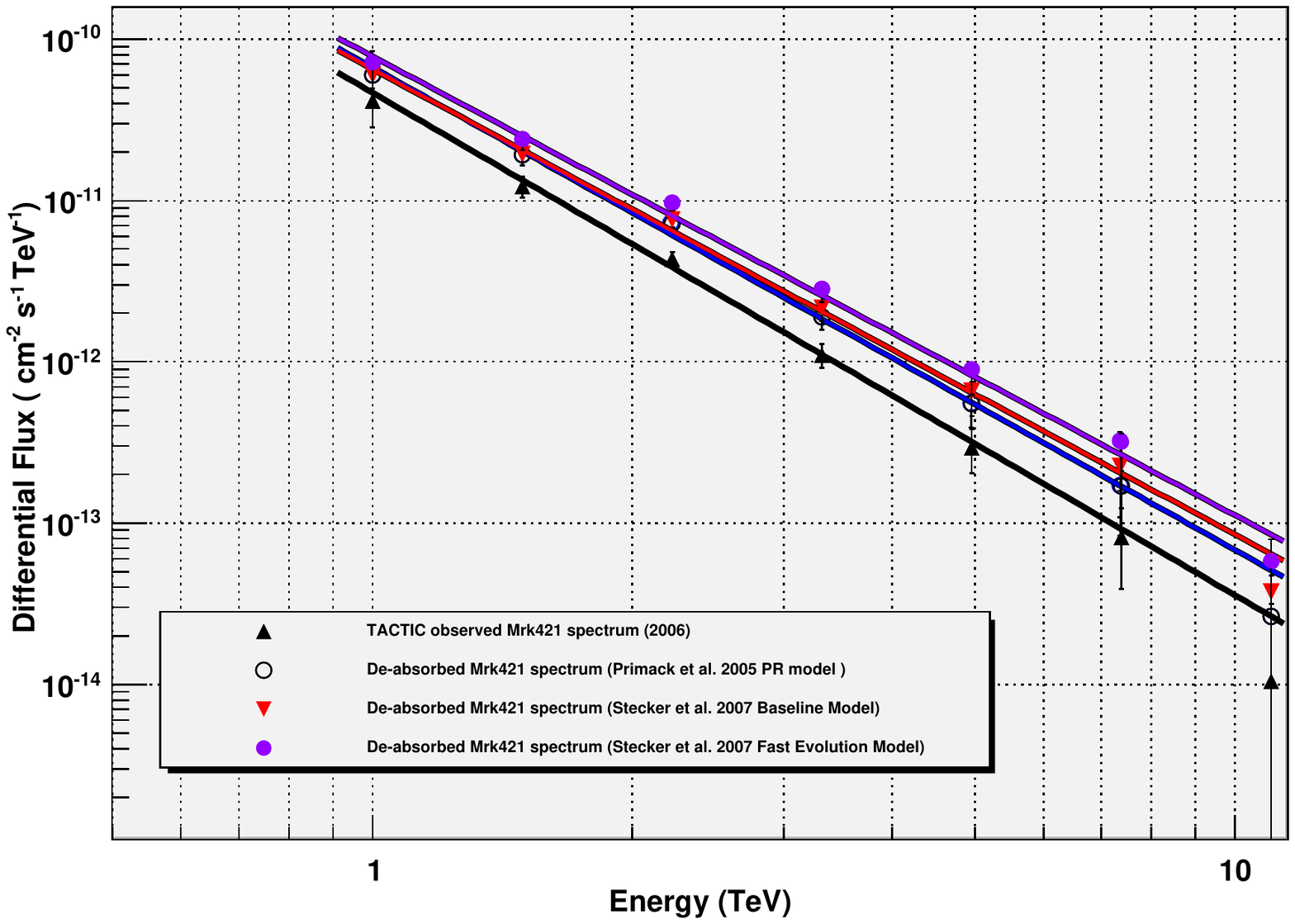}
\caption{\label{fig:es}TACTIC observed differential energy spectrum of Mrk421 derived using the data collected between 27 December 2005 - 07 February 2006 when the source was in a high state (bottom base filled triangle). De-absorbed source spectrum estmated by using the models (i) Primack given in \cite{pri05} (hollow circle), (ii) baseline model given in~\cite{ste07} (top base filled triangle) and (iii) fast evolution model given in \cite{ste07} (filled circle).}
\end{figure} 
\begin{table}[h]
\caption{Estimated intrinsic differential energy spectrum fitted parameters for three models.}\label{tab:fits}
\begin{center}
\begin{tabular}{|c|c|c|c|c|}
\hline
$\Gamma$  &Error in $\Gamma$     &$f_{0}$    &Error in $f_{0}$&Model\\
 &    &   $\gamma$ cm$^{-2}$ s$^{-1}TeV^{-1}$ &  $\gamma$ cm$^{-2}$ s$^{-1}TeV^{-1}$&  \\
\hline 
2.99    &0.101  &6.69$\times$ 10$^{-11}$   &6.38 $\times $10$^{-12}$&Primack \cite{pri05} \\
\hline
2.88   &0.108  &6.53 $\times$ 10$^{-11}$ &6.19 $\times$ 10$^{-12}$&baseline\cite{ste07} \\
\hline
2.84   &0.093  &7.84 $\times$ 10$^{-11}$ &6.20 $\times$ 10$^{-12}$&fast evolution \cite{ste07}\\
\hline
\end{tabular} 
\end{center}
\end{table}
\subsection{Energy spectrum of Mrk421}
We have estimated $TeV$ energy spectrum of Mrk421 using the excess of \gam like events obtained during the high state of this source. The relevant details regarding the estimation of energy spectrum have already been discussed in the last chapter. The resulting differential flux values for various energy bins are given in Table~\ref{flux-value}. Since we know that the source was in a high state during the spell II and spell III of 2005-06 observations, we have used this data alone to determine the time averaged  energy spectrum of Mrk421. Fig.~\ref{m421spectrum} shows the  differential energy  spectrum  after  applying the Dynamic Supercuts to the combined  data of spell II and spell III. A power law fit of the form
\begin{equation}
d\Phi/dE= (4.66\pm0.46)\times 10^{-11}\,E^{(-3.11\pm0.11)}\,\,cm^{-2}\,s^{-1}\,TeV^{-1}
\end{equation} 
in the energy range $1-11\,TeV$ with a $\chi^2/dof= 2.45/5$ (probability = 0.78).\\
A power law with an exponential cutoff 
\begin{equation}
d\Phi/dE=f_0E^{-\Gamma}exp(-E/E_0)
\end{equation} 
was also tried and the result yields the following parameters of the fit $f_0=(4.88\pm0.38)\times10^{-11}\,cm^{-2}\,s^{-1}\,TeV^{-1}$, $\Gamma=2.51\pm0.26$ and  $E_0=(4.7\pm2.1)TeV$ with a $\chi^2/dof= 0.88/4$ (probability = 0.93). The errors in the flux constant, the spectral index and cutoff energy are again standard errors. While work on  understanding the telescope systematics is still in progress,  our preliminary   estimates for the Crab Nebula  spectrum indicate   that   the   systematic errors   in flux   and  the  spectral index  are $<$ $\pm$ 40 $\%$ and $<$ $\pm$ 0.42, respectively.
\par
The energy spectrum of Mrk421 as measured by the TACTIC imaging telescope in the energy  region  $1-11\,TeV$ is compatible with both a pure power law fit and a power law  with an exponential cutoff. However,  a systematic deviation from a pure power law, although not statistically very significant, is evident in Fig.~\ref{m421spectrum} at the energy values of $7.4\,TeV$ and $11.0\,TeV$.  Difficulties like limited \gam event statistics  coupled with rather large  error bars do not allow us to claim the cutoff feature at a high confidence level. The cutoff energy $E_0=(4.7\pm2.1)TeV$  inferred from  our  observations  is fairly consistent with the cutoff values of  $3.6(+0.4-0.3)_{stat}(+0.9-0.8)_{sys} TeV $ and $(4.3\pm0.3)TeV$ reported by  the HEGRA \cite{Aharonian2002-mrk421} and the VERITAS \cite{Krennrich02}
groups, respectively. Furthermore, the results of the HESS group \cite{Aharonian05}, based on 9 nights in April and May 2004, also indicate that the time averaged energy spectrum of Mrk 421  is well described by a power law with an index $\Gamma=2.1\pm0.1_{stat}\pm0.3_{sys}$ and an exponential cutoff at  $3.1(+0.5-0.4)_{stat}\pm0.9_{sys}TeV$. The HESS results also indicate that the cutoff signature in the energy  spectrum  is intrinsic  to the source.
\begin{figure}[t]
\centering
\includegraphics*[width=0.95\textwidth,height=0.35\textheight,angle=0,bb=0 0 311 149]{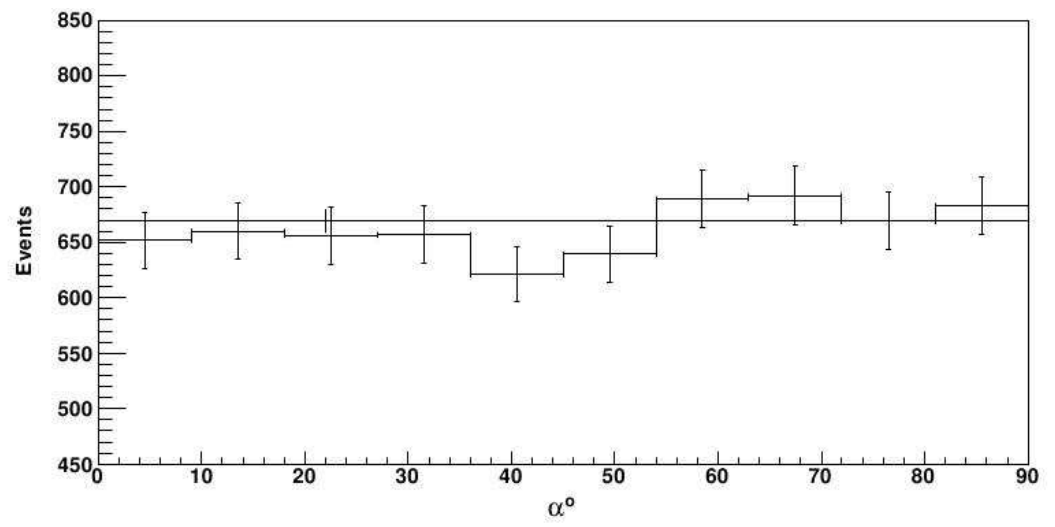}
\caption{\label{alpha-0607}Distributions of image parameter $\alpha$ for the data, taken during 2006-07 observations. Horizontal line represents the mean background level per $9^{\circ}$ bin derived by using resonably flat $\alpha$ region of  $27^{\circ}$ $\leq$ $\alpha$ $\leq 81^{\circ}$. Error bars shown  are for statistical errors only.}
\end{figure} 
\par
We have also attempted to estimate the source intrinsic $\gamma$-ray spectrum by using the SED of the EBL provided by three models given in~\cite{pri05} and~\cite{ste07} (Primack, baseline and fast evolution models). The resulting intrinsic spectra  are shown in Figure \ref{fig:es}  for these three models.  A power law fit to the estimated intrinsic data of the type $d\Phi/dE=f_0 E^{-\Gamma}$ in the energy range $1-11\,TeV$ yields values of $\Gamma$ and $f_{0}$ listed in Table \ref{tab:fits}. As is clear from the Fig.~\ref{fig:es}, there is no significant difference between the measured source spectrum and the three de-absorbed spectra, which is not surprising as the source is associated with a low redshift value. 
\begin{table}[t]
\caption{Monthly spell wise analysis of Mrk421 2006-07 data with only statistical errors.}
\begin{center}
\begin{tabular}{|c|c|c|c|}
\hline 
Spell &$\gamma$-ray &  Significance ($\sigma$) \\
     &photons detected &     \\
\hline
I   & 1.67 $\pm$ 20 &  0.08\\\hline
II  & 49.33 $\pm$ 23 & 2.11\\ \hline
III & 8.33 $\pm$ 17.43 &  0.48\\ \hline
IV  &-56.00 $\pm$ 21.00  & 2.70\\ \hline
V   &-13.00 $\pm$ 8.00  &  -1.64\\ \hline
I+II+III+IV+V &-10.00$\pm$ 42.00 &  -0.25 \\ 
\hline
\end{tabular} 
\label{tab:obs07}
\end{center}
\end{table}
\begin{figure}[h]
\begin{center}
\mbox{\hspace{0cm}\includegraphics[scale=0.85,bb= 0 0 510 360]{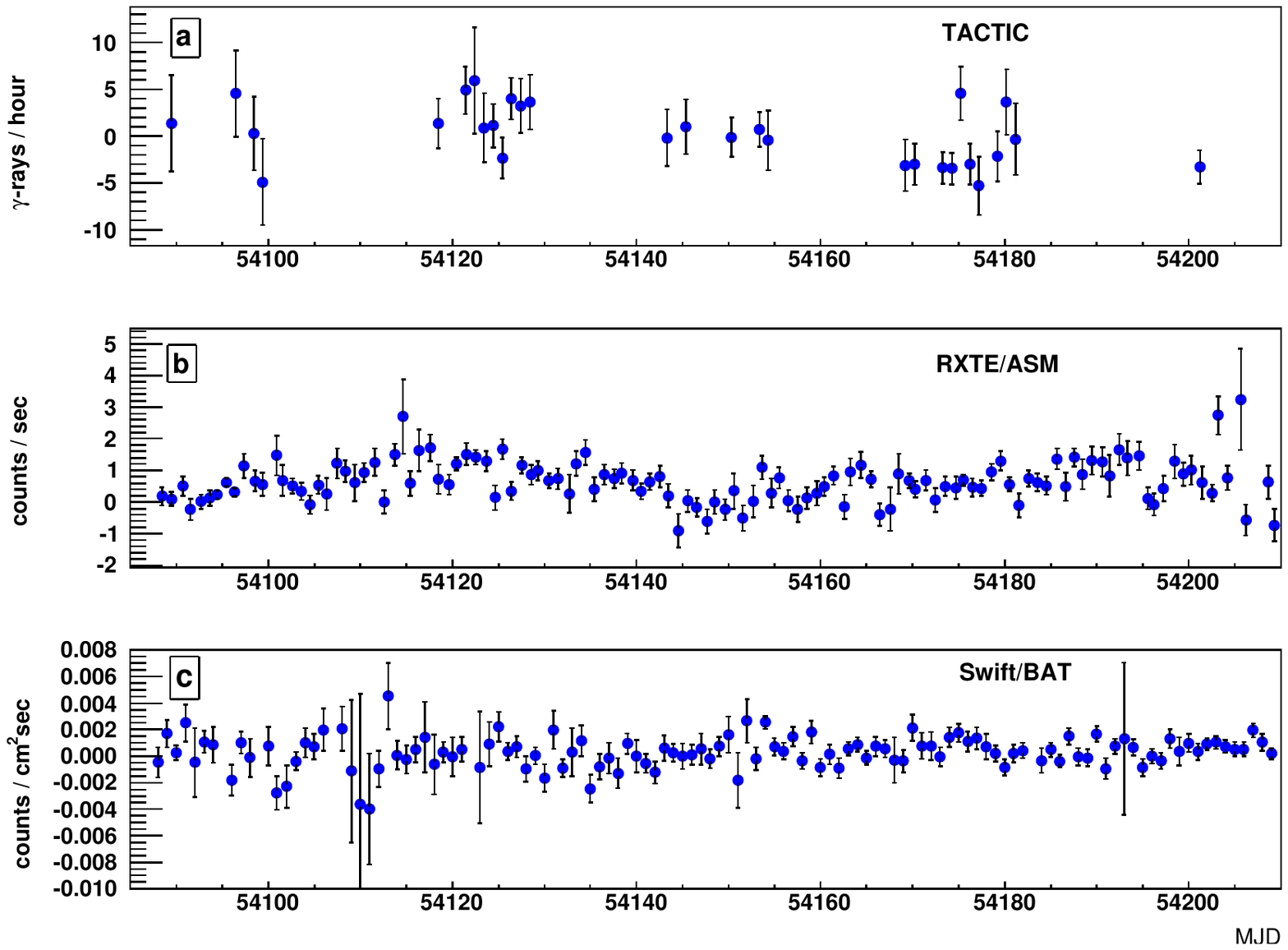}}
\caption{Mrk421 light curves for 2006-07 observations. }
\label{fig:lc07}
\end{center}
\end{figure}
\subsection{Results of 2006-07 data analysis}
In this subsection, we present data analysis results of 2006-07 observations. When all the  data are analyzed together, the corresponding results obtained are shown in Fig.~\ref{alpha-0607}. In this figure the histogram of the $\alpha$ parameter has been shown after having applied shape and orientation related imaging cuts given in Table~\ref{M421cuts}. This distribution  is almost flat and the number of \gam like events  are $-10\pm42$ with a statistical significane of -0.25$\sigma$.  
Thereby indicating that the source was possibly in a low $TeV$ emission state (below TACTIC sensitivity level) during the period of these observations.
We have also carried out an exercise  in which the 2006-07 data are divided into five monthly spells I, II, III, IV and V of December 2006, January, February, March and April 2007 observations respectively and  analyzed by using the same  data analysis procedure. The number of \gam like events  obtained  are 1.67$\pm $0.2, 49.33$\pm $ 23.0, 8.33$\pm $ 17.43, -56.0$\pm $ 21.0 and -13.0$\pm $ 8.0 for  the respective spells, indicating that the source $TeV$ \gam signal level has remained below TACTIC sensitivity during the monthly spells of 2006-07 observations. These results have been given in the Table \ref{tab:obs07}. Further, these data have also been analyzed on nightly basis in order to explore the possibility of very strong episodic $TeV$ emission. The results so obtained are depicted in Fig.~\ref{fig:lc07}a, which shows the night to night variations of the \gam rate (\gams/$hour$) for Mrk421 during 2006-07 observations. This light curve is characterised with  a reduced $\chi^2$ value of 37.59/28 with respect to the mean level of -0.50$\pm$0.47 photon events, with corresponding probability of 0.10 which is consistent with the no variability hypothesis. The magnitude of an excess or deficit recorded on different nights is within $\pm$ 2 $\sigma$  level for these observations and hence indicates the absence of a statistically significant episodic $TeV$ \gam signal from the source direction.
\par
Next we compare the observed $TeV$ light curve which is shown in the Fig.~\ref{fig:lc07}a with those of the source with the RXTE/ASM \cite{ASM} (in the energy range $2-10\,keV$) and \textit{Swift}/BAT \cite{swift} (in the energy range $15-50\,keV$) detectors. First we start with the RXTE/ASM contemporary light curve shown in Fig.~\ref{fig:lc07}b which has been plotted by using the daily average count rates of the ASM from its archived data \cite{ASM}. The ASM light curve is characterised with  a reduced $\chi^2$ value of 300.5/120 (probability obtained is very low which  is consistent with the  variability hypothesis) with respect to the mean level of 0.58 $\pm$ 0.03 $counts/sec$. Next the \textit{Swift} BAT contemporary light curve of the  source obtained from its archived data \cite{swift} is shown in Fig.~\ref{fig:lc07}c. It is characterised with a reduced $\chi^2$ value of 158/116 (probability = 0.0058  which is consistent with the variability hypothesis) with respect to the mean level of 0.00042 $\pm$ 0.00000071 $counts\,cm^{-2}\,s^{-1}$.


\par
We do not find any evidence for the presence of a statistically significant VHE $\gamma$-ray signal  either in the overall data or  when the data are analyzed on a month to month basis or a night to night basis  during 2006-07 observations. We derive  an upper limit of $1.44 \times10^{-12}\, photons\,cm^{-2}\,s^{-1}$ on the VHE $\gamma$-ray emission at 3$\sigma$ confidence  level on the steady state flux from the  source direction  which  is about $9\%$ of the TACTIC detected integrated flux of the Crab Nebula above  $E_{\gamma}\geq1\,TeV$ by using the method of Helene\cite{Helene83}. This result suggests that the BL Lac object Mrk421 was possibly in a quiescent $TeV$ state (below TACTIC sensitivity level) during  2006-07 observations. 
\chapter{TeV observations of H1426+428}
\section{Introduction}
H1426+428 is an established member of the class of extreme synchrotron BL Lac objects detected at $TeV$ energies. Generally, objects of this class are characterized by a flat X-ray spectrum ($F_\nu \propto \nu^\Gamma$ with $\Gamma<1$) extending upto hard X-ray band. In the framework of SSC models, these objects are interesting for observations at $TeV$ energies, as the IC peak of the SED is expected to be at higher energies. The work targeted to measure the IC peak position is hampered by the effect of pair production of VHE photons with low energy (optical and infrared) photons of the EBL. This source is located at a redshift of $z=0.129$ for which the optical depth exceeds unity even for energies of a few $100\,GeV$.
\par
As already discussed in the last chapter, the SED of these objects extend from radio to \gam energies with the double humped  shape  feature in a $\nu F_\nu$ representation \cite{bed97,Ulrich97,Urry1995}. The broadband  SED of the BL Lac objects  identified in X-ray surveys are significantly different compared to those identified in radio surveys and have been classified  into HBL and LBL, depending on their X-ray to radio flux densities \cite{Padovani95}. The first hump in SED of these objects is expected at X-ray energies and generally accepted as originating from synchrotron radiation of relativistic electrons in the magnetic field around the object. The second peak is expected at \gam energies and in the framework of leptonic models \cite{Bloom1996,Konigl81,Maraschi1992,Sikora01,Blandford1995,Dermer97,Ghisellini96} is thought to be stemming from the IC scattering of low energy photons to \gam energies by the same population of relativistic electrons which produces the synchrotron radiation. Whereas, the hadron based  models \cite{Bednarek93,DarLaor97,Aharonian00,Mucke2001} attribute the second hump and X-ray emission to some extent  to processes involving  protons which are also accelerated along with the leptons  in the  relativistic jets. These objects  are extremely variable and are characterized by low emission quiescent states with occasional flaring when the flux can increase by several orders of magnitude. For example, the HESS group  detected the  activity of the  object  PKS2155 -304 in July 2006 when its emission reached a flux level more than two orders of magnitude higher than its quiescent flux \cite{Benbow07}.
\par
H1426+428 was discovered in the 2-6 $keV$ energy band by HEAO 1 \cite{Wood84} and was classified as a BL Lac object in 1989 \cite{Remillard89}. In an observation campaign performed by BeppoSAX in 1998-99, this source was identified as an extreme HBL because of its high synchrotron peak frequency, therefore a potential source of VHE \gams \cite{Costamante2001}.
\par
The  Whipple group reported the first VHE detection of this source in 2002, using the data recorded during   1999 -2001, at a statistical significance of about  $6\sigma$ \cite{Horan02}. Subsequently it was confirmed by the CAT~\cite{Djannati02} and HEGRA groups \cite{Aharonian02H,Aharonian2003-flare} at  statistical significance of $5.2\sigma$ and $7.5\sigma$ respectively. The CELESTE and STACEE groups, both with non-imaging detectors have also observed the source and  reported  upper limits  during  the periods 2002- 2004   \cite{Smith06} and  2001- 2002 \cite{stacee} respectively. However, the Crimean group has reported the detection of VHE $\gamma$-rays from this source using their GT-48 system in April 2004 with  a statistical significance of 5.8$\sigma$\cite{GT-48}. More recently, the VERITAS group  observed this source for $12.5\,hours$ using their  stereoscopic telescope  system at Mt. Hopkins. Preliminary results obtained by them indicate a marginal excess of VHE photons  at 3.2 $\sigma$ statistical level  and they have placed an upper limit of about 3$\%$ of the Crab Nebula flux \cite{kraw07}. In order to draw unbiased conclusions concerning the spectral shape of radiation relevant to the well defined states of the source, it is necessary to derive the energy spectrum in the desired energy range by using a single instrument during a relatively short observation period. The low fluxes and significant variability of such sources make this a rather difficult task. Hence, care should be taken while drawing conclusions about the energy spectrum based on information of fluxes obtained by different groups at different observation epochs.
\begin{figure}[t]
\centering
\includegraphics*[width=0.9\textwidth,height= 0.5\textwidth,angle=0,bb=0 0 1274 633]{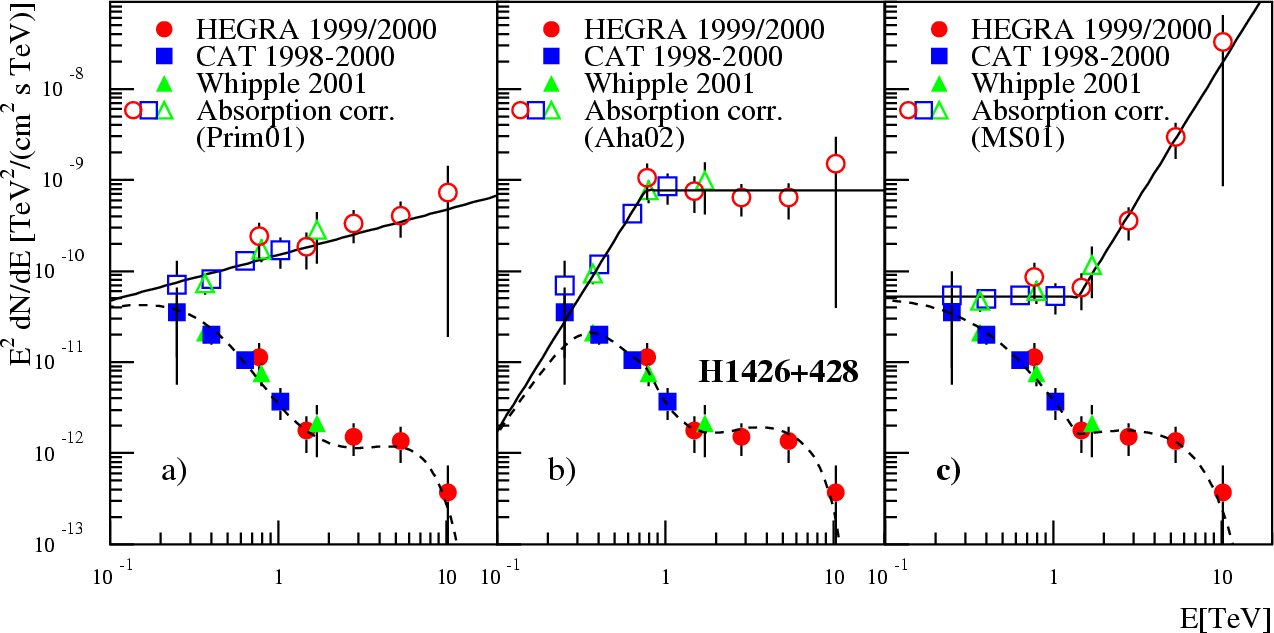}
\caption {Energy spectra of H1426+428. While the filled symbols indicate the measured fluxes reported by Whipple~\cite{Prtry02}, HEGRA~\cite{Aharonian2003-flare} and CAT~\cite{Djannati02} groups, the open symbols represent intrinsic fluxes after EBL correction based on three models. This figure has been adapted from ~\cite{Aharonian2003-flare}.} \label{H1426spect-ebl}
\end{figure}
\par
VHE \gam photons are believed to be absorbed by their interactions with the EBL leading to an energy dependent horizon for viewing of VHE \gam sources. As a result of this absorption the energy spectrum of such a source may show a cut-off feature or steepening of the spectrum, but at the same time such features can also be attributed to the source itself. The interpretation of such features can be used as a probe for EBL measurements \cite{Stecker1992}. Interestingly, the recent detections of the AGN 3C 279 (z = 0.536) by the MAGIC group \cite{Albert2008-ebl}, and H2356-309 (z = 0.165) and 1ES 1101-232 (z = 0.186) by the HESS group \cite{Aharonian-2006ebl} have infused a lot of interest in understanding of the \gam horizon in the VHE region.
\par
The limits on EBL density can be derived from two approaches: (\textit{i}) assuming a limit to the hardness of the intrinsic spectrum of the AGN and deriving limits which assume very little about the EBL spectrum~\cite{Biller1998} and (\textit{ii}) assuming some shape of the EBL spectrum, based on theoretical modeling, and adjusting the normalization of the EBL density to match the VHE spectra~\cite{deJager1994}. The latter can be more stringent but is necessarily more model dependent~\cite{Weekes2003}. Being at very large redshift, the $TeV$ radiation from this extreme BL Lac object suffers severe intergalactic absorption and the measured spectrum can be used to derive limits on EBL. For any resonable EBL model, the absorption effect is at least a factor of 10 or, more likely, factor of 100~\cite{Aharonian-book2004}. This implies not only significant deformation of the primary spectrum, but also very large $TeV$ luminosity of the source. It can be seen from Fig.~\ref{H1426spect-ebl}, the observed \gam spectra of the source, reported by Whipple~\cite{Prtry02}, HEGRA~\cite{Aharonian2003-flare} and CAT~\cite{Djannati02} have a quite unusual shape i.e. the spectra are steep below $1\,TeV$ but become flat above $1.5\,TeV$. However, the source intrinsic spectra after correction for the intergalactic absorption looks significantly different as compared to those observed. While for the EBL model suggested by Primack et al.~\cite{Primack2001}, the intrinsic \gam spectrum has a power law form with a photon index $\sim1.5$ as shown in Fig.~\ref{H1426spect-ebl}(a), for other two models shown in Fig.~\ref{H1426spect-ebl}(b) and (c) the reconstructed intrinsic source spectra have unusual forms with very sharp positive slopes ($E^2 dN/dE \propto E^3$) below $1\,TeV$ for a model discussed in~\cite{Aharonian02H} or above $1\,TeV$ suggested in~\cite{Malkan2001}.
\par
The deformation of the primary $TeV$ \gam spectrum due to the intergalactic absorption is characterised by the factor $e^{-\tau (z,E)}$, where $\tau(z,E)$ is the energy-dependent optical depth for a \gam photon of energy $E$ emitted by a source at a redshift $z$. The spectral deformation factors calculated for the above mentioned 3 different EBL models show that above $300\,GeV$ the intergalactic absorption leads to a strong steepening of the spectrum of \gams from H1426+428, but from $1\,TeV$ to several $TeV$ the spectrum is deformed only slightly, although the suppression of the absolute flux may be as large as a factor of 100. Therefore the intrinsic source spectrum of the \gams i.e. spectrum after correction for the intergalactic absorption 
\begin{equation}
 J_i(E)=J_{obs}(E)\,e^{\tau(z,E)}
\end{equation} 
is expected to be quite different from the observed spectrum $J_{obs}(E)$.
\par
The sharp positive slopes of the \gam spectra ($E^2dN/dE\propto E^3$) in Fig.~\ref{H1426spect-ebl}(b) and (c) can not be easily explained by conventional models invoking either a leptonic or a hadronic origin of \gam emission of BL Lac objects. Even a decent power law intrinsic $TeV$ spectrum of H1426+428 ($E^2dN/dE\propto E^{0.5}$), obtained for a model of EBL spectrum shown in Fig.~\ref{H1426spect-ebl}(a), does not yet guarantee an easy explanation of the broad-band SED of this source. It may require a significant revision of the current conceptual view according to which the synchrotron (X-ray) peak in the SED dominates over the IC ($TeV$) peak. In the case of H1426+428, the detected $TeV$ flux after correcting for intergalactic absorption becomes at least an order of magnitude larger than the X-ray energy flux~\cite{Costamante2003}. Thus, the $TeV$ lumonosity significantly exceeds the X-ray luminosity and it is clear that this should have an impact on the $TeV$ blazar models~\cite{Aharonian-book2004}. Generally, the two bumps in the SED of $TeV$ blazars are always called ``synchrotron'' and ``IC'' peaks which tacitly implies that the leptonic origin of the $TeV$ radiation is accepted as an undisputed fact, and all possible difficulties with the interpretation of the data within this concept can be blamed on the imperfectness of the existing models. However, the $TeV$ observations of sources like H1426+428, indicate that it is quite important to develop in parallel with improvement of leptonic models, alternative scenarios, attributing, for example, the $TeV$ emission to the synchrotron radiation of protons, and the X-ray emission to the synchrotron radiation of primary (directly accelerated) or secondary (pair produced) electrons~\cite{Aharonian-book2004}.
\par
The apparent luminosity of \gams after correcting for the intergalactic absorption appears at least an order of magnitude larger than the synchrotron luminosity.  Within the one-zone SSC model, this would imply a very low magnetic field and a very large density of relativistic electrons. The EC models and the models of electromagnetic cascade in radiation-dominated environments allow more flexibility in modeling the unusual SED of this source. Alternatively, the hard intrinsic $TeV$ \gam spectrum of H1426+428 and high $L_\gamma/L_X$ ratio ($\ge10$) can be explained by the proton-synchrotron model (assuming $\sim \frac{1}{10}$ fraction of the proton synchrotron radiation suffers internal absorption). This scenario couples X-rays to $TeV$ \gams through synchrotron radiation of secondary electrons, therefore we must expect very tight X-rays/$TeV$ \gam correlations. The secondary electrons are promptly cooled in the strong magnetic field, so one may assume that the X-ray will arrive almost simultaeously with the bulk of the $TeV$ \gams. However, more complicated correlations can not be excluded, especially if the magnetic field drops significantly at the outskirts of the emitting region, so an essential part of the secondary electrons are produced and cooled in the low field environment. This unfortunately makes the predictions about the X-ray/$TeV$ \gam correlations less robust.
\par
In this chapter, we present TACTIC results obtained from our observation compaign on this source during the period 2004-07. In addition, we also compare the TACTIC $TeV$ light curves with those obtained with the RXTE/ASM~\cite{ASM} at X-ray energies for the contemporary periods.   
\section{TACTIC observations of H1426+428 (2004-07)}
We have observed the BL Lac object H1426+428 using the imaging element of the TACTIC telescope, during the period 2004-07. The imaging camera was configured with inner 225 pixels for recording ACR images, while innermost 121 pixels were used for the trigger generation. The trigger is based on the 3NCT configuration for 2004 and 2007 data while 2006 data were collected with NNP. The single pixel threshold was set to $\geq$ 8/25 $pe$ for 3NCT/NNP topological trigger logic.
\par
In order to maximize the on-source observation time and to increase  the possibility of recording  flaring activity from the source, we have made these observations in the continuous source tracking  mode. In this mode, the source was always placed at the centre of the camera while observing it. The zenith angle range covered  during these on-source observations  was from 18$^\circ$ to  45$^\circ$. Details of these observations are given in Table \ref{tab:obs}.
\begin{table}[t]
\caption{ Details of TACTIC observations on H1426+428.}
\centering
\begin{tabular}{|c|c|c|c|c|c|c|}
\hline 
Spell &Year & Months  & Observation & Total data & Data selected \\
     &      &         & Dates  & ($hours$) & ($hours$)\\
\hline
I & 2004 & March & 22-26, 28,29,31 & 71.0 & 44.8\\
  &      & April & 12-15, 18, 20-22, 24-26 & &\\ 
\hline
II & 2006 & March & 26-31 &  &  \\
   &      & April & 1, 2, 4, 21, 23-30 &34.9 & 18.4\\
   &      & May  & 01, 02, 19, 21, 23-24, 29-30 & & \\
\hline
III & 2007 & March & 16-27 & & \\
    &      & April & 12-15, 18-20, 22-26 &137.7 & 102.5\\
    &      & May  & 06-15, 17-19, 21-22  & & \\
    &      & June & 07-11  & & \\
\hline
\end{tabular} 
\label{tab:obs}
\end{table}   
\section{Data analysis and results}
While analysing, the data were subdivided into 3 spells each  corresponding  to  the observations made in a particular year.  Standard data quality checks, e.g. compatibility of the PCR and CCR with Poissonian statistics, have been performed in order to evaluate the system behavior and the general quality of the recorded data. As a result of this exercise, $77.9\,hours$ of data were rejected and the rest of the data were used for detailed analysis. While performing detailed analysis of the data,  firstly all the data were  subjected to pedestal removal by using the standard two-level ``image cleaning'' procedure \cite{Konopelko96}. The image cleaning threshold (picture and boundary) levels were first optimized on the Crab Nebula data \cite{Godambe08} and then applied to the data presented here. The digital counts of each pixel in the clean images were then corrected for inter-pixel gain variation.  The clean Cherenkov images were characterized by calculating their standard image parameters like $length$, $width$, $distance$, $alpha$, $size$ and $frac2$ \cite{Hillas1985,Konopelko96,Weekes1989}. The standard extended supercuts \cite{Mohanty1998} procedure was then used to separate $\gamma$-ray like images from those  due to  the huge background of CRs. The $\gamma$-ray selection criteria based on the imaging cuts, given in Table \ref{tab:cuts} used in the analysis have been obtained on the basis of Monte Carlo simulations carried out for  the TACTIC telescope. It may be noted that slightly different \gam domain cuts (Table~\ref{tab:cuts}) have been used for the data collected on H1426+428 as compared to those used for Mrk421 (Table~\ref{M421cuts}). This is done because H1426+428 is normaly observed between March to June (see Table~\ref{tab:obs}) during which period the telescope operates at slightly higher threshold as compared to its winter (October to Februbary) operation due to variations in sky transparency between the two seasons.
\begin{table}[t]
\caption{ Extended Supercuts  selection  criteria used for analyzing the TACTIC data.}
\centering
\begin{tabular}{|c|c|}
\hline 
Parameter  & Cut Values\\
\hline
$length$ (L) & $0.11^\circ\leq L \leq(0.235+0.0265 \times \ln S)^\circ$\\
\hline
$width$ (W) & $0.065^\circ \leq W \leq (0.085+0.012 \times \ln S)^\circ$\\
\hline
$distance$ (D) & $0.5^\circ\leq D \leq 1.27^\circ$ \\
\hline
$size$ (S)  & $S \geq 350  d.c$; (6.5 digital counts $\equiv$ 1.0 pe )\\
\hline
$alpha$ ($\alpha$) &  $\alpha \leq 18^\circ$\\
\hline
$frac2$ (F2) &  $F2 \geq 0.38$ \\
\hline
\end{tabular} 
\label{tab:cuts}
\end{table}  
\par
The next step of the data analysis deals with estimating  the number of $\gamma$- ray events from the source direction in the presence of an overwhelming background of CR events. This is done by plotting the frequency distribution of the $\alpha$ parameter of $length$, $width$, $size$ and $distance$ selected events. The \gams from a point source have smaller values of the $\alpha$ parameter as mentioned earlier in this thesis, whereas the CR events, because of their isotropic nature, are uniformly distributed in all $\alpha$ values ranging  from 0$^\circ$  to 90$^\circ$. For the TACTIC system we find that the $\alpha$ range for \gam events is found to be from  $0^\circ\  to \leq18^\circ$.
The flat distribution of $\alpha$ in the range $27^\circ\leq\alpha\leq81^\circ$ has been used to determine the background events present in the signal region \cite{Yadav07}. The number of \gam events is then calculated by subtracting the expected number of background events (calculated on the basis of the background region) from the \gam domain events and the significance of the excess events has been finally calculated by using the maximum likelihood ratio method of Li $\&$ Ma \cite{Li1983}.
\begin{figure}[t]
\centering
\includegraphics*[width=1.0\textwidth, angle=0,bb=0 0 510 359]{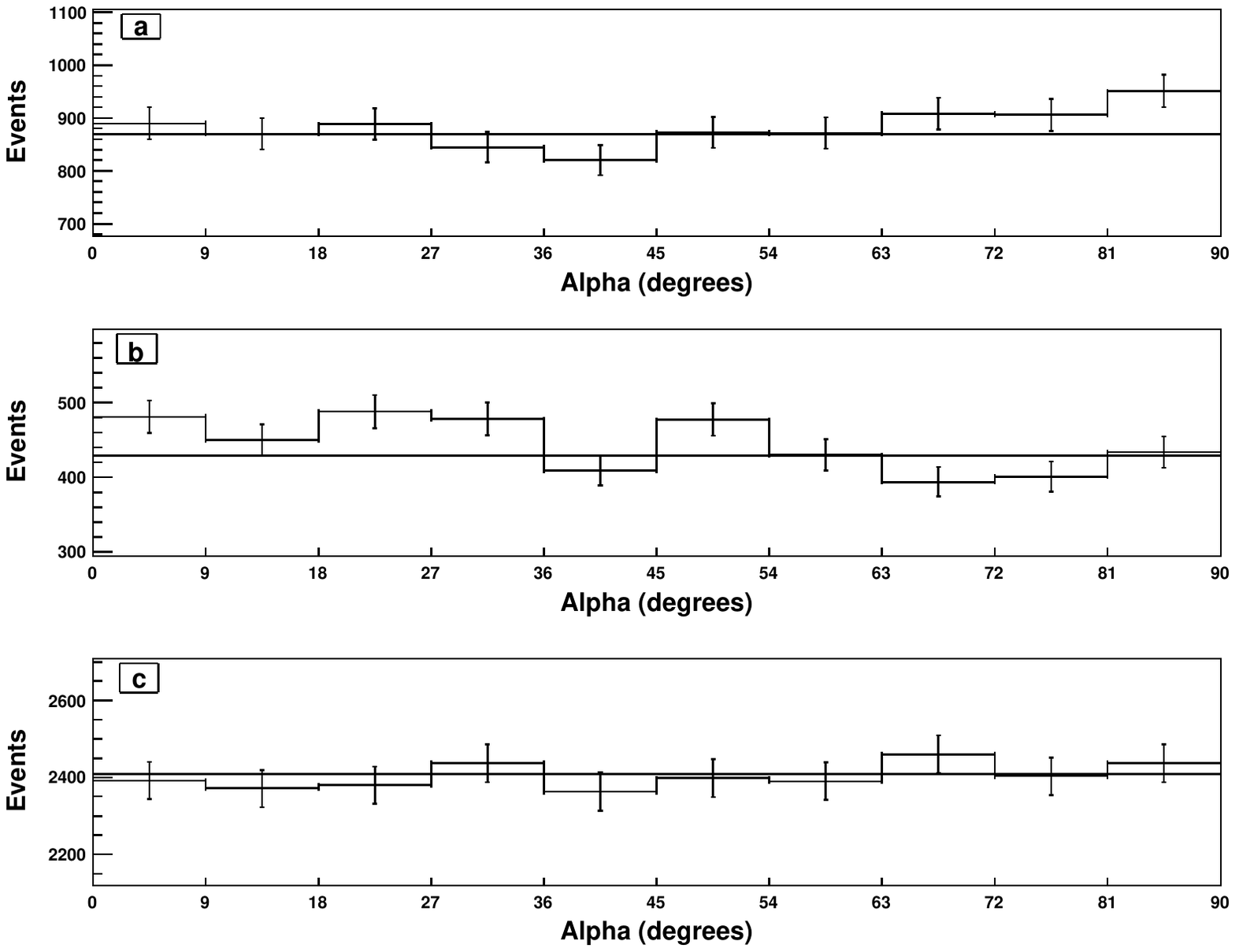} 
\caption {(a) Distribution of image parameter $\alpha$ after applying the cuts given in Table \ref{tab:cuts} for $44.8\,hours$, $18.4\,hours$ and $102.5\,hours$ of data taken on H1426+428 during (a) 2004, (b) 2006 and (c) 2007 observations respectively. The horizontal lines in the figure indicate the expected background in the $\gamma$-domain obtained by using the background region ($27^\circ\leq\alpha\leq81^\circ$). The error bars shown are for statistical errors only.} \label{fig:alpha467}
\end{figure}
\subsection{TeV results}
\begin{figure}[t]
\centering
\includegraphics*[width=1.0\textwidth,angle=0,bb= 0 0 567 366]{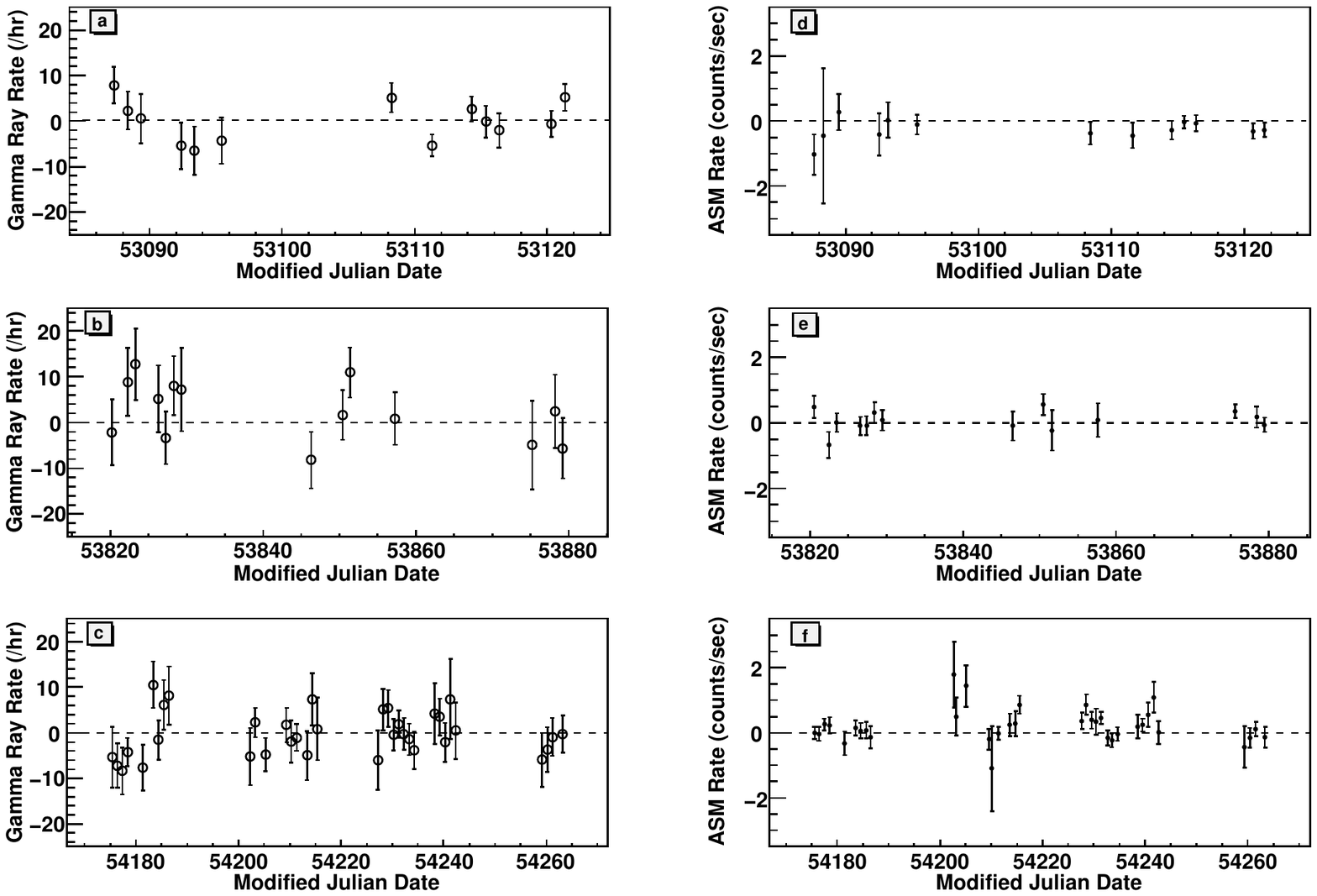}
\caption {TACTIC light curves for (a) 2004, (b) 2006 and (c) 2007 observations  and  corresponding ASM  light curves are shown in  (d), (e) and (f)  respectively.} \label{fig:lc}
\end{figure}

When all the  data recorded during the year  2004 are analysed together, the corresponding results obtained are shown in Fig.~\ref{fig:alpha467}a, wherein the histogram  of the $\alpha$ parameter has been plotted after having applied $length$, $width$, $size$ and $distance$ imaging cuts given in Table~\ref{tab:cuts}.  
As is clear from this figure, the  distribution is almost flat and the number of \gam like events within the \gam domain of the distribution are 19$\pm$48, thereby indicating that  the source was possibly in a low $TeV$ emission state (below TACTIC sensitivity level) during the period of these observations.
Further the same data were also analysed on a nightly basis to explore the possibility of  very strong episodic $TeV$ emissions. The corresponding results obtained are depicted in Fig.~\ref{fig:lc}a, which shows the day-to-day variations of the $\gamma$-ray rate ($\gamma$-rays/$hour$) for 2004  observations. This light curve is characterised with  a reduced $\chi^2$ value of  19.48/12 with respect to the zero degree polynomial fitted  constant value of 0.28$\pm$0.97 photon events, with corresponding probability of  0.11  which is consistent with the no- variability hypothesis. 
The magnitude of an excess or deficit recorded on different nights is within $\pm$ 2 $\sigma$  level for 2004 observations and hence indicates the absence of a statistically significant episodic $TeV$ gamma-ray signal from the H1426+428 direction.
\par
We have followed the same data analysis methodology while analysing the data recorded during the years 2006 and 2007 and  the corresponding $\alpha$ histograms are shown in Fig.~\ref{fig:alpha467}b and  \ref{fig:alpha467}c respectively. As is clear from these figures the  distributions  are almost flat and the number of \gam like events within the \gam domain of the distributions are 68$\pm$35 and -54$\pm$80   
for the 2006 and 2007 observations  
respectively, thereby indicating that  the source was possibly  in a low $TeV$ emission state (below TACTIC sensitivity level) during the period of these observations  as well.
\begin{figure}[t]
\centering
\includegraphics*[width=1.0\textwidth,height= 0.5\textwidth,angle=0,bb=0 0 510 359]{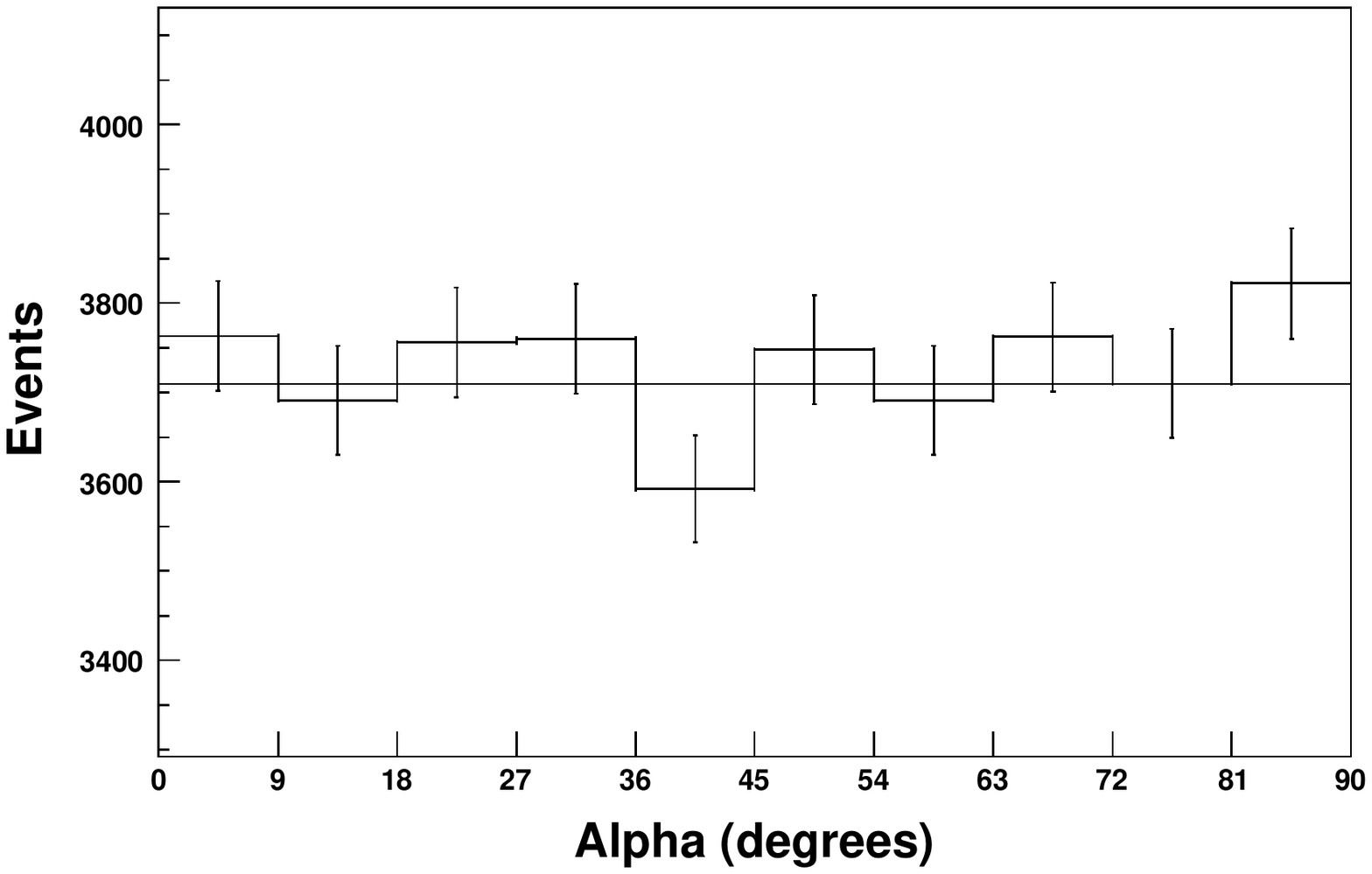}
\caption {Distribution of image parameter $\alpha$ after applying cuts given in Table \ref{tab:cuts} for the entire $165.7\,hours$ of data on H1426+428. The horizontal line in the figure indicates the expected background in the $\gamma$-domain obtained by using the background region ($27^\circ\leq\alpha\leq81^\circ$). The error bars shown are for statistical errors only.} \label{fig:alpha_total}
\end{figure}
Further these data were also analysed on a nightly basis and  the corresponding results obtained are depicted in Fig.~\ref{fig:lc}b and \ref{fig:lc}c respectively, again showing  the day-to-day variations of the  hourly \gam rate with time. Light curves are characterised with the reduced $\chi^2$ values of  12.66 /13  and 31.31 /34  respectively, with respect to the zero degree polynomial fitted  constant value of 2.22 $\pm$1.79 and -0.49$\pm$0.74  photon events with corresponding probabilities of  0.47 and 0.69 respectively which are consistent with the no-variability hypothesis. The magnitude of an excess or deficit recorded on different nights is also within $\pm$ 2 $\sigma$  level for 2006 and 2007  observations thereby again  indicating the absence of a statistically significant episodic $TeV$ \gam signal during these observations as well.
\par
In Fig.~\ref{fig:alpha_total}, we have shown the $\alpha$ histogram obtained when the data for all the years were combined together which is also consistent with the flat distribution. There is thus no evidence for the presence of a  statistically significant $\gamma$-ray signal from the source direction during the period of our observations. We have  placed an   upper limit of $\leq1.18\times10^{-12}$ $photons$ $cm^{-2}$ $s^{-1}$  at 3$\sigma$ level ($13.1\%$ of the TACTIC detected Crab Nebula integrated $\gamma$-ray flux) on the integrated $TeV$ $\gamma$-ray flux above $1\,TeV$ from the source direction, using the method of Helene \cite{Helene83}. In Table \ref{tab:results}, we have compiled all results obtained on a yearly spell basis as well as for the case  when  all the data are considered together. 
\begin{table}[t]
\caption{Yearly spell wise and consolidated data analysis results obtained on  H1426+428 with the TACTIC telescope.}
\centering
\begin{tabular}{|c|c|c|c|c|c|c|}
\hline 
Year & Obs. time & Excess/Deficit of events   &  Significance  & Upper Limit at 3$\sigma$ level     \\ 
&  ( h )  & from source direction       &  ( $\sigma$ )  & photons cm$^{-2}$ s$^{-1}$                \\
\hline
 2004    &  44.8  & 19$\pm$48    & 0.39       & $\leq2.13\times 10^{-12}$          \\
\hline 
 2006  &   18.4   & 68$\pm$35  & 1.95            & $\leq3.72\times10^{-12} $         \\
\hline
 2007  &  102.5   & -54$\pm$80   & -0.68          &$\leq1.54\times10^{-12}$       \\
\hline
 Total  &  165.7   & 33$\pm$100   & 0.33         &$ \leq1.18\times10^{-12} $          \\
\hline
\end{tabular}
\label{tab:results}
\end{table}   
\subsection{ASM X-ray light curves}
In order to examine any correlation between VHE $\gamma$-ray and X-ray ($2-10\,keV$) source emissions, the  Whipple group who discovered this source has done correlation studies for this source and no evidence for significant correlation was found, rather their 2001 data show some evidence for an anticorrelation between the average monthly rates\cite{Horan02}. In the present work we also compare the TACTIC light curves with those of RXTE/ASM X-ray observations in the energy range of $2-10\,keV$. For this we have used the daily average count rate of ASM from its archived data\cite{ASM} to obtain the light curves for the contemporary periods. These light curves are  shown in Fig.~\ref{fig:lc}d, \ref{fig:lc}e and  \ref{fig:lc}f for the corresponding TACTIC observation periods of 2004, 2006 and 2007 respectively. ASM light curves are characterised with  the reduced $\chi^2$ values of 5.27/12, 11.03/13 and 43.78/34  with respect to the zero degree polynomial fitted constant values of -0.2$\pm$0.088, 0.095$\pm$0.082,  and  0.152$\pm$0.046   counts  and corresponding probabilities of 0.96, 0.60 and 0.17 respectively, thereby  indicating that these light curves are  consistent with the constant flux hypothesis and no variability is observed  in a time scale of a day or more in the RXTE/ASM energy band also. We have not done  any correlation studies between the TACTIC and RXTE/ASM observations, mainly because of non-detection of the source in the present work. It may be noted here that we have not used the EBL absorption effect while deriving   the upper limit. We have used ASM data \cite{ASM} in the presented work just to inspect the contemporary RXTE source light curves in the energy band of $2-10\,keV$ and to compare with those obtained with the TACTIC observations. In the earlier reported VHE detections, no evidence for the significant correlation between  X-ray and VHE \gam was found \cite{Horan02} as has been mentioned earlier.

\chapter{Discussion and conclusions}
The aim of this work was to study the VHE \gam emission from two extragalactic sources (of type blazar) Mrk421 and H1426+428. While Mrk421 is the closest ($z=0.030$) known VHE \gam emitter, H1426+428 was the farthest ($z=0.129$) known source at the time of start of this work. However, the most distant source as of now is the FSRQ 3C279 ($z=0.5362$) detected by the MAGIC telescope~\cite{Albert2008-ebl}. The detection of this source, signifies the importance of having a detector with lower energy threshold as the universe is comparatively more transparent to low energy photons.
\par
Blazars are the most powerful variable sources of electromagnetic emission ranging from the radio to the \gam regimes. Their SED show a two peak structure and both of them are found to vary in intensity as well as in peak frequency with the emission level of the blazars. The first peak is commonly believed to originate from synchrotron emission both in leptonic and hadronic models for blazar emission. The origin of the second peak (usually called IC peak) is less well determined and is still being debated. Presently, the leptonic models represent the most preferred concept of \gam emission from blazars and are divided into the following two sub-classes. (i) In SSC models~\cite{Bloom1996,Konigl81,Maraschi1992,Sikora01,Ghisellini1998ssc} the seed photons are synchrotron photons emitted by electrons which are then upscattered with the same electron population via IC process. (ii) In EC models~\cite{Blandford1995,Dermer97,Ghisellini96,Wagner95,Sikora1994,Dermer1992}, the seed photons are ambient infrared or optical photons, photons of CMBR or thermal radiations either directly from the accretion disk or first scattered by the surrounding gas and dust clouds. On the other hand, hadronic models assume that the observed \gam emission is from the accelerated protons interacting with ambient matter~\cite{Pohl2000}, photon fields~\cite{Mannheim1993,Mannheim1996,Mannheim1998}, magnetic fields~\cite{Aharonian2000-hadron} or both magnetic and photon fields~\cite{Mucke2001,Mucke2003}.
\par
In order to understand the properties of blazars and the underlying emission mechanisms responsible for observed variability over the entire electromagnetic spectrum, it is necessary to detect more and more sources. Till date 42 extragalactic sources have been detected~\cite{Wagner}. Most of these detected extragalactic VHE sources belong to the blazars category of AGN except two radio galaxies, two starbust galaxies and three unidentified sources. Further, majority of the detected blazars belongs to HBL type, while LBL and FSRQ count only three sources each. Four recently detected sources belong to type IBL.
\section{Mrk421}
Using TACTIC \gam telescope, we have observed Mrk421 during two spells: (\textit{i}) 07 December 2005 to 30 April 2006 and (\textit{ii}) 18 December 2006 to 18 April 2007, with a total on-source observation time of $\sim$$202\,hours$ and $\sim$$123\,hours$ respectively. The source was detected in a bright state with a statistical significance of $\sim$$12.0 \sigma$ during the period 27 December 2005 to 09 January 2006  and 23 January 2006 to 07 February 2006 in about $\sim$$97\,hours$ of observation time. Total number of $TeV$ photons detected were $951\pm82$ during these observations. The combined average flux was measured to be $\sim$(1.04$\pm$0.14)CU. The Whipple group has also detected this object in a high state during following periods: (\textit{i}) 27 December 2005 - 07 January 2006 and (\textit{ii}) 23 January 2006 - 03 February 2006~\cite{horan09}. Furthermore, our results  for 19 March 2006 to 30 March 2006 are also consistent with the Whipple observations  from 23 March 2006 to 04 April 2006 when the source was observed at a lower flux level. These observations clearly  indicate  that, despite  difference in the observation  time  between the two telescope systems, Mrk421 was observed to be in a high state by both the systems for  a prolonged duration, somewhat  similar  to the flaring  episodes during January - February, 2001 and April - May, 2004~\cite{Blazejowski05}.
\par
We have also determined the observed time-averaged differential \gam spectrum of the source in the energy range $1-11\,TeV$. This spectrum when fitted with a pure power law of the form $d\phi(E)/dE= K\,E^{-\Gamma}$ yields $\Gamma=3.11\pm0.11$ and $K=(4.66\pm0.46)\times 10^{-11}cm^{-2}s^{-1}TeV^{-1}$ with ($\chi^2/dof= 2.45/5$ ). In addition, we have also tried another power law function with an exponential cutoff of the form $d\phi(E)/dE=K\,E^{-\Gamma}exp(-E/E_0)$ which yields $K=(4.88\pm0.38)\times10^{-11}cm^{-2}s^{-1}TeV^{-1}$, $\Gamma=2.51\pm0.26$ and  $E_0=(4.7\pm2.1)TeV$ with a ($\chi^2/dof= 0.88/4$). We find that the later function fits the observed spectrum well with lower value of reduced $\chi^2$. The cutoff energy $E_0=(4.7\pm2.1)TeV$  inferred from our observations is fairly consistent with the cutoff values of  $3.6(+0.4-0.3)_{stat}(+0.9-0.8)_{sys} TeV $ and $(4.3\pm0.3)TeV$ reported by the HEGRA~\cite{Aharonian2002-mrk421} and the VERITAS~\cite{Krennrich02} groups, respectively. Furthermore, the results of the HESS group~\cite{Aharonian05}, based on 9 night observations in April and May 2004, also indicate that the time averaged energy spectrum of Mrk421  is well described by a power law with index $\Gamma=2.1\pm0.1_{stat}\pm0.3_{sys}$ with  an exponential cutoff at  $3.1(+0.5-0.4)_{stat}\pm0.9_{sys}TeV$. The HESS results also indicate that the cutoff signature in the energy  spectrum  is intrinsic  to the source. However, during the observation period from 18 December 2006 to 18 April 2007, the source was found to be in a low emission state and we have placed  an upper limit of $1.44\times\,10^{-12}\,photons\,cm^{-2}\,s^{-1}$ at  $3\sigma$ level on the integrated $TeV$ \gam flux above $1\,TeV$ from the source direction.
\begin{figure}[t]
\centering
\includegraphics*[width=1.05\textwidth,height=0.5\textheight,angle=0,bb= 0 0 567 364]{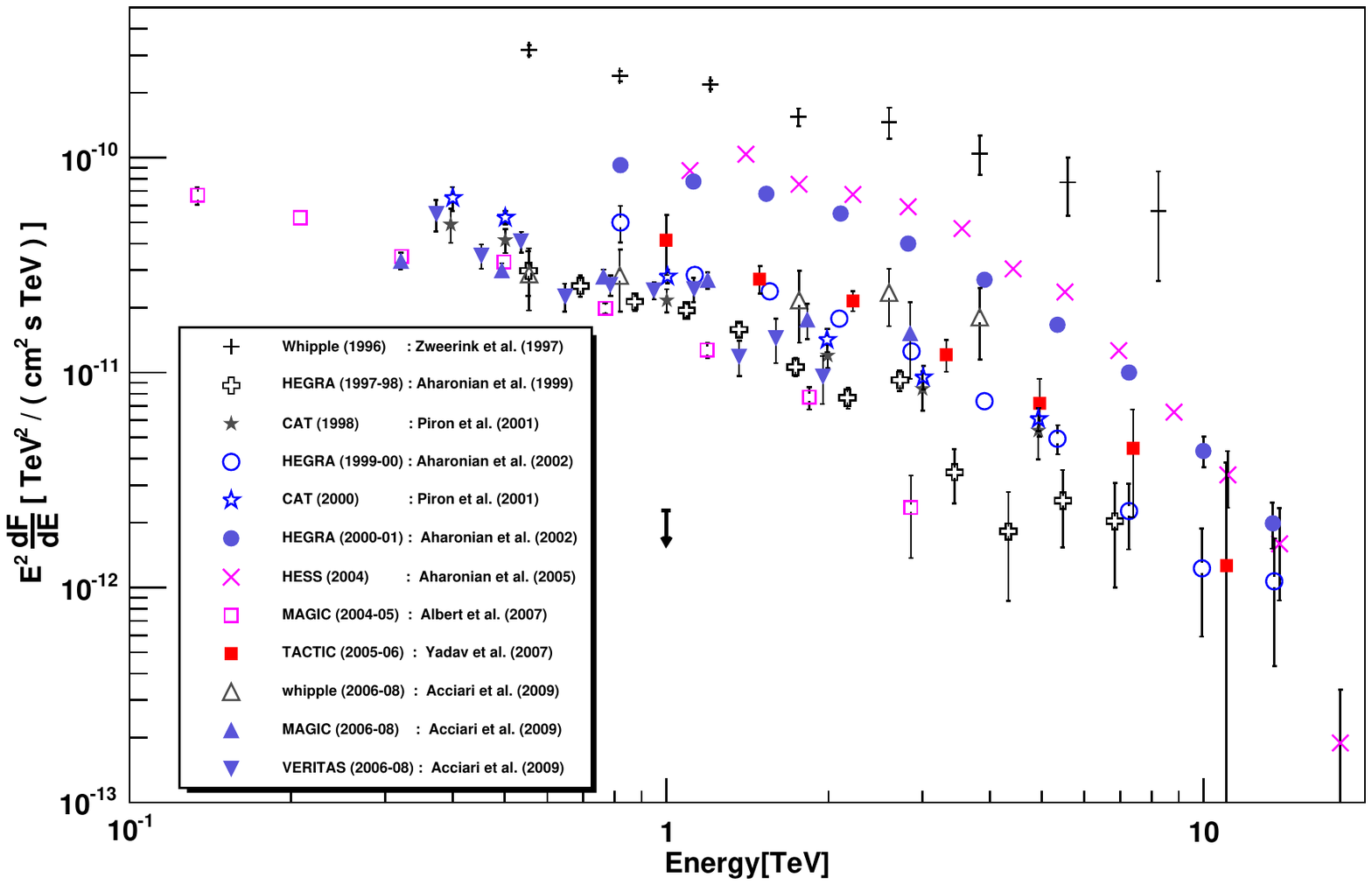}
\caption{\label{sed}Comparison of the  observed  spectral energy   distribution  of Mrk 421 in the VHE range obtained by the HESS, MAGIC, VERITAS, Whipple, HEGRA, CAT and TACTIC~\cite{Aharonian05,Albert07-m421,Zweerink97,acc09,ah99,Aharonian2002-mrk421,Piron01,Yadav07}. The TACTIC observed differential energy spectrum during 2005-06 observations is in the energy range $1-11\,TeV$. The arrow represents  an upper limit obtained at a 3$\sigma$ confidence level on the integrated \gam flux above $1\,TeV$ from this source direction during 2006-07 observations.}
\end{figure} 
\par
In addition, we have compared the TACTIC $TeV$ light curves with those obtained with the RXTE/ASM~\cite{ASM} and \textit{Swift}/BAT~\cite{swift} satellite-based experiments which are sensitive in the energy ranges $2-10\,keV$ and $15-150\,keV$ respectively. Due to the large error bars associated with TACTIC light curves coupled with non-simultaneous observations with RXTE/ASM and \textit{Swift}/BAT, we could not obtain any statistically significant correlation among these data.
\par
From the TACTIC observations of Mrk421 presented in this thesis, we conclude that during 27 December 2005 to 09 January 2006  and 23 January 2006 to 07 February 2006 the source was in a high emission state characterized by a spectral index of $\Gamma=2.51\pm0.26$ and cutoff energy of $E_0=(4.7\pm2.1)TeV$. During rest of the observation period the source was possibly in a $TeV$ quiescent state.
\par
Further, we have obtained the SED of Mrk421 in the VHE range using TACTIC telescope ~\cite{Yadav07} and compared it with those obtained by the HESS, MAGIC, VERITAS, Whipple, HEGRA and CAT telescopes\cite{Aharonian05,Albert07-m421,Zweerink97,acc09,ah99,Aharonian2002-mrk421,Piron01}. This comparision is shown in Fig.~\ref{sed}. The source 2006 observations (28 April to 14 June) reported in~\cite{ale10} have confirmed the indications of rising peak energy with increasing flux, as expected in leptonic acceleration models. It may however be noted that the TACTIC observations for  the year  2006-07 reported in this thesis do not overlap with those reported in the~\cite{ale10}. Therefore there is no conflict between these two observations.
\par
If we compare the earlier high states of this source  the flare of May 1996 detected by the Whipple group remains  the most intense one so far at the level of about 10 times of the Crab Nebula flux. Therefore, in order  to increase the probability of recording  a similar intense flare  more consistent multiwavelength observations are needed with the present age detectors. It seems that such   flares are very rare in nature  but have  a great potential to  provide a  deeper insight into the astrophysics of AGN.
\section{H1426+428}
The BL Lac object H1426+428 is an established source of $TeV$ \gams and detections of these photons from this object also have important implications for estimating the EBL, in addition to the understanding of the  particle acceleration and \gam production mechanisms in the AGN jets. We have observed this source for about $244\,hours$ in 2004, 2006 and 2007 with the TACTIC $\gamma$-ray telescope. Detailed analysis of these data do not indicate the presence of any statistically significant $TeV$ \gam signal from this source in the total data set or when the data set is divided into monthly subsets. Accordingly, we have placed an upper limit of $\leq1.18\times10^{-12}\,photons\,cm^{-2}\,s^{-1}$ on the integrated  $\gamma$-ray flux at 3$\sigma$ significance level~\cite{Yadav2009}.
\begin{figure}[t]
\centering     
\includegraphics*[width=1.0\textwidth,angle=0,bb= 0 0 567 365]{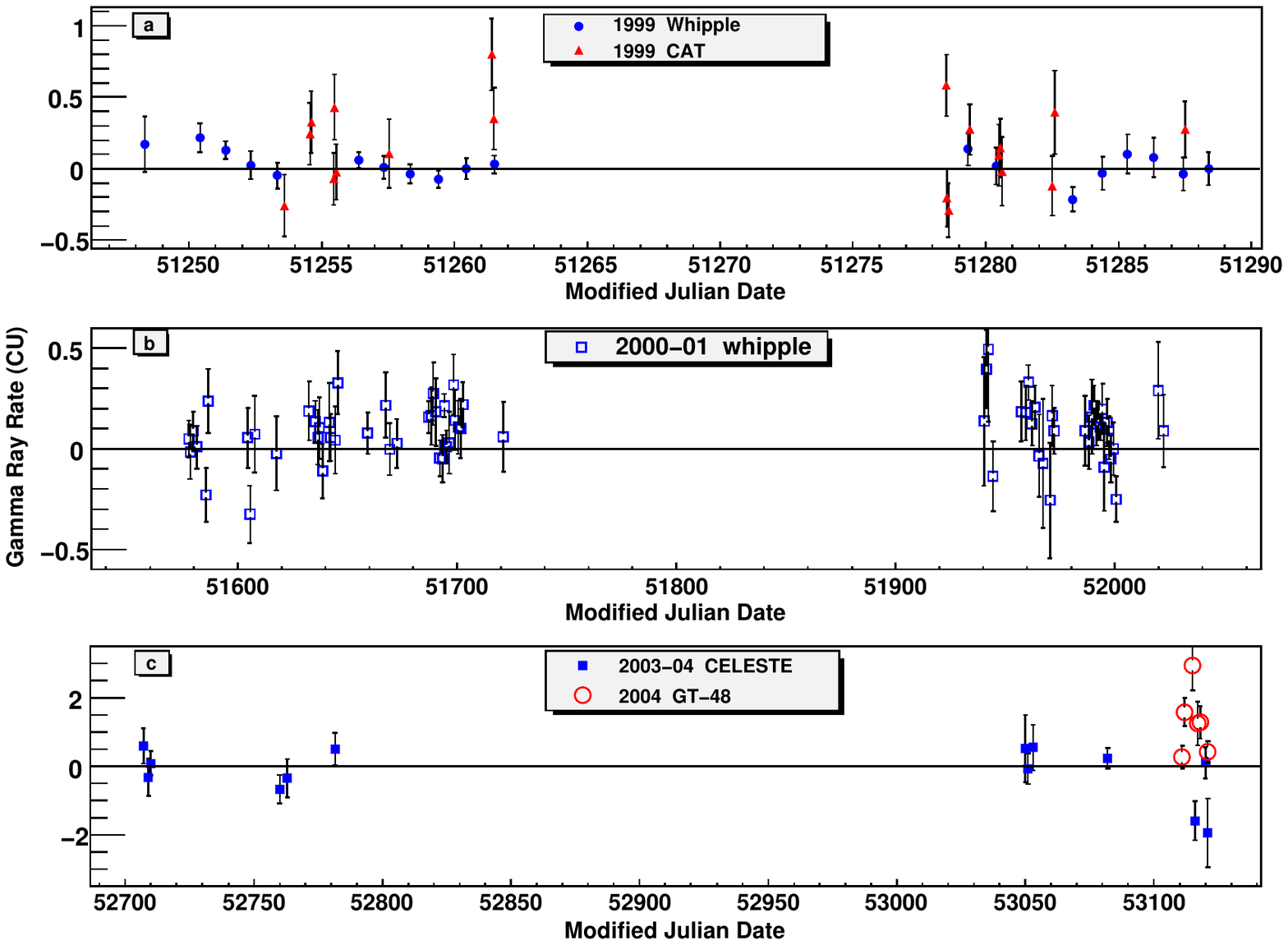}
\caption { H1426+428 $TeV$ light curves   obtained by various groups, namely  Whipple~\cite{Horan02}, CAT\cite{Djannati02}, CELESTE\cite{Smith06} and GT-48\cite{GT-48}.}  \label{fig:tev}
\end{figure} 
\begin{figure}[t]
\centering     
\includegraphics*[width=1.0\textwidth,angle=0,bb=0 0 567 365]{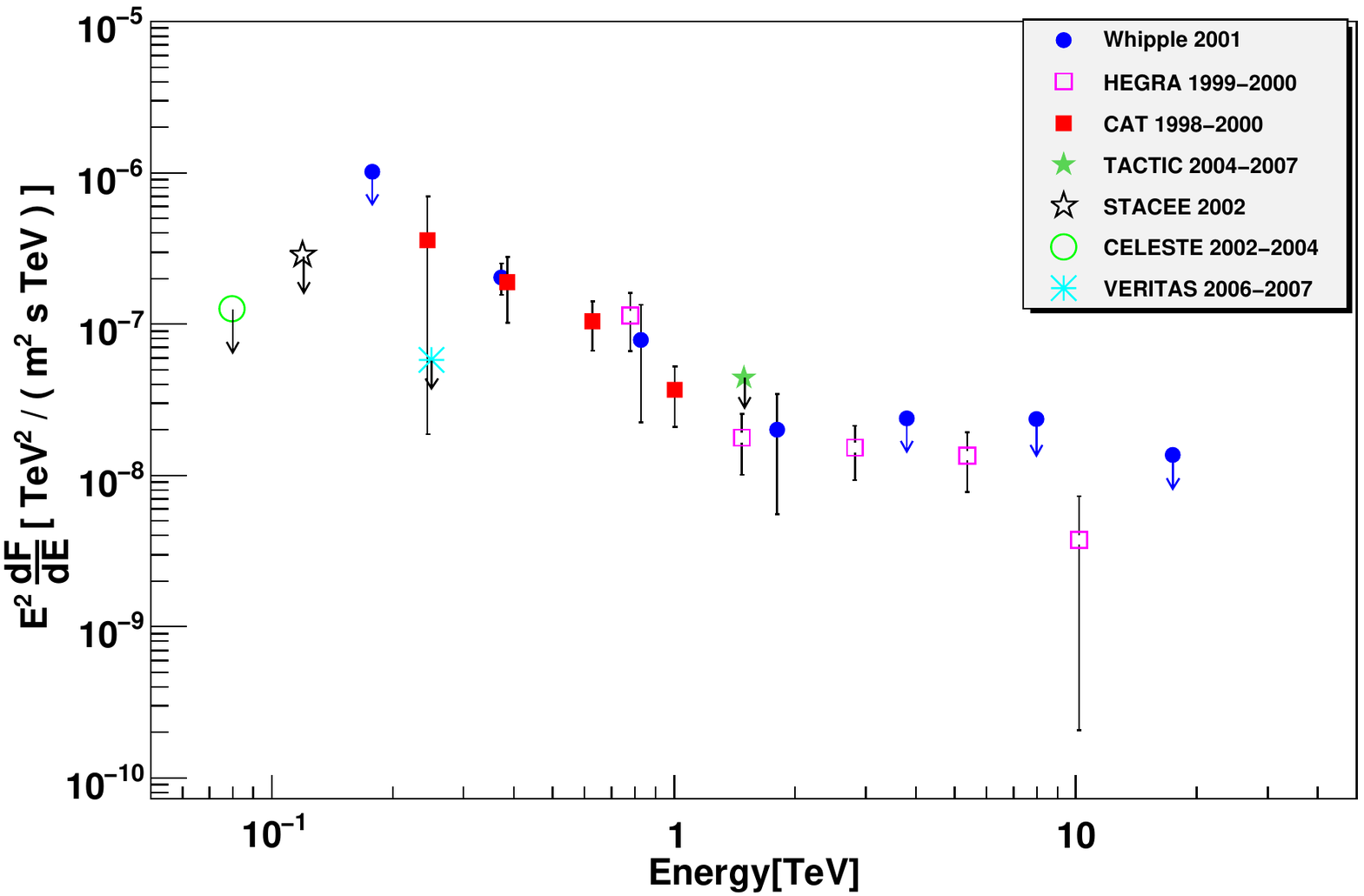}
\caption {TACTIC upper limit on H1426+428 derived using $165.7\,hours$ of data has been shown along with results obtained by Whipple~\cite{Prtry02}, HEGRA~\cite{Aharonian2003-flare}, CAT~\cite{Djannati02} CELESTE~\cite{Smith06}, STACEE~\cite{stacee} and VERITAS~\cite{kraw07} groups.} \label{fig:comp}
\end{figure} 
\par
In Fig.~\ref{fig:tev}, we show the available light curves of the source from 1999 to 2004   observed by the Whipple, CAT, CELESTE and GT-48~\cite{Horan02,Djannati02,Smith06,GT-48} groups. As is clear from this figure, the GT-48 telescope has recorded the source at more than  200$\%$ of the GT-48 detected Crab level on April 19, 2004 (MJD 53115) at 3.9 $\sigma$ level~\cite{GT-48,gt}. It may be noted here that  the TACTIC 2004  results discussed above on this source are in  agreement with those obtained with the CELESTE system \cite{Smith06} around $100\,GeV$ during the same period. However,  our results are in conflict with those  of GT-48 telescope  above $1\,TeV$ during the period from 15 April to 25 April, 2004, wherein a \gam signal at 5.8$\sigma$  level was reported~\cite{GT-48}. 
\par
In addition, we have compared RXTE/ASM~\cite{ASM} X-ray light curves ($2-10\,keV$) with those obtained with the TACTIC observations. We do not find any enhanced activity of the source at X-ray energies too. Further, we have also compared the available $TeV$ light curves of the source from 1999 to 2004  observed by the Whipple, CAT, CELESTE and GT-48  groups~\cite{Horan02,Djannati02,Smith06,GT-48}. It is interesting to point out here that the TACTIC 2004 results are in close agreement with those obtained by the CELESTE group above $100\,GeV$  during the same period. We conclude that the emission level of H1426+428 was below the TACTIC sensitivity with the estimated flux level $\sim13.1\%$ of the Crab Nebula. In Fig.~\ref{fig:comp}, we have  compared the TACTIC  upper limit obtained during 2004-2007 with  the source   VHE spectra results  obtained during different epochs  by the   Whipple~\cite{Horan02}, HEGRA~\cite{Aharonian2003-flare} and CAT~\cite{Djannati02} alongwith the upper limits provided by the CELESTE~\cite{Smith06}, STACEE~\cite{stacee} and VERITAS~\cite{kraw07} collaborations. Further, those groups who have reported VHE $\gamma$-ray detections have derived source spectra which are consistent with  the expected signature of absorption of $TeV$ $\gamma$-rays by EBL \cite{Prtry02,Aharonian02H,Aharonian2003-flare,Djannati02}. 
\par
We  feel that  the long term  multi-wavelength  observations of the source  are required   to possibly  record more  flaring episodes of the source by using the  technologically advanced telescopes.  These  observations in the VHE regime would also possibly record a better statistics of VHE photons to derive the source related observed  and intrinsic spectra corresponding to various  high states. Indeed this would also help  in   understanding   the particle acceleration and $\gamma$-ray production mechanisms in the AGN jets, in addition to the  VHE photon horizon  and EBL related  quest.
\section{Future outlook}
Exploring the \gam sky in the VHE energy range is expected to lead to a potentially rich harvest of astrophysical discoveries. This has been amply demonstrated in the last few years by ground based IACTs such as HESS, MAGIC and VERITAS at \gam energies $\ge100\,GeV$. Further, at still lower energies, the new generation of satellite based \gam telescopes particularly Fermi which was launched in 2008 has heralded a new era in \gam astronomy~\cite{Fermi-conf}. The Fermi LAT which is sensitive to \gams with energies in the range from $20\,MeV$ to $100\,GeV$ has to date detected about 1500 objects. However, due to low flux, instruments for detection of VHE \gams of more than few tens of $GeV$, require a large effective detection area which is not possible with space based instruments. On the other hand, ground based instruments allow much larger detection area. The recent rapid growth in the field of VHE \gam astronomy was therefore primarily driven by IACTs which reach a sensitivity of $\sim1\%$ of the Crab Nebula flux for typical observation time of $25\,hours$. The sensitivity degrades towards lower energies due to threshold effects and towards higher energies due to the limited detection area~\cite{CTA}.
\par
For many unanswered questions in the field of \gam astronomy, low energy threshold instruments are expected to be promising. For example, VHE \gam photons from distant sources interact in the intergalactic space with infrared and other low energy photon fields leading to an energy dependent \gam horizon for vewing VHE \gam sources. The observations at lower energy threshold give access to \gam sources at high redshifts. The low threshold energy can be attained by increasing the light collector area of the telescopes and installing them at higher altitudes where the photon density of the atmospheric Cherenkov events is higher~\cite{Aharonian2001-55}.
\par
As part of augmentation of our activities in \gam astronomy we are setting up a large area imaging \gam telescope MACE (Major Atmospheric Cherenkov Experiment) at Hanle (32.8$^\circ$ N, 78.9$^\circ$ E, $4200\,m$ asl) in the Ladakh region of India. This high altitude site offers an average of about 260 uniformly distributed spectroscopic nights per year which is a major advantage in terms of sky coverage for source observations. Located closer to the shower maximum the Cherenkov photon density at Hanle is substantially high as compared to that at sea level~\cite{Cowsik2001}. The higher photon density along with the low background light level at this site helps in lowering the energy threshold of the Cherenkov telescope being setup there.
\par 
The  MACE telescope with high resolution imaging camera is designed to operate in the sub-$TeV$ energy range. 
The altitude-azimuth mounted telescope will deploy a $21\,m$ diameter parabolic light collector made of 356 panels of $984\,mm\times984\,mm$ size with each panel consisting of 4 spherical mirror facets of $488\,mm\times488\,mm$ size. Each facet is diamond turned to a mirror finish yielding a reflectivity of $\geq$ 85$\%$ in the visible band. The telescope with the F/D of $\sim1.2$ will use graded focal length ($25\,m$ to $26.2\,m$) mirrors in order to reduce the D$_{80}$ spot size (defined as the diameter of the circle within which 80$\%$ of the reflected rays lie) of the light collector to $\sim15\,mm$ for on-axis incidence. Each mirror panel will be equipped with motorized orientation controllers for aligning them to form a single parabolic light collector.
\par
The focal plane instrumentation will have a photomultiplier tube based imaging camera covering a field of view of 4$^\circ\times$4$^\circ$. The imaging camera will comprise of 1088 pixels arranged in a square matrix with uniform pixel resolution of $0.125^\circ$. The inner 576 pixels with a field of view of $2.4^\circ\times2.4^\circ$  will be used for generating the event trigger. The PMTs will be provided with acrylic front-aluminized light cones for enhancing the light collection efficiency of the camera. The signal processing instrumentation will also be housed within the camera and the acquired data will be sent to the control room over the computer network for processing and archiving. Detailed Monte Carlo simulation studies have been carried out using CORSIKA~\cite{corsika}code and the results suggest that using a pixel threshold of $\geq$4pe and a 4 nearest neighbour pixel trigger, gamma-ray energy threshold of $\sim$30 GeV is achievable by the MACE telescope. 
When operational by 2013 the MACE telescope will have the distinction of being the largest high altitude \gam telescope in the world. Operating at a flux sensitivity of a few milli Crab units the MACE will augment international efforts in the field of ground based \gam astronomy.


\newpage


\bibliographystyle{plain}
\bibliography{kky_thesis}

\end{document}